\documentclass[12pt, a4paper]{article}
\usepackage{graphicx}
\usepackage{amssymb}
\usepackage{amsmath}
\usepackage{bm}
\usepackage{color}
\usepackage{theorem}
\usepackage{subcaption}
\usepackage{multirow}

\usepackage[sort&compress,numbers, merge]{natbib}

\definecolor{Orange}{cmyk}{0,0.61,0.87,0}
\definecolor{JungleGreen}{cmyk}{0.99,0,0.52,0}
\definecolor{OliveGreen}{cmyk}{0.64,0,0.95,0.40}
\definecolor{Brown}{cmyk}{0,0.81,1,0.60}
\definecolor{RoyalBlue}{cmyk}{0.71,0.53,0,0.12}

\usepackage[colorlinks=true, linkcolor=OliveGreen, citecolor=RoyalBlue,
urlcolor=Brown]{hyperref}

\begin{document}

\begin{titlepage}

\begin{flushright}
{\tt 
UT-18-03\\
KYUSHU-RCAPP-2018-01 
}\\
March, 2018
\end{flushright}

\vskip 1.35cm
\begin{center}

%\textcolor{RoyalBlue}
{\large
{\bf
Searching for Metastable Particles with Sub-Millimeter \\[3pt]
Displaced Vertices at Hadron Colliders
}
}

\vskip 1.2cm

Hayato Ito$^a$,
Osamu Jinnouchi$^b$,
Takeo Moroi$^{a}$,
Natsumi Nagata$^a$,
and
Hidetoshi Otono$^c$

\vskip 0.8cm

{\it $^a$Department of Physics, University of Tokyo, Tokyo
 113-0033, Japan}\\[3pt]
{\it $^b$Department of Physics, Tokyo Institute of Technology, Tokyo 152-8551,
 Japan}\\[3pt]
{\it $^c$Research Center for Advanced Particle Physics, Kyushu University,\\
 Fukuoka 819-0395, Japan}

\date{\today}

\vskip 1.5cm

\begin{abstract}

  A variety of new-physics models predict metastable particles whose
  decay length is $\lesssim 1$~mm. Conventional displaced-vertex
  searches are less sensitive to this sub-millimeter decay range, and
  thus such metastable particles have been looked for only in usual
  prompt decay searches. In this paper, we show that an additional
  event-selection cut based on the vertex reconstruction using charged
  tracks considerably improves the sensitivity of ordinary searches
  which rely only on kinematic selection criteria, for particles with
  a decay length of $\gtrsim 100~\mu \text{m}$. To that end, we
  consider a metastable gluino as an example, and study the impact of
  this new event-selection cut on gluino searches at the LHC by
  simulating both the signal and Standard Model background
  processes. Uncertainty of the displaced-vertex reconstruction due to
  the limited resolution of track reconstruction is taken into
  account. We also discuss possibilities for optimization of the
  kinematic selection criteria, which takes advantage of significant
  reduction of background through the requirement of displaced
  vertices. In addition, we demonstrate that using the method
  discussed in this paper it is possible to measure the lifetime of
  metastable particles with an ${\cal O}(1)$ accuracy at the
  high-luminosity LHC. Implications for a future 100~TeV collider are
  also studied, where produced particles tend to be more boosted and
  thus it is easier to detect the longevity of metastable particles.

\end{abstract}

\end{center}

\end{titlepage}

\renewcommand{\thefootnote}{\#\arabic{footnote}}
\setcounter{footnote}{0}

%%%%%%%%%%%%%%%%%%%%%%%%%%%%%%%%%%%%%%%%%%
\section{Introduction}
\label{sec:intro}
%%%%%%%%%%%%%%%%%%%%%%%%%%%%%%%%%%%%%%%%%%

Many models beyond the Standard Model (SM) predict the existence of
long-lived particles~\cite{Fairbairn:2006gg}. For instance, in
supersymmetric (SUSY) models, gluino tends to be long-lived if sfermion
masses are in the multi-TeV range \cite{Toharia:2005gm, Gambino:2005eh,
Sato:2012xf}. Such a long-lived particle is also found in the
gauge-mediation models \cite{Giudice:1998bp, Draper:2011aa,
Evans:2016zau, Allanach:2016pam}, in $R$-parity violating models
\cite{Barbier:2004ez, Graham:2012th, Ghosh:2017yeh}, in SUSY axion
models \cite{Brandenburg:2005he, Hamaguchi:2006vu, Freitas:2011fx,
Barenboim:2014kka, Redino:2015mye, Co:2016fln, Co:2017orl}, in the
stealth SUSY scenario \cite{Fan:2011yu, Fan:2012jf, Fan:2015mxp}, in the
SUSY relaxion models \cite{Batell:2015fma, Evans:2016htp}, and in
a degenerate SUSY spectrum \cite{Feng:1999fu, Ibe:2006de, Asai:2007sw,
Asai:2008sk, Asai:2008im, Liu:2015bma, Nagata:2015hha, Nagata:2015pra,
Rolbiecki:2015gsa, Nagata:2017gci, Fukuda:2017jmk}. Other well-motivated
scenarios such as Neutral Naturalness \cite{Chacko:2005pe,
Burdman:2006tz, Cai:2008au, Burdman:2015oej, Chacko:2015fbc}, 
hidden-valley models
\cite{Strassler:2006im, Strassler:2006ri, Strassler:2006qa,
Nakai:2015swg, Knapen:2017kly}, composite Higgs models
\cite{Barnard:2015rba}, dark matter models \cite{Chang:2009sv,
Co:2015pka, Buchmueller:2017uqu, Acharya:2017kfi}, and neutrino models
\cite{Basso:2008iv, Helo:2013esa, Izaguirre:2015pga, Maiezza:2015lza,
Antusch:2016vyf, Antusch:2016ejd, Accomando:2016rpc, Nemevsek:2016enw,
Dev:2016vle} also provide metastable particles, and the searches of
these particles offer a promising way of testing these models. 
The predicted lifetime of these particles spreads over a wide range;
some particles are predicted to decay much before they reach the
detector region, while others can be regarded as stable particles at
collider experiments. Therefore it is important to develop various
strategies to cover the whole potential signatures of long-lived
particles, which depend on the lifetime of these particles as well as
their decay properties.  
Indeed, a lot of efforts have been dedicated to searching for long-lived
particles at the LHC in a variety of search channels, including
displaced-vertex searches \cite{Aad:2014yea, CMS:2014wda, CMS:2014hka,
Aaij:2014nma, Aad:2015uaa, Aad:2015rba, Khachatryan:2016unx,
Aaboud:2017iio}, disappearing track searches \cite{Aad:2013yna,
CMS:2014gxa, Aaboud:2017mpt}, the searches for
large energy-loss signatures \cite{ATLAS:2014fka, Aad:2015qfa,
Aaboud:2016dgf, Aaboud:2016uth, Khachatryan:2016sfv}, and so on. Null
results in these searches have imposed stringent limits on long-lived
particles. 

Most of these searches, however, have sensitivities only to particles
with a proper decay length of $c\tau \gtrsim 1$~mm, where $\tau$ is the
proper lifetime and $c$ is the speed of light.\footnote{We however note
that there are several exceptions, \textit{i.e.}, those which are
sensitive to $c\tau \lesssim 1$~mm \cite{CMS:2014hka,
Khachatryan:2016unx, Khachatryan:2014mea, CMS-PAS-EXO-16-022}; these
searches are focused on rather specific signatures and, in particular,
unable to be directly applied to the setup we consider in this paper. } 
Particles with a shorter decay length have been probed only with ordinary
prompt-decay searches. Although these prompt searches do give strong
limits on such particles, the ignorance of the lifetime information
leaves room for improvement in these searches. Given the null result
so far, it is desirable to pursue every possibility of potential
improvements by fully utilizing the ability of the detectors at the
LHC.

Motivated by this observation, in Ref.~\cite{Ito:2017dpm}, we proposed a
new event-selection criterion based on the reconstruction of displaced
vertices using charged tracks associated with the decay points. These
vertices are reconstructed in the same way as those used for the primary
vertex reconstruction, which makes it possible to determine the vertex
position with a resolution of ${\cal O}(10)~\mu\text{m}$. 
We then found that the reach of ordinary prompt-decay searches could be
significantly extended for metastable particles with $c\tau \gtrsim
100~\mu\text{m}$ when we require this new event-selection cut in
addition to other existing selection criteria. Moreover, the
reconstruction of displaced vertices allows us to measure the
lifetime of metastable particles. 

The aim of the present paper is to elaborate on this idea with the help
of detailed Monte Carlo (MC) collider simulations. To be concrete, we here focus on
metastable gluino searches, but a similar study can also be performed
for other metastable particles. With the generation of SM background
processes, we reevaluate the efficiency of the cut proposed in
Ref.~\cite{Ito:2017dpm} and confirm that this is able to separate signal
events from background quite efficiently for particles with $c\tau
\gtrsim 100~\mu\text{m}$. We also discuss optimization of
a kinematical selection cut in order to make the most of the
background-reduction ability of this new cut. It is then found that the reach
of prompt-decay searches can considerably be extended with this method
for particles with $100~\mu\text{m} \lesssim c\tau \lesssim 100$~mm. The
improvement in the sensitivity is especially significant when the parent
and daughter particles have masses close to each other. We also find that the
lifetime of metastable particles can be measured with an ${\cal O}(1)$
accuracy at the high-luminosity LHC. 

The selection method discussed in this paper can in principle be applied
to any collider experiments where high tracking performance is
implemented. We thus study the prospects of our method for a future
100~TeV $pp$ collider \cite{Arkani-Hamed:2015vfh, Golling:2016gvc} on
the assumption that charged tracks are reconstructed with
an accuracy similar to that obtained at the LHC. We see that the extent
of the resultant improvement in the sensitivity in this case is rather
large compared with the 13~TeV LHC case, since gluinos at a 100~TeV
collider are produced in a highly boosted state so that their flight
distance tends to be prolonged. This study suggests that it is desirable
to develop a detector system at a future 100~TeV collider which has
a good capability for track and vertex reconstruction.

This paper is organized as follows. In the next section, we briefly
review the vertex reconstruction method used in our analysis, which is
discussed in more detail in Ref.~\cite{Ito:2017dpm}. As we mentioned
above, in our analysis we consider metastable gluinos as a concrete
example; we thus summarize the properties of metastable gluinos in
Sec.~\ref{sec:gluino} with some discussions on theoretical motivations
for such particles. Then, we show the results of our analysis for the
13~TeV LHC and a future 100~TeV collider in Sec.~\ref{sec:lhc} and
Sec.~\ref{sec:100tev}, respectively. Section~\ref{sec:conclusion} is
devoted to conclusion and discussions.

%%%%%%%%%%%%%%%%%%%%%%%%%%%%%%%%%%%%%%%%%%
\section{Vertex reconstruction method}
\label{sec:vertex}
%%%%%%%%%%%%%%%%%%%%%%%%%%%%%%%%%%%%%%%%%%

To begin with, we describe the vertex reconstruction procedure we use
in the following analysis. We also discuss how to take account of the
deterioration in the vertex reconstruction caused by the limited
resolution of track reconstruction. See Ref.~\cite{Ito:2017dpm} for
more detailed explanation of this procedure and the validation of our
method.

In the following discussion, we focus on the event topology in which
metastable particles are pair-produced and their decay product
contains a large number of charged particles as well as a stable
neutral massive particle, which yields a large missing transverse
energy. Such an event topology is realized in gluino decays, where
each decay vertex is accompanied with two hard jets as we see in
Sec.~\ref{sec:gluino}. By using the information of tracks associated
with the charged particles emitted from the decay points (such as
those in the jets in the case of gluino decay), we can determine the
positions of the decay points of metastable particles.  In particular,
if a metastable particle is pair-produced, two decay vertices exist.
%Observation of the distance between two decay vertices 
Observation of two distinct decay vertices
is a strong
evidence of a new metastable particle like long-lived gluino.  In the
following analysis, we propose to use the distance between two
reconstructed decay vertices as a discriminator to reduce SM
backgrounds.

The procedure to reconstruct vertices emitting hadronic activities has
been well established, as it is used for the primary-vertex
reconstruction by both the ATLAS \cite{ATLAS:2010lca, Aaboud:2016rmg,
  ATL-PHYS-PUB-2015-026} and CMS \cite{Chatrchyan:2014fea,
  CMS-DP-2016-041} collaborations.  We propose to apply such a
procedure to the search of metastable particles.  Typical resolution
of the primary vertex position is ${\cal O}(10)~\mu\text{m}$, which
indicates that the decay positions of metastable particles can also be
determined with a similar precision if the number of charged tracks
associated with the decay vertices is sufficiently large, which is
expected to be realized in gluino decay events.

In our analysis, we reconstruct vertices from charged tracks using the
adaptive vertex fitting algorithm~\cite{Fruhwirth:2007hz}, which is
adopted in Refs.~\cite{ATLAS:2010lca, Aaboud:2016rmg}. In this
algorithm, we first obtain a set of crossing points of the tracks;
each crossing point is defined as the midpoint between the two points
of closest approach of two tracks. A vertex seed is then determined
from these crossing points using a method called the fraction of
sample mode with weights \cite{Bickel20063500}, where the vertex
position is set to be the point of highest density of the crossing
points (this is performed separately in each spatial coordinate). We
denote the position vector of this vertex by $\bm{v}$. Next, we assign
a weight $w_i$ to each track (labeled by $i$) such that tracks which
lie distant from the vertex seed $\bm{v}$ are down-weighted. This
weight contains a parameter that controls the reduction factor of
weights for distant tracks. See Ref.~\cite{Ito:2017dpm} for the
explicit form of $w_i$. We then find another vertex which minimizes
the sum of the squared standardized distances from the vertex seed
$\chi_i^2 (\bm{v})$,
\begin{align}
  \chi^2_i (\bm{v}) \equiv 
  \frac{d_i^2 (\bm{v})}{\sigma_{d_0}^2+\sigma_{z_0\sin\theta}^2},
  \label{eq:chi_sq}
\end{align}
with $d_i (\bm{v})$ being the distance from the vertex $\bm{v}$,
 multiplied by the weights $w_i$ (we refer to this
as the weighted vertex chi square value, $\chi^2_{w} \equiv \sum_{i}
w_i \chi^2_i (\bm{v})$), and regard this vertex as the new vertex
seed. This process is iterated while changing the parameter in the
weights so that the distant tracks are more highly down-weighted,
until the change in the parameter is stopped and the convergence of
the vertex position is reached within a certain accuracy (we take this
to be $1~\mu\text{m}$). With the help of the parameter in the weight,
we can gradually narrow down the range of the tracks we use to obtain
a vertex candidate, through which we can avoid being stuck at a local
minimum of the weighted vertex chi square value. For the choice of the
parameters and the procedure of the iteration, we follow
Refs.~\cite{Fruhwirth:2007hz, Ito:2017dpm}.

The resolution of the vertex reconstruction strongly depends on that of
the track reconstruction. To take this effect into account, we smear 
tracks obtained from the MC-truth information by shifting each track in
parallel by impact parameters which are randomly chosen from 
Gaussian distributions with the variances set by impact parameter
resolutions. To be specific, we refer to the performance of the
ATLAS detector in what follows. The tracking performance of the ATLAS
inner detector for the 13~TeV run is given in Refs.~\cite{ATLAS:trk,
ATL-PHYS-PUB-2015-018}. Here we only consider the resolution of the
transverse and longitudinal impact parameters, $d_0$ and
$z_0\sin\theta$, respectively, as the resolution for the track direction
is sufficiently small \cite{PERF-2007-01}.\footnote{
The polar angle $\theta$ is defined with respect to the $z$-axis, which
lies in the direction of the beam axis. $d_0$ and $z_0$ denote
the distance of the closest approach between the track and the beam axis
and its $z$-coordinate, respectively. } The impact parameter resolutions
are parametrized as functions of the transverse momentum $p_{\text{T}}$
\cite{PERF-2007-01}: 
\begin{equation}
 \sigma_X (p_{\text{T}}) = \sigma_X (\infty) (1\oplus p_X/p_{\text{T}})
  ~,
\label{eq:sigmax}
\end{equation}
for $X = d_0, z_0\sin\theta$, where $\sigma_X(\infty)$ and $p_X$ are
some constant parameters determined via the fitting of this expression
onto the experimental result obtained by the ATLAS collaboration
\cite{ATLAS:trk,ATL-PHYS-PUB-2015-018}. 
We here neglect the $\eta$-dependence of $\sigma_X$ as
it turns out to be sufficiently small for $p_{\text{T}} \gtrsim\text{a
few GeV}$ \cite{PERF-2007-01, impres2015}. 

The validation of the above procedure was performed in
Ref.~\cite{Ito:2017dpm} using minimum-bias event samples, where the
position of primary vertices are reconstructed with our method using
smeared tracks. The vertex resolution obtained in this way was found to
be in good agreement with those given in
Ref.~\cite{ATL-PHYS-PUB-2015-026}. We thus use this prescription to
reconstruct displaced vertices associated with the decay of metastable
particles in what follows.

%%%%%%%%%%%%%%%%%%%%%%%%%%%%%%%%%%%%%%%%%%%%%%%%%%%%%%%%%%
\begin{figure}
  \centering
  \includegraphics[width=0.7\columnwidth]{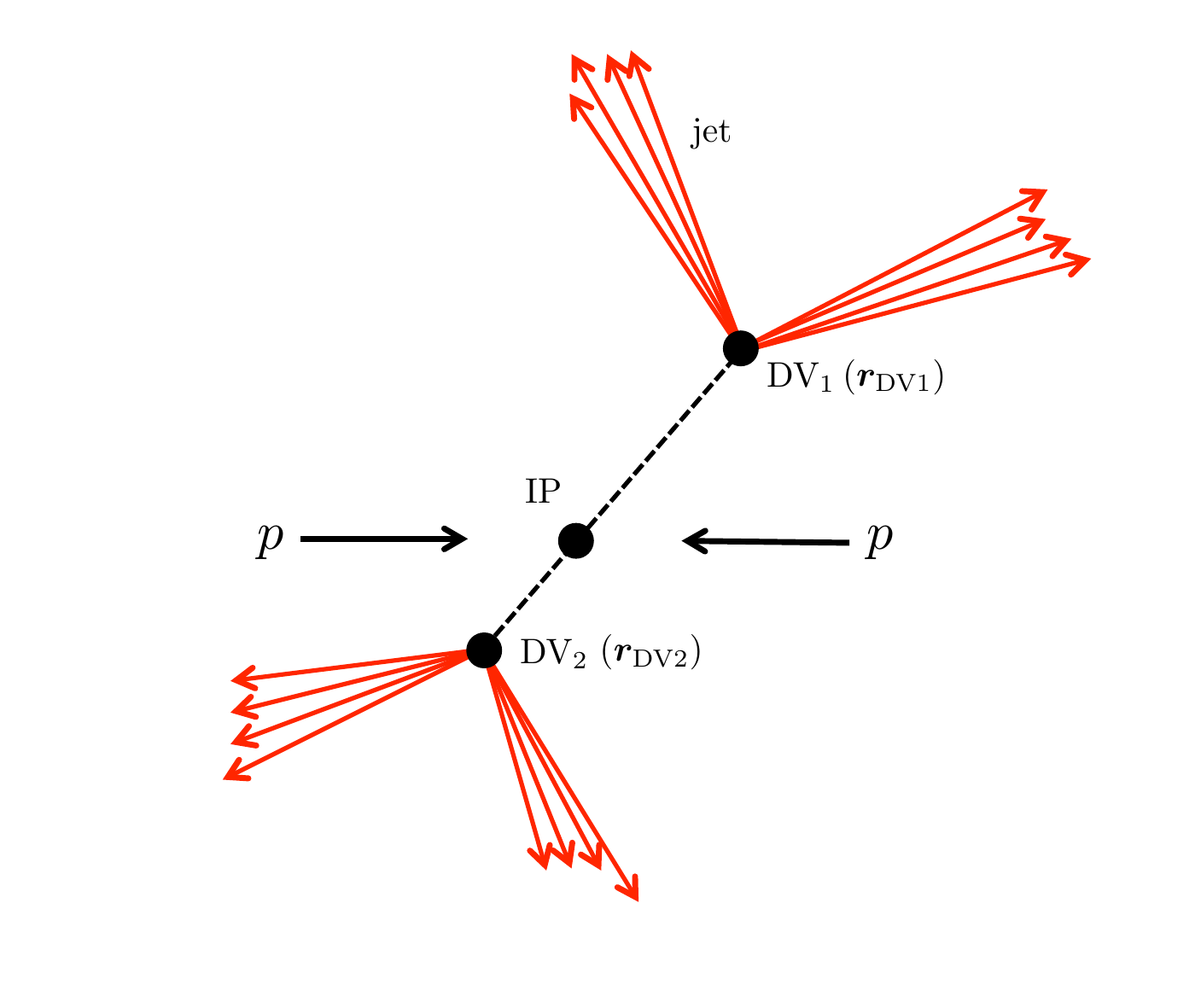}
 \caption{The event topology of metastable gluino decay. 
 Here, ${\rm DV}_1$ and ${\rm DV}_2$ are two decay vertices of gluino,
while IP is the
interaction point.}
  \label{fig:dv_signature}
\end{figure}
%%%%%%%%%%%%%%%%%%%%%%%%%%%%%%%%%%%%%%%%%%%%%%%%%%%%%%%%%%%

As mentioned above, gluinos are pair produced and the decay of each
gluino gives rise to two hard jets; this event topology is illustrated
in Fig.~\ref{fig:dv_signature}. Considering this, we reconstruct
displaced vertices in the metastable gluino decay events by using only
tracks associated with four high-$p_{\text{T}}$ jets. The way of
choosing these four high-$p_{\text{T}}$ jets, as well as the track
conditions, will be given in Sec.~\ref{sec:lhc}.  The reconstruction
of the decay vertices in gluino decay events is complicated due to the
fact that it is unclear which pair of jets is associated with the same
vertex. In our analysis, we consider all three possible patterns of
pairings out of the four jets and reconstruct a vertex for each pair
using the method described above.  For each pairing, we determine the
positions of two decay vertices assuming that the jets in a pair
originate from the same decay vertex; 
%the decay vertex is determined
%with the tracks associated to the jets using the procedure of the
%vertex reconstruction explained above.  
each decay vertex is determined
with the tracks associated to the jets constituting the pair.
We then choose the pairing
that has the smallest value of $\chi^2$ defined by
\begin{equation}
 \chi^2 \equiv \frac{
\sum_{i\in \text{trk}(\bm{v}_1) } w_i  \chi^2_i (\bm{v}_1) 
+\sum_{j\in \text{trk}(\bm{v}_2) } w_j \chi^2_j (\bm{v}_2)
}{
\sum_{i\in \text{trk}(\bm{v}_1) } w_i   (\bm{v}_1) 
+\sum_{j\in \text{trk}(\bm{v}_2) } w_j (\bm{v}_2)} ~,
\label{eq:chisq}
\end{equation}
where $\text{trk}(\bm{v}_{1,2})$ denotes the set of tracks associated
with the vertex reconstructed at the position $\bm{v}_{1,2}$ for each
pair of jets.  
Note that 
$\chi_i^2 (\bm{v})$ is
the squared standardized distance from the vertex $\bm{v}$,
which is defined by Eq.~\eqref{eq:chi_sq}.
%with $d_i (\bm{v})$ being the distance from the vertex $\bm{v}$, while
%and
%$w_i$ is the weight factor of the track $i$ (see
%Ref.~\cite{Ito:2017dpm} for more detail). 
We employ the vertices
obtained with this pairing, and denote their positions by
$\bm{r}_{\text{DV1,2}} \equiv \bm{v}_{1,2}$. This new information 
of the position of the displaced vertices
will be used to improve the metastable gluino searches in the following
analysis.

%%%%%%%%%%%%%%%%%%%%%%%%%%%%%%%%%%%%%%%%%%%%%
\section{Gluino phenomenology}
\label{sec:gluino}
%%%%%%%%%%%%%%%%%%%%%%%%%%%%%%%%%%%%%%%%%%%%%

As mentioned in the introduction, we consider a metastable gluino as a
concrete example and study the effect of the new selection cut on the
gluino searches. Here we assume that all of the squarks are much heavier
than gluino and there is at least one neutralino which is lighter than
the gluino. In this case, gluinos $\tilde{g}$ decay either in the
tree-level three-body decay processes with the emission of a pair of
quarks and a neutralino $\tilde{\chi}^0$ or chargino $\tilde{\chi}^\pm$,
$\tilde{g} \to q \bar{q} \tilde{\chi}^0, q\bar{q}^\prime
\tilde{\chi}^\pm$, or in the one-loop two-body decay process where 
a gluino and a neutralino is emitted: $\tilde{g} \to g
\tilde{\chi}^0$. It is found that if the gluino mass is $\gtrsim 2$~TeV,
then the three-body decay processes dominate the two-body decay
\cite{Toharia:2005gm, Gambino:2005eh, Sato:2012xf}. As the three-body
decay processes are induced by the exchange of virtual heavy squarks,
the amplitudes of these processes are suppressed by the squared masses
of these heavy squarks, $m_{\tilde{q}}^2$. As a consequence, gluinos
become long-lived if the squark masses are sufficiently heavy. To see
this more quantitatively, we show an approximate formula for the decay
length of gluinos:
\begin{equation}
 c\tau_{\tilde{g}} \simeq 200\times 
\biggl(\frac{m_{\tilde{q}}}{10^3~\text{TeV}}\biggr)^4
\biggl(\frac{m_{\tilde{g}}}{2~\text{TeV}}\biggr)^{-5} ~\mu\text{m} ~,
\label{eq:ctaugapp}
\end{equation} 
where $m_{\tilde{g}}$ is the mass of gluino. In deriving this
approximate formula, we have assumed that the gluino mainly decays into
the first-generation quarks, the masses of bino and wino are much
smaller than the gluino mass, and higgsinos are heavier than the
gluino. In this case, $m_{\tilde{q}}$ corresponds to the mass of the
first-generation squarks. From Eq.~\eqref{eq:ctaugapp}, we see
that metastable gluinos with a decay length of $\gtrsim 100~\mu\text{m}$
are obtained if squarks lie around the PeV scale or higher. 

In fact, the PeV-scale squarks are phenomenologically well motivated. First
and foremost, with such heavy squarks (especially heavy stops), the
observed value of the SM-like Higgs boson mass \cite{Aad:2015zhl} can
be explained within the minimal SUSY SM \cite{Okada:1990vk,
Okada:1990gg, Ellis:1990nz, Ellis:1991zd, Haber:1990aw}. In addition, 
the contributions of SUSY particles to 
the flavor-changing neutral current
processes, the electric dipole moments, and the
dimension-five proton decay processes are all suppressed by heavy
sfermion masses, and thus the experimental limits on such dangerous processes can
easily be evaded \cite{Gabbiani:1996hi, Moroi:2013sfa, McKeen:2013dma,
Altmannshofer:2013lfa, Fuyuto:2013gla, Hisano:2013exa, Nagata:2013sba,
Evans:2015bxa}. Even though squarks are as heavy as ${\cal O}(1)$~PeV,
fermionic SUSY particles can naturally be around the TeV scale since
their masses are protected by chiral symmetry. A simple framework to
realize this split spectrum \cite{Giudice:1998xp, Wells:2003tf,
ArkaniHamed:2004fb, Giudice:2004tc, ArkaniHamed:2004yi, Wells:2004di} is
provided by the so-called anomaly-mediated SUSY-breaking mechanism
\cite{Giudice:1998xp, Randall:1998uk},\footnote{In the anomaly
mediation, the gluino mass is proportional to the gravitino mass with
its proportional constant given by the strong gauge coupling beta
function. Under this relation, if both gravitino and squarks are around
the PeV scale, then the gluino mass tends to be much larger
than the TeV scale. This consequence can be evaded if the SUSY-breaking
effects are transmitted to the visible sector below the Planck scale; in
this case, squark masses are larger than the gravitino mass, and thus it
is possible to obtain a TeV-scale gluino and PeV-scale squarks
simultaneously, by assuming a gravitino mass lower than the PeV
scale. Keeping this in mind, we consider both the gluino mass and 
squark masses, and thus the gluino lifetime as well, to be free
parameters in the following discussion. } 
where gaugino masses are induced by quantum corrections and thus
suppressed by a loop factor compared with the gravitino mass. 
This gives theoretical support for the split SUSY mass spectrum. A
distinguished feature of this type of 
spectrum is that the lightest SUSY particle (LSP), which is one of
the TeV-scale fermionic SUSY particles such as the neutral wino and
higgsino, can be a promising candidate for dark matter in the Universe 
\cite{Gherghetta:1999sw, Moroi:1999zb}. These advantages have thus
attracted a lot of attentions \cite{Hall:2011jd, Ibe:2011aa, Ibe:2012hu,
Arvanitaki:2012ps, Hall:2012zp, ArkaniHamed:2012gw, Evans:2013lpa,
Evans:2013dza} especially after the early stage of the LHC run. In such
a split spectrum, gluino may also lie around the TeV scale and be within
the reach of the LHC and/or a future hadron collider. 
Gluinos in this case are expected to have a sizable decay length according to
Eq.~\eqref{eq:ctaugapp}---in this sense, a metastable gluino may be
regarded as a smoking-gun signature of the split-type SUSY
spectrum and the detection of such a gluino enables us to
confirm this scenario experimentally. Moreover, since the lifetime of
metastable gluinos reflects the squark mass scale, we may probe this
via the precise measurements of the gluino decay length, even though we
cannot produce squarks directly at colliders. We discuss this
possibility in more detail in Sec.~\ref{sec:lifetime}.

If the gaugino masses are induced mainly by the anomaly mediation, the
gaugino mass spectrum is determined by the gauge-coupling beta
functions; in particular, one finds that the LSP is the neutral wino in
this case. On the other hand, if
there are other contributions which are comparable to the
anomaly-mediation contribution in size, the gaugino mass spectrum can
deviate from the anomaly-mediation relation and, in fact, 
%be almost arbitrary. 
the spectrum could take any form.
For example, threshold corrections by heavy Higgs
bosons/higgsinos \cite{Pierce:1996zz, Giudice:1998xp} and some extra
matter fields \cite{Pomarol:1999ie, Nelson:2002sa, Hsieh:2006ig,
Gupta:2012gu, Nakayama:2013uta, Harigaya:2013asa, Evans:2014xpa}
give rise to such contributions. Considering this, we regard both the
gluino and LSP masses as free parameters in what follows.

An interesting possibility of gaugino mass spectrum is that the LSP
(say bino) is degenerate with gluino in mass with a mass splitting of
$\sim 100$~GeV. In this case, the annihilation cross section of the
LSP in the early Universe is strongly enhanced by the coannihilation
effect \cite{Griest:1990kh}, which has significant implications for
the relic abundance of the LSP. In fact, it is found that the relic
abundance of the bino LSP agrees to the observed dark matter density
if the mass difference between bino and gluino is $\lesssim 100$~GeV
\cite{Profumo:2004wk, Feldman:2009zc, deSimone:2014pda,
  Harigaya:2014dwa, Ellis:2015vaa, Nagata:2015hha, Ellis:2015vna,
  Liew:2016hqo}. Motivated by this observation, we will also consider
such a degenerate mass spectrum in our analysis.

%%%%%%%%%%%%%% FIGURE %%%%%%%%%%%%%%%%%%%%%%%%%%%%%%%%%%%%
\begin{figure}
  \centering
  \subcaptionbox{\label{fig:prodcr13TeV} 13 TeV}{
  \includegraphics[width=0.48\columnwidth]{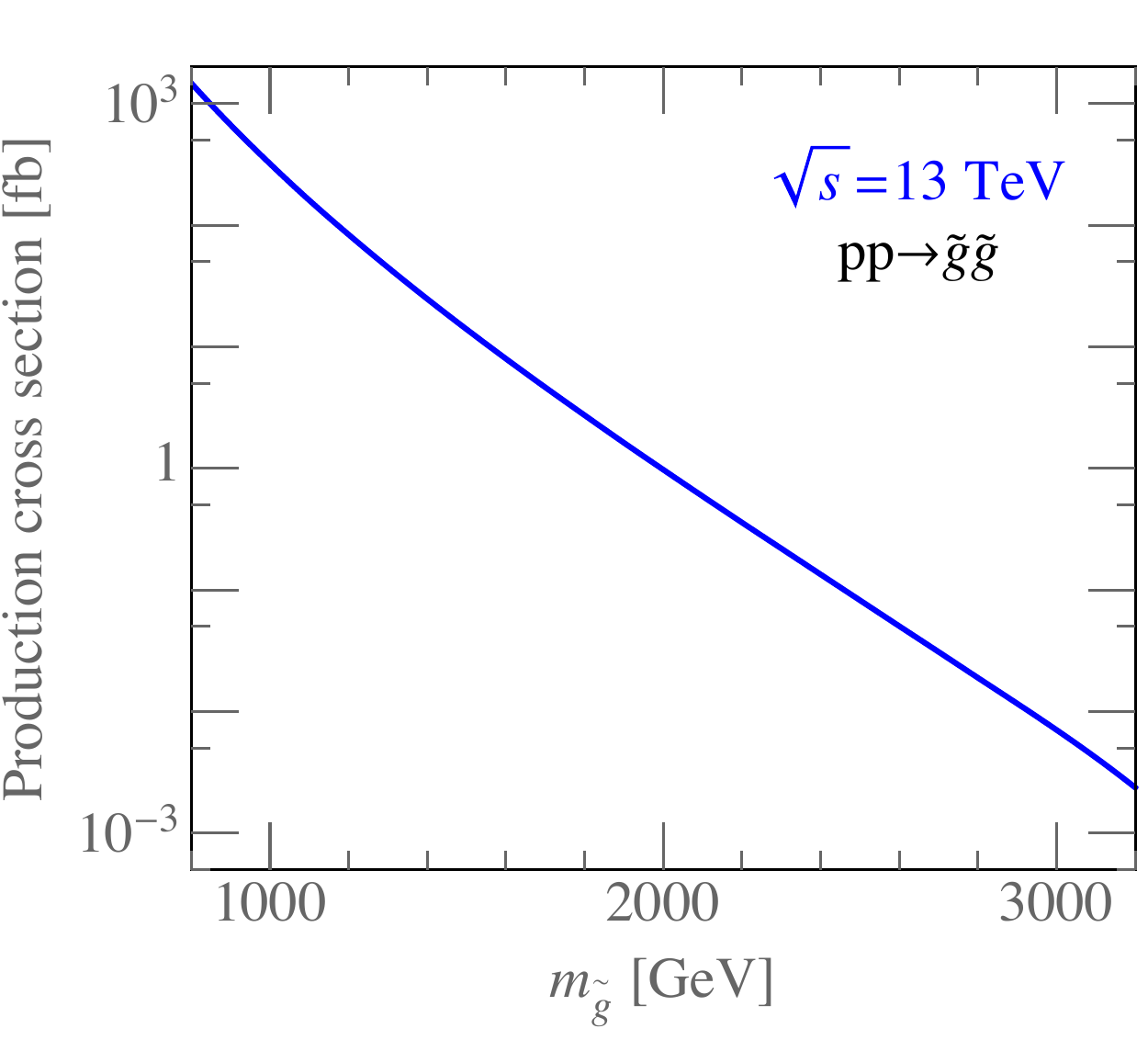}}
  \subcaptionbox{\label{fig:prodcr100TeV} 100 TeV}{
  \includegraphics[width=0.48\columnwidth]{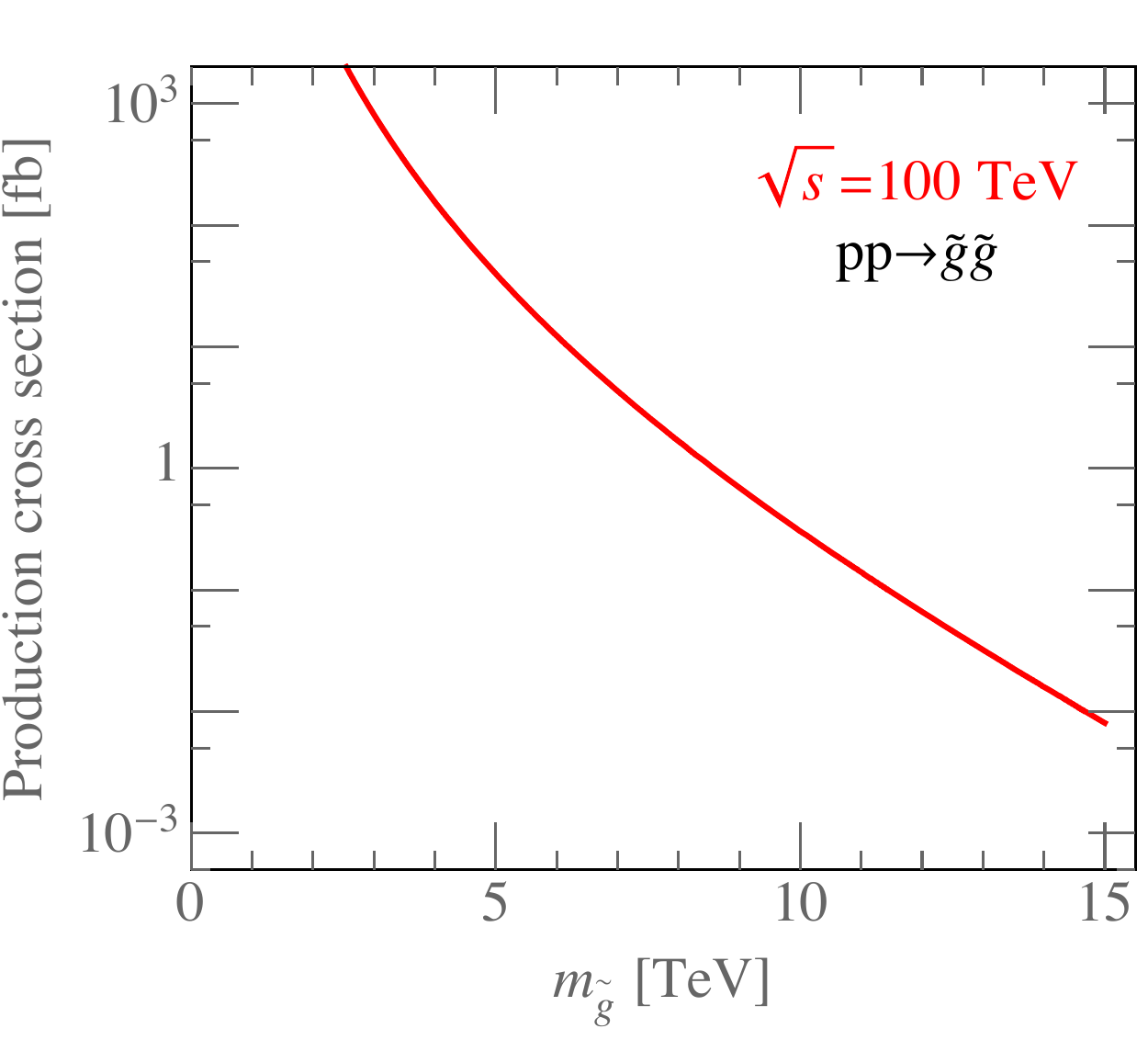}}
\caption{The pair-production cross sections of gluinos as functions of
 the gluino mass.} 
  \label{fig:gluinopr}
\end{figure}
%%%%%%%%%%%%%%%%%%%%%%%%%%%%%%%%%%%%%%%%%%%%%%%%%%%%%%%%%%

At hadron colliders, gluinos are pair-produced through the strong
interactions, and the cross section of this process is unambiguously
determined if squarks are heavy enough. We show the pair-production
cross sections of gluinos as functions of the gluino mass for $\sqrt{s}
= 13$ and 100~TeV in Fig.~\ref{fig:prodcr13TeV} and
Fig.~\ref{fig:prodcr100TeV}, respectively. They are computed using the
\texttt{CTEQ6.6} PDF set \cite{Nadolsky:2008zw} with the next-to-leading
order QCD corrections included and soft gluon emission resummed at the
next-to-leading logarithmic accuracy \cite{Borschensky:2014cia}. We will
use these production cross sections in the following analyses.

%%%%%%%%%%%%%%%%%%%%%%%%%%%%%%%%%%%%%%%%%%%%%%%%%%
\section{13 TeV LHC}
\label{sec:lhc}
%%%%%%%%%%%%%%%%%%%%%%%%%%%%%%%%%%%%%%%%%%%%%%%%%%

%Now we study how much extent we may extend the reach of the prompt-decay
Now we study how far we may extend the reach of the prompt-decay
gluino searches at the 13~TeV LHC using a new cut based on the
reconstruction of displaced vertices. We first describe our MC
simulation procedure in Sec.~\ref{sec:mc}, which is followed by the
summary of the event selection criteria used in this analysis in
Sec.~\ref{sec:eventsel}. We then show the prospects of the new
event-selection cut in Sec.~\ref{sec:prospects}. Finally, in
Sec.~\ref{sec:lifetime}, we discuss the possibility of measuring the
lifetime of gluino in the future LHC experiments.

%%%%%%%%%%%%%%%%%%%%%%%%%%%%%
\subsection{MC simulation}
\label{sec:mc}
%%%%%%%%%%%%%%%%%%%%%%%%%%%%%

%-------------------------------
\begin{table}
  \begin{center}
    \begin{tabular}{c|c}
      \hline\hline
      $m_{\tilde{g}}$ & 
      $1000,\,1200,\,\ldots, \,3200$ GeV\\ \hline
      \multirow{3}{*}{$m_{\tilde{\chi}_1^0}$} &
      $100$ GeV, \\
      &
      $( 0.2,\, 0.4,\, 0.6,\, 0.7,\, 0.8 )\times m_{\tilde{g}}$,  \\ 
      & 
      $m_{\tilde{g}}-(150\,{\rm GeV},\, 100\,{\rm GeV},\, 50\,{\rm GeV},\, 25 \,{\rm GeV})$ \\
      \hline
      \multirow{2}{*}{$c\tau_{\tilde{g}}$} &   
      $0,\, 50,\,100,\, 200,\, 500,\, 1000,\, 3000$ ${\rm \mu m}$, \\
      &
      $1\times10^4,\, 3\times10^4,\, 1\times10^5,\, 3\times10^5,\, 1\times10^6$ ${\rm \mu m}$\\
      \hline\hline
    \end{tabular}
    \caption{Sample points for signal events. We generate 50000 events
      for each sample point.
    }
    \label{table:sample_point}
  \end{center}
\end{table}
% -------------------------------

In the following analysis, both signal and background events are
generated with MC simulations.  For signal events, we use {\tt
  MadGraph5\_aMC@NLO\,v2}~\cite{Alwall:2014hca} and {\tt PYTHIA\,v8.2}
\cite{Sjostrand:2007gs}, and generate 50000 events for each model
point categorized in terms of the gluino mass $m_{\tilde{g}}$, the LSP
mass $m_{\tilde{\chi}_1^0}$, and the gluino decay length
$c\tau_{\tilde{g}}$; sample points with this categorization are
summarized in Table~\ref{table:sample_point}.\footnote{Among these
  points, we do not generate events for $m_{\tilde{g}} \leq 1800$~GeV
  and $m_{\tilde{\chi}^0_1} \leq 600$~GeV, and for $m_{\tilde{g}} =
  3200$~GeV and $m_{\tilde{\chi}^0_1} \neq 100$~GeV. }  In generating
the signal events, (i) the gluinos are forced to decay at the
interaction point, then (ii) the decay points are spatially shifted by
the flight lengths of the parent gluinos, whose probability
distribution is determined from the lifetime of gluino
$\tau_{\tilde{g}}$.  All the trajectories of the decay products of the
parent gluinos are shifted accordingly.  The number of signal event
samples is normalized according to the production cross section given
in Fig.~\ref{fig:prodcr13TeV}.  In our analysis, gluino is assumed to
decay only into the first-generation quarks. In reality, gluino may
also have sizable decay fractions into the other two generations of
quarks, especially top quark when the stop mass is similar to or
smaller than the first-generation squark masses.  We will briefly
discuss such cases in Sec.~\ref{sec:conclusion}. Moreover, we neglect
the hadronization effect of gluinos as it is known that the energy and
momentum of the resultant $R$-hadrons are almost the same as those of
the produced gluinos, for a gluino mass of ${\cal O}(1)$~TeV
\cite{Fairbairn:2006gg, Bjorken:1977md}.

As for the SM background events, we consider the production processes
of the electroweak gauge bosons ($W^\pm$ and $Z$) and top-quark pairs
($t \bar{t}$). We do not consider the fully hadronic decay processes
of these particles since we require a large missing energy and thus
such events are expected to be efficiently eliminated. In addition, we
do not take account of diboson production and multi-jet processes as
they are found to be subdominant \cite{Aaboud:2017vwy}. For the $Z$
boson production channel, the matrix elements are calculated with up
to four additional partons. Due to the limitation of our available
computational resources, we include only up to three and one extra
partons for the $W$ and $t\bar{t}$ production processes, respectively,
which are subdominant compared with the $Z$ boson production
process. 
%(We also studied these processes with one more parton.
%Although the number of generated events is not enough for our analysis
%of the gluino search, we checked that the number of SM background is
%expected not to be significantly changed even if we add one more
%parton to these samples.  \rem{TM: Is this comment OK?})  
(We also studied these processes with one more parton.
Although the number of generated events is not enough for our analysis
of the gluino search, we observed that the number of SM background 
does not seem to 
 be significantly changed even if we add one more
parton to these samples.) 
%\rem{TM: Is this comment OK? 
%-- The statement is made weakened (Ito).}
The samples
with different number of additional partons are merged using the five
flavor MLM matching scheme with $k_t$ jets \cite{Alwall:2007fs}. Here
we adopt the shower-$k_t$ scheme and set the matching parameter ${\rm
  QCUT} = {\rm XQCUT}$ to $40$~GeV for the $Z$ and $W$ production
processes and $80$~GeV for the $t\bar{t}$ production process,
following Ref.~\cite{Avetisyan:2013onh}. All generated samples are
passed to {\tt PYTHIA\,v8.2}~\cite{Sjostrand:2007gs} and then {\tt
  DELPHES\,v3} \cite{deFavereau:2013fsa} for the purpose of parton
showering and fast detector simulation, respectively.

%------------------------------------------------------------------
% Z
\begin{table}
  \centering
  \subcaptionbox{$Z \rightarrow \nu\bar{\nu} + 4j$. \label{table:bg_table_z4}}[.9\linewidth]{
    \begin{tabular}{cc|ccccc}
      \hline\hline
      & bin $\alpha$ &
      0 & 1 & 2 & 3 & 4 \\ \hline
      $H_{T,0}^{\rm max}(\alpha)$ & (GeV) & 
      $600$ & $1200$ & $2000$ & $2900$ & $\infty$ \\
      %$\sigma_{\text{no-matching}}(\alpha)$ & (fb) &
      %$1.91{\scriptstyle \times10^3}$ & $1.33{\scriptstyle \times10^3}$ & $1.83{\scriptstyle \times10^2}$ & $19.7$ & $2.60$ \\
      $\sigma_{\text{matched}}(\alpha)$ & (fb) &
      $8.85{\scriptstyle \times10^2}$ & $5.50{\scriptstyle \times10^2}$ & $66.5$ & $6.6$ & $0.83$ \\
      $N_{\text{matched}}(\alpha)$ & $(\times 10^3)$ &
      %$10$ & $130$ & $180$ & $590$ & $90$ \\
      $4.6$ & $53.6$ & $65.3$ & $198$ & $28.8$ \\
      $\mathcal{L}_{\rm gen}(\alpha)$ & (${\rm ab}^{-1}$) &
      $5.23{\scriptstyle \times10^{-3}}$ & $0.0975$ & $0.982$ & $29.9$ & $34.6$ \\
      % $\mathcal{L}_{\rm gen}(\alpha)$ & (${\rm fb}^{-1}$) &
      % $5.23$ & $97.5$ & $9.82\times10^2$ & $2.99\times10^4$ & $3.46\times10^4$ \\
      \hline\hline
    \end{tabular}
  }
  
  \vspace{3mm}
  \centering
  \subcaptionbox{$Z \rightarrow \nu\bar{\nu} + 0,1,2,3\,j$. \label{table:bg_table_z3}}[.9\linewidth]{
    \begin{tabular}{cc|ccccc}
      \hline\hline
      & bin $\alpha$ &
      0 & 1 & 2 & 3 & 4 \\ \hline
      $H_{T,0}^{\rm max}(\alpha)$ & (GeV) & 
      $600$ & $1200$ & $2000$ & $2900$ & $\infty$ \\
      %$\sigma_{\text{no-matching}}(\alpha)$ & (fb) &
      %$2.84{\scriptstyle \times10^4}$ & $2.42{\scriptstyle \times10^3}$ & $1.94{\scriptstyle \times10^2}$ & $17.2$ & $2.03$ \\
      $\sigma_{\text{matched}}(\alpha)$ & (fb) &
      $1.22{\scriptstyle \times10^4}$ & $7.85{\scriptstyle \times10^2}$ & $55.7$ & $4.65$ & $0.54$ \\
      $N_{\text{matched}}(\alpha)$ & $(\times 10^3)$ &
      %$500$ & $1000$ & $2000$ & $500$ & $500$ \\
      $214$ & $325$ & $574$ & $135$ & $133$ \\
      $\mathcal{L}_{\rm gen}(\alpha)$ & (${\rm ab}^{-1}$) &
      $0.0176$ & $0.413$ & $10.3$ & $29.1$ & $247$ \\
      % $\mathcal{L}_{\rm gen}(\alpha)$ & (${\rm fb}^{-1}$) &
      % $17.6$ & $4.13\times10^2$ & $1.03\times10^4$ & $2.91\times10^4$ & $2.47\times10^5$ \\
      \hline\hline
    \end{tabular}
  }
  
  \vspace{3mm}
  \centering
  \subcaptionbox{$Z \rightarrow \ell\bar{\ell} + 0,1,2,3\,j$ \label{table:bg_table_zll}}[.9\linewidth]{
    \begin{tabular}{cc|ccccc}
      \hline\hline
      & bin $\alpha$ &
      0 & 1 & 2 & 3 & 4 \\ \hline
      $H_{T,0}^{\rm max}(\alpha)$ & (GeV) & 
      $600$ & $1200$ & $2000$ & $2900$ & $\infty$ \\
      %$\sigma_{\text{no-matching}}(\alpha)$ & (fb) &
      %$1.12{\scriptstyle \times10^5}$ & $2.34{\scriptstyle \times10^3}$ & $1.75{\scriptstyle \times10^2}$ & $14.6$ & $1.67$ \\
      $\sigma_{\text{matched}}(\alpha)$ & (fb) &
      $6.28{\scriptstyle \times10^4}$ & $1.04{\scriptstyle \times10^3}$ & $74.4$ & $6.05$ & $0.69$ \\
      $N_{\text{matched}}(\alpha)$ & $(\times 10^3)$ &
      %$500$ & $500$ & $500$ & $500$ & $500$ \\
      $280$ & $223$ & $212$ & $207$ & $205$ \\
      $\mathcal{L}_{\rm gen}(\alpha)$ & (${\rm ab}^{-1}$) &
      $4.46{\scriptstyle \times10^{-3}}$ & $0.214$ & $2.85$ & $34.1$ & $300$ \\
      % $\mathcal{L}_{\rm gen}(\alpha)$ & (${\rm fb}^{-1}$) &
      % $4.46$ & $2.14\times10^2$ & $2.85\times10^3$ & $3.41\times10^4$ & $3.00\times10^5$ \\
      \hline\hline
    \end{tabular}
  }
  \caption{
    The upper edge value of $H_{T,0}$, $H_{T,0}^{\rm max}(\alpha)$,
    the leading-order matched cross section,
    %the cross section after matching, 
    $\sigma_{\text{matched}}$,
    the number of samples after matching,
    $N_{\text{matched}}$, and the equivalent
    luminosity, $\mathcal{L}_{\rm gen}$, in each $H_{T,0}$ bin for the
    $Z$-boson production process. 
    %\TODO{
    %Isn't the cross section in the bin 0 in
      %Table~\ref{table:bg_table_z4} smaller than expected by an order
      %of magnitude?? 
      %Do we want to show the matched quantities rather than
      %non-matched ones??}
  }
  \label{table:bg_table_z}
\end{table}
% ------------------------------------------------------------------

% ------------------------------------------------------------------
% W
\begin{table}
  \begin{center}
    \begin{tabular}{cc|cccccc}
      \hline\hline
      & bin $\alpha$ &
      0 & 1 & 2 & 3 & 4 & 5\\ \hline
      $H_{T,0}^{\rm max}(\alpha)$ & (GeV) & 
      $400$ & $900$ & $1600$ & $2500$ & $3400$ & $\infty$ \\
      %$\sigma_{\text{no-matching}}(\alpha)$ & (fb) &
      %$9.01{\scriptstyle \times10^5}$ & $6.75{\scriptstyle \times10^4}$ &
      %$4.82{\scriptstyle \times10^3}$ & $4.15{\scriptstyle \times10^2}$ & $34.2$ & $4.66$ \\
      $\sigma_{\text{matched}}(\alpha)$ & (fb) &
      $5.15{\scriptstyle \times10^5}$ & $3.15{\scriptstyle \times10^4}$ &
      $2.10{\scriptstyle \times10^3}$ & $1.74{\scriptstyle \times10^2}$ & $14.2$ & $1.92$ \\
      $N_{\text{matched}}(\alpha)$ & $(\times 10^3)$ &
      %$500$ & $500$ & $5000$ & $4000$ & $1000$ & $500$ \\
      $286$ & $233$ & $2180$ & $1680$ & $414$ & $206$ \\
      $\mathcal{L}_{\rm gen}(\alpha)$ & (${\rm ab}^{-1}$) &
      $5.5{\scriptstyle \times10^{-4}}$ & $7.41{\scriptstyle \times10^{-3}}$ &
      $1.04$ & $9.64$ & $29.2$ & $107$ \\
      % $\mathcal{L}_{\rm gen}(\alpha)$ & (${\rm fb}^{-1}$) &
      % $0.55$ & $7.41$ & $1.04{\scriptstyle \times10^3}$ & 
      % $9.64{\scriptstyle \times10^3}$ & $2.92{\scriptstyle \times10^4}$ & $1.07{\scriptstyle \times10^5}$ \\
      \hline\hline
    \end{tabular}
    \caption{
      Same as in Table~\ref{table:bg_table_z}, but for the $W$ production process.
    }
    \label{table:bg_table_w}
  \end{center}
\end{table}
% ------------------------------------------------------------------

% ------------------------------------------------------------------
% tt
\begin{table}
  \centering
  \subcaptionbox{$t\bar{t}\rightarrow$ (semi-leptonic)$+ 0,1 j$. \label{table:bg_table_tt_jl}}[.9\linewidth]{
    \begin{tabular}{cc|cccccc}
      \hline\hline
      & bin $\alpha$ &
      0 & 1 & 2 & 3 & 4 & 5\\ \hline
      $H_{T,0}^{\rm max}(\alpha)$ & (GeV) & 
      $600$ & $1100$ & $1700$ & $2400$ & $3200$ & $\infty$ \\
      %$\sigma_{\text{no-matching}}(\alpha)$ & (fb) &
      %$9.62{\scriptstyle \times10^4}$ & $6.96{\scriptstyle \times10^3}$ &
      %$5.01{\scriptstyle \times10^2}$ & $41.6$ & $3.78$ & $0.362$ \\
      $\sigma_{\text{matched}}(\alpha)$ & (fb) &
      $7.43{\scriptstyle \times10^4}$ & $5.64{\scriptstyle \times10^3}$ &
      $4.26{\scriptstyle \times10^2}$ & $36.8$ & $3.43$ & $0.34$ \\
      $N_{\text{matched}}(\alpha)$ & $(\times 10^3)$ &
      %$200$ & $1500$ & $1500$ & $200$ & $200$ & $200$ \\
      $154$ & $1220$ & $1280$ & $177$ & $182$ & $186$ \\
      $\mathcal{L}_{\rm gen}(\alpha)$ & (${\rm ab}^{-1}$) &
      $2.08{\scriptstyle \times10^{-3}}$ & $0.216$ &
      $2.99$ & $4.81$ & $53.0$ & $552$ \\
      \hline\hline
    \end{tabular}
  }
  
  \vspace{3mm}
  \centering
  \subcaptionbox{$t\bar{t}\rightarrow$ (leptonic)$+ 0,1 j$.  \label{table:bg_table_tt_ll}}[.9\linewidth]{
    \begin{tabular}{cc|cccc}
      \hline\hline
      & bin $\alpha$ &
      0 & 1 & 2 & 3 \\ \hline
      $H_{T,0}^{\rm max}(\alpha)$ & (GeV) & 
      $800$ & $1400$ & $2100$ & $\infty$ \\
      %$\sigma_{\text{no-matching}}(\alpha)$ & (fb) &
      %$510$ & $40.9$ & $3.85$ & $0.342$ \\
      $\sigma_{\text{matched}}(\alpha)$ & (fb) &
      $418$ & $38.9$ & $3.74$ & $0.34$ \\
      $N_{\text{matched}}(\alpha)$ & $(\times 10^3)$ &
      %$50$ & $50$ & $100$ & $50$ \\
      $41$ & $48$ & $97$ & $49$ \\
      $\mathcal{L}_{\rm gen}(\alpha)$ & (${\rm ab}^{-1}$) &
      $0.0980$ & $1.22$ & $26.0$ & $146$ \\
      \hline\hline
    \end{tabular}
  }
  \caption{
    Same as in Table~\ref{table:bg_table_z}, but for the $t\bar{t}$
    production process. 
  }
  \label{table:bg_table_tt}
\end{table}
% ------------------------------------------------------------------

In new physics searches with a large integrated luminosity, it is the
distribution tail of discriminating variables from the SM background
processes that affects the sensitivities of these searches. A
simulation of such tails in general requires a huge amount of
background samples. To deal with this difficulty, we use the technique
described in Ref.~\cite{Avetisyan:2013onh}. In this method, the
generator-level phase space is divided into several regions, and MC
samples are separately generated and matched for each region. The
generated samples are then combined with each other after multiplied
by some weight factor. In our analysis, we divide the generator-level
phase space into bins numbered by $\alpha = 0,1,2,\dots$, in terms of
the scalar sum of $p_{\text{T}}$ of partons in each event: 
\begin{align}
  H_{T,0} \equiv \sum_{\text{partons}} p_{\text{T}}.
\end{align}
We show the values of the upper edge of each bin $\alpha$, denoted by
$H_{T,0}^{\text{max}} (\alpha)$, for the $Z$, $W$, and $t\bar{t}$
production processes in Tables~\ref{table:bg_table_z},
\ref{table:bg_table_w}, and \ref{table:bg_table_tt},
respectively. Here, the width of the bin $\alpha + 1$ is determined
such that the corresponding cross section is $\simeq 1/10$ of that of
the bin $\alpha$. 
\footnote{
For $Z$ production processes, 
the widths of the bins are taken universal; the width of
the bin $\alpha+1$ is determined such that the corresponding cross section
becomes $\sim 1/10$ of that of the bin $\alpha$ for
$Z\rightarrow\nu\bar{\nu}+0,1,2,3j$.
}
We also give 
the leading-order matched cross section
%the cross sections after the matching procedure 
$\sigma_{\text{matched}}$, 
the number of samples after matching
%the number of generated event samples 
$N_{\text{matched}}$, and the equivalent luminosity $\mathcal{L}_{\rm
  gen} =N_{\text{matched}}/\sigma_{\text{matched}}$ in these
tables. 
\footnote{We do not include $K$-factors in our analysis.}
When generating these samples, we impose a generator-level cut
of $| \sum_{{\rm all}\, \nu} \bm{p}_{\rm T}| > 200$~GeV for $Z
\rightarrow \nu\bar{\nu}$ and $| \sum_{{\rm all}\, l,\nu} \bm{p}_{\rm
  T}| > 100$~GeV for the other processes. 
 We then define the weight
$w_\alpha$ for the samples in the bin $\alpha$ by
\begin{equation}
 w_\alpha \equiv \frac{\sigma_{\text{matched}}
  (\alpha)}{N_{\text{matched}} (\alpha)} ~,
\label{eq:weight}
\end{equation}
%where $\sigma_{\text{matched}} (\alpha)$ denotes the leading-order
%matched cross section and $N_{\text{matched}} (\alpha)$ is the number of
%samples after matching.\footnote{We do not include $K$-factors in our
%analysis. %\TODO{Do we want to say this??}
%}
and
all samples in the bins are
added by using these weights.

%%%%%%%%%%%%%% FIGURE %%%%%%%%%%%%%%%%%%%%%%%%%%%%%%%%%%%%
\begin{figure}
  \centering
  \subcaptionbox{\label{fig:meff_valid_z} $Z$ boson production}{
  \includegraphics[width=0.48\columnwidth]{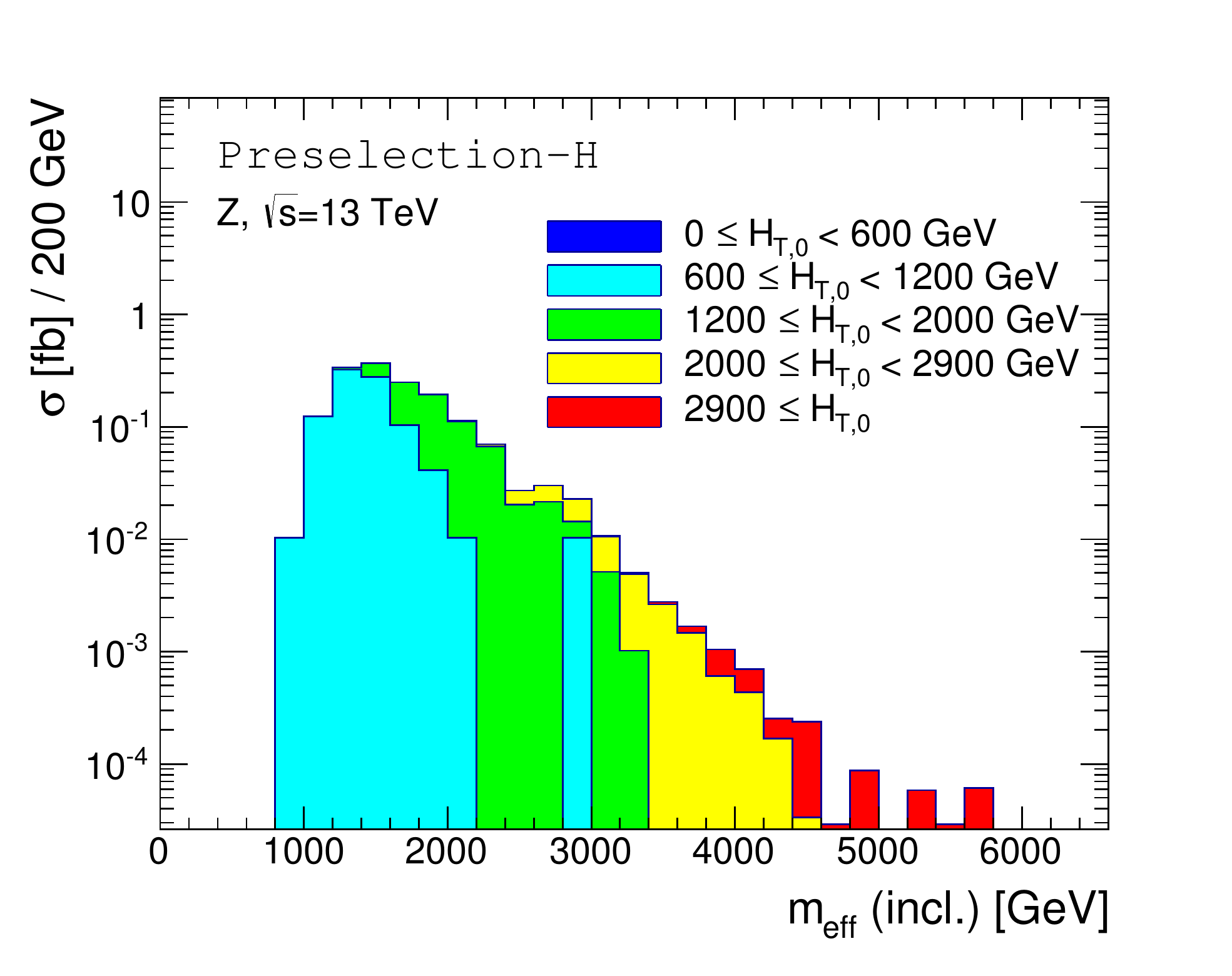}}
  \subcaptionbox{\label{fig:meff_valid_w} $W$ boson production}{
  \includegraphics[width=0.48\columnwidth]{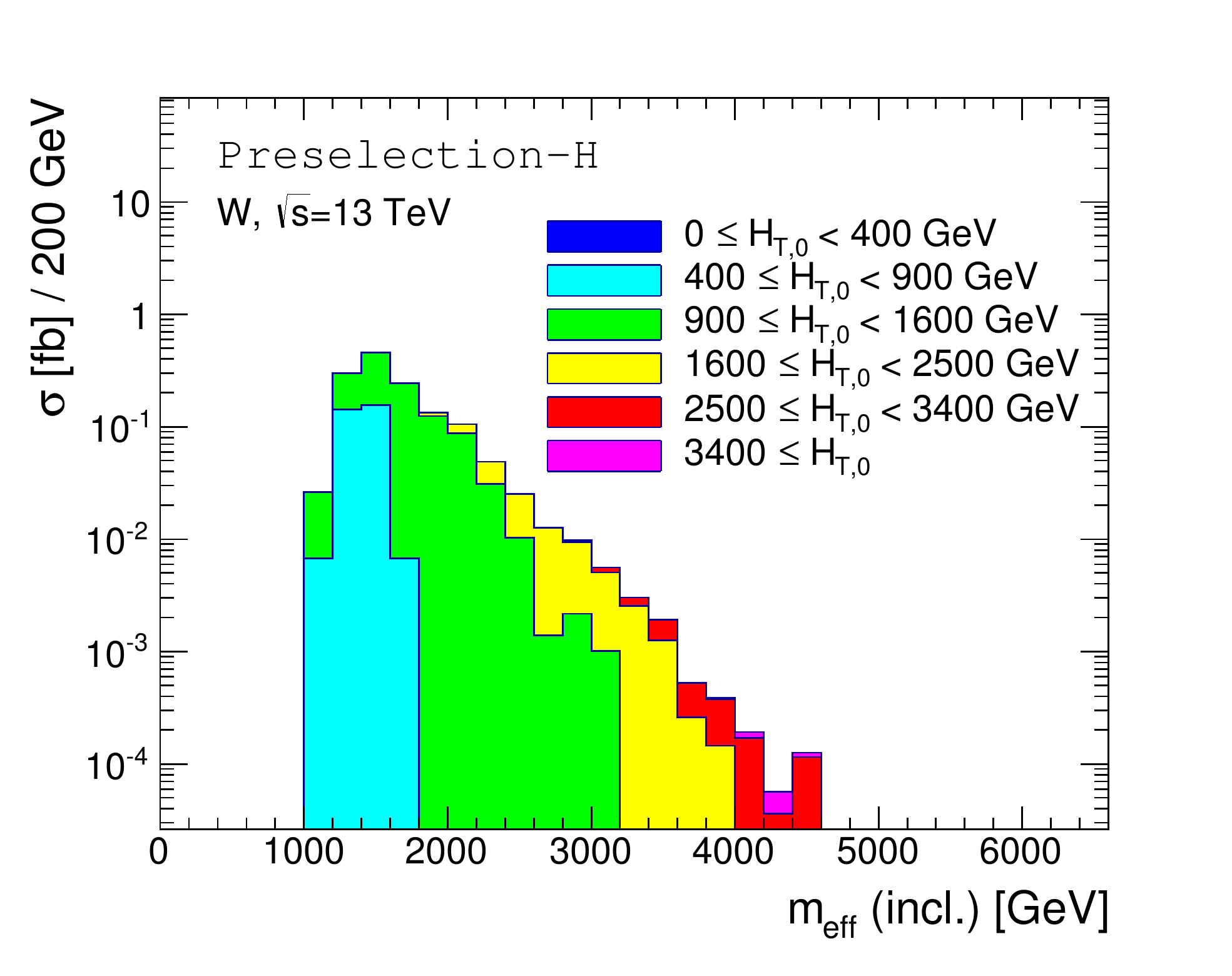}}
\caption{Distributions of $m_{\text{eff}} (\text{incl.})$ with {\tt
 preselection-H} required. Contributions from different $H_{T,0}$ bins
 are filled with different colors.} 
  \label{fig:meff_valid_wz}
\end{figure}
%%%%%%%%%%%%%%%%%%%%%%%%%%%%%%%%%%%%%%%%%%%%%%%%%%%%%%%%%%

%%%%%%%%%%%%%% FIGURE %%%%%%%%%%%%%%%%%%%%%%%%%%%%%%%%%%%%
\begin{figure}
  \centering
  \subcaptionbox{\label{fig:meff_valid_tt_semil} Semi-leptonic $t\bar{t}$ decay}{
  \includegraphics[width=0.48\columnwidth]{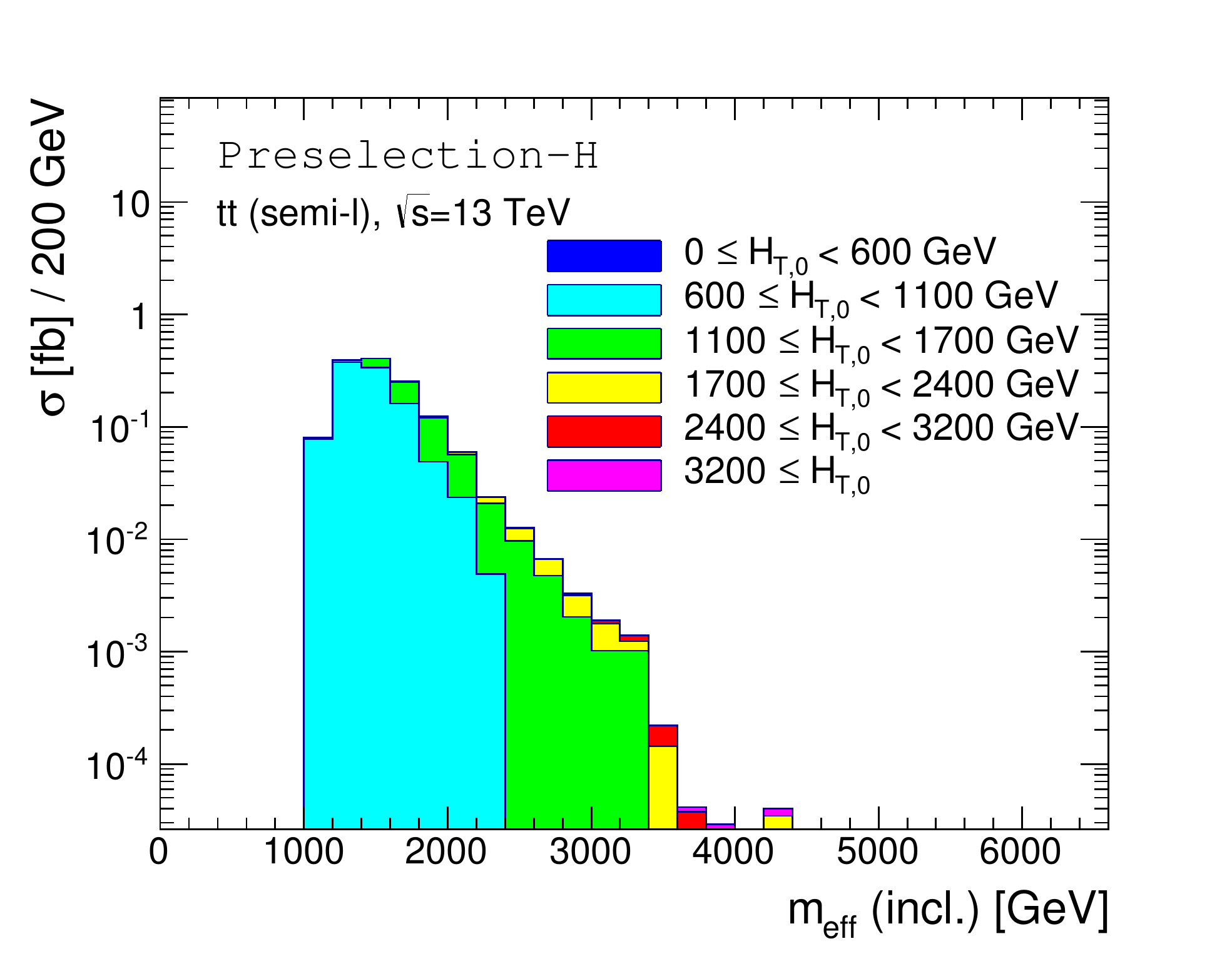}}
  \subcaptionbox{\label{fig:meff_valid_tt_ll} Leptonic $t\bar{t}$ decay}{
  \includegraphics[width=0.48\columnwidth]{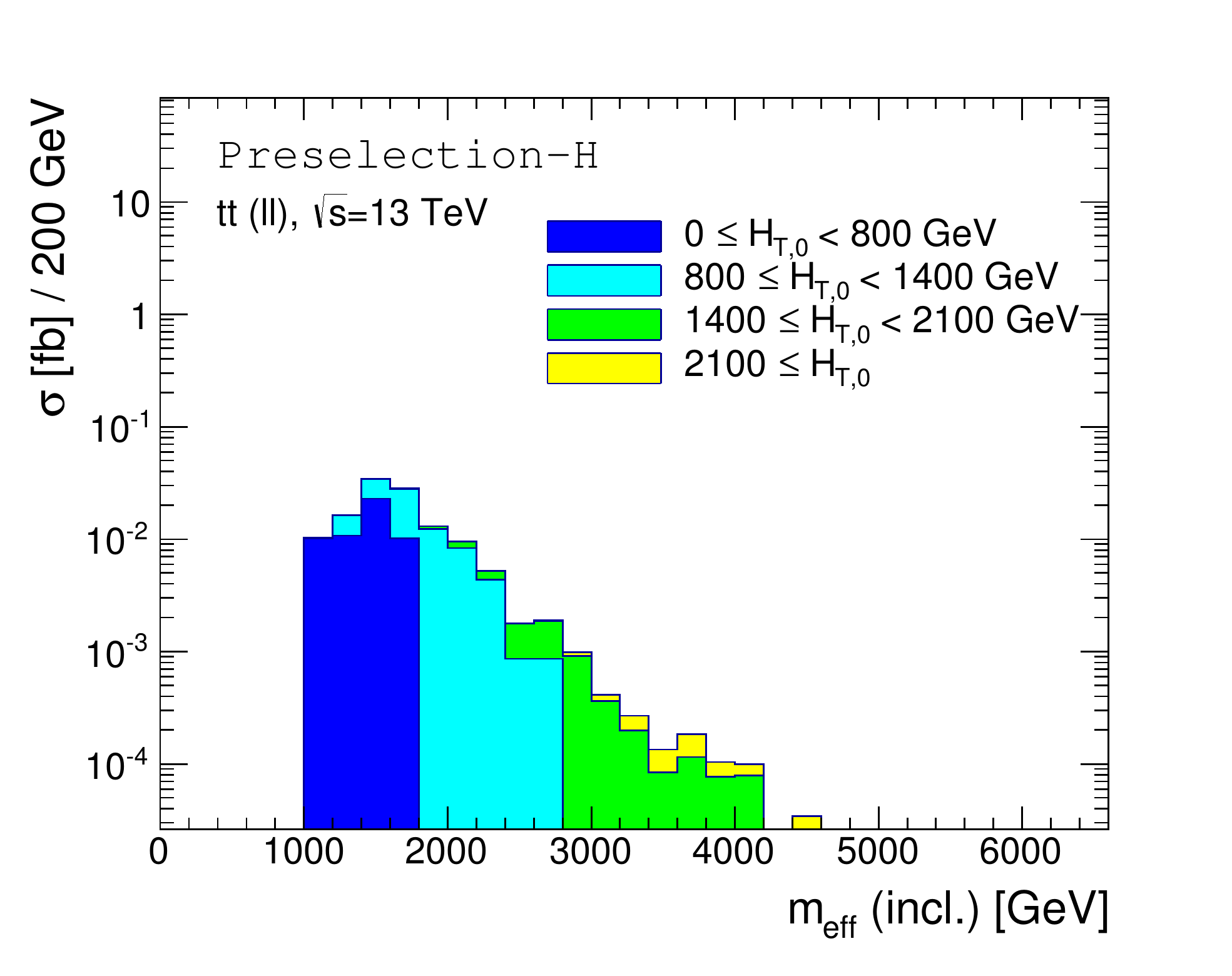}}
\caption{Same as in Fig.~\ref{fig:meff_valid_wz}, but for the $t\bar{t}$
 production process.} 
  \label{fig:meff_valid_tt}
\end{figure}
%%%%%%%%%%%%%%%%%%%%%%%%%%%%%%%%%%%%%%%%%%%%%%%%%%%%%%%%%%

By using the samples obtained above, we generate the distribution of
$m_{\text{eff}} (\text{incl.})$ for each process, where
$m_{\text{eff}} (\text{incl.})$ is defined as the scalar sum of the
missing transverse energy $E_{\rm T}^{\rm (miss)}$ and the transverse
momenta of all jets with $p_{\rm T}>50$~GeV; the results are given in
Figs.~\ref{fig:meff_valid_wz} and \ref{fig:meff_valid_tt}. These
distributions are obtained for events passing a selection requirement
{\tt Preselection-H}, whose definition is found in
Table~\ref{table:preselection} in the next section. We will use
$m_{\text{eff}} (\text{incl.})$ to define signal regions in the
following analyses. From these figures, we find that there is a %tight
correlation between $H_{T, 0}$ and $m_{\text{eff}} (\text{incl.})$. 
%\rem{TM: tight correlation?}
In
addition, we find a sizable number of events in the tails of the
distributions, which smoothly spread out from the bulk of the
distributions.  These observations indicate that the division of the
generator-level phase space in terms of $H_{T, 0}$ offers an adequate
way to estimate the number of events in signal regions which are
defined by the values of $m_{\text{eff}} (\text{incl.})$.

For the definition of the objects such as charged tracks, jets, and
charged leptons, we basically follow Ref.~\cite{Aaboud:2017vwy}. We
require $p_{\text{T}} > 1$~GeV and $|\eta| < 2.5$ for the reconstruction
of a charged track, and reject it if the production point of the
particle associated with the charged track is outside the innermost
pixel layer, which is located at $|\bm{r}_{\text{T}}| =33.25$~mm. These
requirements are intended to remove tracks with poor measurement
quality. Jets are clustered using {\tt  FastJet\,v3.1}~\cite{Cacciari:2011ma} 
with a jet radius parameter of $0.4$ and required to satisfy $p_{\rm T}
> 20$ GeV and $|\eta|<2.8$. For leptons, we require $p_{\rm T} > 7$~GeV,
and $|\eta|<2.47$ $(2.7)$ for electrons (muons). 

As we will mention in the next section, we require neither electrons nor
muons be reconstructed in our analysis. In this case, the SM background
processes listed above can contribute to the signal regions only if
all of the leptons in the processes fail to be reconstructed. To take
account of this, we introduce the reconstruction probability of an
electron (muon), $p^{e(\mu)}_{\text{reco.}} (p_{\text{T}}, \eta)$,
which we take from {\tt DELPHES\,v3}~\cite{deFavereau:2013fsa}. We then
multiply the weight $w_\alpha$ in Eq.~\eqref{eq:weight} by a factor of 
$\prod_{{\rm all}\, e_i} (1-p_{{\rm reco.}}^{e_i}) \times \prod_{{\rm
all}\, \mu_i} (1-p_{ {\rm reco.}}^{\mu_i})$ when we require the absence
of electrons and muons.

To validate our MC simulations, we compute the expected number of
background events as well as the expected $95$\% confidence level (CL)
exclusion limit on gluino mass for an integrated luminosity of
$36.1\,{\rm fb}^{-1}$ at the 13~TeV LHC. We use the kinematical
selection category called {\tt Meff-4j-3000} given in
Ref.~\cite{Aaboud:2017vwy}.\footnote{When we require a large
$m_{\text{eff}} (\text{incl.})$ as in {\tt Meff-4j-3000}, the diboson
production channel may become comparable to the $W$ and $t\bar{t}$
production channels \cite{Aaboud:2017vwy}, though this is still
subdominant compared with the $Z$-boson production process.} 
We then find the expected number of background events to be
$1.4$ and the exclusion limit on the gluino mass to be $1960$~GeV.
Comparing these results with the ones reported by the ATLAS
collaboration~\cite{Aaboud:2017vwy}---$2.0$ and $2030$~GeV,
respectively---we can safely conclude that our MC simulation
satisfactorily reproduces the ATLAS results. 
%\TODO{Check the ATLAS
%values. Also, shouldn't we use the values given in
%Ref.~\cite{Aaboud:2017vwy}, rather than those in
%Ref.~\cite{ATLAS-CONF-2017-022}?}

%%%%%%%%%%%%%%%%%%%%%%%%%%%%%%%%%%%%%%%%
\subsection{Event selection criteria}
\label{sec:eventsel}
%%%%%%%%%%%%%%%%%%%%%%%%%%%%%%%%%%%%%%%%

%%%%%%%%%%%%%%%%%%%%%%%%%%%%%%%%%%%%%%%%%%%%%%%%%%%%%%%%%%%%%%%%%%%%%

\begin{table}
  \begin{center}
    \begin{tabular}{c|ccc}
      \hline\hline
      Requirement & {\tt L} & {\tt M} & {\tt H} \\ \hline
      Number of jets with $p_{\rm T}>50$ GeV & 
       \multicolumn{3}{c}{$\ge4$} \\ \hline
      $E_{\rm T}^{\rm (miss)}$ [GeV] &
      \multicolumn{3}{c}{$>250$} \\ \hline
      $p_{\rm T} (j_1)$ [GeV] &
       \multicolumn{3}{c}{$>200$} \\ \hline 
      $p_{\rm T} (j_4)$ [GeV] & 
      $>100$ & $>100$ & $>150$ \\ \hline
      % \multicolumn{2}{c}{$>100$} & $>150$ \\ \hline
      $|\eta(j_{i=1,2,3,4})|$ &
      $<1.2$ &$<2.0$ & $<2.0$ \\ \hline
      %$<1.2$ & \multicolumn{2}{c}{$<2.0$}  \\ \hline
      $\Delta\phi (j_{p_{\rm T}>50\,{\rm GeV}},E_{\rm T}^{\rm (miss)})_{\rm min}$ &
      \multicolumn{3}{c}{$>0.4$}  \\ \hline
      $E_{\rm T}^{\rm (miss)}/m_{\rm eff}(4)$ &
      $>0.3$ & $>0.25$ & $>0.2$ \\ \hline
      Aplanarity &
      \multicolumn{3}{c}{$>0.04$}
      \\ \hline\hline
    \end{tabular}
    \caption{ 
    Definition of {\tt preselection-L, M, H}.
    Besides these criteria, we require that neither electrons nor muons
   be reconstructed. 
    }
    \label{table:preselection}
  \end{center}
  \end{table}
%%%%%%%%%%%%%%%%%%%%%%%%%%%%%%%%%%%%%%%%%%%%%%%%%%%%%%%%%%%%%%%%%%%%%%%%

%%%%%%%%%%%%%% FIGURE %%%%%%%%%%%%%%%%%%%%%%%%%%%%%%%%%%%%
\begin{figure}
  \centering
  \subcaptionbox{\label{fig:meff_L} \texttt{Preselection-L}}{
  \includegraphics[width=0.48\columnwidth]{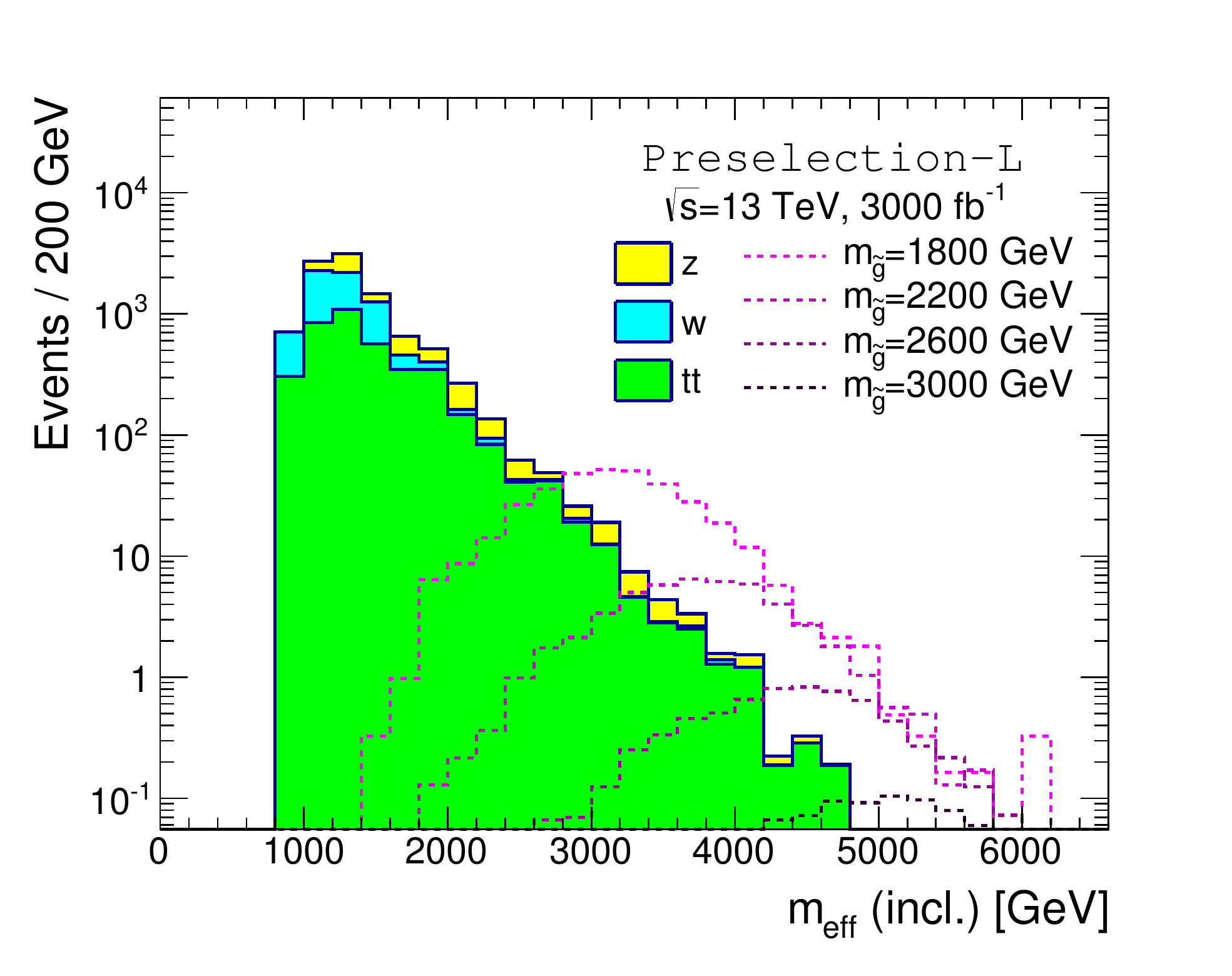}}
  \subcaptionbox{\label{fig:meff_M} \texttt{Preselection-M}}{
  \includegraphics[width=0.48\columnwidth]{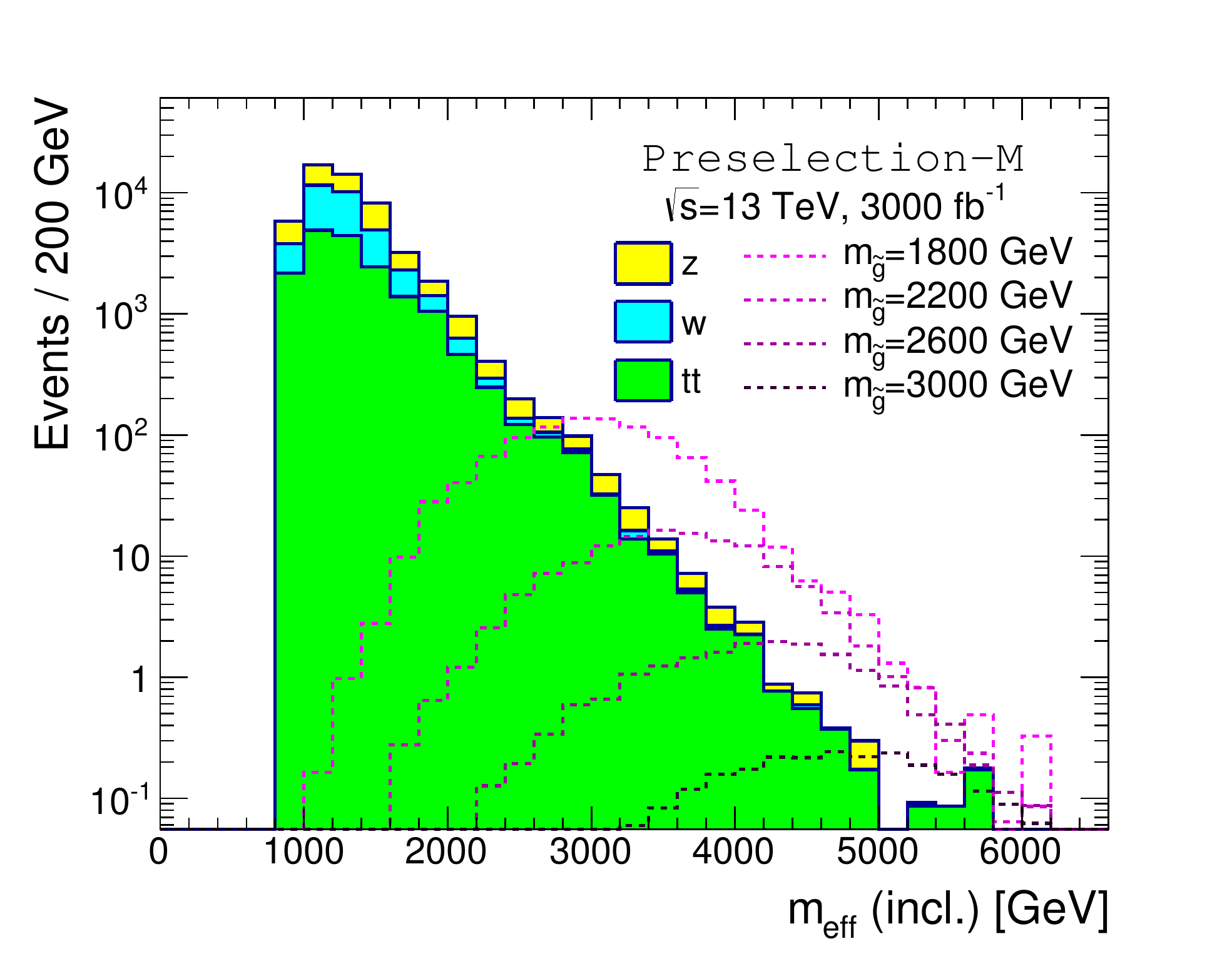}}
  \subcaptionbox{\label{fig:meff_H} \texttt{Preselection-H}}{
  \includegraphics[width=0.48\columnwidth]{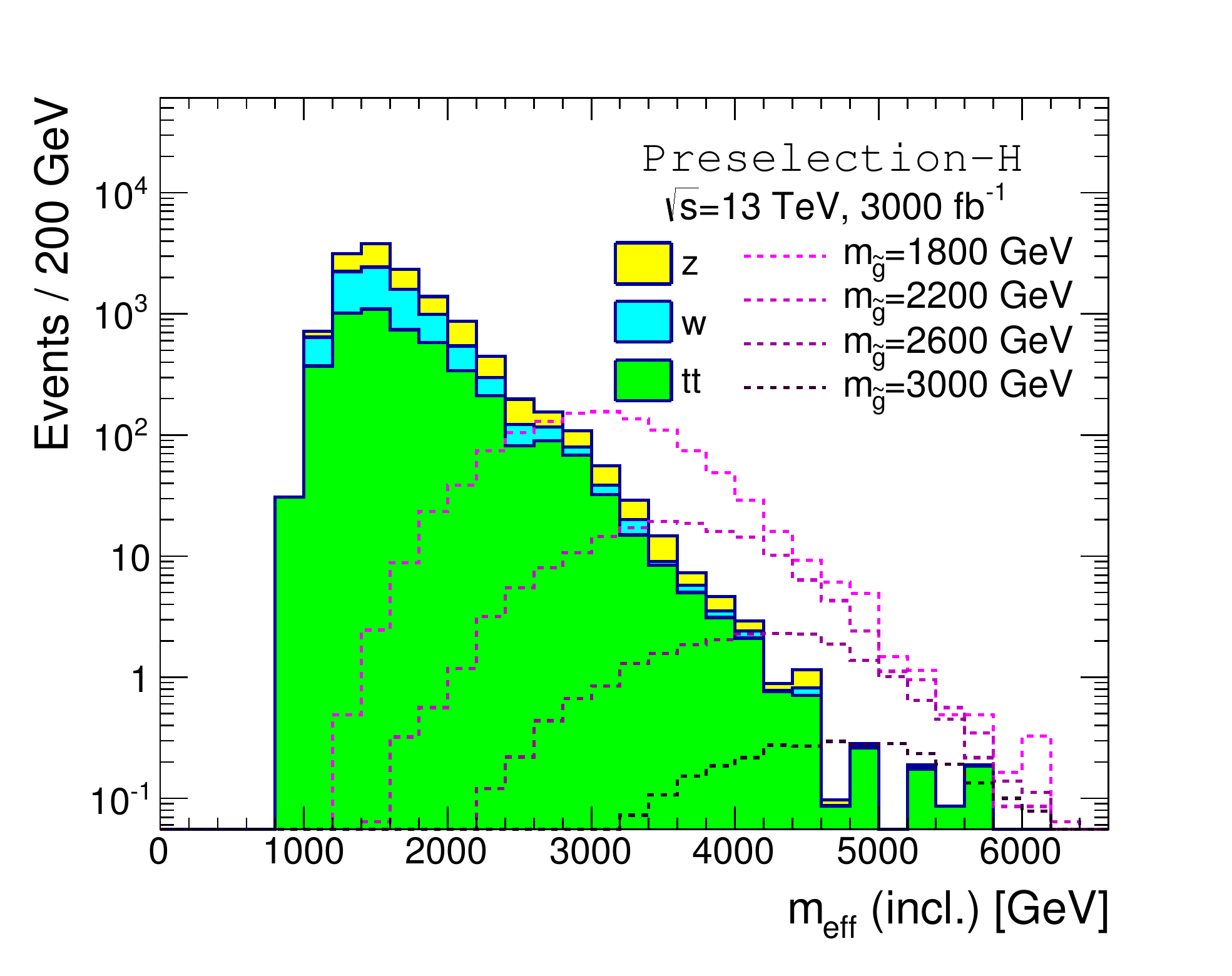}}
\caption{Distributions of $m_{\text{eff}} (\text{incl.})$ for the SM
 background and gluino events. We set the LSP mass to be 100~GeV. } 
  \label{fig:meff_LMH}
\end{figure}
%%%%%%%%%%%%%%%%%%%%%%%%%%%%%%%%%%%%%%%%%%%%%%%%%%%%%%%%%%

In our analysis, we impose a new event-selection cut using the
information of displaced vertices in addition to the ordinary
selection criteria based on kinematics. We further try to optimize the
kinematic-based cuts as the new selection cut is expected to reduce
the SM background efficiently. To that end, we first divide the
kinematic-based selection criteria into two classes; $m_{\text{eff}}
(\text{incl.})$, which is to be varied to optimize the event
selection, and the other criteria adopted in
Ref.~\cite{Aaboud:2017vwy}, which we call the preselection. Moreover,
we divide the preselection into three classes, {\tt preselection-L},
{\tt -M}, and {\tt -H}, and use one of them for each sample point so
that the sensitivity is maximized. We summarize the preselection
criteria in Table~\ref{table:preselection}. Here, $p_{\rm T} (j_i)$
denotes the transverse momentum of $i$-th jet, $\Delta\phi$ is the
azimuthal angle between the jet and the missing energy, and $m_{\rm
  eff}(4)$ is the scalar sum of $E_{\rm T}^{\rm (miss)}$ and the
transverse momenta of the leading $4$-jets. The requirements on
$\Delta\phi$ and $E_{\rm T}^{\rm (miss)}/m_{\rm eff}(4)$ were imposed
to reduce the contributions from QCD multi-jet processes in
Ref.~\cite{Aaboud:2017vwy}---although such multi-jet processes are not
included in the present analysis, we still impose these requirements
since they are also able to reject background events from the
semi-leptonically decaying $t$ and $\bar{t}$ efficiently. The
definition of aplanarity can be found in Refs.~\cite{Bjorken:1969wi,
  Chen:2011ah}; if a distribution of jets is highly directional, then
the value of aplanarity gets close to zero, while if it is completely
isotropic then this is equal to $1/2$. In addition to these
kinematical selection criteria, we require that neither electrons nor
muons be reconstructed.  In Fig.~\ref{fig:meff_LMH}, we give the
distributions of $m_{\text{eff}} (\text{incl.})$ for the SM background
and gluino events with each preselection imposed. Here, we set the LSP
mass to be 100~GeV. These figures show that we may effectively select
gluino events while eliminating the SM background if we require a
large $m_{\text{eff}} (\text{incl.})$. The size of this cut will be
chosen so that the sensitivity is maximized, as we see in the
subsequent section.

Now we discuss a selection criterion based on the reconstruction of
displaced vertices. As described in Sec.~\ref{sec:vertex}, we
reconstruct vertices using charged tracks associated with the decay
product. The resolutions of impact parameters, which determine the
resolution of vertex reconstruction, are obtained by fitting
Eq.~\eqref{eq:sigmax} onto the experimental results given by the ATLAS
collaboration \cite{ATLAS:trk,ATL-PHYS-PUB-2015-018}. 
As we mentioned in Sec.~\ref{sec:mc},
only the tracks with $p_{\text{T}} > 1$~GeV are used in this analysis.
Performing the fit in this range, we determine the parameters in
Eq.~\eqref{eq:sigmax} as $\sigma_{d_0} (\infty) = 23~\mu\text{m}$,
$\sigma_{z_0 \sin \theta} (\infty) = 78~\mu\text{m}$, $p_{d_0}
=3.1$~GeV, and $p_{z_0 \sin\theta} =1.6$~GeV.

As we discussed in Sec.~\ref{sec:vertex}, we use four
high-$p_{\text{T}}$ jets to reconstruct displaced vertices generated
in gluino-decay events. For this purpose, we basically choose the
four-highest $p_{\text{T}}$ jets. We further require that the tracks
in these jets satisfy $d_0 < 10$~mm and $|z_0| < 320$~mm in order to
focus on decays that occur inside the inner detector. However, if one
of these four jets contains no track which satisfies the above
conditions, then we add the fifth-highest $p_{\rm T}$ jet to the
vertex reconstruction analysis. 
%If more than one jets among these five
%jets do not offer any tracks which meet the track criteria, 
%then we conclude that the reconstruction of vertices fails in such an event.
If the number of jets which contain tracks satisfying the criteria 
is smaller than four among these five jets,
then we conclude that the reconstruction of vertices fails in such an event.

As our main focus is on displaced vertices located inside the
innermost pixel layer, vertices generated by hadronic interactions in
the detector materials rarely contribute to the background. To assure
this, we reject events in which a vertex is reconstructed inside the
detector materials: {\it i.e.}, $22~{\rm mm}\le |(\bm{r}_{\rm
  DV1,2})_{\rm T}| \le 25$~mm, $29~{\rm mm}\le |(\bm{r}_{\rm
  DV1,2})_{\rm T}| \le 38$~mm, $46~{\rm mm} \le |(\bm{r}_{\rm
  DV1,2})_{\rm T}| \le 73$~mm, $84~{\rm mm} \le |(\bm{r}_{\rm
  DV1,2})_{\rm T}| \le 111$~mm, or $|( \bm{r}_{\rm DV1,2})_{\rm T}|\ge
120~{\rm mm}$ \cite{PERF-2007-01, Capeans:1291633,
  1748-0221-9-02-C02018, 1748-0221-11-11-P11020}. In the absence of
this possibility, the displaced vertices from the SM processes are
mainly caused by the misinterpretation of non-displaced vertices due
to the limited resolution of track impact parameters, which we take
into account in a manner discussed in Sec.~\ref{sec:vertex}.

In the metastable gluino production processes under consideration, we do
not expect that particles with hard momenta are generated at the primary
interaction point (except those from initial state radiation), as the
metastable gluinos decay after they flew away from the primary
interaction point. For this reason, we do not try to
determine the position of the primary interaction point in each
event.\footnote{We however note that it may be possible to reconstruct
the primary vertex by using the initial state radiation and/or soft
products generated by remnants of the $pp$ collision. Implications of
this possibility will be discussed in Sec.~\ref{sec:conclusion}. }
Instead, we use the distance between the two reconstructed vertices as a
discriminator. As for the definition of the distance, we try the
following three candidates and adopt the one which yields the best
sensitivity for each sample point: $|\bm{r}_{\rm DV1}-\bm{r}_{\rm
DV2}|$, $|(\bm{r}_{\rm DV1}-\bm{r}_{\rm DV2})_{\rm T}|$, and
$|(\bm{r}_{\rm DV1}-\bm{r}_{\rm DV2})_z|$, where $\bm{r}_{\rm DV1,2}$
are the position vectors of the displaced vertices defined in
Sec.~\ref{sec:vertex}, and T and $z$ stand for the transverse and $z$
directions, respectively.

%%%%%%%%%%%%%% FIGURE %%%%%%%%%%%%%%%%%%%%%%%%%%%%%%%%%%%%
\begin{figure}
  \centering
  \subcaptionbox{\label{fig:rDV_pdf} $|\bm{r}_{\rm DV1} -\bm{r}_{\rm
 DV2}|$ distribution}{
  \includegraphics[width=0.48\columnwidth]{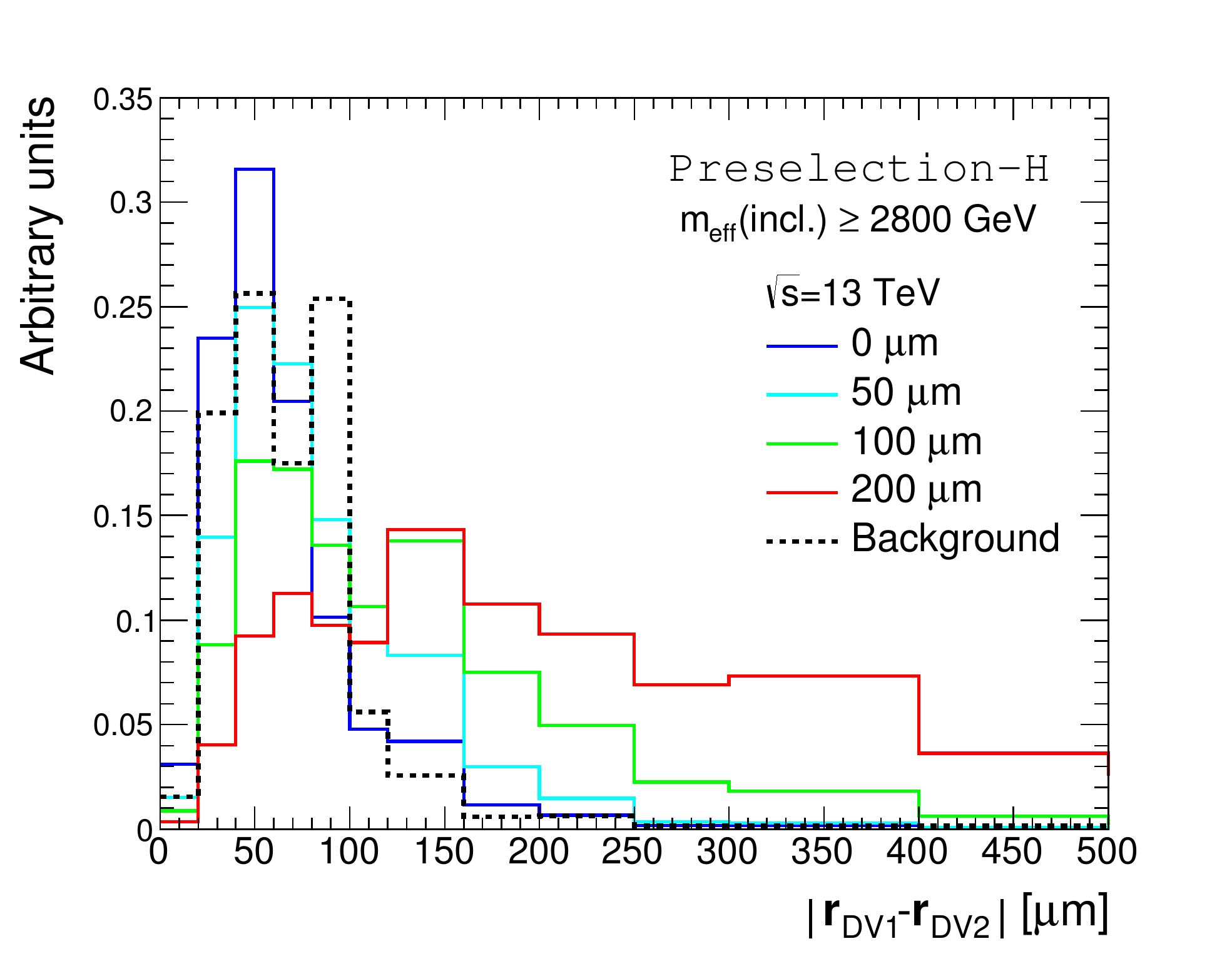}}
  \subcaptionbox{\label{fig:rDV_cdf} Efficiency}{
  \includegraphics[width=0.48\columnwidth]{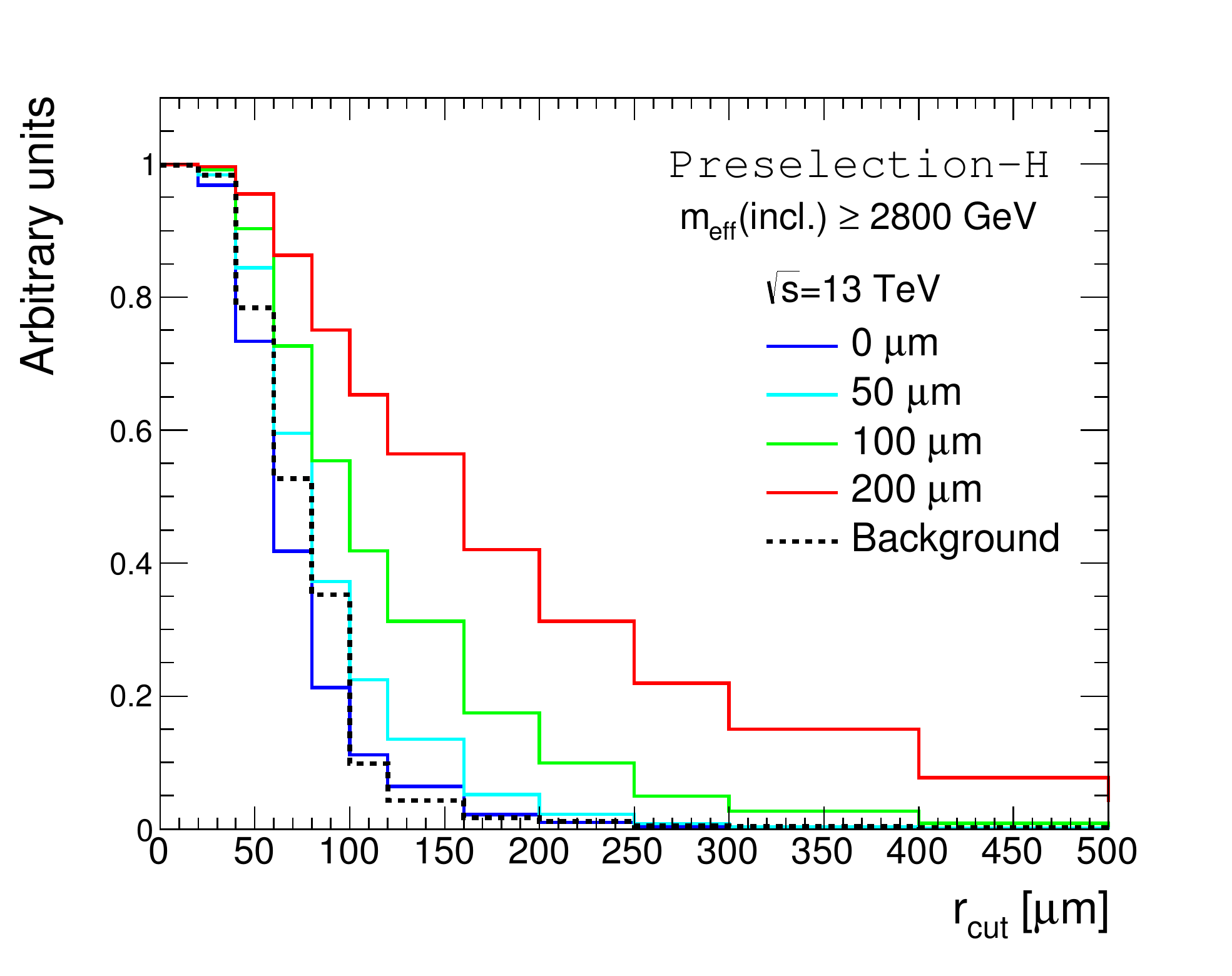}}
\caption{The distributions of $|\bm{r}_{\rm DV1} - \bm{r}_{\rm DV2}|$
 and the fractions to pass the selection cut $|\bm{r}_{\rm DV1} -
 \bm{r}_{\rm DV2}| > r_{\rm cut}$ for a gluino with
 $m_{\tilde{g}}=3$~TeV and different values of
 $c\tau_{\tilde g}$, shown in the solid lines. The distributions for the
 SM background events are also shown in the dotted lines. We have
 imposed {\tt preselection-H} and $m_{\rm eff}({\rm incl.}) > 2800
 \,\,{\rm GeV}$. }  
  \label{fig:rDV}
\end{figure}
%%%%%%%%%%%%%%%%%%%%%%%%%%%%%%%%%%%%%%%%%%%%%%%%%%%%%%%%%%

In Fig.~\ref{fig:rDV_pdf}, we show the $|\bm{r}_{\rm DV1} -
\bm{r}_{\rm DV2}|$ distribution of signal events 
for a gluino with $m_{\tilde{g}}=3$~TeV and different values of the
decay distance $c\tau_{\tilde g}$ in the solid lines.
The distribution for the SM background events is also shown in the
dotted line. Figure~\ref{fig:rDV_cdf} shows a fraction of events
passing a selection cut of $|\bm{r}_{\rm DV1} - \bm{r}_{\rm DV2}| >
r_{\rm cut}$ as a function of $r_{\rm cut}$. 
For these simulations, we have imposed {\tt preselection-H} and
$m_{\rm eff}({\rm incl.}) > 2800 \,\,{\rm GeV}$.
Notice that the distribution of the SM background deviates from the signal
distribution with $c\tau_{\tilde g} = 0$. This is because the flavor
content of jets in the SM background sample is different from that in
the signal event sample. In the signal events, gluinos are forced to
decay into only the first-generation quarks as we mentioned above.  
In the SM background events, on the other hand, jets may also be 
induced by heavy flavor quarks, which then contain metastable hadrons; 
for example, the typical decay length of $B$ mesons is about $400\,{\rm
\mu m}$ and these mesons can fly over a few mm when they are highly boosted.
Decay products of such metastable hadrons may form a secondary vertex
and deteriorate resolution of the vertex reconstruction position.
We however note that our vertex reconstruction method is actually robust
against the presence of metastable hadrons. A jet originating from
a heavy flavor quark contains not only metastable hadrons such as $B$ mesons
but also other many hadrons emitted during hadronization.
Since our vertex reconstruction algorithm chooses as the vertex position
a point at which tracks are most densely concentrated, 
it is less likely to identify a secondary vertex caused by a metastable hadron
as the reconstructed vertex.
For the same reason, our vertex reconstruction method is less
affected by pile-up events as well, especially if we reconstruct
vertices using only tracks in high-$p_{\rm T}$ jets as we do in our
analysis. The effect of pile-up events on kinematical
selection cuts is also expected to be
small~\cite{Cohen:2013xda}. Considering these, we do not include the
pile-up effects in our analysis.

The plots in Fig.~\ref{fig:rDV} show that if we set $r_{\rm cut}$ to
be $\gtrsim 100~\mu{\rm m}$, then a significant fraction of the SM
background fails to pass the selection cut while a sizable number of
signal events for $c\tau_{\tilde{g}} \gtrsim 100~\mu{\rm m}$ still
remain after the selection cut. This observation indicates that
this cut may be useful to probe a gluino with a decay length of
$c\tau_{\tilde{g}} \gtrsim 100~\mu{\rm m}$, which we demonstrate in the
subsequent section.

%%%%%%%%%%%%%%%%%%%%%%%%%%%%%%%%%
\subsection{Prospects}
\label{sec:prospects}
%%%%%%%%%%%%%%%%%%%%%%%%%%

%-------------------------------
\begin{table}
  \begin{center}
    \begin{tabular}{c|c}
    \hline\hline
     	{\bf Selection} & {\bf Requirements} \\ \hline\hline
 	{ Preselection} & $\in$ ({\tt preslection-L, M, H}) \\ %\hline
	{ Lepton veto} & No reconstructed electrons and muons \\ \hline
	\multirow{3}{*}{Material veto} & No vertices reconstructed in material regions:\\
	& {\it i.e.}  (in units of mm) \\ 
	& $|(\bm{r}_{\rm DV})_{\rm T}|\notin$ $(22,25)$, $(29,38)$, $(46,73)$, $(84,111)$, $(120,\infty)$  \\
	\hline
	\multirow{3}{*}{$m_{\rm eff}({\rm incl.})$ cut} & 
	Require $m_{\rm eff}({\rm incl.}) > \left(m_{\rm effcut}\right)_{\rm optimal}$ \\ 
	& over\\
	& $m_{\rm effcut} \in (1000~{\rm GeV}, 10^4~{\rm GeV})$ \\ \hline
	\multirow{4}{*}{$\Delta r_{\rm DV}$ cut} & Require $\Delta r_{\rm DV} > \left(r_{\rm cut}\right)_{\rm optimal}$ \\ 
	& over\\
	& $\Delta r_{\rm DV}\in(
		\left| \Delta \bm{r}_{\rm DV} \right|,
		\left| \left(\Delta \bm{r}_{\rm DV}\right)_{\rm T} \right|,
		\left| \left(\Delta \bm{r}_{\rm DV}\right)_{\rm z} \right|
	)$ \\
	& $r_{\rm cut} \in (0, 2\times10^5 {\rm \mu m})$
	\\ \hline\hline
    \end{tabular}
    \caption{Summary of event selection criteria.
    }
    \label{table:selections}
  \end{center}
  \end{table}
%-------------------------------

Let us summarize the event selection criteria we use in the following
analysis. After we apply one of the preselections, {\tt
preselection-L,-M,-H}, we further require 
\begin{align}
m_{\rm eff}({\rm incl.}) >  \left(m_{\rm effcut}\right)_{\rm optimal}
 \quad \text{and} \quad
\Delta r_{\rm DV} > \left(r_{\rm cut}\right)_{\rm optimal} 
\label{eq:cutpara}
\end{align}
where $\Delta r_{\rm DV}$ is one of the discriminators proposed in the
previous section, {\it i.e.}, $\Delta r_{\rm DV} = | \Delta
\bm{r}_{\rm DV}|$, $|(\Delta \bm{r}_{\rm DV})_\text{T}|$, or $|(\Delta
\bm{r}_{\rm DV})_{z}|$, where $\Delta \bm{r}_{\rm DV} \equiv
\bm{r}_{\rm DV1}-\bm{r}_{\rm DV2}$.  We adopt the one which leads to
the best sensitivity for a given sample point. We vary the cut
parameters $m_{\rm effcut}$ and $r_{\rm cut}$ in
Eq.~\eqref{eq:cutpara} from $1000$~GeV to $10^4$~GeV and from
$0~\mu\text{m}$ to $2\times10^5~{\rm \mu m}$, respectively, and employ
the values which maximize the performance. These event selection
criteria are listed in Table~\ref{table:selections}.

Now we study the performance of the new selection cut based on the
reconstruction of displaced vertices. To that end, we evaluate the
discovery reach and exclusion limit for metastable gluino searches and
compare them with the existing results from the prompt decay searches. 
For exclusion limits, we compute the expected 95\% confidence level
(CL) limits on the gluino mass using the $CL_{s}$ prescription
\cite{Read:2002hq, Junk:1999kv}. For the discovery reach, we compute
the expected significance of discovery $Z_0$ \cite{Cowan:2010js}:
\begin{align}
Z_0 = \sqrt{2\left\{ \left(S + B\right) \log \left( 1 + S/B \right) -S
 \right\} }~,
 \label{eq:significance}  
\end{align}
where $S$ and $B$ are the expected numbers of signal and background
events, respectively. We then require both $Z_0$ and $S$ to be larger
than $5$ for the discovery.

%%%%%%%%%%%%%% FIGURE %%%%%%%%%%%%%%%%%%%%%%%%%%%%%%%%%%%%
\begin{figure}
  \centering
  \subcaptionbox{\label{fig:optimal_0mu} $c\tau_{\tilde{g}}=0~{\rm \mu m}$}{
  \includegraphics[width=0.48\columnwidth]{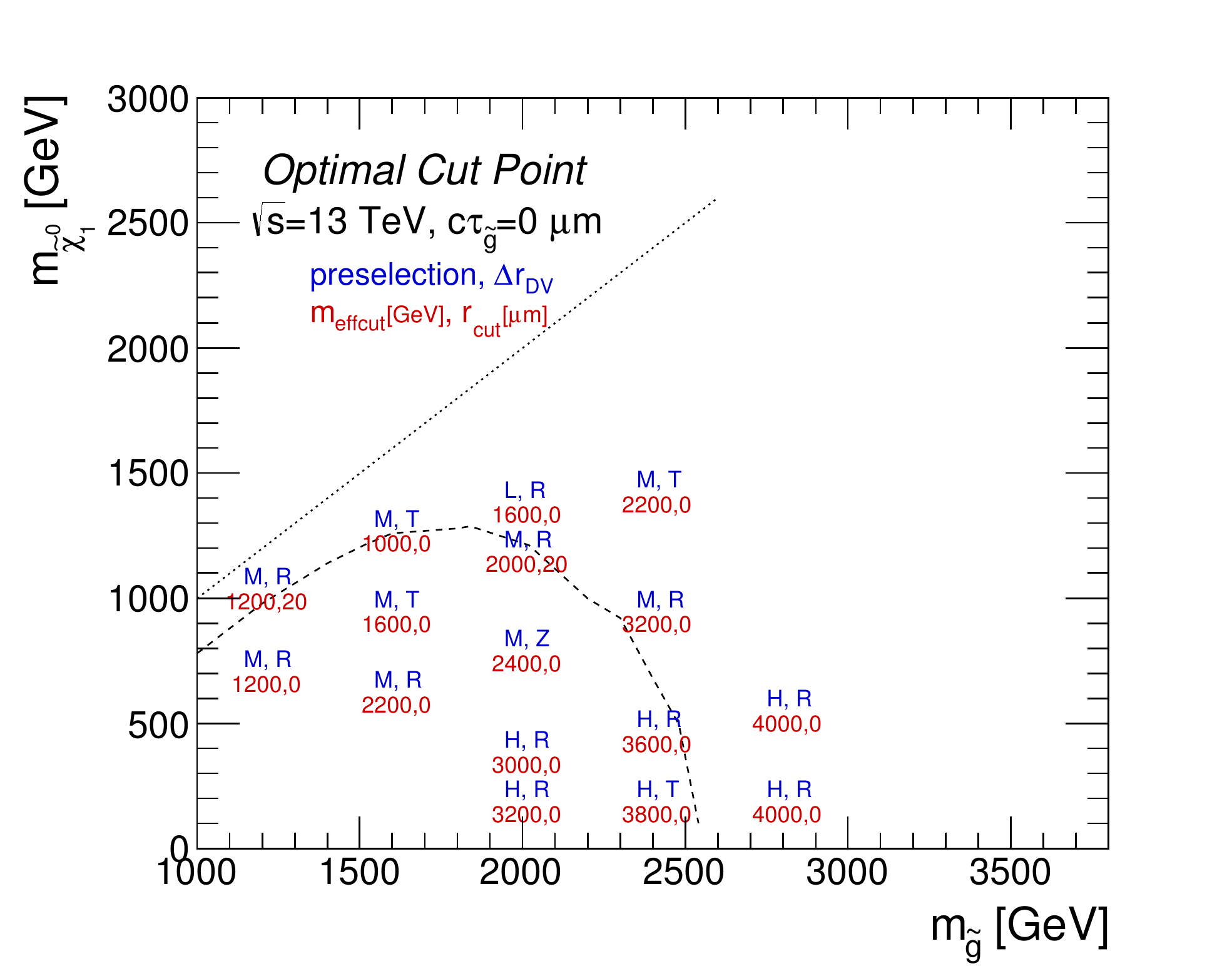}}
  \subcaptionbox{\label{fig:optimal_100mu} $c\tau_{\tilde{g}}=100~{\rm \mu m}$}{
  \includegraphics[width=0.48\columnwidth]{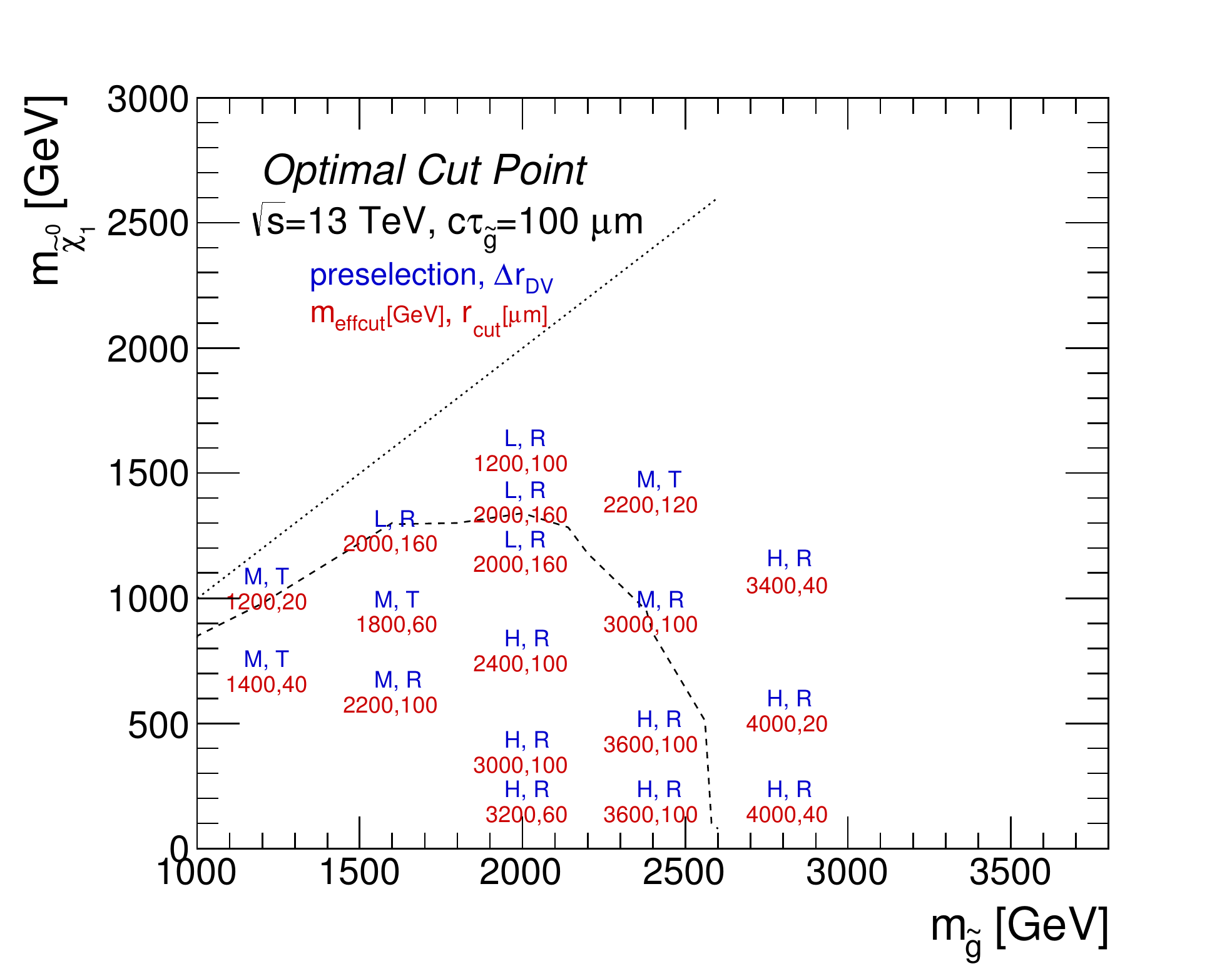}}
  \subcaptionbox{\label{fig:optimal_200mu} $c\tau_{\tilde{g}}=200~{\rm \mu m}$}{
  \includegraphics[width=0.48\columnwidth]{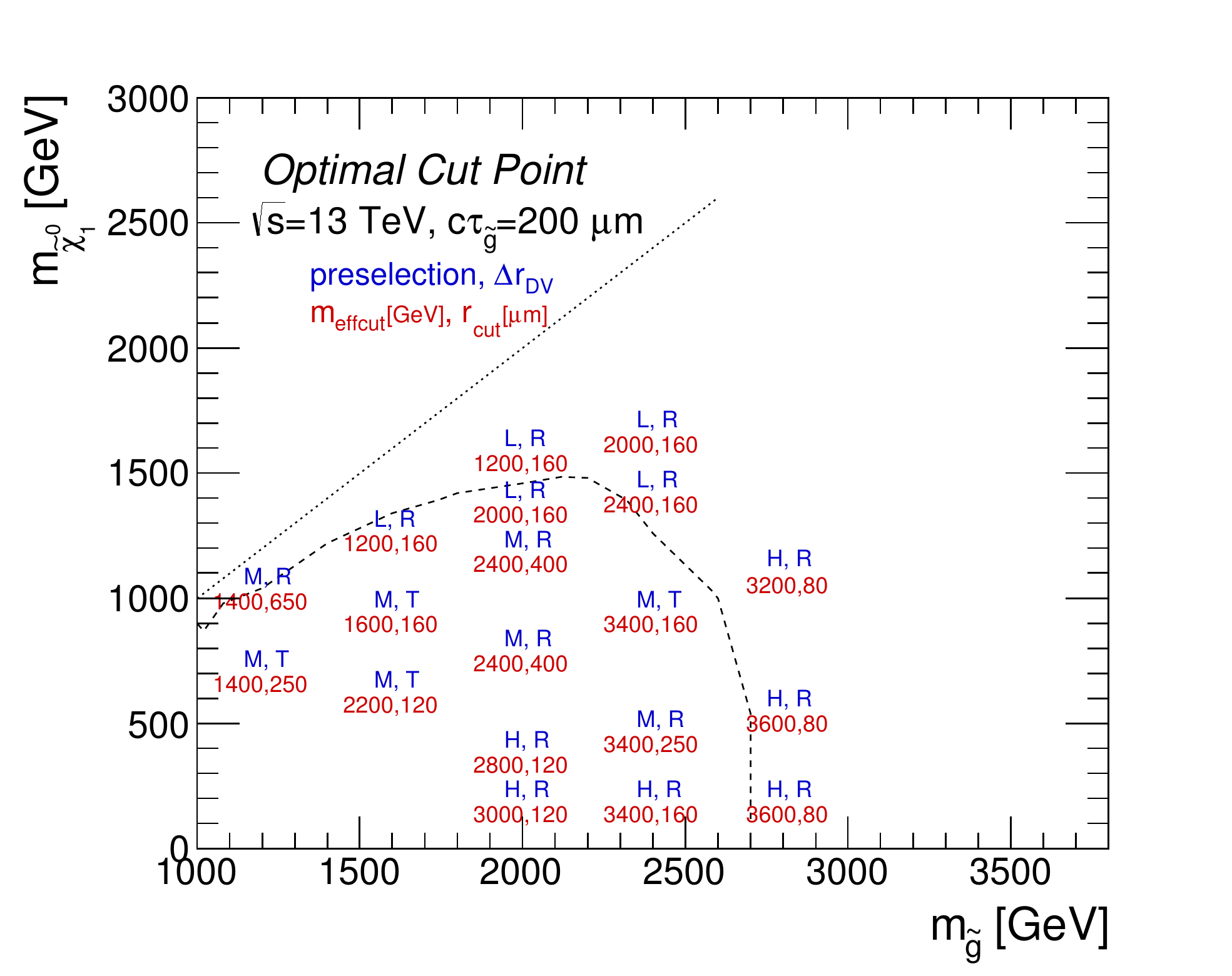}}
  \subcaptionbox{\label{fig:optimal_500mu} $c\tau_{\tilde{g}}=500~{\rm \mu m}$}{
  \includegraphics[width=0.48\columnwidth]{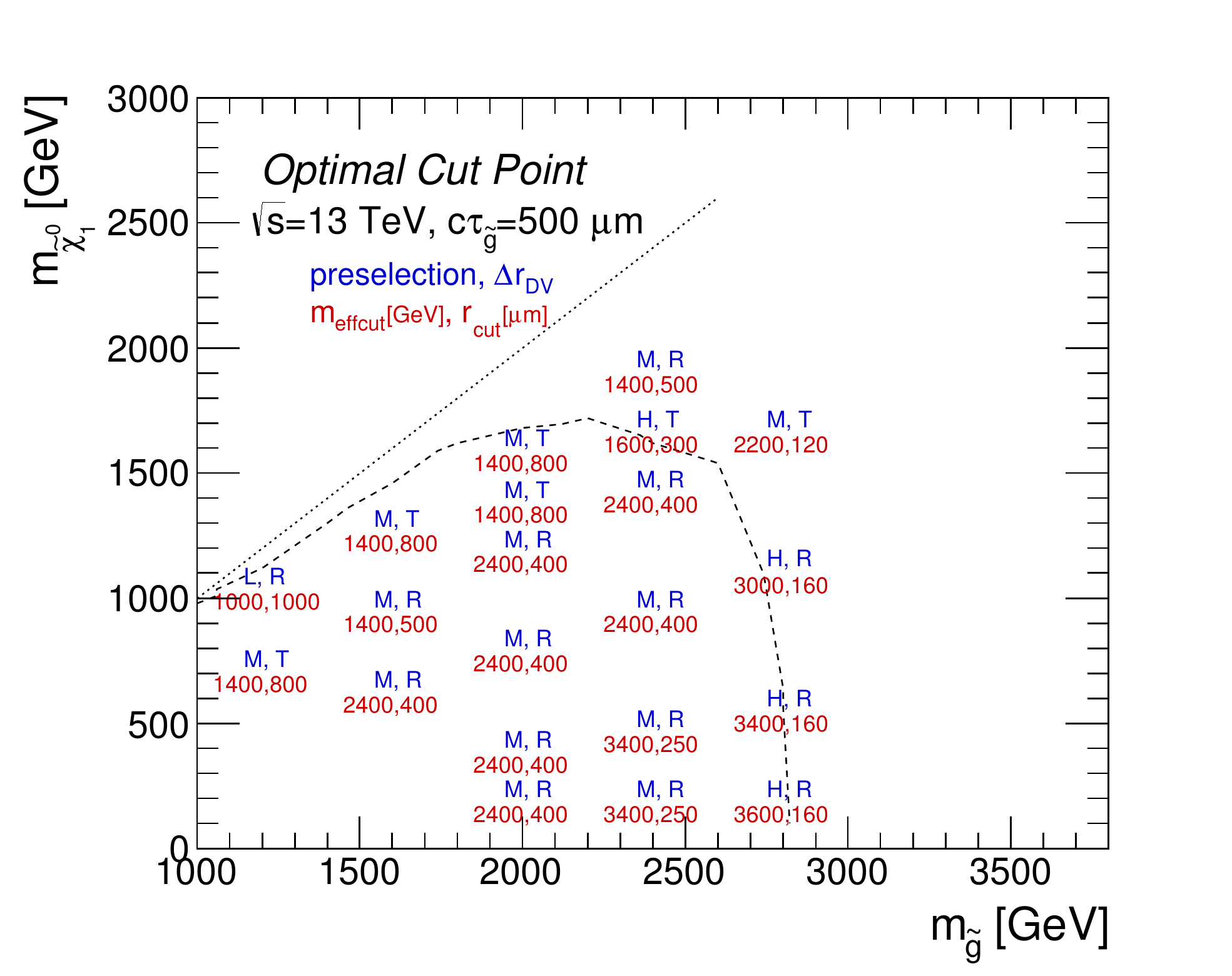}}
  \subcaptionbox{\label{fig:optimal_1000mu} $c\tau_{\tilde{g}}=1~{\rm mm}$}{
  \includegraphics[width=0.48\columnwidth]{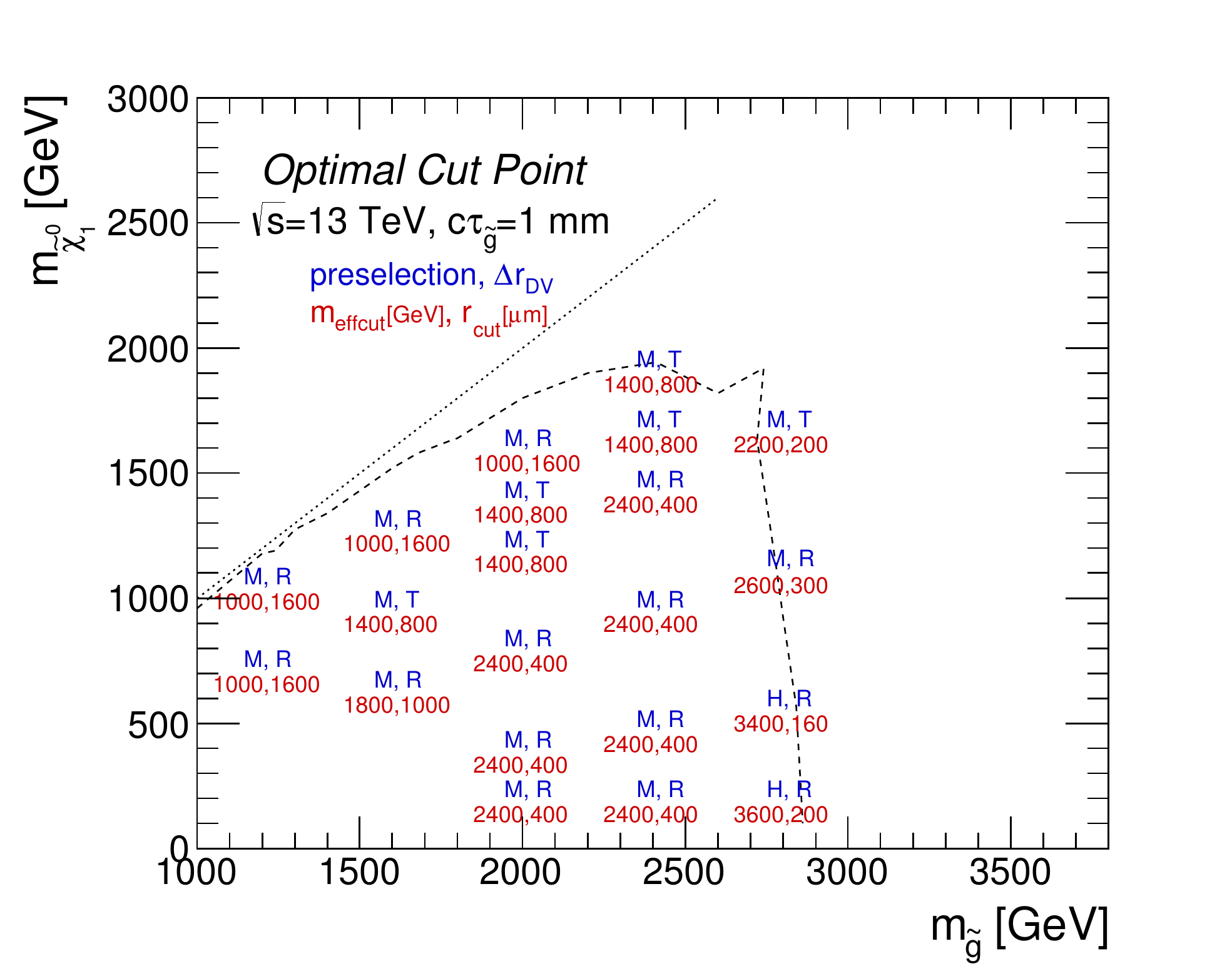}}
  \subcaptionbox{\label{fig:optimal_10000mu} $c\tau_{\tilde{g}}=10~{\rm mm}$}{
  \includegraphics[width=0.48\columnwidth]{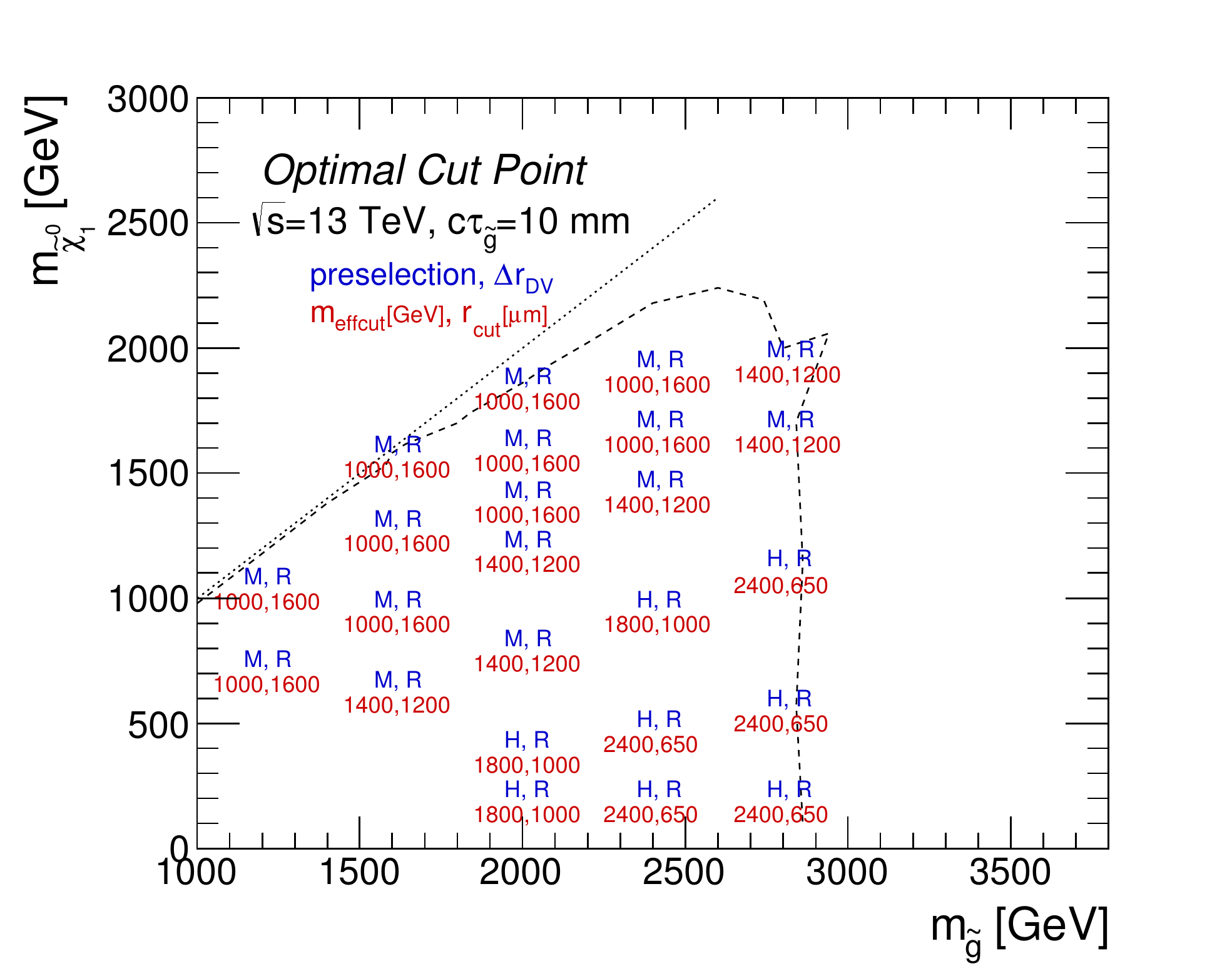}}
\caption{Optimal cut values ($(m_{\rm effcut})_{\rm optimal}$,
  $(r_{\rm cut})_{\rm optimal}$), the preselection ({\tt L}, {\tt M},
  {\tt H}), and the discriminator $\Delta r_{\rm DV}$ ({\tt R}, {\tt
    T}, {\tt Z}) for each sample point with various
  $c\tau_{\tilde{g}}$, with an integrated luminosity of $3000~{\rm
    fb}^{-1}$.  The expected $5\sigma$ discovery reaches for gluinos
  are also shown in the dashed lines.}
  \label{fig:optimal}
\end{figure}
%%%%%%%%%%%%%%%%%%%%%%%%%%%%%%%%%%%%%%%%%%%%%%%%%%%%%%%%%%

In Fig.~\ref{fig:optimal}, we show optimal cut values ($(m_{\rm
  effcut})_{\rm optimal}$, $(r_{\rm cut})_{\rm optimal}$), the
preselection ({\tt L}, {\tt M}, or {\tt H}), and the discriminator
$\Delta r_{\rm DV}$ ({\tt R}, {\tt T}, or {\tt Z}) for each sample
point with various $c\tau_{\tilde{g}}$, with an integrated luminosity
of $3000~{\rm fb}^{-1}$. 
Here, for the discriminator $\Delta r_{\rm
  DV}$, {\tt R}, {\tt T}, and {\tt Z} represent $| \Delta \bm{r}_{\rm
  DV}|$, $|(\Delta \bm{r}_{\rm DV})_\text{T}|$, and $|(\Delta
\bm{r}_{\rm DV})_{z}|$, respectively.
They are obtained so that the expected
significance $Z_0$ is maximized.  The expected $5\sigma$ discovery
reaches for gluinos are also shown in the dashed lines, which we
discuss in more detail below. 
%For the discriminator $\Delta r_{\rm
%  DV}$, {\tt R}, {\tt T}, and {\tt Z} represent $| \Delta \bm{r}_{\rm
%  DV}|$, $|(\Delta \bm{r}_{\rm DV})_\text{T}|$, and $|(\Delta
%\bm{r}_{\rm DV})_{z}|$, respectively. 
It is found that the new
selection cut allows us to relax the kinematic selection cut on
$m_{\rm eff}({\rm incl.})$ considerably, especially for gluinos with a
long lifetime. As for the optimization of $(r_{\rm cut})_{\rm
  optimal}$, $(r_{\rm cut})_{\rm optimal} \simeq c\tau_{\tilde{g}}$
tends to be favored for gluinos with a sub-millimeter decay length.

%-------------------------------
\begin{table}
  \centering
\subcaptionbox{The expected number of background events. }[.9\linewidth]{
    \begin{tabular}{l|ccc|c}
    \hline\hline
    & $Z$ & $W$ & $t\bar{t}$ & total \\
    \hline
    {\tt preselection-H}~~$(\times10^3)$& 
    $4.7{\scriptstyle~ \pm 0.3}$ &
    $4.1{\scriptstyle~ \pm 0.6}$ &
    $4.5{\scriptstyle~ \pm 0.2}$ &
    $13.3{\scriptstyle~ \pm 0.7}$\\
    %{\tt preselection-H} & 
    %$( 4.68{\scriptstyle~ \pm 0.30} ){\scriptstyle \times10^{3} }$ &
    %$( 4.13{\scriptstyle~ \pm 0.58} ){\scriptstyle \times10^{3} }$ &
    %$( 4.45{\scriptstyle~ \pm 0.22} ){\scriptstyle \times10^{3} }$ &
    %$( 1.33{\scriptstyle~ \pm 0.07} ){\scriptstyle \times10^{4} }$\\
    $m_{\rm eff}({\rm incl.}) > 3600$ GeV & 
    $12.5{\scriptstyle~ \pm 1}$ &
    $4.1{\scriptstyle~ \pm 0.7}$ &
    $1.6{\scriptstyle~ \pm 0.4}$ &
    $18.2{\scriptstyle~ \pm 1}$\\
    \hline
    $ \bigl| \bm{r}_{\rm DV1}-\bm{r}_{\rm DV2} \bigr| > 80~{\rm \mu m}$ & 
    $2.1{\scriptstyle~ \pm 0.4}$ &
    $0.8{\scriptstyle~ \pm 0.4}$ &
    $0.2{\scriptstyle~ \pm 0.1}$ &
    $3.2{\scriptstyle~ \pm 0.6}$\\
    $ \bigl| \bm{r}_{\rm DV1}-\bm{r}_{\rm DV2} \bigr| > 160~{\rm \mu m}$ & 
    $0.1{\scriptstyle~ \pm 0.1}$ &
    $< 0.1$ &
    $< 0.1$ &
    $0.1{\scriptstyle~ \pm 0.1}$\\
    $ \bigl| \bm{r}_{\rm DV1}-\bm{r}_{\rm DV2} \bigr| > 200~{\rm \mu m}$ & 
    $0.1{\scriptstyle~ \pm 0.1}$ &
    $< 0.1$ &
    $< 0.1$ &
    $0.1{\scriptstyle~ \pm 0.1}$\\
    \hline\hline
    \end{tabular}
}

\vspace{3mm}
\centering
\subcaptionbox{The expected number of signal events for 
    $m_{\tilde{g}}=2800$~GeV and $m_{\tilde{\chi}_1^0}=100$~GeV with
    different values of $c\tau_{\tilde{g}}$.}[.9\linewidth]{
    \begin{tabular}{l|cccc}
    \hline\hline
    & $c\tau_{\tilde{g}}=0$ 
    & $c\tau_{\tilde{g}}=200~{\rm \mu m}$ & $c\tau_{\tilde{g}}=500~{\rm \mu m}$ & $c\tau_{\tilde{g}}=1~{\rm  mm}$ \\
    \hline
    {\tt preselection-H} & 
    \multicolumn{4}{c}{$8.2{\scriptstyle~\pm 0.1}$} \\
    $m_{\rm eff}({\rm incl.}) > 3600$ GeV & 
     \multicolumn{4}{c}{$6.9{\scriptstyle~\pm 0.1}$} \\
     \hline
    $ \bigl| \bm{r}_{\rm DV1}-\bm{r}_{\rm DV2}\bigr| > 80~{\rm \mu m}$ & 
    $1.6{\scriptstyle~ \pm 0.04}$ &
    $5.3{\scriptstyle~ \pm 0.1}$ &
    $6.3{\scriptstyle~ \pm 0.1}$ &
    $6.6{\scriptstyle~ \pm 0.1}$\\
    $ \bigl|\bm{r}_{\rm DV1}-\bm{r}_{\rm DV2} \bigr| > 160~{\rm \mu m}$ & 
    $0.2{\scriptstyle~ \pm 0.01}$ &
    $2.9{\scriptstyle~ \pm 0.1}$ &
    $5.1{\scriptstyle~ \pm 0.1}$ &
    $5.9{\scriptstyle~ \pm 0.1}$\\
    $ \bigl| \bm{r}_{\rm DV1}-\bm{r}_{\rm DV2} \bigr| > 200~{\rm \mu m}$ & 
    $< 0.1$ &
    $2.2{\scriptstyle~ \pm 0.1}$ &
    $4.5{\scriptstyle~ \pm 0.1}$ &
    $5.6{\scriptstyle~ \pm 0.1}$\\
    \hline\hline
    \end{tabular}
}
    \caption{
    The expected number of background (a) and signal (b) events for an
   integrated luminosity of $\mathcal{L}=3000~{\rm fb}^{-1}$. We set
    the masses of gluino and the LSP to be $2800$~GeV and $100$~GeV, respectively.
    }
    \label{table:cut_flow}
  \end{table}
\begin{table}
  \centering
\subcaptionbox{The expected number of background events. }[.9\linewidth]{
    \begin{tabular}{l|ccc|c}
    \hline\hline
    & $Z$ & $W$ & $t\bar{t}$ & total \\
    \hline
    {\tt preselection-M}~~$(\times10^4)$& 
    $1.7{\scriptstyle~ \pm 0.10}$ &
    $1.7{\scriptstyle~ \pm 0.22}$ &
    $1.8{\scriptstyle~ \pm 0.05}$ &
    $5.2{\scriptstyle~ \pm 0.25}$\\
    $m_{\rm eff}({\rm incl.}) > 2400$ GeV& 
    $357{\scriptstyle~ \pm 41}$ &
    $148{\scriptstyle~ \pm 13}$ &
    $37{\scriptstyle~ \pm 5}$ &
    $542{\scriptstyle~ \pm 44}$\\
    \hline
    $ \bigl| \bm{r}_{\rm DV1}-\bm{r}_{\rm DV2} \bigr| > 160~{\rm \mu m}$ & 
    $8.2{\scriptstyle~ \pm 4}$ &
    $3.6{\scriptstyle~ \pm 1}$ &
    $0.5{\scriptstyle~ \pm 0.2}$ &
    $12.3{\scriptstyle~ \pm 4}$\\
    $ \bigl| \bm{r}_{\rm DV1}-\bm{r}_{\rm DV2} \bigr| > 320~{\rm \mu m}$ & 
    $0.2{\scriptstyle~ \pm 0.1}$ &
    $< 0.1$ &
    $0.2{\scriptstyle~ \pm 0.2}$ &
    $0.5{\scriptstyle~ \pm 0.2}$\\
    $ \bigl| \bm{r}_{\rm DV1}-\bm{r}_{\rm DV2} \bigr| > 400~{\rm \mu m}$ & 
    $< 0.1$ &
    $< 0.1$ &
    $< 0.1$ &
    $< 0.1$ \\
    \hline\hline
    \end{tabular}
    }

\vspace{3mm}
\centering
\subcaptionbox{The expected number of signal events for 
    $m_{\tilde{g}}=2400$~GeV, $m_{\tilde{\chi}_1^0}=1440$~GeV
    with different values of $c\tau_{\tilde{g}}$.}[.9\linewidth]{
    \begin{tabular}{l|cccc}
    \hline\hline
    & $c\tau_{\tilde{g}}=0$ 
    & $c\tau_{\tilde{g}}=200~{\rm \mu m}$ & $c\tau_{\tilde{g}}=500~{\rm \mu m}$ & $c\tau_{\tilde{g}}=1~{\rm  mm}$ \\
    \hline
    {\tt preselection-M} & 
    \multicolumn{4}{c}{$68.6{\scriptstyle~\pm 0.7}$} \\
    $m_{\rm eff}({\rm incl.}) > 2400$ GeV & 
     \multicolumn{4}{c}{$31.2{\scriptstyle~\pm 0.5}$} \\
     \hline
    $ \bigl| \bm{r}_{\rm DV1}-\bm{r}_{\rm DV2}\bigr| > 160~{\rm \mu m}$ & 
    $0.8{\scriptstyle~ \pm 0.1}$ &
    $12.7{\scriptstyle~ \pm 0.3}$ &
    $21.1{\scriptstyle~ \pm 0.4}$ &
    $24.1{\scriptstyle~ \pm 0.4}$\\
    $ \bigl|\bm{r}_{\rm DV1}-\bm{r}_{\rm DV2} \bigr| > 320~{\rm \mu m}$ & 
    $< 0.1$ &
    $4.6{\scriptstyle~ \pm 0.2}$ &
    $13.9{\scriptstyle~ \pm 0.3}$ &
    $18.9{\scriptstyle~ \pm 0.4}$\\
    $ \bigl| \bm{r}_{\rm DV1}-\bm{r}_{\rm DV2} \bigr| > 400~{\rm \mu m}$ & 
    $< 0.1$ &
    $2.5{\scriptstyle~ \pm 0.1}$ &
    $10.2{\scriptstyle~ \pm 0.3}$ &
    $16.2{\scriptstyle~ \pm 0.4}$\\
    \hline\hline
    \end{tabular}
}
    \caption{The same as in Table.~\ref{table:cut_flow} but {\tt
   preselection-M} is imposed and the masses of gluino and the LSP are
   set to be $2400$~GeV and $1440$~GeV, respectively.
    }
    \label{table:cut_flow_mod_degen}
  \end{table}
%-------------------------------

%%%%%%%%%%%%%% FIGURE %%%%%%%%%%%%%%%%%%%%%%%%%%%%%%%%%%%%
\begin{figure}
  \centering
  \subcaptionbox{\label{fig:meff_w_rdv_100mu} $| \bm{r}_{\rm
 DV1}-\bm{r}_{\rm DV2}| > 100~{\rm \mu
 m}$}{\includegraphics[width=0.48\columnwidth]{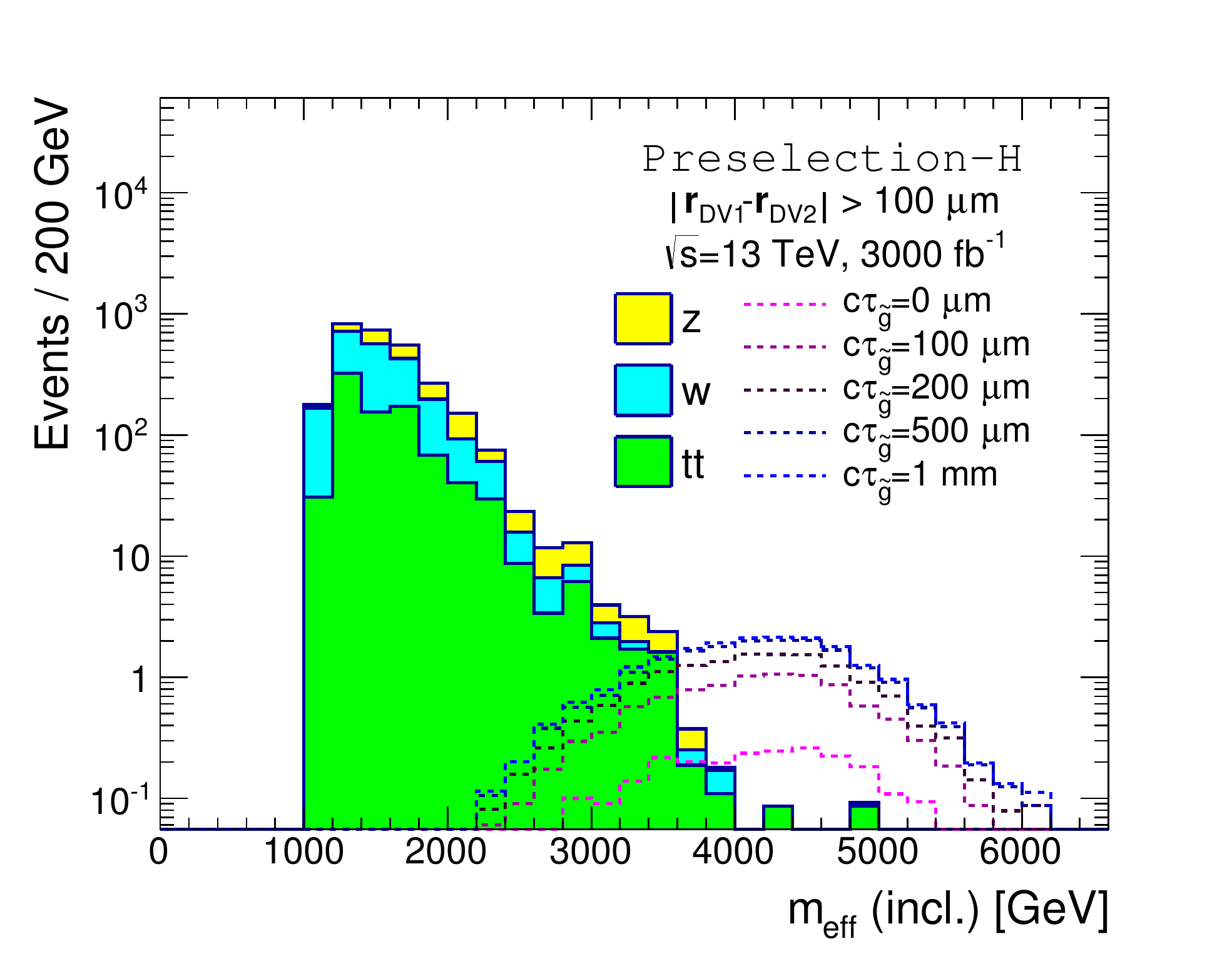}} 
  \subcaptionbox{\label{fig:meff_w_rdv_200mu} $| \bm{r}_{\rm
 DV1}-\bm{r}_{\rm DV2}| > 200~{\rm \mu
 m}$}{\includegraphics[width=0.48\columnwidth]{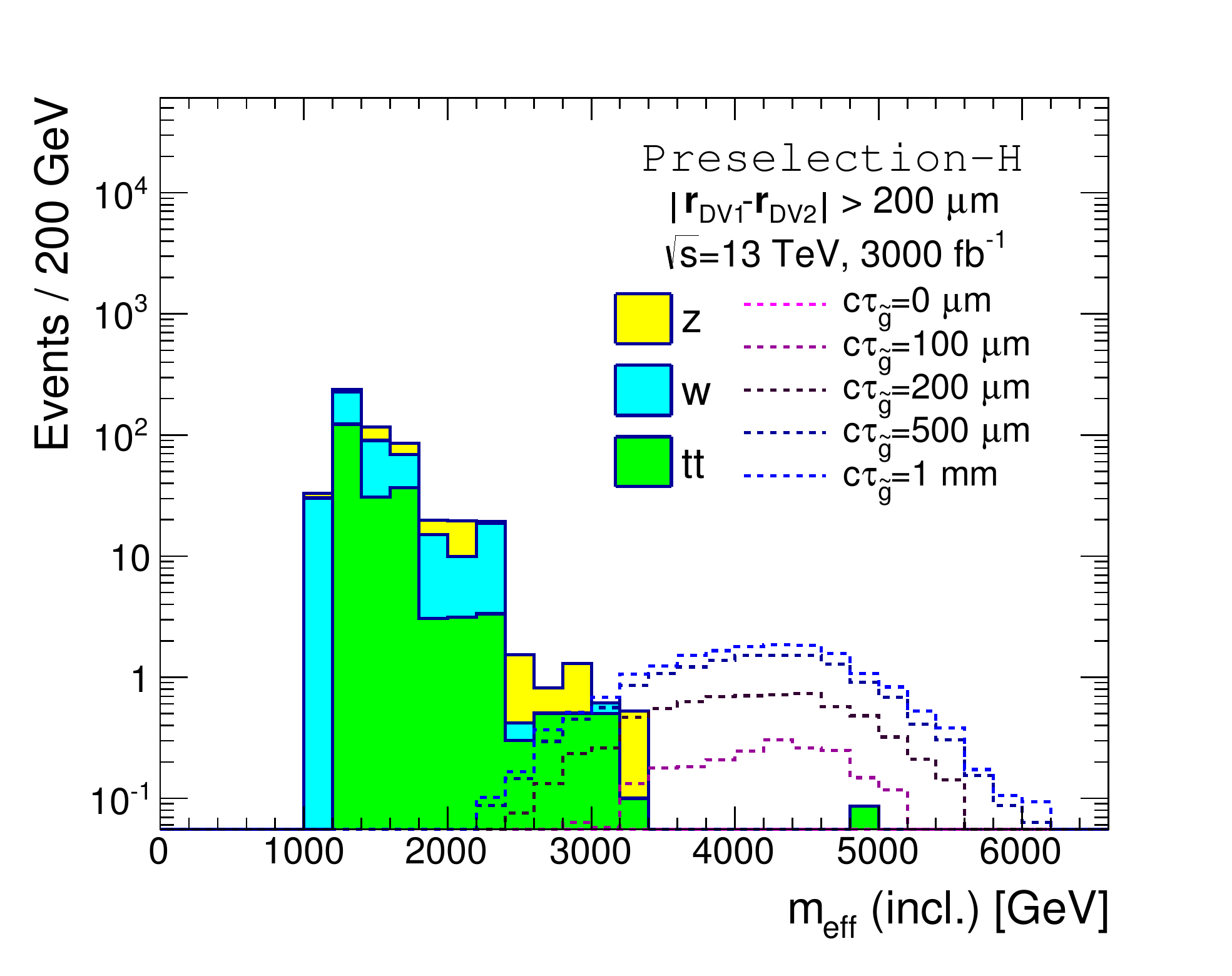}} 
  \subcaptionbox{\label{fig:meff_w_rdv_500mu} $| \bm{r}_{\rm
 DV1}-\bm{r}_{\rm DV2}| > 500~{\rm \mu
 m}$}{\includegraphics[width=0.48\columnwidth]{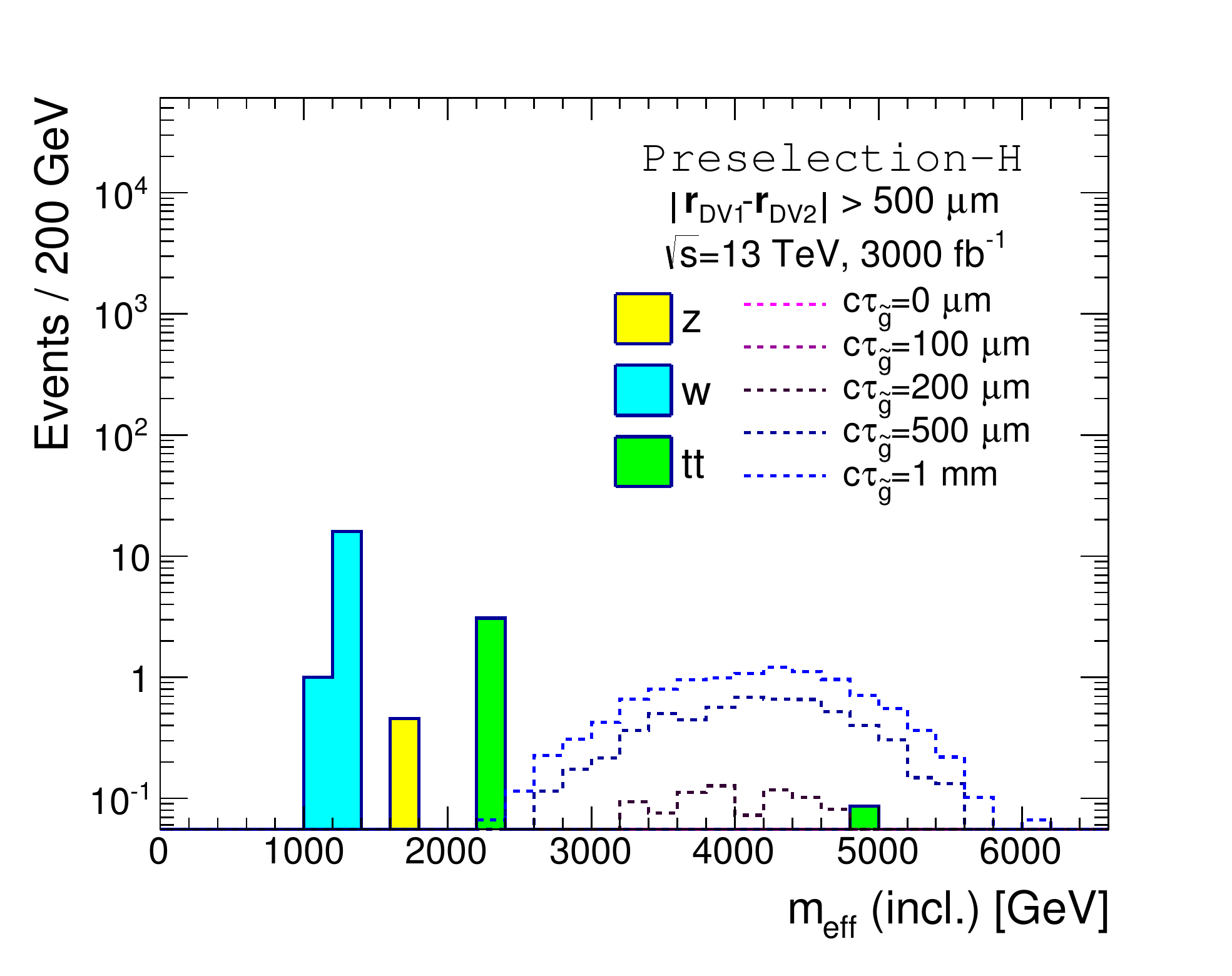}} 
\caption{Distributions of $m_{\rm eff}({\rm incl.})$ for the SM background and
signal events with different values of $c\tau_{\tilde{g}}$. The masses
of gluino and the LSP are set to be $2600$~GeV and $100$~GeV, 
respectively. We have imposed {\tt preselection-H} and the vetoes given in
Table~\ref{table:selections}. }  
  \label{fig:meff_w_rdv}
\end{figure}
%%%%%%%%%%%%%%%%%%%%%%%%%%%%%%%%%%%%%%%%%%%%%%%%%%%%%%%%%%

In Tables~\ref{table:cut_flow} and \ref{table:cut_flow_mod_degen}, we
present the expected number of background and signal events for an
integrated luminosity of $\mathcal{L}=3000~{\rm fb}^{-1}$.\footnote{We
also show the statistical uncertainties of MC simulation, which we
estimate as 
\begin{align*}
\Delta N = \mathcal{L}\times \sqrt{ \sum w_{\rm sample}^2 } ~,
\end{align*}
where $\mathcal{L}$ is an integrated luminosity, $w_{\rm sample}$ is the
MC sample weight given by Eq.~\eqref{eq:weight}, and the summation is
taken over all MC samples which have passed the selection cut.
}
In Table~\ref{table:cut_flow} (\ref{table:cut_flow_mod_degen}),
we consider the case of a light (heavy) LSP with
$m_{\tilde{g}}=2800~(2400)$~GeV and 
$m_{\tilde{\chi_1^0}}=100~(1440)$~GeV, where \texttt{preselection-H}
(\texttt{-M}) is used for the event preselection. We see that in both
cases the new selection cut efficiently removes the SM background while
maintaining a sizable amount of signal events for metastable gluinos. To
see this more clearly, in Fig.~\ref{fig:meff_w_rdv}, we show the 
distributions of $m_{\rm eff}({\rm incl.})$ for the SM background and
signal events with different values of $c\tau_{\tilde{g}}$, with {\tt
preselection-H} and the vetoes in Table~\ref{table:selections} imposed. The masses
of gluino and the LSP are set to be $2600$~GeV and $100$~GeV, 
respectively.  As seen in these plots, the new selection
cut considerably reduces the SM background especially if we require a large
separation between the reconstructed displaced vertices, which allows us
to relax the cut on $m_{\rm eff}({\rm incl.})$ to keep a large number of
signal events. 

%------------------------------------------------------------------------------
\begin{figure}
  \centering
  \includegraphics[width=0.7\columnwidth]{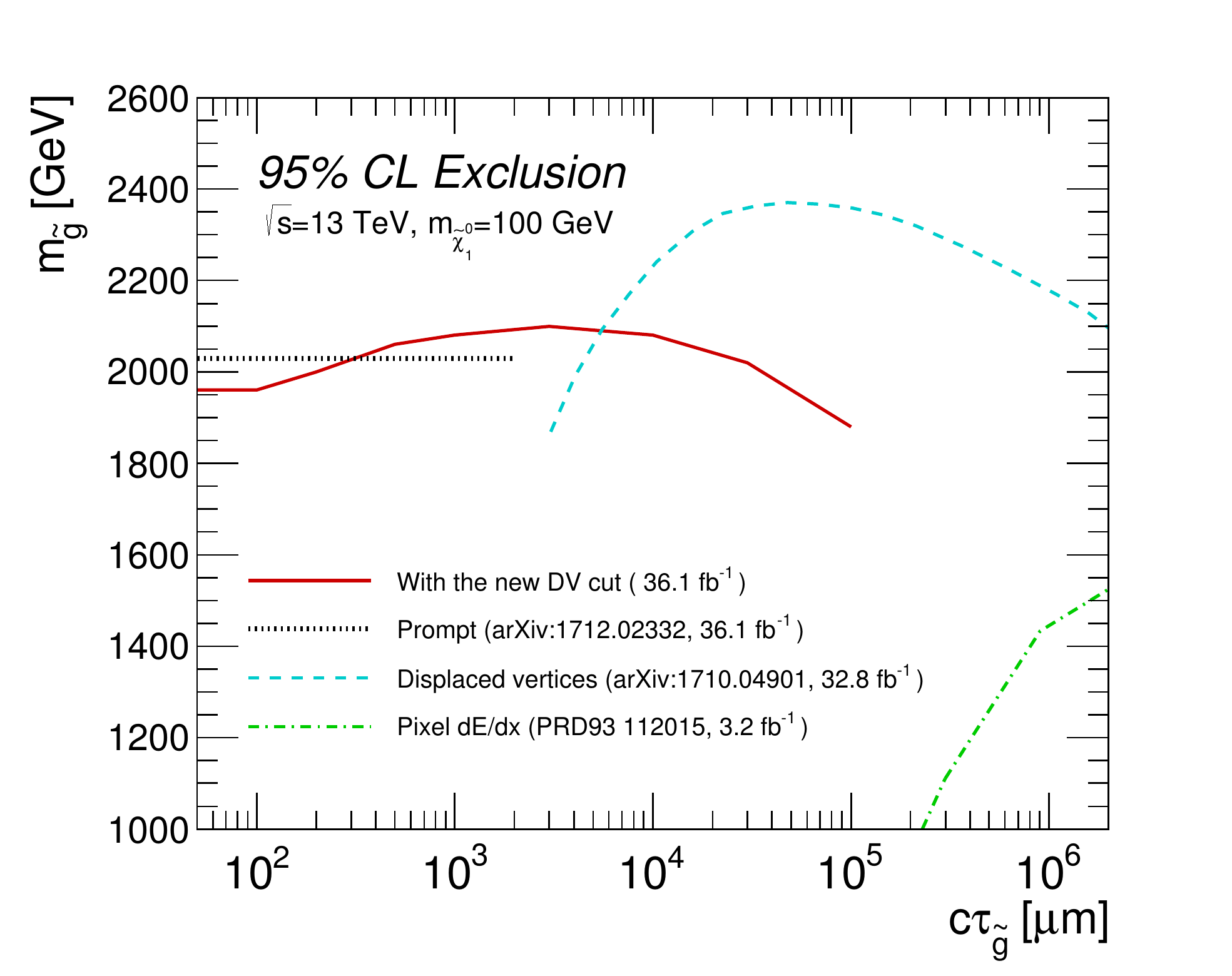}
 \caption{The 95\% CL expected exclusion limit on the gluino
    mass with ${\cal L}=36.1\ {\rm fb}^{-1}$ at the 13~TeV LHC run as
    a function of $c\tau_{\tilde g}$ (red solid line). For
    comparison, we also show the 95\% CL exclusion limits given by the
    ATLAS prompt-decay gluino search (black dotted line)
    \cite{Aaboud:2017vwy}, the ATLAS displaced-vertex search (blue dashed line)
    \cite{Aaboud:2017iio}, and the ATLAS search of large ionization
    energy loss in the Pixel detector (green dot-dashed line)
    \cite{Aaboud:2016dgf}. }
  \label{fig:exclusionNOW}
\end{figure}
%------------------------------------------------------------------------------

Now we show in Fig.~\ref{fig:exclusionNOW} the expected limit on the gluino
mass as a function of $c\tau_{\tilde g}$ based on the currently
available luminosity of $36.1\ {\rm fb}^{-1}$ at the 13~TeV LHC (red solid line). 
Here the mass of LSP is set to be 100~GeV. The improvement in reach
because of the new selection cut may be seen by comparing the reach for each
$c\tau_{\tilde g}$ with that for $c\tau_{\tilde g} = 0$, for which the
new selection cut is ineffective as seen from Fig.~\ref{fig:optimal_0mu}. 
It turns out that, even with the current data, the exclusion limit can
be improved by about $100$ and $120\ {\rm GeV}$ for
$c\tau_{\tilde{g}}=0.5$ and $1\ {\rm mm}$, respectively.
We also find that the sensitivity gets worse for $c\tau_{\tilde{g}}
\gtrsim\mathcal{O}(10)$ mm. This is because the signal efficiency is
decreased due to the requirements on the production point and impact
parameters of tracks as well as the detector material veto on the
position of reconstructed vertices. 
To compare the result with the current sensitivities of
other gluino searches, we also show the 95\% CL exclusion limits given
by the ATLAS prompt-decay gluino search with the 13~TeV 36.1~fb$^{-1}$
data (black dotted line) \cite{Aaboud:2017vwy}, the ATLAS
displaced-vertex search with the 13~TeV 32.8~fb$^{-1}$ data (blue dashed line)
\cite{Aaboud:2017iio}, and the ATLAS search of large ionization energy
loss in the Pixel detector with the 13~TeV 3.2~fb$^{-1}$ data (green dot-dashed line) 
\cite{Aaboud:2016dgf}. 
Note that we extend the black dotted line for the ATLAS prompt-decay
gluino search up to $c\tau_{\tilde{g}}\sim\mathcal{O}(1)$~mm just for
comparison; the reach of the prompt-decay gluino search is expected to
become worse for $c\tau_{\tilde{g}}\gtrsim\mathcal{O}(1)$~mm
\cite{ATLAS-CONF-2014-037}.\footnote{In the gluino search performed by the
CMS in Ref.~\cite{Sirunyan:2018vjp}, the sensitivity is maximized for
$c\tau_{\tilde{g}}\simeq \mathcal{O}(1)$~mm and then gets worse for
larger values of $c\tau_{\tilde{g}}$. The good sensitivity at
$c\tau_{\tilde{g}}\simeq \mathcal{O}(1)$~mm is due to the CSV tag
\cite{Sirunyan:2017ezt} for the $b$ tagging. This result indeed
indicates that an elaborated vertex reconstruction algorithm can improve
the sensitivity of metastable gluino searches, as we discuss in the
present paper.} 
We see that the existing metastable gluino
searches are insensitive to a gluino with $c\tau_{\tilde{g}} \lesssim
1$~mm, where our event-selection criterion may offer
a good sensitivity.  In this sense, this new search strategy plays a
complementary role in probing metastable gluinos.

%%%%%%%%%%%%%% FIGURE %%%%%%%%%%%%%%%%%%%%%%%%%%%%%%%%%%%%
\begin{figure}
  \centering
  \subcaptionbox{\label{fig:future} $m_{\tilde{\chi}^0_1} =
 100$~GeV}{\includegraphics[width=0.48\columnwidth]{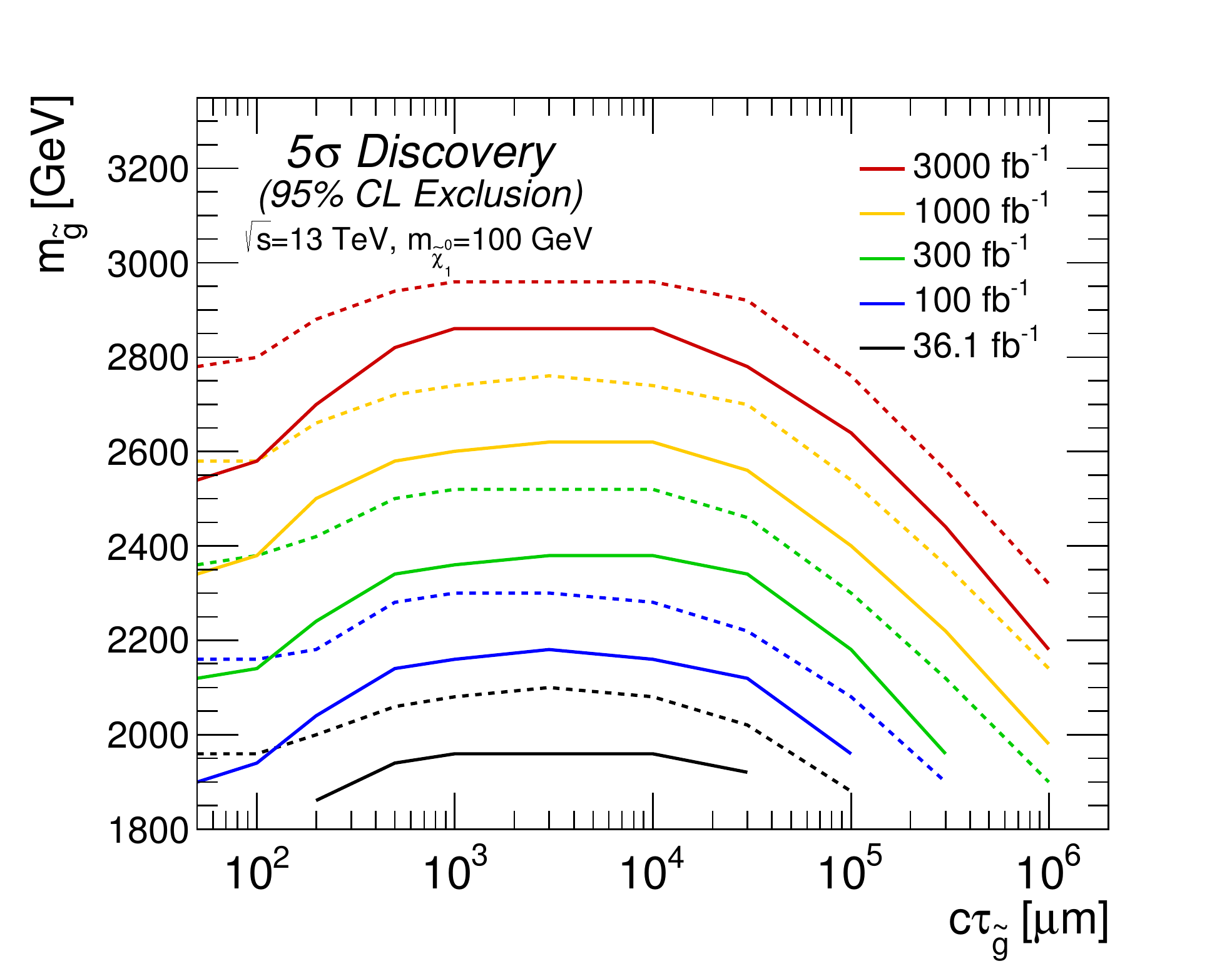}}  
  \subcaptionbox{\label{fig:future_diff100} $|m_{\tilde{g}} -m_{\tilde{\chi}^0_1}| =
 100$~GeV}{\includegraphics[width=0.48\columnwidth]{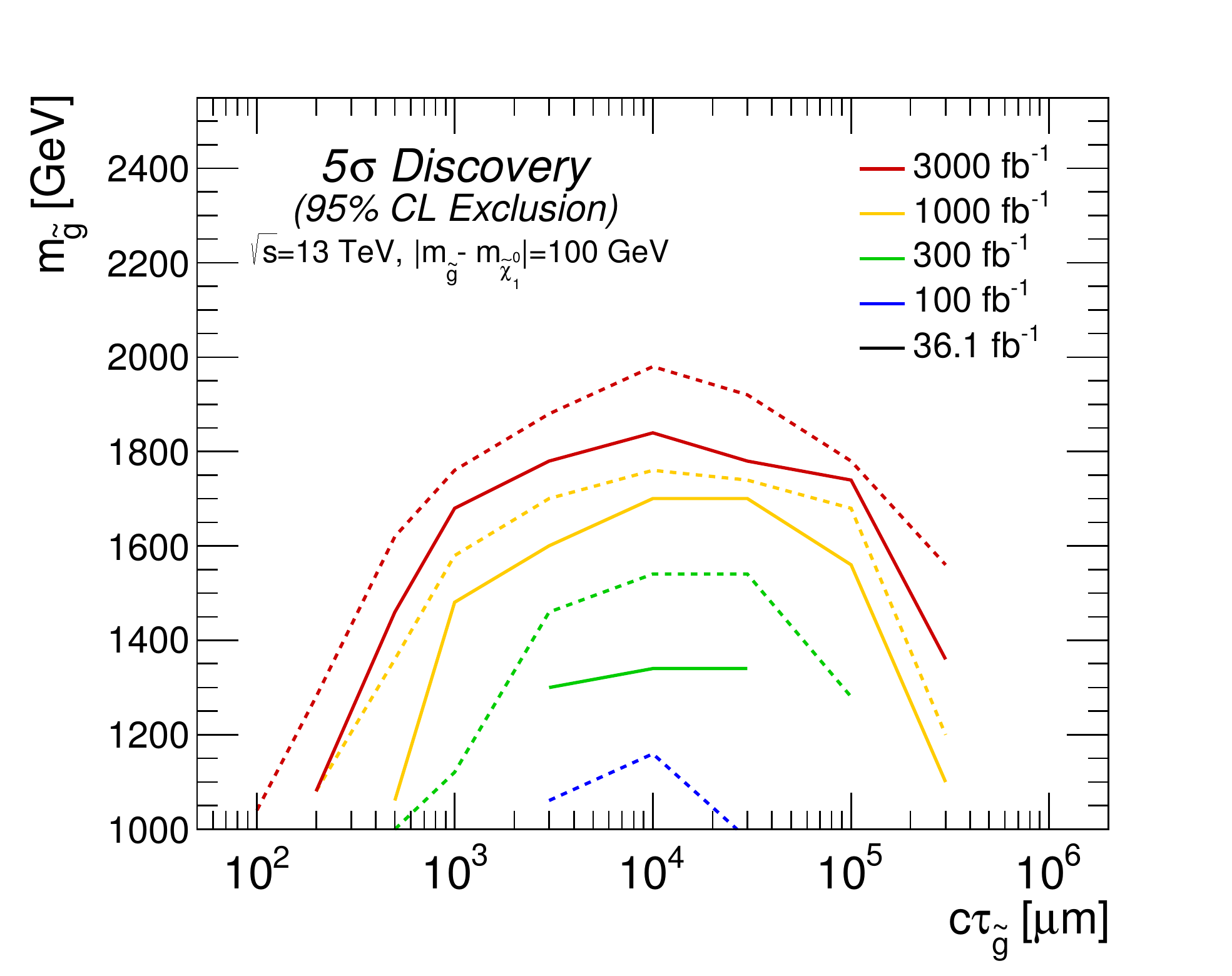}}  
\caption{The expected 95\% CL exclusion limits (dotted) and
    $5\sigma$ discovery reaches (solid) as functions of $c\tau_{\tilde
      g}$ for different values of integrated luminosity at the 13~TeV
    LHC run.}  
  \label{fig:fu}
\end{figure}
%%%%%%%%%%%%%%%%%%%%%%%%%%%%%%%%%%%%%%%%%%%%%%%%%%%%%%%%%%

In Fig.~\ref{fig:future}, we show the expected 95\% CL exclusion limits 
(in dotted lines) and $5\sigma$ discovery reaches (in solid lines) for
gluino as functions of $c\tau_{\tilde  g}$ for different values of
integrated luminosity at the 13~TeV LHC run, where the LSP mass is set to be 100~GeV.
Notice that the expected reaches for an extremely small
$c\tau_{\tilde{g}}$ should correspond to those for the prompt-decay
gluino search with the same data set since the new selection cut plays
no role in this case.  
As can be seen from this figure, the reach for the gluino
can be extended with the help of the new selection cut for
$c\tau_{\tilde g} \gtrsim 100~\mu{\rm m}$; \textit{e.g.}, for a gluino
with $c\tau_{\tilde g} \sim \mathcal{O}$(1--10)~mm, the expected
discovery reach for the gluino mass can be extended by as large as
$\sim 240$~GeV ($320$~GeV) with an integrated luminosity of 
${\mathcal  L}=300~{\rm fb}^{-1}$ (3000~fb$^{-1}$).
These reaches for a gluino with 
$c\tau_{\tilde g} \sim \mathcal{O}$(1--10)~mm
are obtained with {\tt preselection H},
$\left(m_{\rm effcut}\right)_{\rm optimal}=2400$ GeV ($3400$ GeV), and 
$\left( r_{\rm cut} \right)_{\rm optimal}\sim300\,\mu{\rm m}$ ($160\,\mu{\rm m}$) for 
${\mathcal L}=300\,{\rm fb}^{-1}$ ($3000\,{\rm fb}^{-1}$).
Compared to a promptly decaying gluino, for which
$\left(m_{\rm effcut}\right)_{\rm optimal}=3400$~GeV ($4200$ GeV),
the $m_{\rm eff}({\rm incl.})$ selection cut is found to be
significantly loosened, which implies that the new selection cut plays
an important role in background rejection.
As we already mentioned,
because charged tracks with
$|d_0|>10$~mm are not included in the analysis, and also because we
reject all events in which a vertex is reconstructed with a position radius
larger than $120$~mm, the expected exclusion limits decrease for
$c\tau_{\tilde{g}} \gtrsim 100$~mm. Such a larger $c\tau_{\tilde{g}}$
region can however be covered by other long-lived gluino searches.

We also study the case where gluino and the LSP are degenerate in mass,
which is motivated by the coannihilation scenario as we mentioned in
Sec.~\ref{sec:gluino}.
The result is shown in Fig.~\ref{fig:future_diff100}. Here, the mass
difference between gluino and the LSP is set to be $100$~GeV. 
We find that the limits and the reaches are strongly enhanced especially for 
$c\tau_{\tilde{g}}\sim {\mathcal O}$(1--100)~mm. Contrary to the
previous case, we are unable to set a reach or limit higher than $1000$~GeV for 
low luminosities. 
A caveat here is that we impose relatively tight requirements on jet
$p_{\rm T}$ and thus this analysis is not optimized for the degenerate
mass region. Indeed, according to the analysis done by the ATLAS
collaboration~\cite{ATLAS:2016kts}, the event selection category called
{\tt Meff-5j-1400} provides the best sensitivity for the degenerate mass
region, where conditions on $p_{\rm T}$ for 2nd--4th jets are relaxed
and another 5th jet is required instead. We however do not try to
further explore such an optimization in this paper---as the number of
additional partons in our MC simulation is restricted to less than
five---and defer it to another occasion.

%%%%%%%%%%%%%% FIGURE %%%%%%%%%%%%%%%%%%%%%%%%%%%%%%%%%%%%
\begin{figure}
  \centering
  \subcaptionbox{\label{fig:mg_mneu1_36p1_disc} $5\sigma$ discovery}{\includegraphics[width=0.48\columnwidth]{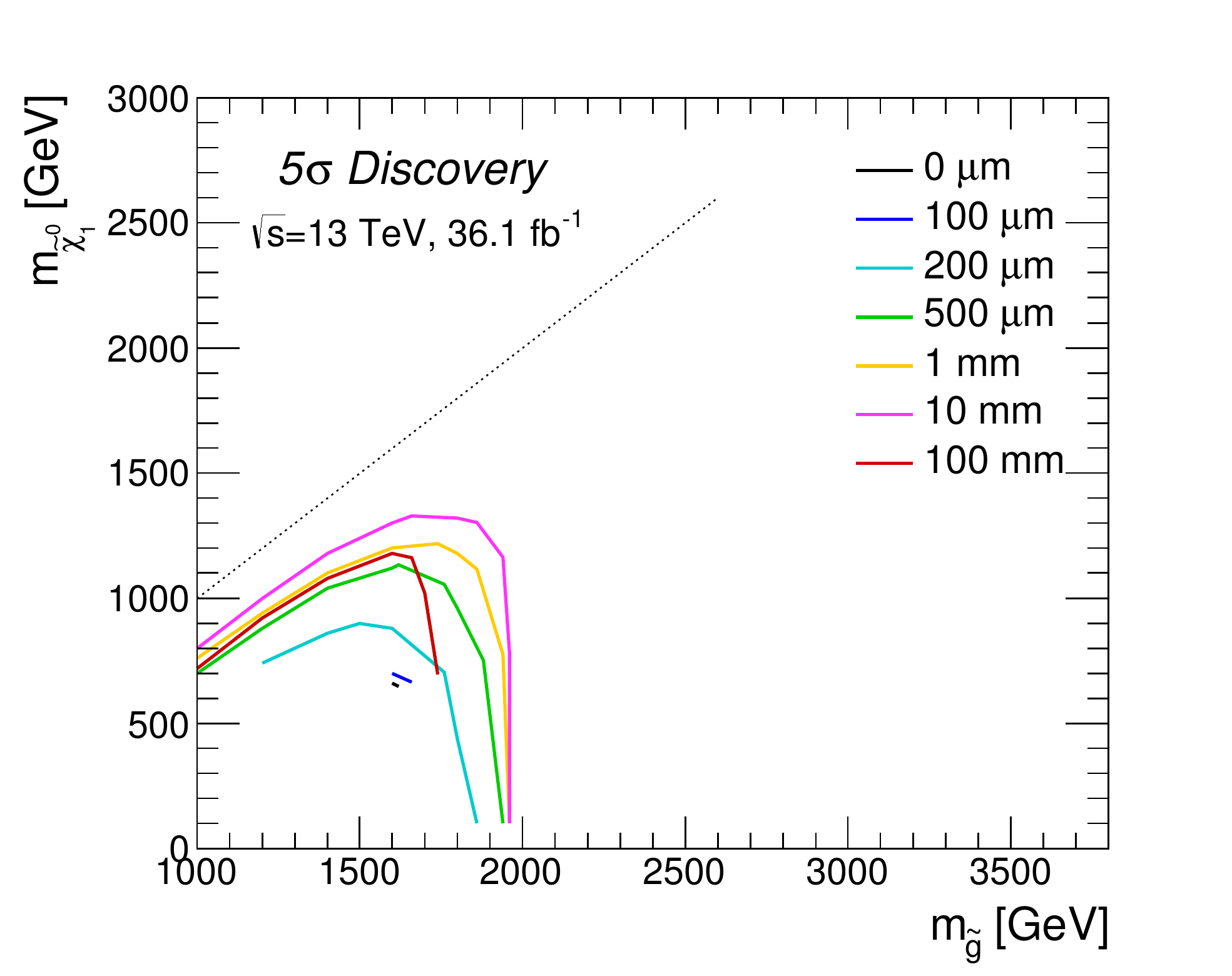}}  
  \subcaptionbox{\label{fig:mg_mneu1_36p1_excl} Expected 95\% CL exclusion}{\includegraphics[width=0.48\columnwidth]{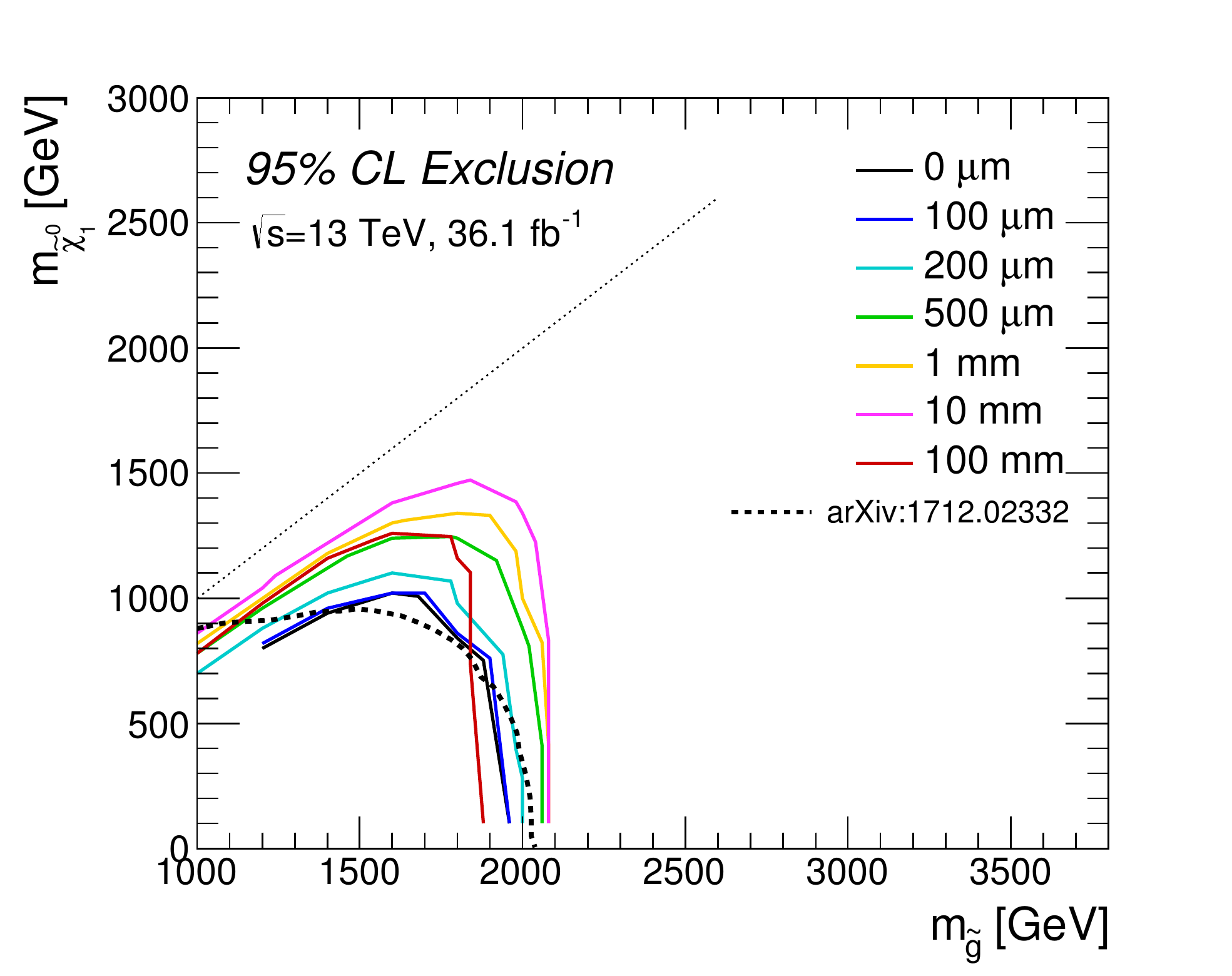}}  
\caption{$5\sigma$ discovery reaches and expected 95\% CL exclusion
 limits for gluinos with different $c\tau_{\tilde{g}}$, for an
 integrated luminosity of $36.1\,{\rm fb}^{-1}$. 
 The expected 95\% CL
 exclusion limit from the ATLAS promptly-decaying gluino
 search~\cite{Aaboud:2017vwy} is also shown in the black dotted
 line. 
 %\TODO{Use the result in Ref.~\cite{Aaboud:2017vwy}
 %instead of \cite{ATLAS-CONF-2017-022}. }
 }  
  \label{fig:mg_mneu1_36p1}
\end{figure}
%%%%%%%%%%%%%%%%%%%%%%%%%%%%%%%%%%%%%%%%%%%%%%%%%%%%%%%%%%

%%%%%%%%%%%%%% FIGURE %%%%%%%%%%%%%%%%%%%%%%%%%%%%%%%%%%%%
\begin{figure}
  \centering
  \subcaptionbox{\label{fig:mg_mneu1_300_disc} $5\sigma$ discovery}{\includegraphics[width=0.48\columnwidth]{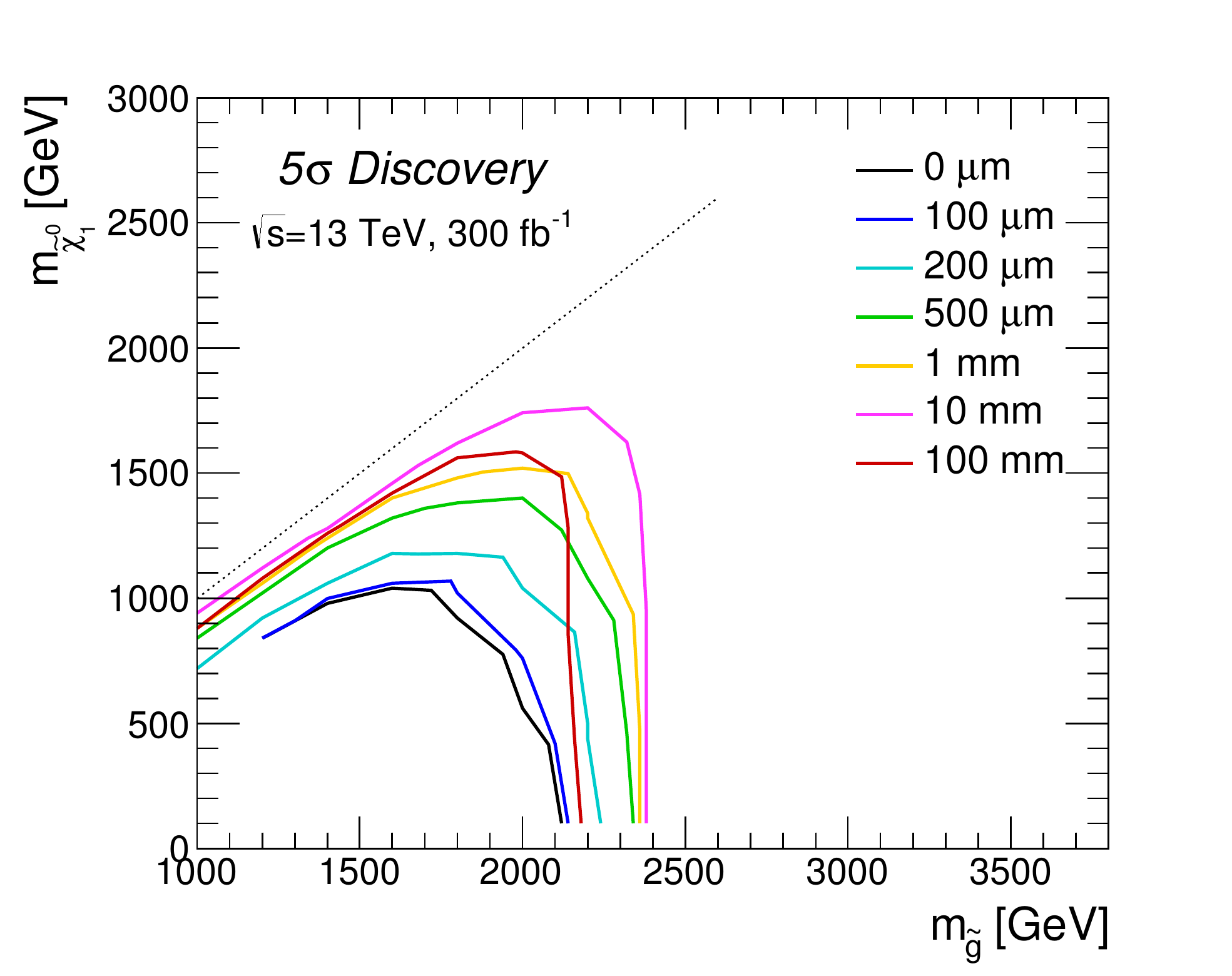}}  
  \subcaptionbox{\label{fig:mg_mneu1_300_excl} Expected 95\% CL exclusion}{\includegraphics[width=0.48\columnwidth]{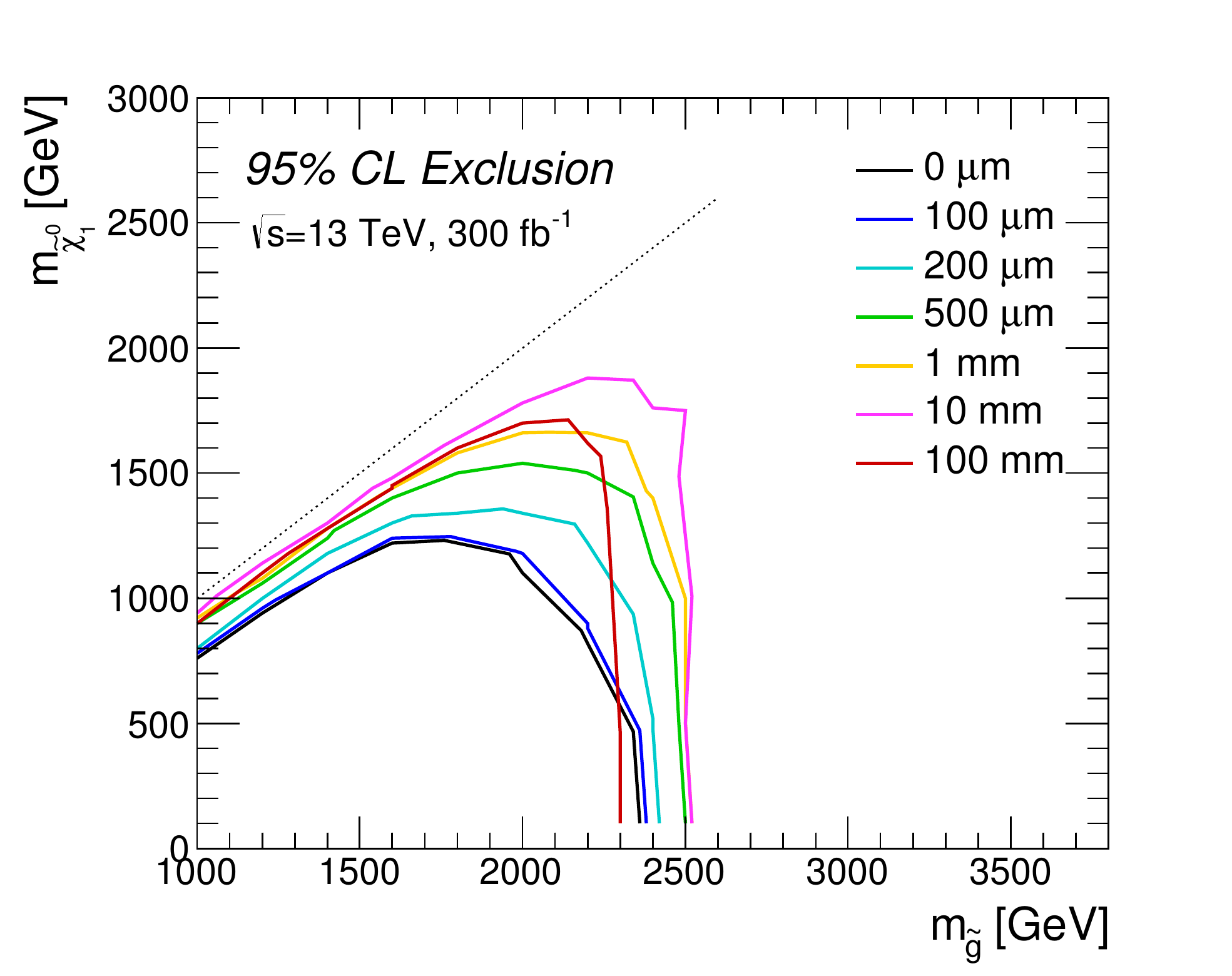}}  
\caption{Same as in Fig.~\ref{fig:mg_mneu1_36p1}, but for an integrated
 luminosity of $300\,{\rm fb}^{-1}$. }  
  \label{fig:mg_mneu1_300}
\end{figure}
%%%%%%%%%%%%%%%%%%%%%%%%%%%%%%%%%%%%%%%%%%%%%%%%%%%%%%%%%%

%%%%%%%%%%%%%% FIGURE %%%%%%%%%%%%%%%%%%%%%%%%%%%%%%%%%%%%
\begin{figure}
  \centering
  \subcaptionbox{\label{fig:mg_mneu1_3000_disc} $5\sigma$ discovery}{\includegraphics[width=0.48\columnwidth]{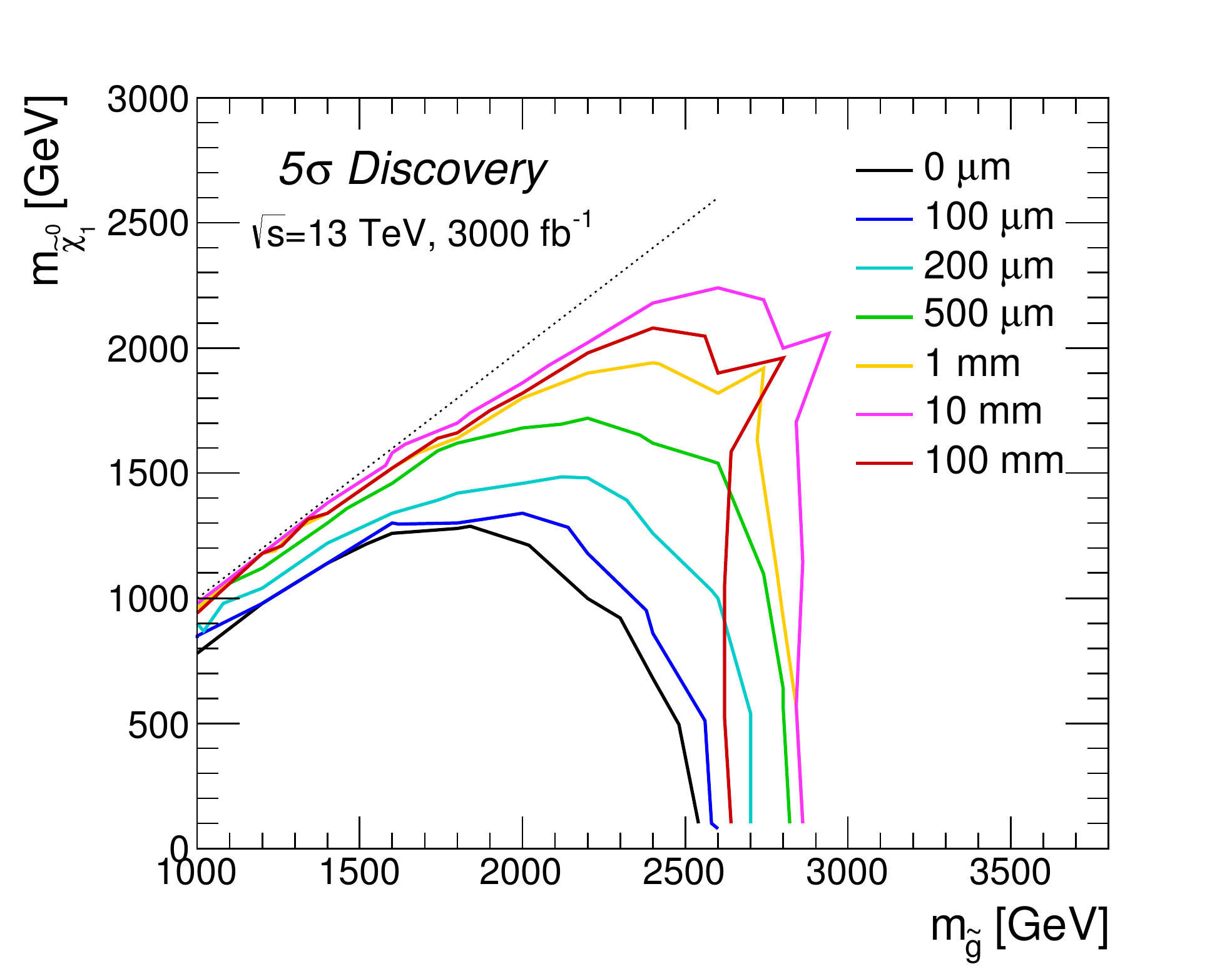}}  
  \subcaptionbox{\label{fig:mg_mneu1_3000_excl} Expected 95\% CL exclusion}{\includegraphics[width=0.48\columnwidth]{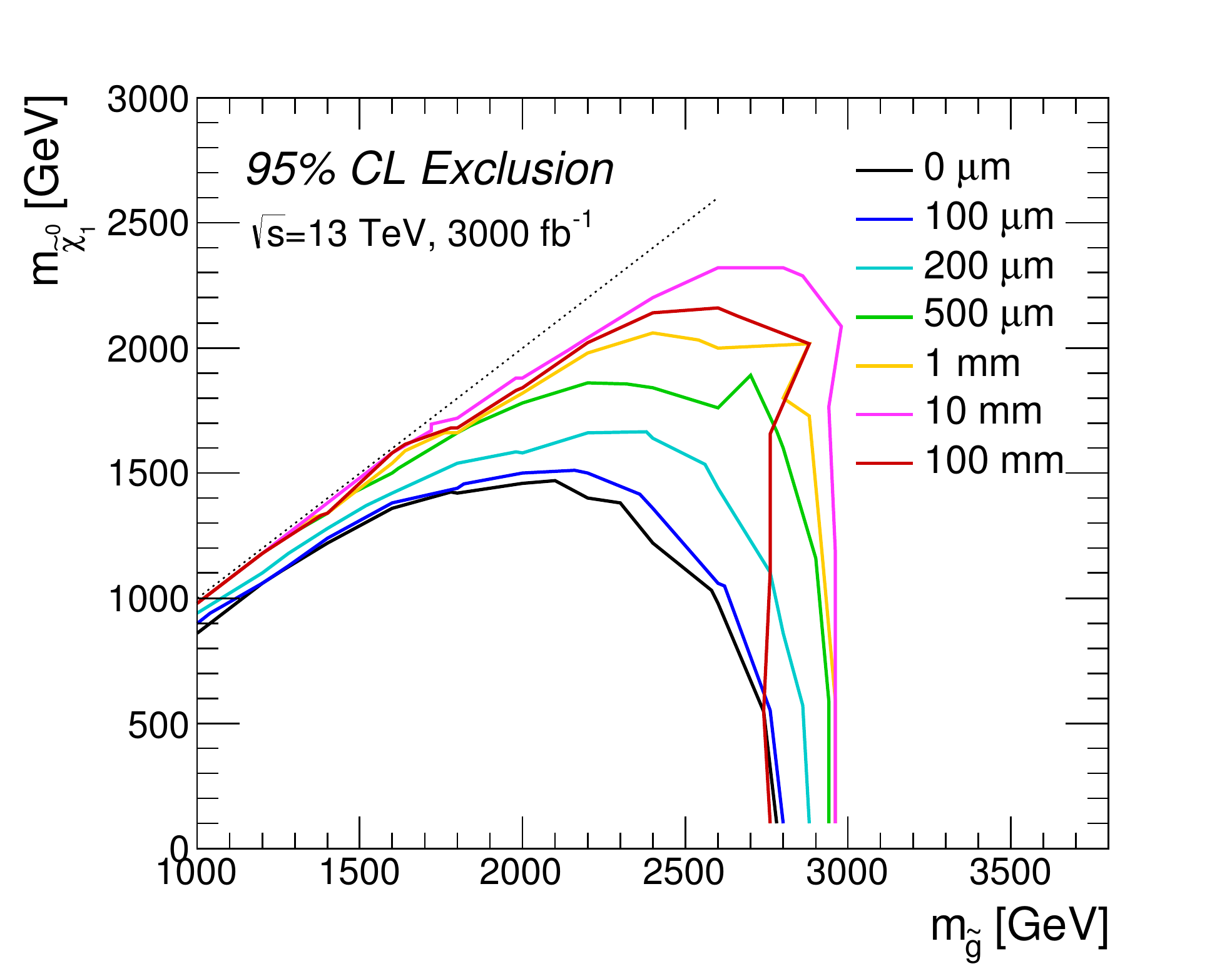}}  
\caption{Same as in Fig.~\ref{fig:mg_mneu1_36p1}, but for an integrated
 luminosity of $3000\,{\rm fb}^{-1}$. }  
  \label{fig:mg_mneu1_3000}
\end{figure}
%%%%%%%%%%%%%%%%%%%%%%%%%%%%%%%%%%%%%%%%%%%%%%%%%%%%%%%%%%

Finally, we present the $5\sigma$ discovery reaches and expected 95\% CL
exclusion limits for various values of $c\tau_{\tilde{g}}$ in
Figs.~\ref{fig:mg_mneu1_36p1}, \ref{fig:mg_mneu1_300}, and
\ref{fig:mg_mneu1_3000} for an integrated luminosity of 36.1, 300, and
3000~fb$^{-1}$, respectively. In Fig.~\ref{fig:mg_mneu1_36p1_excl}, we
also show the expected 95\% CL exclusion limit given by the ATLAS
experiment~\cite{Aaboud:2017vwy} in the black dotted line. We see
that this ATLAS limit is in a fairly good agreement with our limit for
$c\tau_{\tilde{g}}=0~{\rm \mu m}$ shown in the black solid line, besides
the regions where gluino and the LSP are highly degenerate in mass. This
discrepancy is expected since our analysis is not optimized for the
degenerate region as we discussed above. 
From these plots, we see that an implementation of an extra cut
utilizing the vertex reconstruction leads to a significant improvement in
both the discovery reach and the exclusion limit for
$c\tau_{\tilde{g}}\gtrsim200\,\mu{\rm m}$. The extent of the improvement
is maximized for $c\tau_{\tilde{g}}\sim {\mathcal O}$(1--10)~mm and
tends to get larger for a heavier LSP. This feature is also seen in
Fig.~\ref{fig:fu}, where we find that the improvement in the degenerate
case (Fig.~\ref{fig:future_diff100}) is much significant compared with
the light LSP case  (Fig.~\ref{fig:future}). This follows from the fact
that jets and the missing energy in gluino decays for the degenerate
case tend to be soft, and thus traditional kinematical selection cuts
become less powerful in this case. In such a situation, a sizable amount
of the SM background remain after the kinematical selection cuts,
but an additional cut based on vertex reconstruction can remove this
efficiently, which results in a drastic improvement in the sensitivity.

%%%%%%%%%%%%%%%%%%%%%%%%%%%%%%%%%%
\subsection{Lifetime measurements}
\label{sec:lifetime}
%%%%%%%%%%%%%%%%%%%%%%%%%%%%%%%%%%%

Once a new metastable particle is discovered at the LHC, the measurement of
its lifetime is of crucial importance to understand the nature of new
physics behind this metastable particle. For example, by measuring the
lifetime of metastable gluino, we can infer the mass scale of squarks
through Eq.~\eqref{eq:ctaugapp}. In this subsection, we discuss the
prospects of the lifetime measurement by means of the displaced-vertex
reconstruction.

To see this, we study the expected
significance of rejection of a hypothesis that the gluino decay length
is $c\tau_{\tilde{g}}^{\rm (hypo)}$ for gluino samples with a decay
length of $c\tau_{\tilde{g}}$. 
Event samples are binned according to the distance between the two
reconstructed vertices $|\bm{r}_{\rm DV1} - \bm{r}_{\rm DV2}|$ of the events.
Then the expected significance $\langle
Z_{c\tau_{\tilde{g}}^{\rm (hypo)}} \rangle_{c\tau_{\tilde{g}}}$ is
defined by 
\begin{equation}
 \langle Z_{c\tau_{\tilde{g}}^{\rm (hypo)}}
\rangle_{c\tau_{\tilde{g}}}\equiv\sqrt{
\Delta \chi^2 (c\tau_{\tilde{g}}^{\rm (hypo)},c\tau_{\tilde{g}})} ~,
\end{equation}
where
\begin{align}
\Delta \chi^2 (c\tau_{\tilde{g}}^{\rm (hypo)},c\tau_{\tilde{g}})
= 
\sum_{{\rm bin}\,\, i} 
\frac{\left\{ S_i(c\tau_{\tilde{g}}^{\rm (hypo)}) - S_i(c\tau_{\tilde{g}})
 \right\}^2}{S_i(c\tau_{\tilde{g}}^{\rm (hypo)}) +  B_i}\, .
\label{eq:chisq}
\end{align}
Here, $S_i(c\tau)$ is the expected number of signal events in the bin
$i$ on the assumption that gluinos have a decay length of $c\tau$,
while $B_i$ is the number of SM background events.

%%%%%%%%%%%%%% FIGURE %%%%%%%%%%%%%%%%%%%%%%%%%%%%%%%%%%%%
\begin{figure}
  \centering
  \subcaptionbox{\label{fig:ctau_bound300} ${\mathcal L}=300\,{\rm fb}^{-1}$}{\includegraphics[width=0.48\columnwidth]{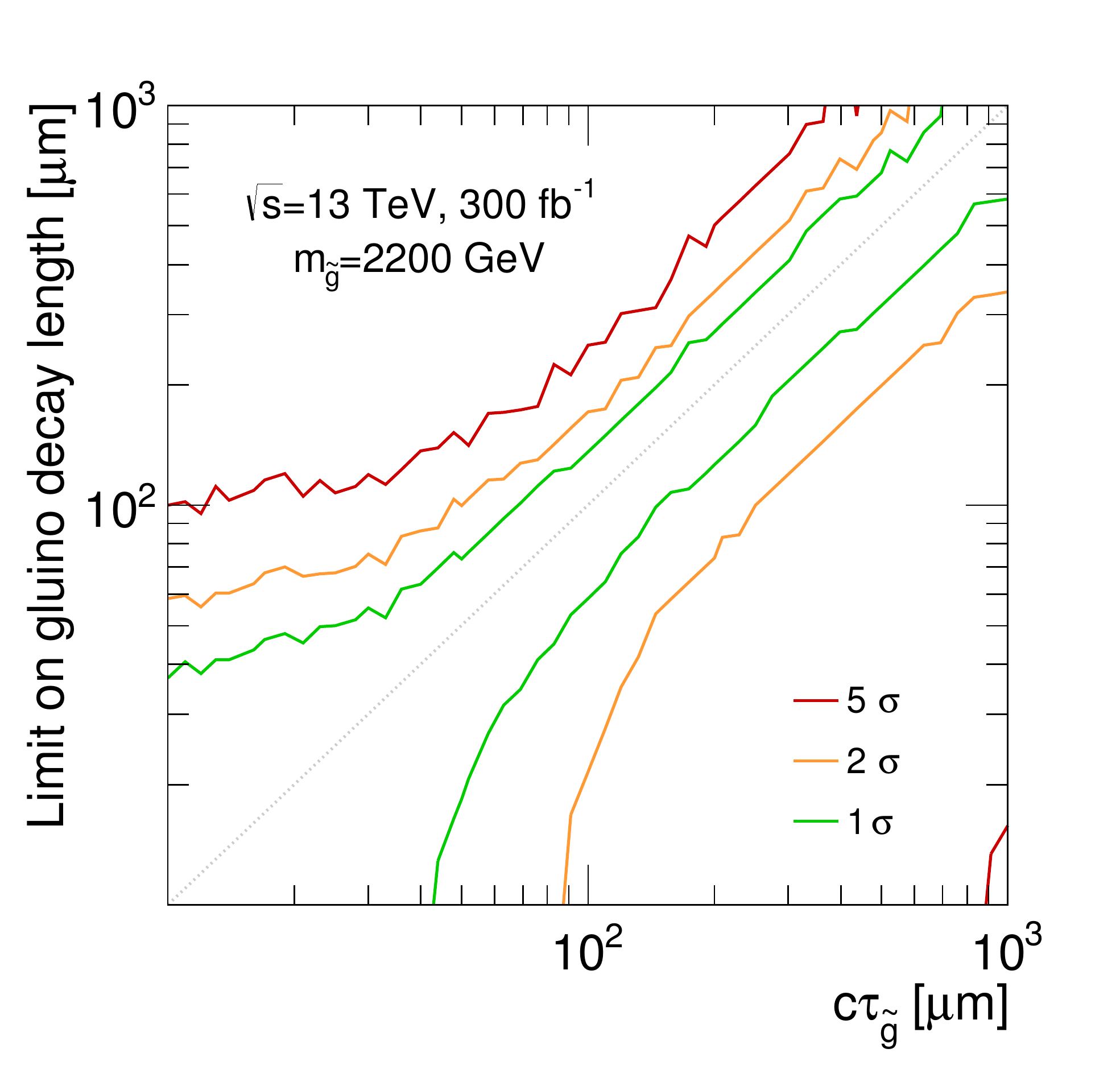}}  
  \subcaptionbox{\label{fig:ctau_bound3000} ${\mathcal L}=3000\,{\rm fb}^{-1}$}{\includegraphics[width=0.48\columnwidth]{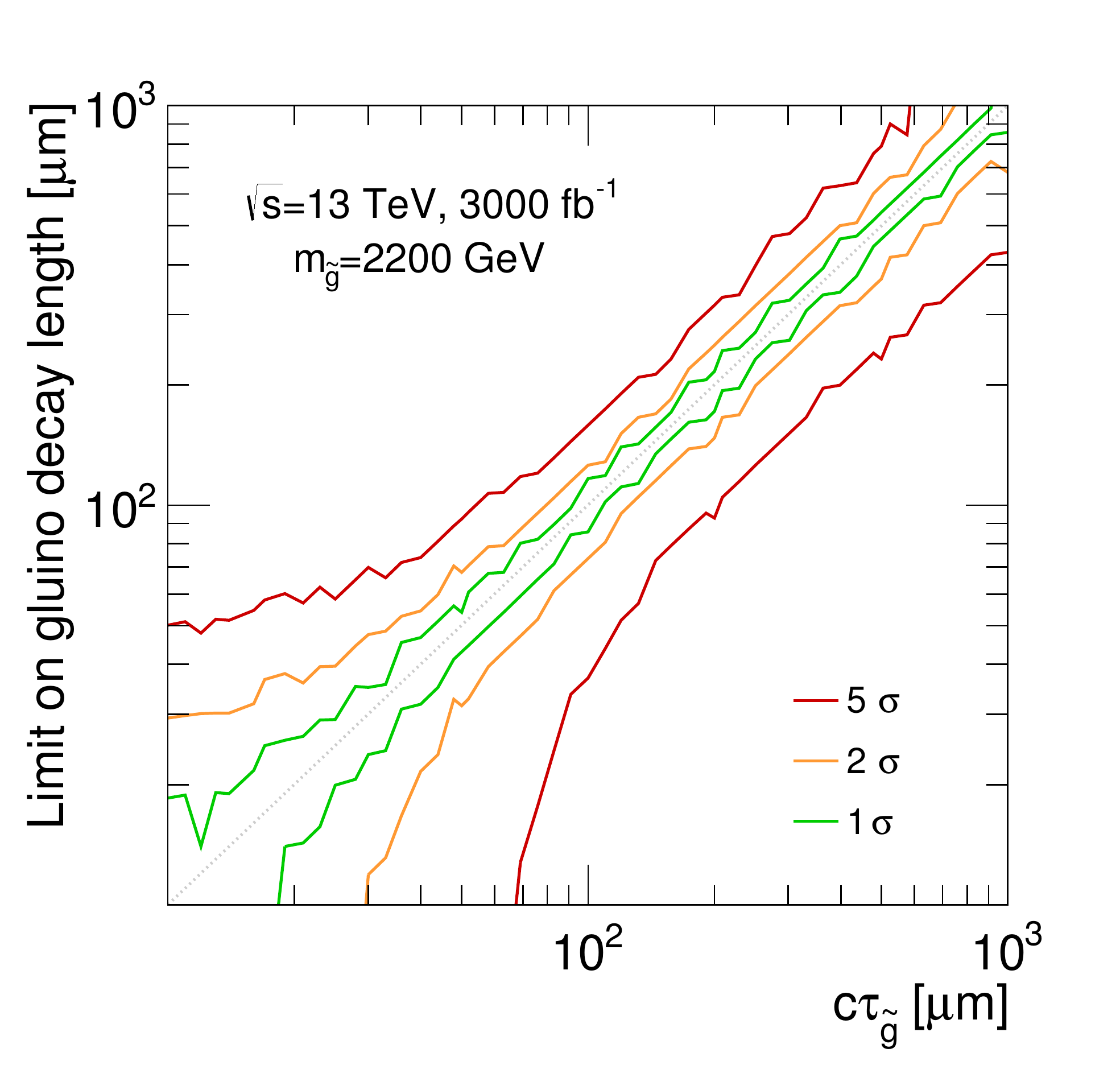}}  
\caption{
The expected upper and lower bounds on the decay length of
    gluino as a function of the underlying value of $c\tau_{\tilde
      g}$. Here, we set $m_{\tilde{g}} = 2.2$~TeV, and impose
       {\tt preselection-H} and a selection cut  
       $m_{\rm eff}({\rm incl.}) > 3500$ GeV.
 }  
  \label{fig:ctau_bound}
\end{figure}
%%%%%%%%%%%%%%%%%%%%%%%%%%%%%%%%%%%%%%%%%%%%%%%%%%%%%%%%%%

In Figs.~\ref{fig:ctau_bound300} and
\ref{fig:ctau_bound3000}, we show the expected upper and lower
bounds on the decay length as a function of 
$c\tau_{\tilde g}$ for a gluino with a mass of 2.2~TeV.
Here we impose {\tt preselection-H} and 
require $m_{\rm eff}({\rm incl.}) > 3500$ GeV.
From these figures, we find that a metastable 
gluino with $c\tau_{\tilde g} \gtrsim 30~(60)~{\rm \mu m}$ can be
distinguished from a promptly decaying one with the significance of
$2\sigma$ ($5\sigma$) with an integrated luminosity of $3000~ {\rm
fb}^{-1}$. Moreover, Fig.~\ref{fig:ctau_bound3000} shows that the decay
length of a gluino with $c\tau_{\tilde{g}} \sim {\cal O}(100)~\mu {\rm
m}$ can be measured with an ${\cal O}(1)$ accuracy at the
high-luminosity LHC. With such a measurement, we may probe the squark
mass scale $m_{\tilde{q}}$ via Eq.~\eqref{eq:ctaugapp} even though
squarks are inaccessible at the LHC. We also note that even if only an
upper limit on the gluino decay length is obtained, this gives
valuable implications for SUSY models, since such a limit
results in an upper bound on the squark mass scale; {\it e.g.}, for
$m_{\tilde{g}} = 2.2$~TeV, an upper limit on the decay length
$c\tau_{\tilde{g}}|_{\text{limit}} $ leads to
\begin{equation}
 m_{\tilde{q}} \lesssim 10^3 \times
  \biggl(\frac{c\tau_{\tilde{g}}|_{\text{limit}}}{100~\mu \text{m}}\biggr)^{\frac{1}{4}}
  ~\text{TeV} ~.
\end{equation}
After all, the
reconstruction of sub-millimeter decay vertices is quite important even
after the discovery of a new particle such as gluino, as its lifetime
contains precious information on the underlying physics.

%%%%%%%%%%%%%%%%%%%%%%%%%%%%%%%
\section{100 TeV collider}
\label{sec:100tev}
%%%%%%%%%%%%%%%%%%%%%%%%%%%%%%%

The vertex-reconstruction method developed above can also be applied to
other collider experiments. Recently, there have been a growing
interest in future collider experiments with a center-of-mass energy
higher than those at the LHC, such as a 100~TeV collider
\cite{Arkani-Hamed:2015vfh, Golling:2016gvc}. Motivated by this, in this
section we apply our new selection cut to the searches for metastable
gluinos at a future $100$ TeV $pp$ collider and study how much this new
selection cut can improve the sensitivity of the prompt-decay searches
in this case. As we have already seen in Sec.~\ref{sec:gluino},
the production cross section of gluinos at a $100$ TeV collider is much
larger than those at the LHC, which drastically extends the reach of
gluino searches \cite{Arkani-Hamed:2015vfh, Golling:2016gvc, Cohen:2013xda,
Ellis:2015xba}. In addition, particles tend to be produced in a highly
boosted state at a 100~TeV collider, which then prolongs the lifetime of
metastable particles---this allows us to probe particles with shorter
decay length, as we actually see in the following analysis.

%%%%%%%%%%%%%%%%%%%%%%%%%%%%%%%%%%%%
\subsection{MC Simulation}
\label{sec:mc_100tev}
%%%%%%%%%%%%%%%%%%%%%%%%%%%%%%%%%%%

%-------------------------------
\begin{table}
  \begin{center}
    \begin{tabular}{c|c}
    \hline\hline
     	$m_{\tilde{g}}$ & 
	$3,\,4,\,\ldots 6,\,8,\,\ldots 16,\, 18$ TeV\\ \hline
	\multirow{3}{*}{$m_{\tilde{\chi}_1^0}$} &
%	$m_{\tilde{\chi}_1^0}$ & 
	$100$ GeV, \\
	&
	 $( 0.2,\, 0.4,\, 0.6,\, 0.8 )\times m_{\tilde{g}}$,  \\ 
	 %\cline{2-2}
	  & 
	  $m_{\tilde{g}}-(150\,{\rm GeV},\, 100\,{\rm GeV},\, 50\,{\rm GeV},\, 25 \,{\rm GeV})$ \\
	 \hline
	 \multirow{2}{*}{$c\tau_{\tilde{g}}$} &   
	 $0,\, 50,\,100,\, 200,\, 500,\, 1000,\, 3000$ ${\rm \mu m}$, \\
	 &
	$1\times10^4,\, 3\times10^4,\, 1\times10^5,\, 3\times10^5,\, 1\times10^6$ ${\rm \mu m}$\\
\hline\hline
    \end{tabular}
    \caption{Sample points for signal events.
    We generate 50000 events for each sample point.
    }
    \label{table:sample_point_100tev}
  \end{center}
  \end{table}
%-------------------------------

%------------------------------------------------------------------
% Z
\begin{table}
\centering
\subcaptionbox{$Z \rightarrow \nu\bar{\nu} + 0,1,2,3\,j$.  \label{table:bg_table_znu3_100tev}}[.9\linewidth]{
    \begin{tabular}{cc|ccccc}
    \hline\hline
	 & bin $\alpha$ &
     	 0 & 1 & 2 & 3 & 4 \\ \hline
	$H_{T,0}^{\rm max}(\alpha)$ & (GeV) & 
	$900$ & $2100$ & $4000$ & $6400$ & $\infty$ \\
	%$\sigma_{\text{no-matching}}(\alpha)$ & (pb) &
	%$1.22{\scriptstyle \times10^3}$ & $91.4$ & $8.56$ & 
	%$0.911$  & $0.166$  \\
	$\sigma_{\text{matched}}(\alpha)$ & (pb) &
	$3.69{\scriptstyle \times10^2}$ & $25.8$ & $2.30$ & 
	$0.23$  & $0.04$  \\
	$N_{\text{matched}}(\alpha)$ & $(\times 10^3)$ &
	%$2400$ & $1840$ & $1720$ & $1840$ & $2400$ \\
	$728$ & $520$ & $462$ & $473$ & $590$ \\
	$\mathcal{L}_{\rm gen}(\alpha)$ & (${\rm ab}^{-1}$) &
	$1.97{\scriptstyle \times10^{-3}}$ & $0.0201$ & 
	$0.201$ & $2.02$ & $14.5$ \\
    \hline\hline
    \end{tabular}
}

    \vspace{3mm}
\centering
\subcaptionbox{$Z \rightarrow \ell\bar{\ell} + 0,1,2,3\,j$.  \label{table:bg_table_zl3_100tev}}[.9\linewidth]{
    \begin{tabular}{cc|ccccc}
    \hline\hline
	 & bin $\alpha$ &
     	 0 & 1 & 2 & 3 & 4 \\ \hline
	$H_{T,0}^{\rm max}(\alpha)$ & (GeV) & 
	$900$ & $2100$ & $4000$ & $6400$ & $\infty$ \\
	%$\sigma_{\text{no-matching}}(\alpha)$ & (pb) &
	%$2.97{\scriptstyle \times10^3}$ & $73.8$ & $6.16$ & 
	%$0.605$  & $0.102$  \\
	$\sigma_{\text{matched}}(\alpha)$ & (pb) &
	$1.20{\scriptstyle \times10^3}$ & $22.4$ & $1.70$ & 
	$0.16$  & $0.03$  \\
	$N_{\text{matched}}(\alpha)$ & $(\times 10^3)$ &
	%$300$ & $300$ & $300$ & $300$ & $300$  \\
	$121$ & $90.9$ & $82.9$ & $79.3$ & $75.2$  \\
	$\mathcal{L}_{\rm gen}(\alpha)$ & (${\rm ab}^{-1}$) &
	$1.01{\scriptstyle \times10^{-4}}$ & $4.06{\scriptstyle \times10^{-3}}$ & 
	$0.0487$ & $0.496$ & $2.95$ \\
    \hline\hline
    \end{tabular}
}
     \caption{
     The upper edge value of $H_{T,0}$,
     the leading-order matched cross section,
    % cross section after matching,
     the number of samples after matching,
     and equivalent
     luminosity in each $H_{T,0}$ bin for the $Z$-boson production process.
     }
     \label{table:bg_table_z_100tev}
  \end{table}
%------------------------------------------------------------------

%------------------------------------------------------------------
% W
\begin{table}
  \begin{center}
    \begin{tabular}{cc|cccccc}
    \hline\hline
	 & bin $\alpha$ &
     	 0 & 1 & 2 & 3 & 4 & 5\\ \hline
	$H_{T,0}^{\rm max}(\alpha)$ & (GeV) & 
	$600$ & $1500$ & $3000$ & $5000$ & $7500$ & $\infty$ \\
	%$\sigma_{\text{no-matching}}(\alpha)$ & (pb) &
	%$2.32{\scriptstyle \times10^4}$ & $1.63{\scriptstyle \times10^3}$ & $1.44{\scriptstyle \times10^2}$ & 
	%$14.5$ & $1.86$ & $0.379$ \\
	$\sigma_{\text{matched}}(\alpha)$ & (pb) &
	$9.62{\scriptstyle \times10^3}$ & $5.19{\scriptstyle \times10^2}$ & $41.7$ & 
	$3.93$ & $0.49$ & $0.09$ \\
	$N_{\text{matched}}(\alpha)$ & $(\times 10^3)$ &
	%$4600$ & $3200$ & $2800$ & $2800$ & $3800$ & $7800$   \\
	$1910$ & $1020$ & $813$ & $759$ & $993$ & $1950$   \\
	$\mathcal{L}_{\rm gen}(\alpha)$ & (${\rm ab}^{-1}$) &
	$1.99{\scriptstyle \times10^{-4}}$ & $1.97{\scriptstyle \times10^{-3}}$ & 
	$0.0195$ & $0.193$ & $2.05$ & $20.6$ \\
    \hline\hline
    \end{tabular}
    \caption{
     Same as in Table~\ref{table:bg_table_z_100tev} but for the $W$-boson
   production process.
     }
     \label{table:bg_table_w_100tev}
  \end{center}
  \end{table}
%------------------------------------------------------------------

%------------------------------------------------------------------
% tt
\begin{table}
\centering
\subcaptionbox{$t\bar{t}\rightarrow$ (semi-leptonic)$+ 0,1 j$.  \label{table:bg_table_tt_jl_100tev}}[1.\linewidth]{
    \begin{tabular}{cc|ccccccc}
    \hline\hline
	 & bin $\alpha$ &
     	 0 & 1 & 2 & 3 & 4 & 5 & 6\\ \hline
	$H_{T,0}^{\rm max}(\alpha)$ & (GeV) & 
	$800$ & $1600$ & $2800$ & $4600$ & $7200$ & $10000$ & $\infty$ \\
	%{\small $\sigma_{\text{no-matching}}(\alpha)$} & (pb) &
	%$6.56{\scriptstyle \times10^3}$ & $4.73{\scriptstyle \times10^2}$ &
	 %$42.4$ & $4.17$ & 
	 %$0.406$ & $0.0359$ & $6.10{\scriptstyle \times10^{-3}}$\\
	 {\small $\sigma_{\text{matched}}(\alpha)$} & (pb) &
	$4.40{\scriptstyle \times10^3}$ & $3.54{\scriptstyle \times10^2}$ &
	 $33.7$ & $3.40$ & 
	 $0.34$ & $0.03$ & $5.18{\scriptstyle \times10^{-3}}$\\
	$N_{\text{matched}}(\alpha)$ & $(\times 10^3)$ &
	%$80$ & $80$ & $80$ & $80$ & $80$ & $80$ & $80$\\
	$53.6$ & $59.8$ & $63.5$ & $65.3$ & $66.2$ & $67.2$ & $68$\\
	$\mathcal{L}_{\rm gen}(\alpha)$ & (${\rm ab}^{-1}$) &
	$1.22{\scriptstyle \times10^{-5}}$ & $1.69{\scriptstyle \times10^{-4}}$ &
	 $1.89{\scriptstyle \times10^{-3}}$ & $0.0192$ & $0.197$ & $2.23$ & $13.1$\\
    \hline\hline
    \end{tabular}
}

    \vspace{3mm}
\centering
\subcaptionbox{$t\bar{t}\rightarrow$ (leptonic)$+ 0,1 j$.  \label{table:bg_table_tt_ll_100tev}}[.9\linewidth]{
    \begin{tabular}{cc|ccccc}
    \hline\hline
	 & bin $\alpha$ &
     	 0 & 1 & 2 & 3 & 4 \\ \hline
	%$H_{T,0}^{\rm max}(\alpha)$ & (GeV) & 
	$H_{T,0}^{\rm max}(\alpha)$ & (GeV) & 
	$1100$ & $2300$ & $3900$ & $6000$ & $\infty$ \\
	%$\sigma_{\text{ no-matching}}(\alpha)$ & (pb) &
	%$72.1$ & $7.23$ & 
	%$0.603$ & $0.0589$ & $7.30{\scriptstyle \times10^{-3}}$ \\
	$\sigma_{\text{matched}}(\alpha)$ & (pb) &
	$54.4$ & $6.57$ & 
	$0.56$ & $0.06$ & $6.90{\scriptstyle \times10^{-3}}$ \\
	$N_{\text{matched}}(\alpha)$ & $(\times 10^3)$ &
	%$80$ & $80$ & $80$ & $80$ & $80$\\
	$60.4$ & $72.6$ & $74.2$ & $74.8$ & $75.7$\\
	$\mathcal{L}_{\rm gen}(\alpha)$ & (${\rm ab}^{-1}$) &
	$1.11{\scriptstyle \times10^{-3}}$ & $1.11{\scriptstyle \times10^{-2}}$ & 
	$0.133$ & $1.36$ & $11.0$\\
    \hline\hline
    \end{tabular}
}
     \caption{\small
     Same as in Table~\ref{table:bg_table_z_100tev} but for the
 $t\bar{t}$ production process. 
     }
     \label{table:bg_table_tt_100tev}
  \end{table}
%------------------------------------------------------------------

For MC simulation of a 100~TeV collider, we basically follow the same
procedure as described in Sec.~\ref{sec:mc}. We list the sample points
for signal events in Table~\ref{table:sample_point_100tev},\footnote{For
$m_{\tilde{g}}=18$ TeV, we only generate samples with
$m_{\tilde{\chi}_1^0} =100$ GeV. }
with the same categorization as in Table~\ref{table:sample_point}. We
again generate 50000 events for each sample point. For the SM background
processes, we again focus on the $Z$, $W$, and $t\bar{t}$ production
processes, which turn out to be dominant~\cite{Cohen:2013xda}. We
carry out simulations with up to three and one additional partons for the
$Z/W$ and $t\bar{t}$ production processes, respectively. 
In generating background samples, we divide the generator-level phase
space in terms of $H_{T,0}$; we show the binning of this division in 
Tables~\ref{table:bg_table_z_100tev}--\ref{table:bg_table_tt_100tev}
for each process with 
the leading-order matched cross section,
%the cross section after matching,
the number of samples after matching, 
%the number of MC samples we generate, 
and the equivalent luminosity.

%%%%%%%%%%%%%% FIGURE %%%%%%%%%%%%%%%%%%%%%%%%%%%%%%%%%%%%
\begin{figure}
  \centering
  \subcaptionbox{\label{fig:meff_valid_z_100tev} $Z$ boson production}{
  \includegraphics[width=0.48\columnwidth]{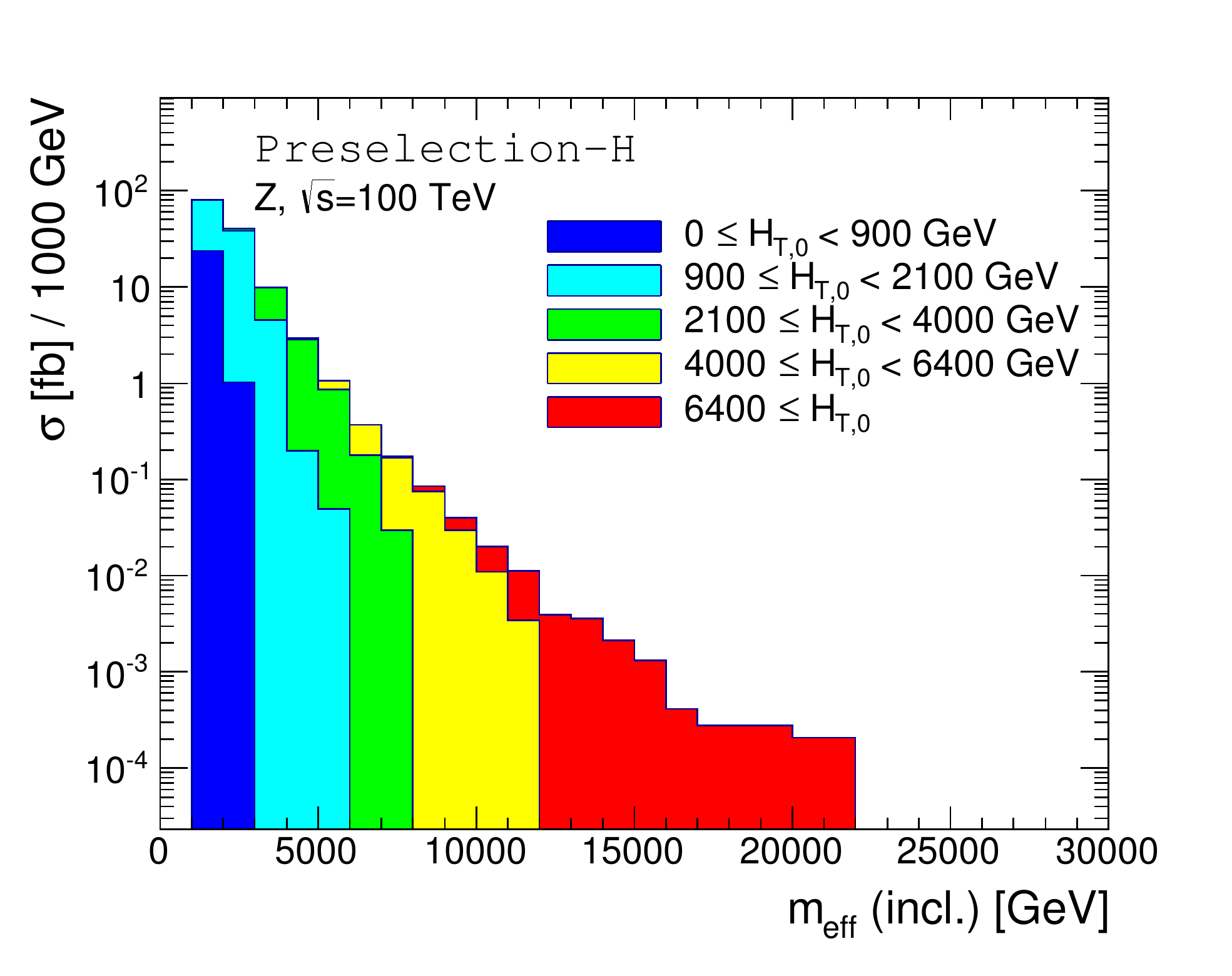}}
  \subcaptionbox{\label{fig:meff_valid_w_100tev} $W$ boson production}{
  \includegraphics[width=0.48\columnwidth]{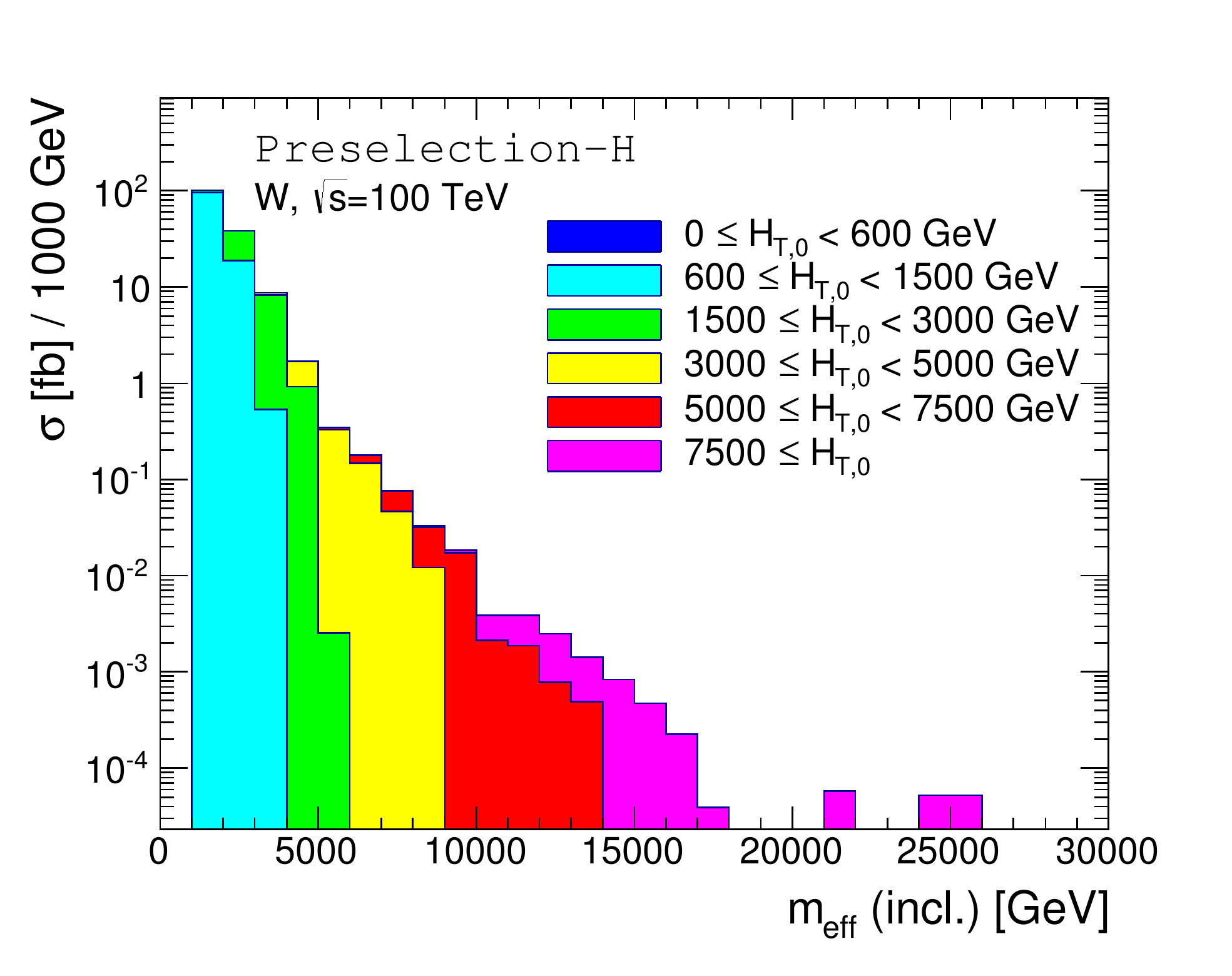}}
\caption{Distributions of $m_{\text{eff}} (\text{incl.})$ with {\tt
 preselection-H} required. Contributions from different $H_{T,0}$ bins
 are filled with different colors.} 
  \label{fig:meff_valid_wz_100tev}
\end{figure}
%%%%%%%%%%%%%%%%%%%%%%%%%%%%%%%%%%%%%%%%%%%%%%%%%%%%%%%%%%

%%%%%%%%%%%%%% FIGURE %%%%%%%%%%%%%%%%%%%%%%%%%%%%%%%%%%%%
\begin{figure}
  \centering
  \subcaptionbox{\label{fig:meff_valid_tt_semil_100tev} Semi-leptonic $t\bar{t}$ decay}{
  \includegraphics[width=0.48\columnwidth]{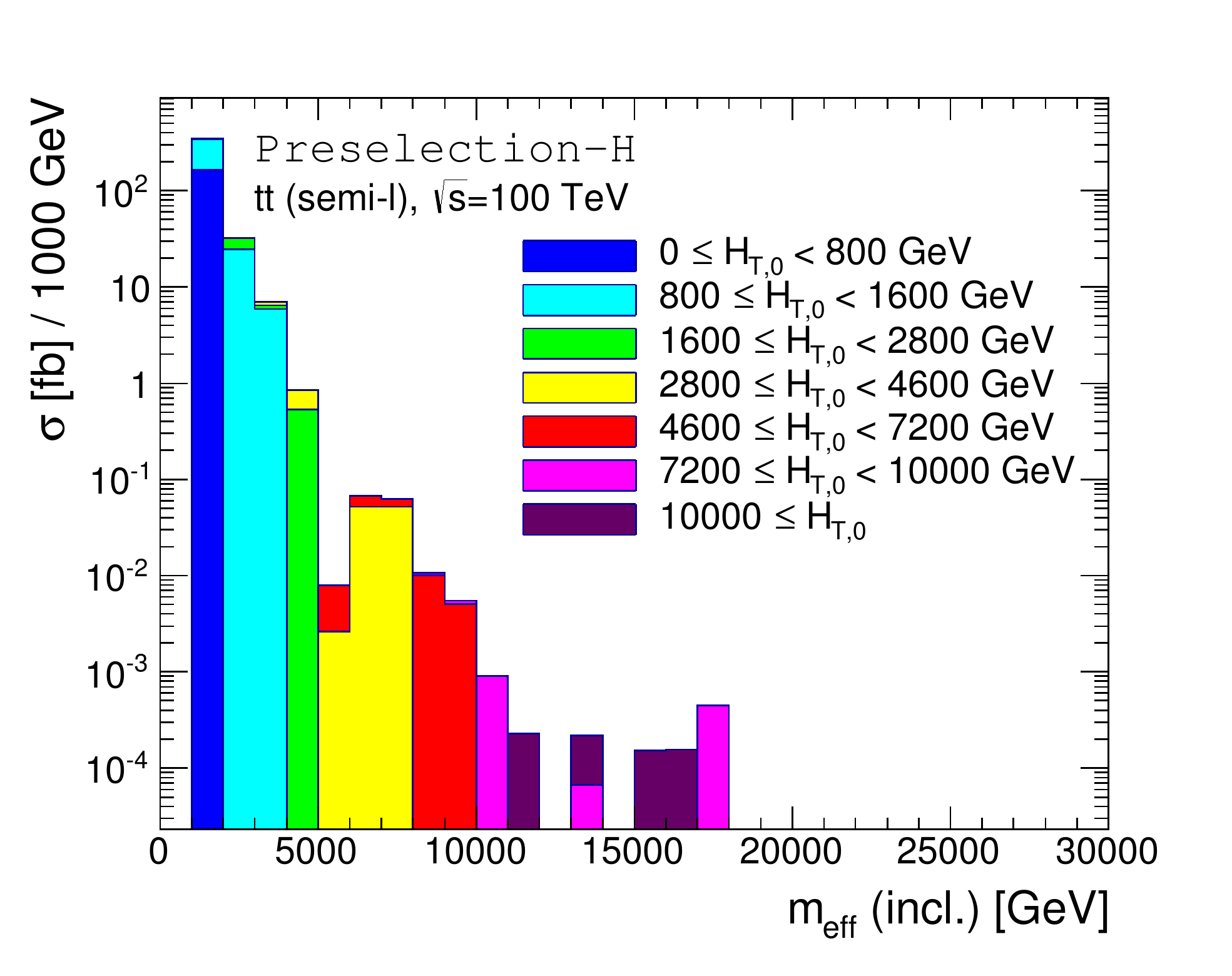}}
  \subcaptionbox{\label{fig:meff_valid_tt_ll_100tev} Leptonic $t\bar{t}$ decay}{
  \includegraphics[width=0.48\columnwidth]{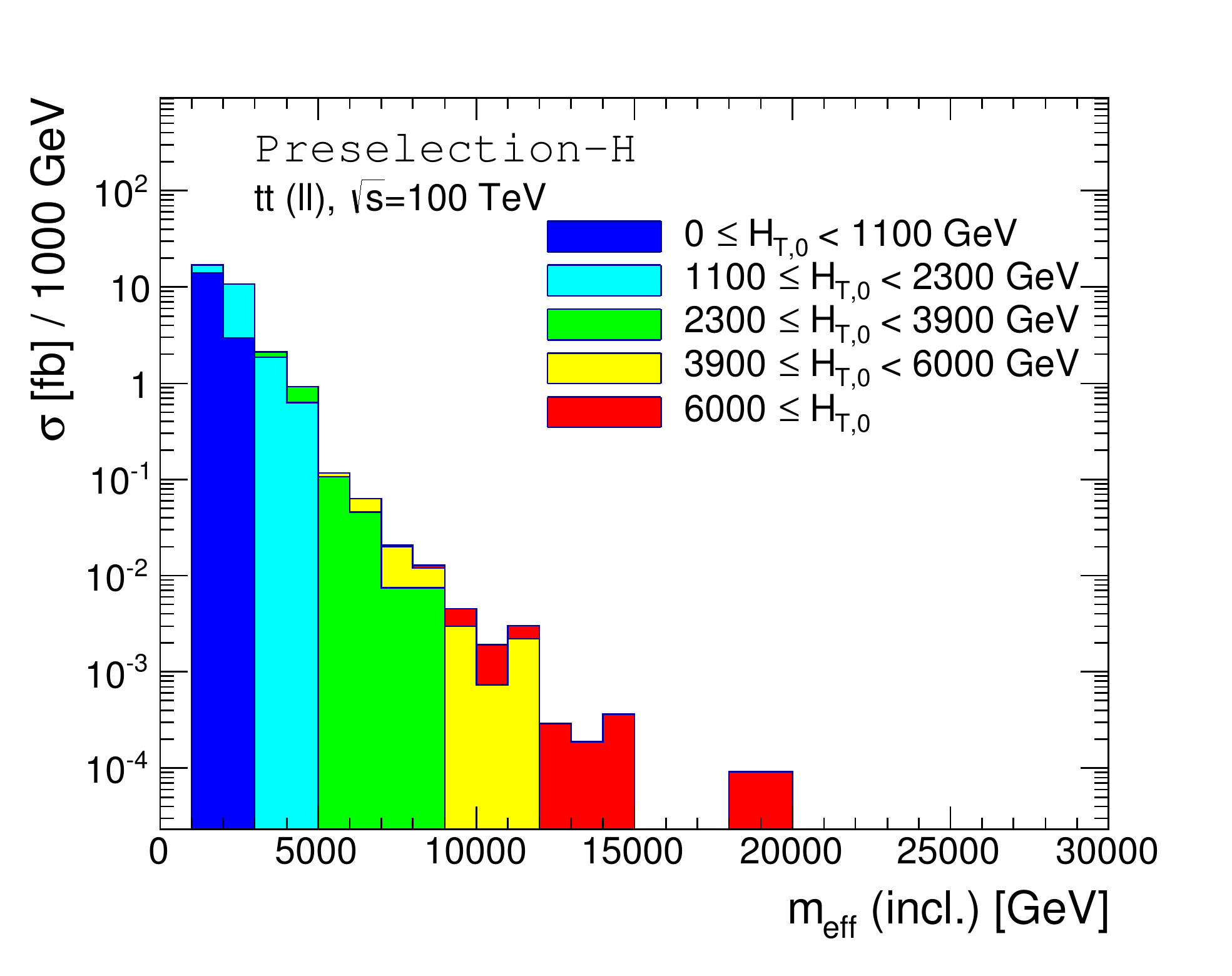}}
\caption{Same as in Fig.~\ref{fig:meff_valid_wz_100tev}, but for the $t\bar{t}$
 production process.} 
  \label{fig:meff_valid_tt_100tev}
\end{figure}
%%%%%%%%%%%%%%%%%%%%%%%%%%%%%%%%%%%%%%%%%%%%%%%%%%%%%%%%%%

Using these samples, we generate distributions of $m_{\rm eff}({\rm
incl.})$ for each process with {\tt Preselection-H} imposed. 
For reconstructed objects such as jets, charged leptons, and charged
tracks, we use the same criteria as in Sec.~\ref{sec:mc}. The resultant
distributions are shown  in
Figs.~\ref{fig:meff_valid_wz_100tev} and \ref{fig:meff_valid_tt_100tev}. 
Again, we see a correlation between $H_{T,0}$ and $m_{\rm eff}({\rm
incl.})$ and a sizable number of events in the tails, which justifies
our way of dividing the phase space in terms of $H_{T,0}$. 

To validate our MC simulation, we have compared the expected number of
events obtained by our MC simulation with that given in
Ref.~\cite{Cohen:2013xda}, with the same selection cuts adopted
there. We have found a fairly good agreement ($\sim 20\%$ level) with
each other over the signal regions.

%%%%%%%%%%%%%%%%%%%%%%%%%%%%%%%%%%%%%%%%%%
\subsection{Event Selection}
\label{sec:event_selection_100tev}
%%%%%%%%%%%%%%%%%%%%%%%%%%%%%%%%%%%%%%%

%%%%%%%%%%%%%% FIGURE %%%%%%%%%%%%%%%%%%%%%%%%%%%%%%%%%%%%
\begin{figure}
  \centering
  \subcaptionbox{\label{fig:meff_L_100tev} \texttt{Preselection-L}}{
  \includegraphics[width=0.48\columnwidth]{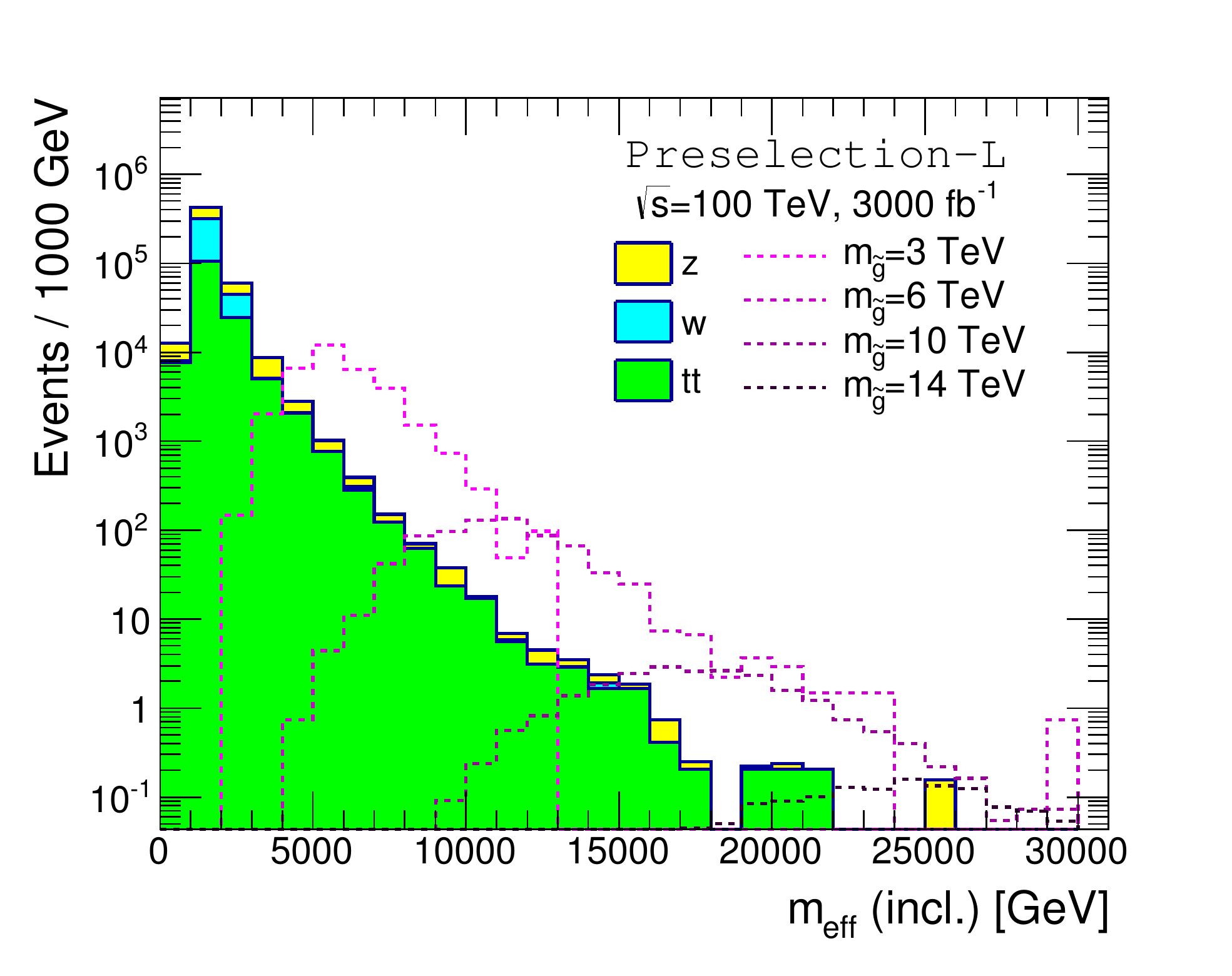}}
  \subcaptionbox{\label{fig:meff_M_100tev} \texttt{Preselection-M}}{
  \includegraphics[width=0.48\columnwidth]{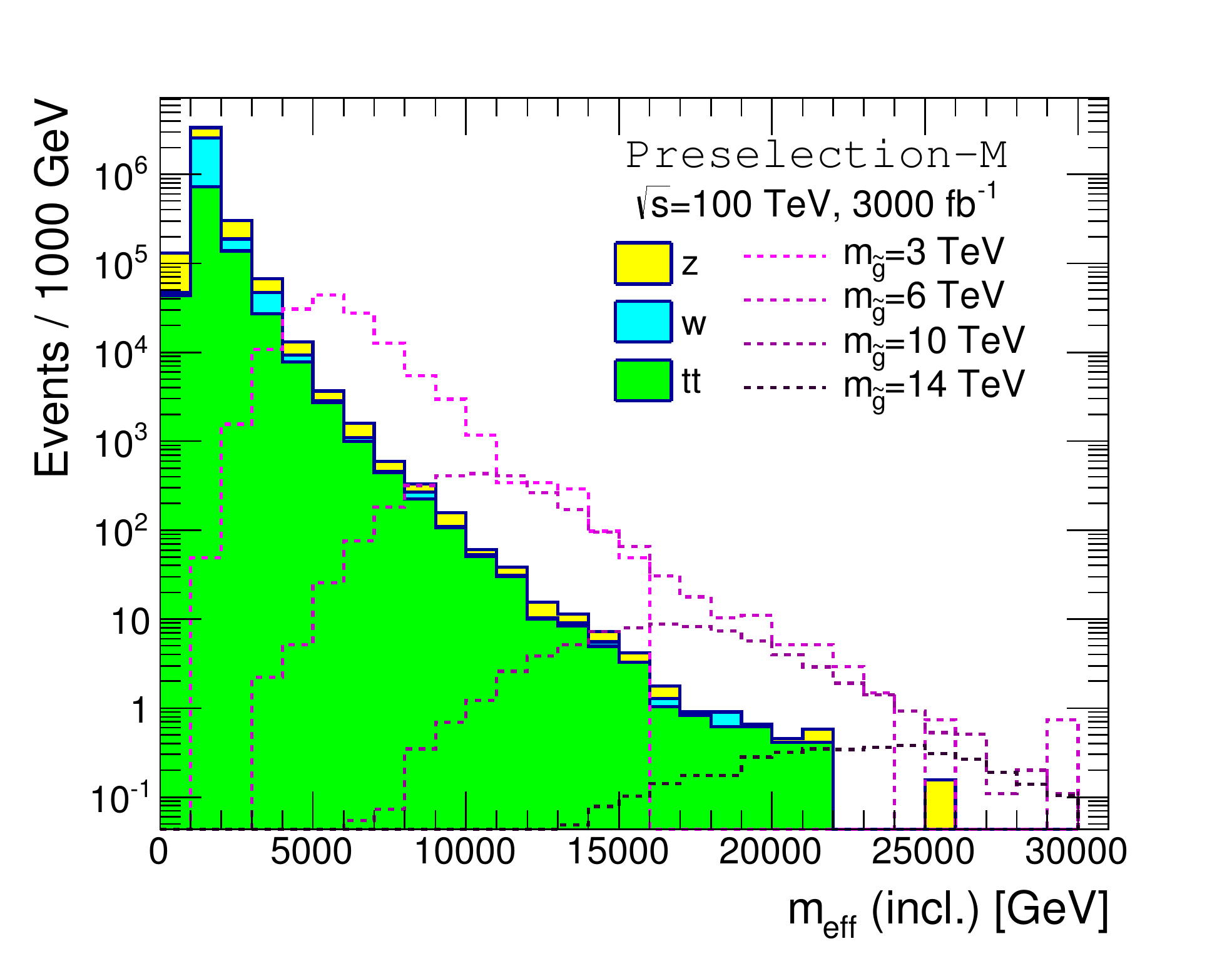}}
  \subcaptionbox{\label{fig:meff_H_100tev} \texttt{Preselection-H}}{
  \includegraphics[width=0.48\columnwidth]{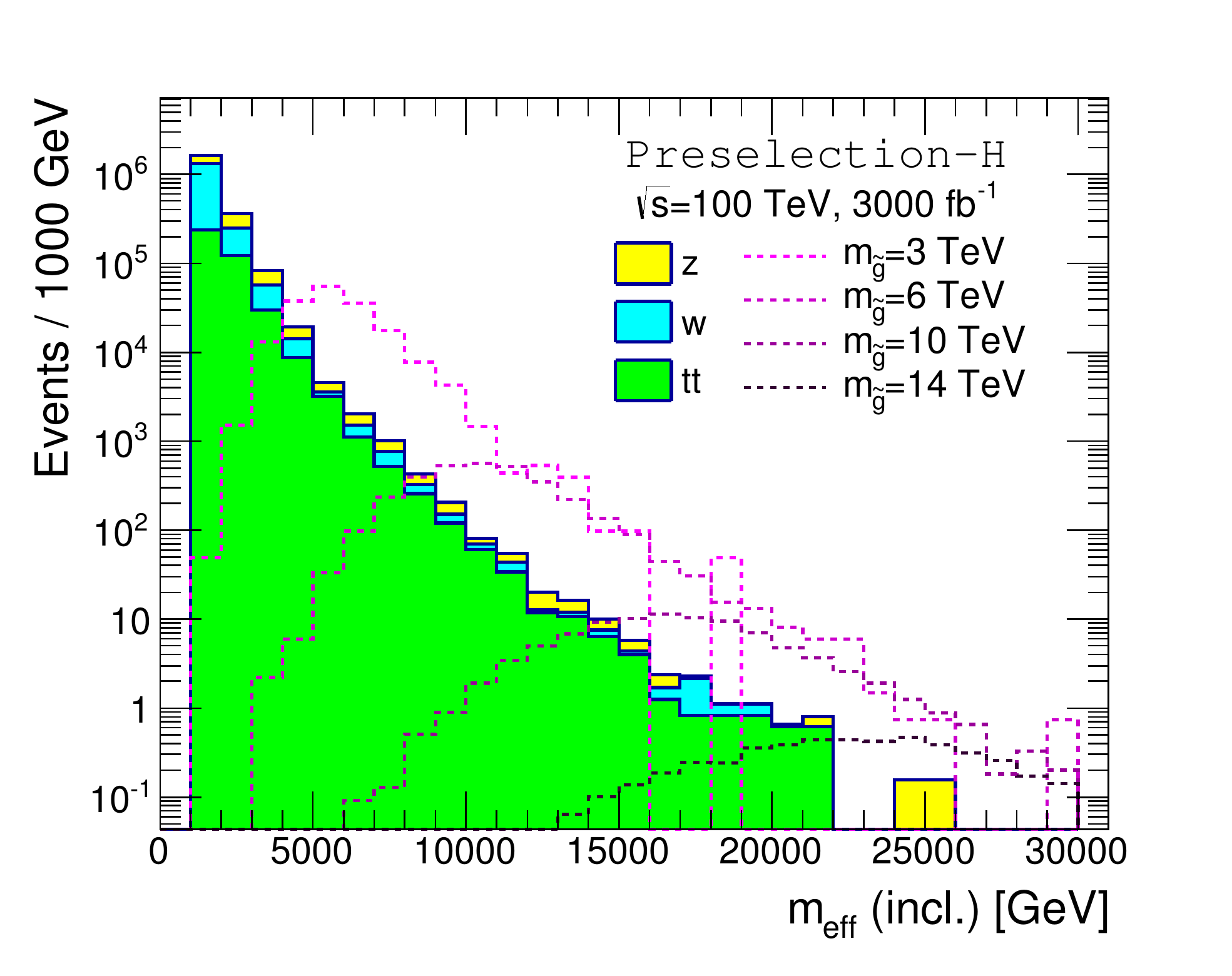}}
\caption{Distributions of $m_{\text{eff}} (\text{incl.})$ for the SM
 background and gluino signal events. We set the LSP mass to be 100~GeV. } 
  \label{fig:meff_LMH_100tev}
\end{figure}
%%%%%%%%%%%%%%%%%%%%%%%%%%%%%%%%%%%%%%%%%%%%%%%%%%%%%%%%%%

For the preselection, we consider the same criteria as in
Table~\ref{table:preselection}. These criteria may be further
optimized for a 100~TeV collider, but we do not discuss this possibility
in this paper. We again impose the lepton and detector-material vetoes. 
In Fig.~\ref{fig:meff_LMH_100tev}, we show the distributions of $m_{\rm
eff}({\rm incl.})$ for the SM background and gluino signal events with
each preselection imposed. We set the LSP mass to be 100~GeV in these
plots.

%%%%%%%%%%%%%% FIGURE %%%%%%%%%%%%%%%%%%%%%%%%%%%%%%%%%%%%
\begin{figure}
  \centering
  \subcaptionbox{\label{fig:rDV_pdf_3tev_100tev} $m_{\tilde{g}} = 3$~TeV
 }{
  \includegraphics[width=0.48\columnwidth]{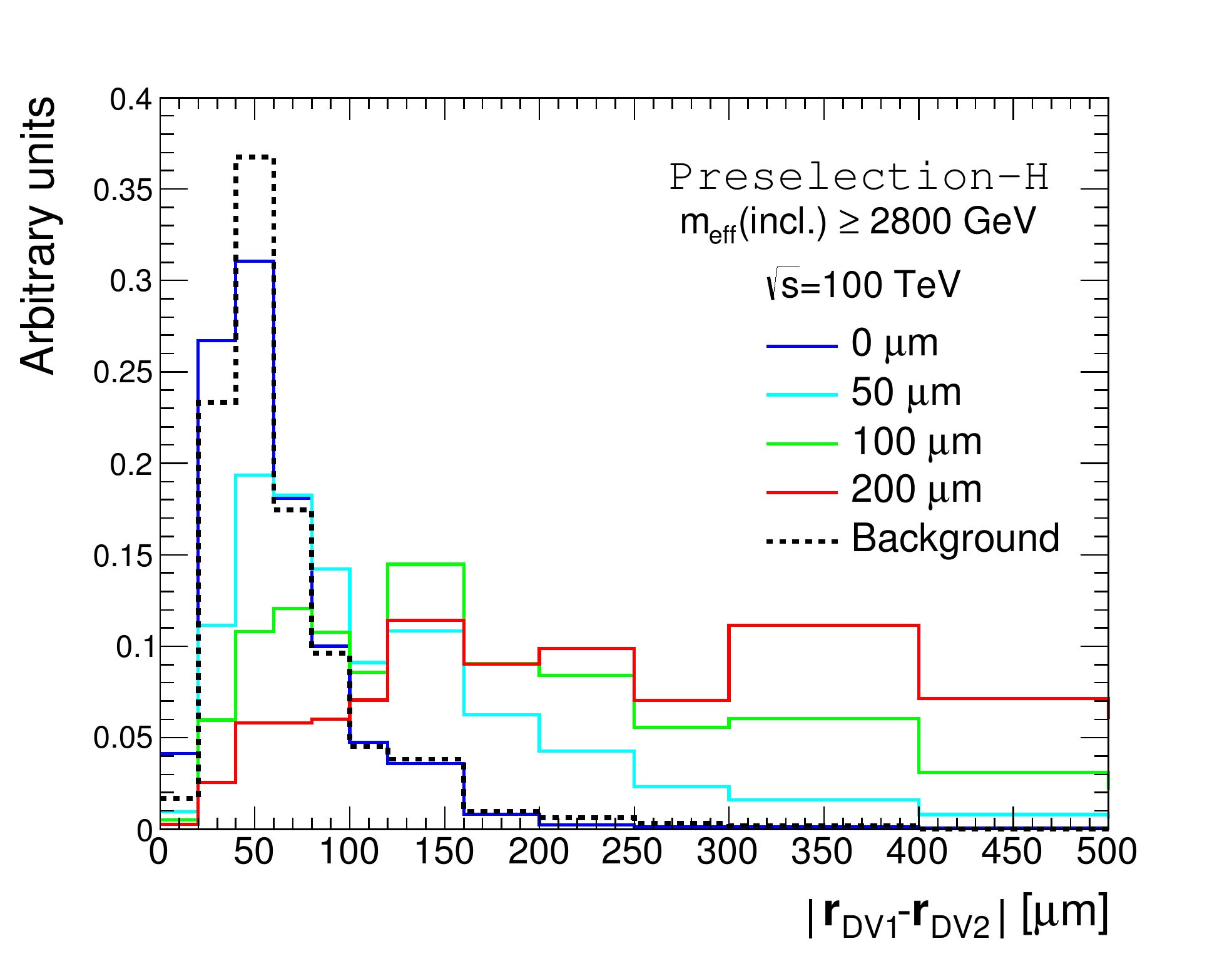}}
  \subcaptionbox{\label{fig:rDV_pdf_14tev_100tev} $m_{\tilde{g}} =
 14$~TeV}{
  \includegraphics[width=0.48\columnwidth]{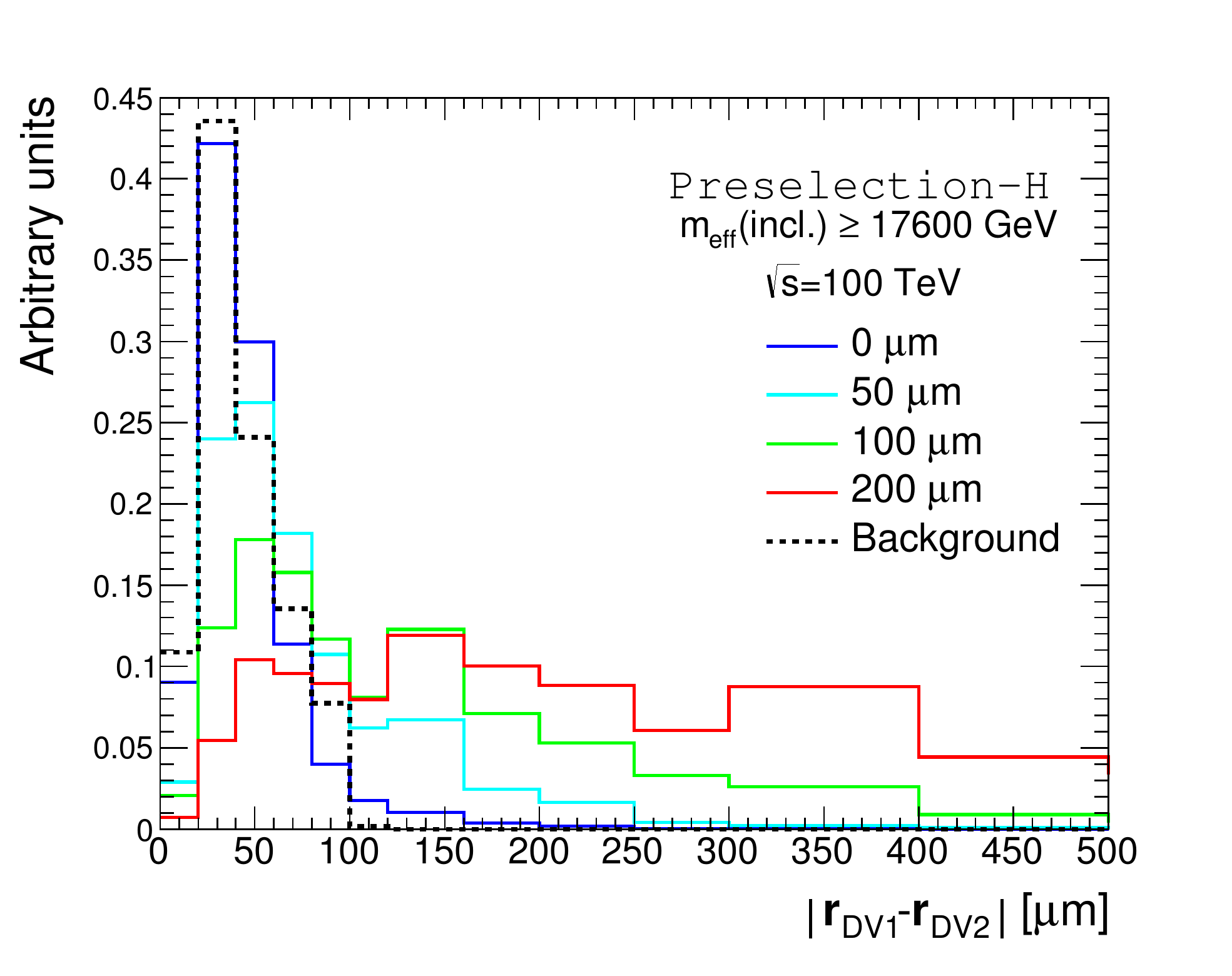}}
\caption{The distributions of $|\bm{r}_{\rm DV1} - \bm{r}_{\rm DV2}|$
 for a gluino with different values of
 $c\tau_{\tilde g}$, shown in the solid lines. The distributions for the
 SM background events are also shown in the dotted lines. We have
 imposed {\tt preselection-H}, and $m_{\rm eff}({\rm incl.}) > 2800$~GeV
 (17600~GeV) in Fig.~\ref{fig:rDV_pdf_3tev_100tev}
 (Fig.~\ref{fig:rDV_pdf_14tev_100tev}).  }  
  \label{fig:rDV_pdf_100tev}
\end{figure}
%%%%%%%%%%%%%%%%%%%%%%%%%%%%%%%%%%%%%%%%%%%%%%%%%%%%%%%%%%

%%%%%%%%%%%%%% FIGURE %%%%%%%%%%%%%%%%%%%%%%%%%%%%%%%%%%%%
\begin{figure}
  \centering
  \subcaptionbox{\label{fig:rDV_cdf_3tev_100tev} $m_{\tilde{g}} = 3$~TeV}{
  \includegraphics[width=0.48\columnwidth]{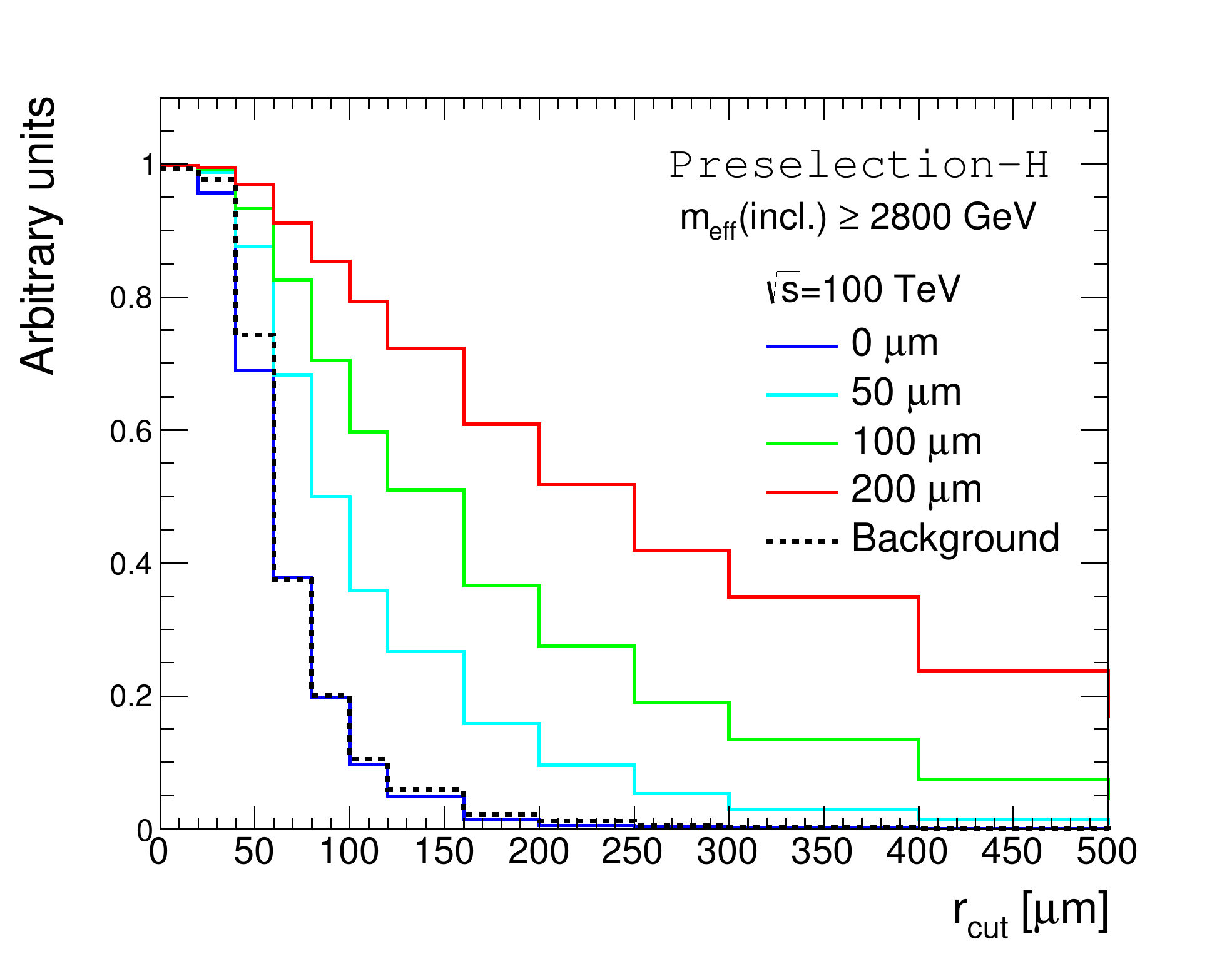}}
  \subcaptionbox{\label{fig:rDV_cdf_14tev_100tev} $m_{\tilde{g}} = 14$~TeV}{
  \includegraphics[width=0.48\columnwidth]{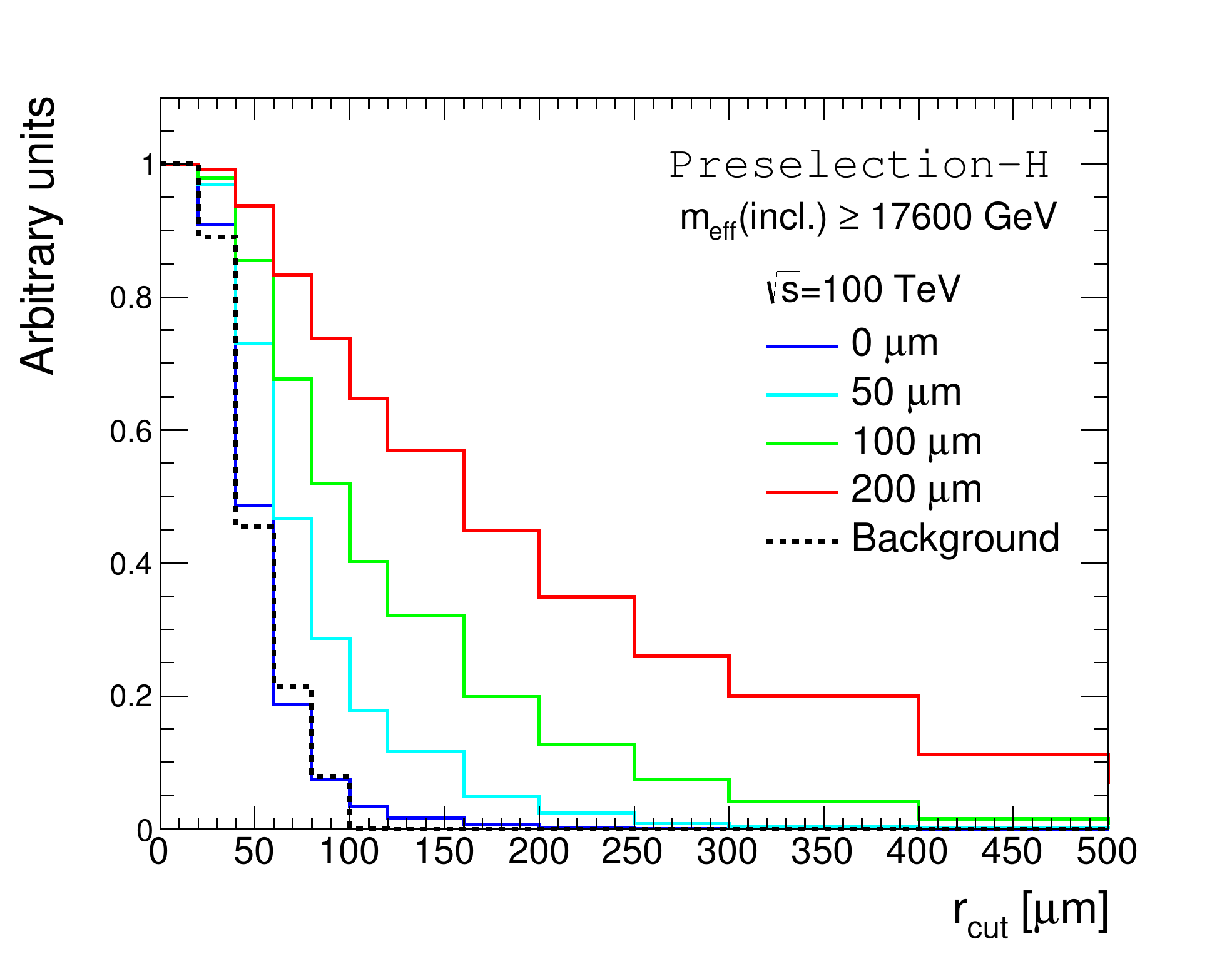}}
\caption{Fractions of events which pass the selection cut $|\bm{r}_{\rm DV1} -
 \bm{r}_{\rm DV2}| > r_{\rm cut}$ for a gluino with different values of
 $c\tau_{\tilde g}$, shown in the solid lines. The distributions for the
 SM background events are also shown in the dotted lines. We have
 imposed {\tt preselection-H}, and $m_{\rm eff}({\rm incl.}) > 2800$~GeV
 (17600~GeV) in Fig.~\ref{fig:rDV_cdf_3tev_100tev}
 (Fig.~\ref{fig:rDV_cdf_14tev_100tev}). }  
  \label{fig:rDV_cdf_100tev}
\end{figure}
%%%%%%%%%%%%%%%%%%%%%%%%%%%%%%%%%%%%%%%%%%%%%%%%%%%%%%%%%%

As we discuss in Sec.~\ref{sec:vertex}, the resolution of the vertex
reconstruction highly depends on the track reconstruction performance of
a detector. Due to a lack of concrete information on detectors at future
100~TeV colliders, in the following analysis, we just assume the same
track-resolution parameters as those given in Sec.~\ref{sec:eventsel},
and reconstruct vertices in the same manner as before. 
We show the distributions of $|\bm{r}_{\rm DV1} - \bm{r}_{\rm DV2}|$ in
the solid lines for a gluino with different values of $c\tau_{\tilde g}$
and a mass of 3~TeV and 14~TeV in Figs.~\ref{fig:rDV_pdf_3tev_100tev}
and ~\ref{fig:rDV_pdf_14tev_100tev}, respectively. Moreover, we show
in Figs.~\ref{fig:rDV_cdf_3tev_100tev} and
\ref{fig:rDV_cdf_14tev_100tev} fractions of events which pass a
selection cut of $|\bm{r}_{\rm DV1} - \bm{r}_{\rm DV2}| > r_{\rm cut}$
as functions of $r_{\rm cut}$.  The
distributions for the SM background events are also shown in the dotted
lines. Here, we have imposed {\tt preselection-H}, and $m_{\rm eff}({\rm
incl.}) > 2800$~GeV (17600~GeV) in the left
(right) panel. By comparing these figures with Fig.~\ref{fig:rDV}, we
clearly see the effect of Lorentz boost of metastable particles on the
displaced-vertex reconstruction. 
In Fig.~\ref{fig:rDV}, the distribution for the $c\tau_{\tilde{g}} =
50~{\rm \mu m}$ case is found to be fairly close to those for $c\tau_{\tilde{g}} =
0~{\rm \mu m}$ and the SM background. On the other hand, as seen in
Figs.~\ref{fig:rDV_pdf_3tev_100tev} and \ref{fig:rDV_cdf_3tev_100tev}, 
we can easily distinguish $c\tau_{\tilde{g}} = 50~{\rm \mu m}$ 
from $c\tau_{\tilde{g}} = 0~{\rm \mu m}$ and the SM background at a
100~TeV collider. For a heavier gluino, however, the separation becomes
less clear due to reduction in the boost factor.

%%%%%%%%%%%%%%%%%%%%%%%%%%%%
\subsection{Prospects}
\label{sec:reach_100tev}
%%%%%%%%%%%%%%%%%%%%%%%%%%%%%

%%%%%%%%%%%%%% FIGURE %%%%%%%%%%%%%%%%%%%%%%%%%%%%%%%%%%%%
\begin{figure}
  \centering
  \subcaptionbox{\label{fig:optimal_0mu_100tev} $c\tau_{\tilde{g}}=0~{\rm \mu m}$}{
  \includegraphics[width=0.48\columnwidth]{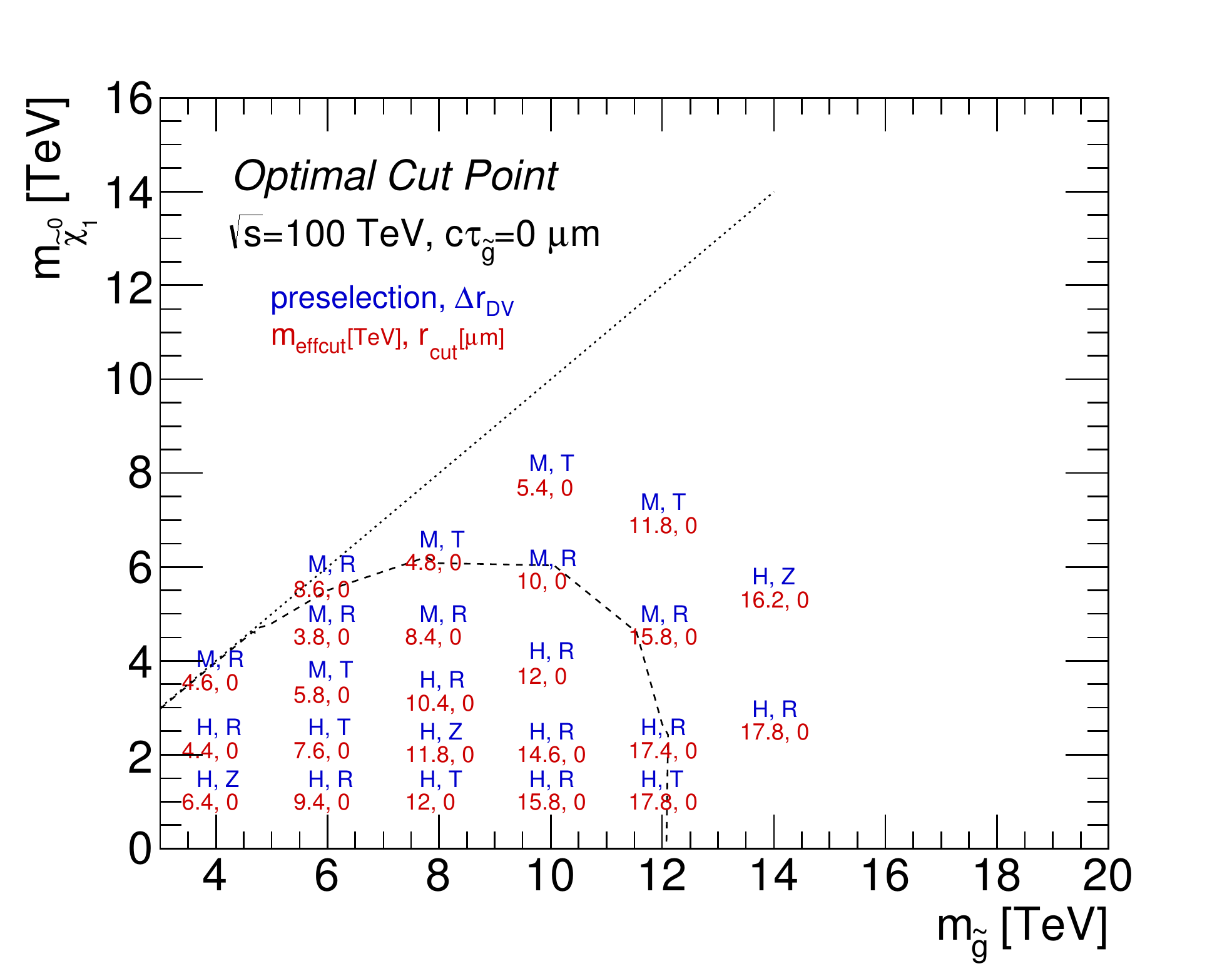}}
  \subcaptionbox{\label{fig:optimal_100mu_100tev} $c\tau_{\tilde{g}}=100~{\rm \mu m}$}{
  \includegraphics[width=0.48\columnwidth]{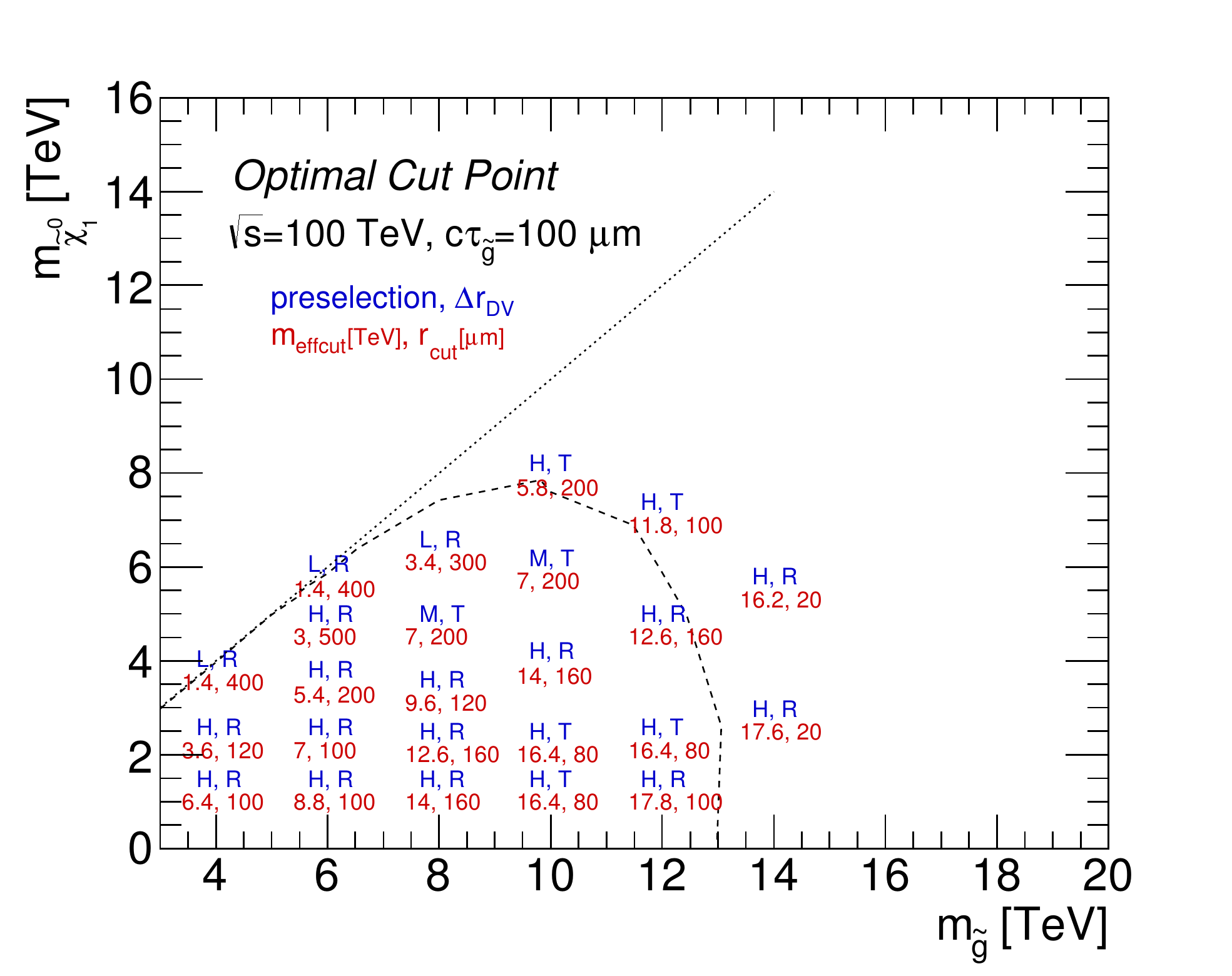}}
  \subcaptionbox{\label{fig:optimal_200mu_100tev} $c\tau_{\tilde{g}}=200~{\rm \mu m}$}{
  \includegraphics[width=0.48\columnwidth]{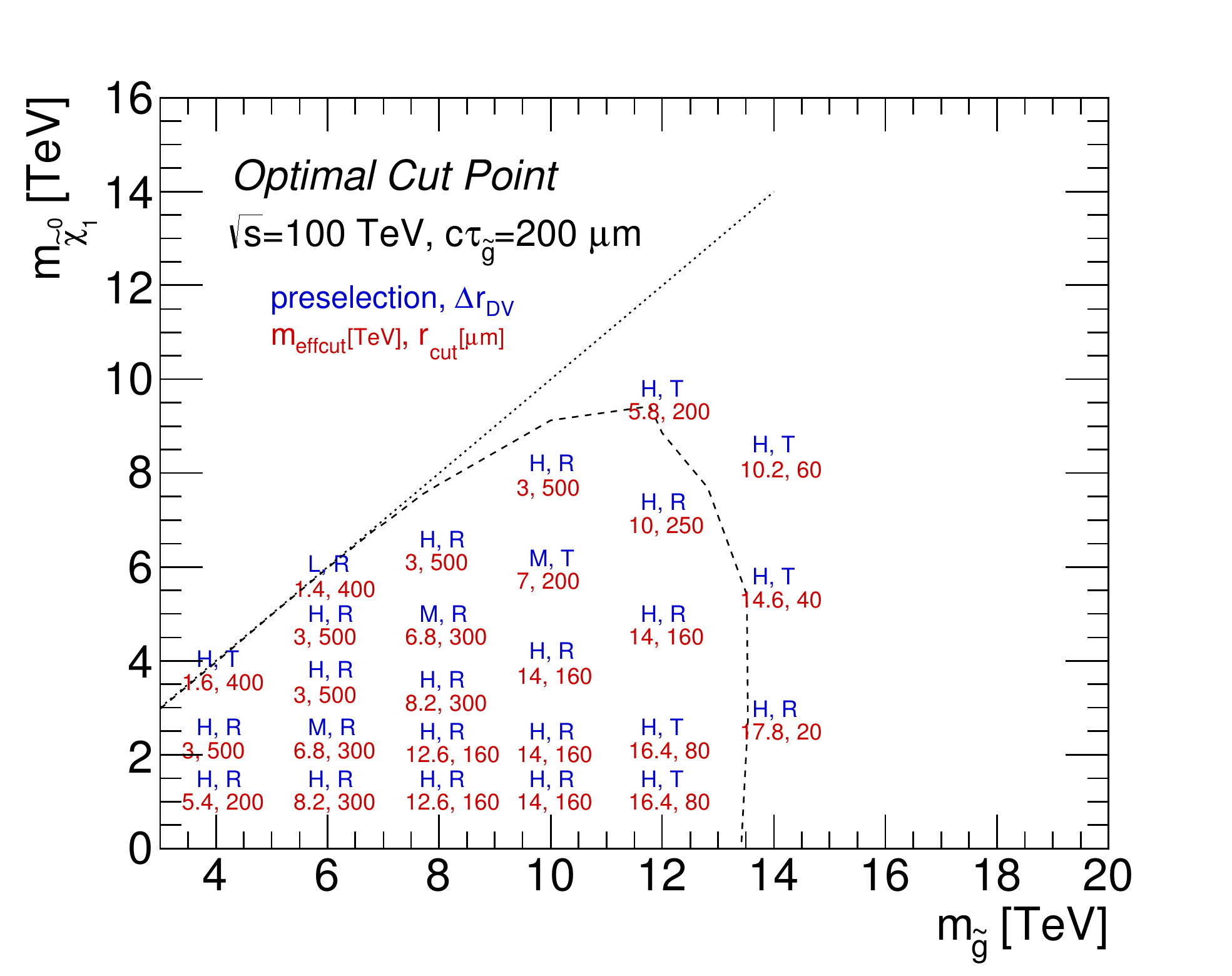}}
  \subcaptionbox{\label{fig:optimal_500mu_100tev} $c\tau_{\tilde{g}}=500~{\rm \mu m}$}{
  \includegraphics[width=0.48\columnwidth]{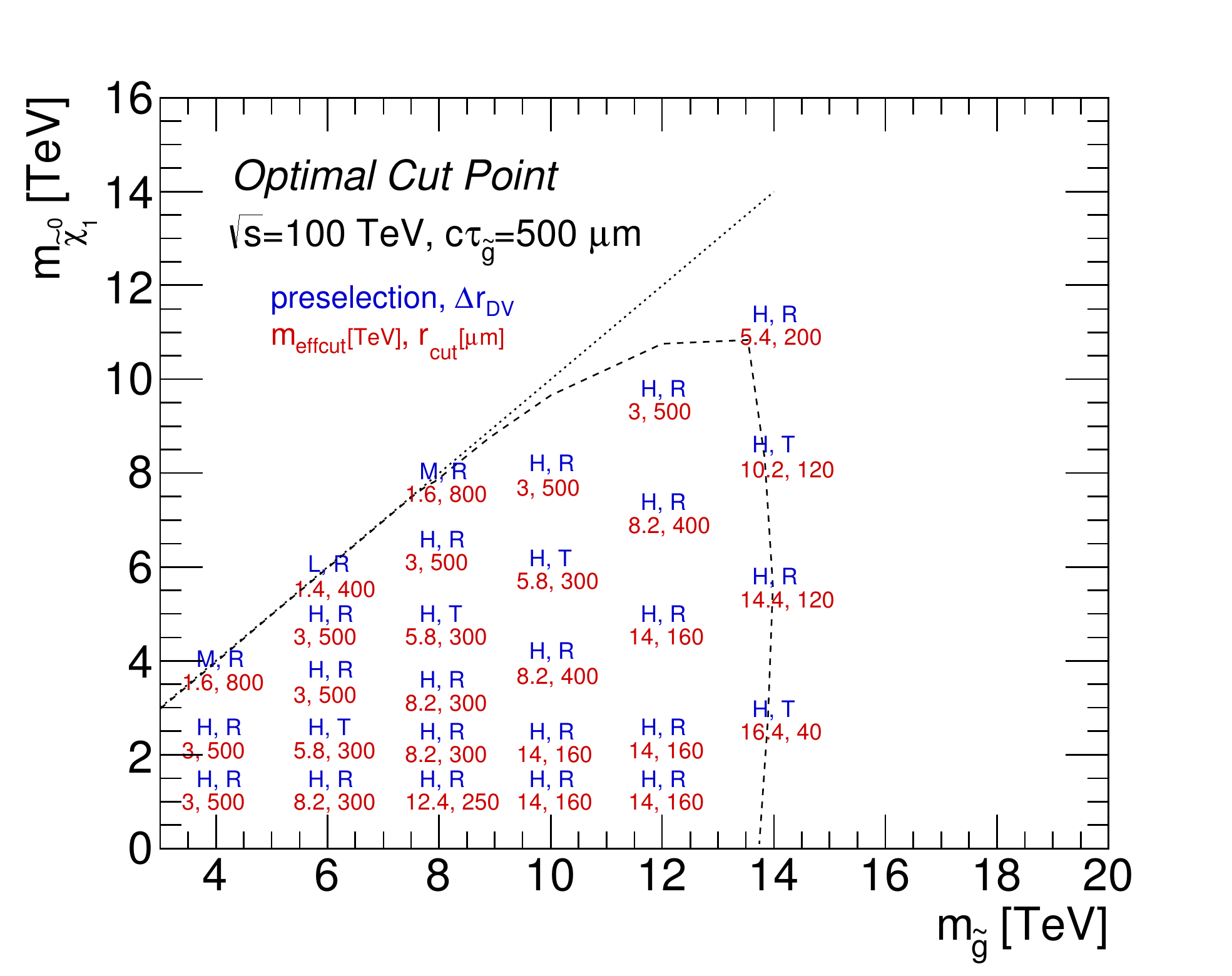}}
  \subcaptionbox{\label{fig:optimal_1000mu_100tev} $c\tau_{\tilde{g}}=1~{\rm mm}$}{
  \includegraphics[width=0.48\columnwidth]{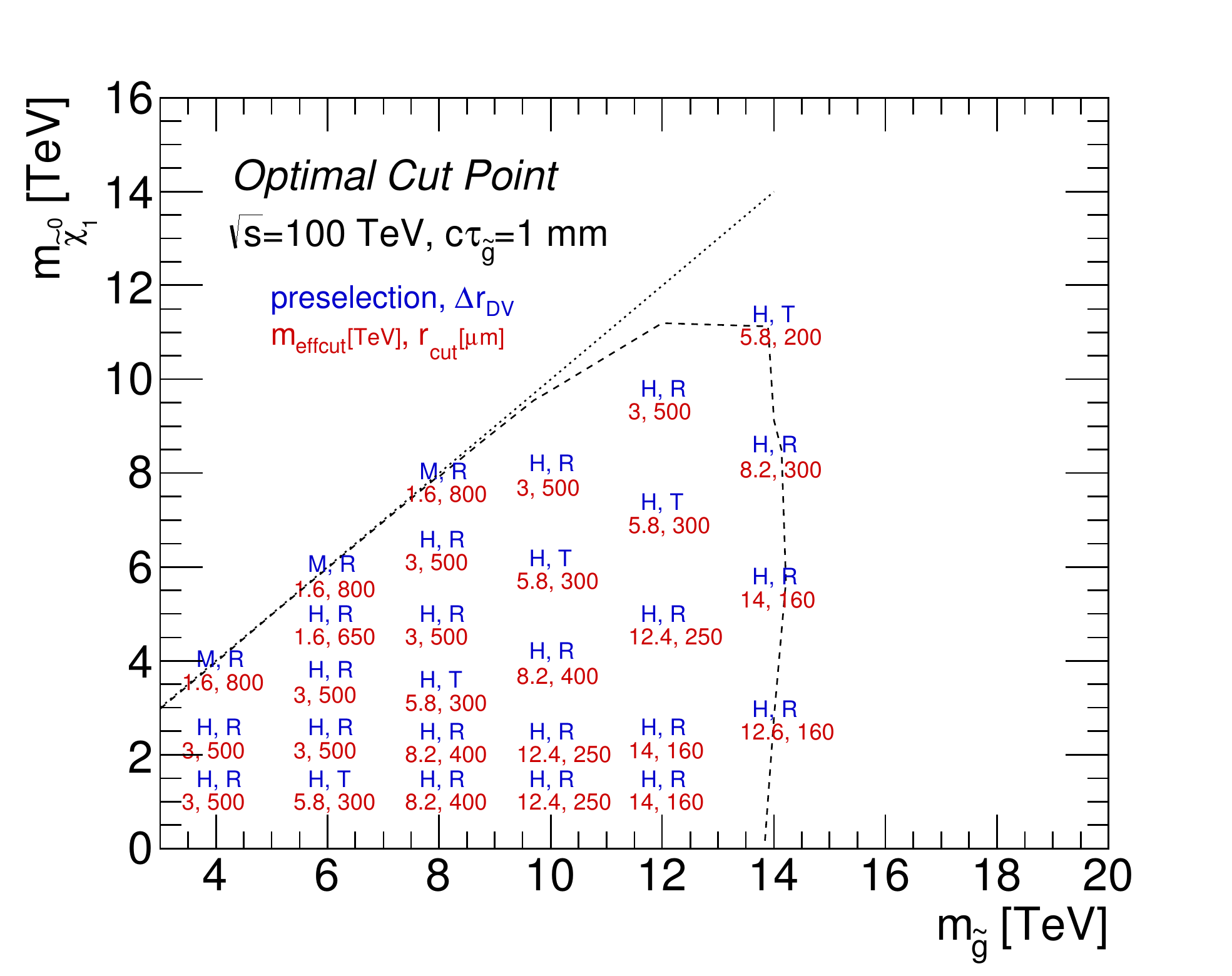}}
  \subcaptionbox{\label{fig:optimal_10000mu_100tev} $c\tau_{\tilde{g}}=10~{\rm mm}$}{
  \includegraphics[width=0.48\columnwidth]{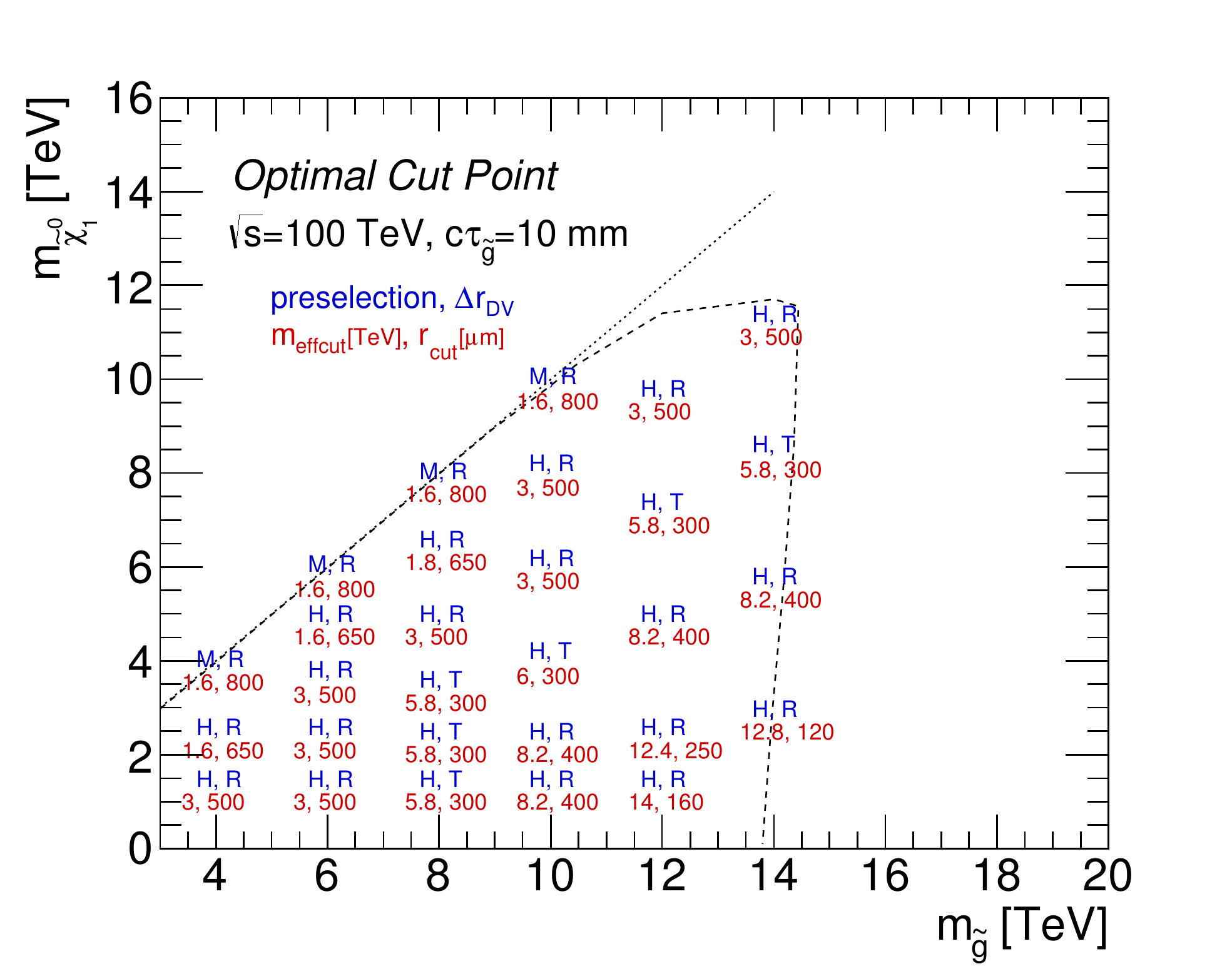}}
\caption{Optimal cut values ($(m_{\rm effcut})_{\rm optimal}$,
  $(r_{\rm cut})_{\rm optimal}$), the preselection ({\tt L}, {\tt M},
  {\tt H}), and the discriminator $\Delta r_{\rm DV}$ ({\tt R}, {\tt
    T}, {\tt Z}) for each sample point with various
  $c\tau_{\tilde{g}}$, with an integrated luminosity of $3000~{\rm
    fb}^{-1}$.  The expected $5\sigma$ discovery reaches for gluinos
  are also shown in the dashed lines.}
  \label{fig:optimal_100tev}
\end{figure}
%%%%%%%%%%%%%%%%%%%%%%%%%%%%%%%%%%%%%%%%%%%%%%%%%%%%%%%%%%

%-------------------------------
\begin{table}
  \centering
\subcaptionbox{The expected number of background events. }[.9\linewidth]{
    \begin{tabular}{l|ccc|c}
    \hline\hline
    & $Z$ & $W$ & $t\bar{t}$ & total \\
    \hline
    {\tt preselection-H}~~$(\times10^6)$& 
    $0.42{\scriptstyle~ \pm 0.01}$ &
    $0.45{\scriptstyle~ \pm 0.02}$ &
    $1.26{\scriptstyle~ \pm 0.4}$ &
    $2.13{\scriptstyle~ \pm 0.4}$\\
    $m_{\rm eff}({\rm incl.}) > 17.8$ TeV & 
    $3.1{\scriptstyle~ \pm 0.8}$ &
    $0.9{\scriptstyle~ \pm 0.3}$ &
    $0.5{\scriptstyle~ \pm 0.4}$ &
    $4.6{\scriptstyle~ \pm 0.9}$\\
    \hline
    $ \bigl| \bm{r}_{\rm DV1}-\bm{r}_{\rm DV2} \bigr| > 100~{\rm \mu m}$ & 
    $< 0.1$ &
    $< 0.1$ &
    $< 0.1$ &
    $< 0.1$ \\
    $ \bigl| \bm{r}_{\rm DV1}-\bm{r}_{\rm DV2} \bigr| > 200~{\rm \mu m}$ & 
    $< 0.1$ &
    $< 0.1$ &
    $< 0.1$ &
    $< 0.1$ \\
    \hline\hline
    \end{tabular}
}

    \vspace{3mm}
  \centering
\subcaptionbox{The expected number of signal events for 
    $m_{\tilde{g}}=12$~TeV, $m_{\tilde{\chi}_1^0}=100$~GeV with
    different values of $c\tau_{\tilde{g}}$. }[.9\linewidth]{
  \begin{tabular}{l|cccc}
    \hline\hline
    & $c\tau_{\tilde{g}}=0$ 
    & $c\tau_{\tilde{g}}=200~{\rm \mu m}$ & $c\tau_{\tilde{g}}=500~{\rm \mu m}$ & $c\tau_{\tilde{g}}=1~{\rm  mm}$ \\
    \hline
    {\tt preselection-H} & 
    \multicolumn{4}{c}{$20.5{\scriptstyle~\pm 0.3}$} \\
    $m_{\rm eff}({\rm incl.}) > 17.8$ TeV &
     \multicolumn{4}{c}{$14.5{\scriptstyle~\pm 0.2}$} \\
     \hline
    $ \bigl| \bm{r}_{\rm DV1}-\bm{r}_{\rm DV2}\bigr| > 100~{\rm \mu m}$ & 
    $0.5{\scriptstyle~ \pm 0.04}$ &
    $9.8{\scriptstyle~ \pm 0.2}$ &
    $12.5{\scriptstyle~ \pm 0.2}$ &
    $12.9{\scriptstyle~ \pm 0.2}$\\
    $ \bigl|\bm{r}_{\rm DV1}-\bm{r}_{\rm DV2} \bigr| > 200~{\rm \mu m}$ & 
    $< 0.1$ &
    $5.5{\scriptstyle~ \pm 0.1}$ &
    $9.6{\scriptstyle~ \pm 0.2}$ &
    $11.3{\scriptstyle~ \pm 0.2}$\\
    \hline\hline
    \end{tabular}
}
    \caption{
    The expected number of background (a) and signal (b) events for an
 integrated luminosity of $\mathcal{L}=3000~{\rm fb}^{-1}$. We set
    the masses of gluino and the LSP to be $12$~TeV and
    $100$~GeV, respectively.
    }
    \label{table:cut_flow_100tev}
  \end{table}
%-------------------------------

%-------------------------------
\begin{table}
  \centering
\subcaptionbox{The expected number of background events. }[.9\linewidth]{
    \begin{tabular}{l|ccc|c}
    \hline\hline
    & $Z$ & $W$ & $t\bar{t}$ & total \\
    \hline
    {\tt preselection-H}~~$(\times10^6)$& 
    $0.42{\scriptstyle~ \pm 0.01}$ &
    $0.45{\scriptstyle~ \pm 0.02}$ &
    $1.26{\scriptstyle~ \pm 0.4}$ &
    $2.13{\scriptstyle~ \pm 0.4}$\\
    $m_{\rm eff}({\rm incl.}) > 6$ TeV~~$(\times10^3)$ & 
    $2.2{\scriptstyle~ \pm 0.1}$ &
    $1.0{\scriptstyle~ \pm 0.1}$ &
    $0.8{\scriptstyle~ \pm 0.2}$ &
    $4.0{\scriptstyle~ \pm 0.3}$\\
    \hline
    $ \bigl| \bm{r}_{\rm DV1}-\bm{r}_{\rm DV2} \bigr| > 200~{\rm \mu m}$ & 
    $5.9{\scriptstyle~ \pm 3}$ &
    $5.7{\scriptstyle~ \pm 2.}$ &
    $2.2{\scriptstyle~ \pm 2}$ &
    $14{\scriptstyle~ \pm 4}$\\
    $ \bigl| \bm{r}_{\rm DV1}-\bm{r}_{\rm DV2} \bigr| > 500~{\rm \mu m}$ & 
    $< 0.1$ &
    $< 0.1$ &
    $< 0.1$ &
    $< 0.1$ \\
    \hline\hline
    \end{tabular}
    }

\vspace{3mm}
  \centering
\subcaptionbox{The expected number of signal events for 
    $m_{\tilde{g}}=10$ TeV, $m_{\tilde{\chi}_1^0}=8$ TeV
    with different values of $c\tau_{\tilde{g}}$. }[.9\linewidth]{
  \begin{tabular}{l|cccc}
    \hline\hline
    & $c\tau_{\tilde{g}}=0$ 
    & $c\tau_{\tilde{g}}=200~{\rm \mu m}$ & $c\tau_{\tilde{g}}=500~{\rm \mu m}$ & $c\tau_{\tilde{g}}=1~{\rm  mm}$ \\
    \hline
    {\tt preselection-H} & 
    \multicolumn{4}{c}{$143{\scriptstyle~\pm 2}$} \\
    $m_{\rm eff}({\rm incl.}) > 6$ TeV & 
     \multicolumn{4}{c}{$70{\scriptstyle~\pm 1}$} \\
     \hline
    $ \bigl|\bm{r}_{\rm DV1}-\bm{r}_{\rm DV2} \bigr| > 200~{\rm \mu m}$ & 
    $0.4{\scriptstyle~ \pm 0.1}$ &
    $26.5{\scriptstyle~ \pm 0.7}$ &
    $45.1{\scriptstyle~ \pm 0.9}$ &
    $53.5{\scriptstyle~ \pm 1}$\\
    $ \bigl|\bm{r}_{\rm DV1}-\bm{r}_{\rm DV2} \bigr| > 500~{\rm \mu m}$ & 
    $< 0.1$ &
    $6.4{\scriptstyle~ \pm 0.3}$ &
    $22.8{\scriptstyle~ \pm 0.6}$ &
    $36.5{\scriptstyle~ \pm 0.8}$\\
    \hline\hline
    \end{tabular}
    }
    \caption{
    Same as Table.~\ref{table:cut_flow_100tev} but  
    the masses of gluino and the LSP are set to be $10$~TeV and
    $8$~TeV, respectively.
    }
    \label{table:cut_flow_mod_degen_100tev}
  \end{table}
%-------------------------------

%%%%%%%%%%%%%% FIGURE %%%%%%%%%%%%%%%%%%%%%%%%%%%%%%%%%%%%
\begin{figure}
  \centering
  \subcaptionbox{\label{fig:meff_w_rdv_100mu_100tev} $| \bm{r}_{\rm
 DV1}-\bm{r}_{\rm DV2}| > 100~{\rm \mu
 m}$}{\includegraphics[width=0.48\columnwidth]{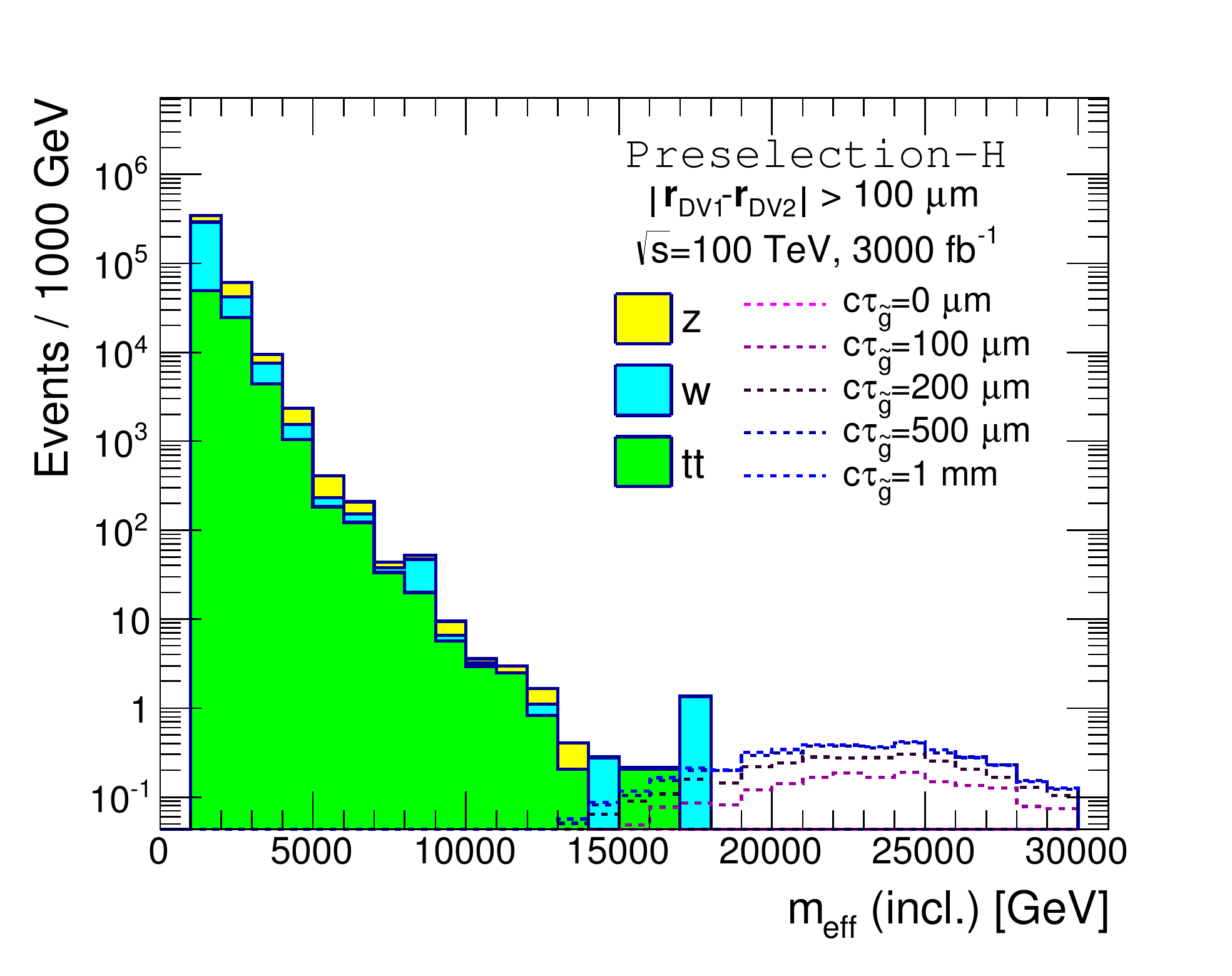}} 
  \subcaptionbox{\label{fig:meff_w_rdv_200mu_100tev} $| \bm{r}_{\rm
 DV1}-\bm{r}_{\rm DV2}| > 200~{\rm \mu
 m}$}{\includegraphics[width=0.48\columnwidth]{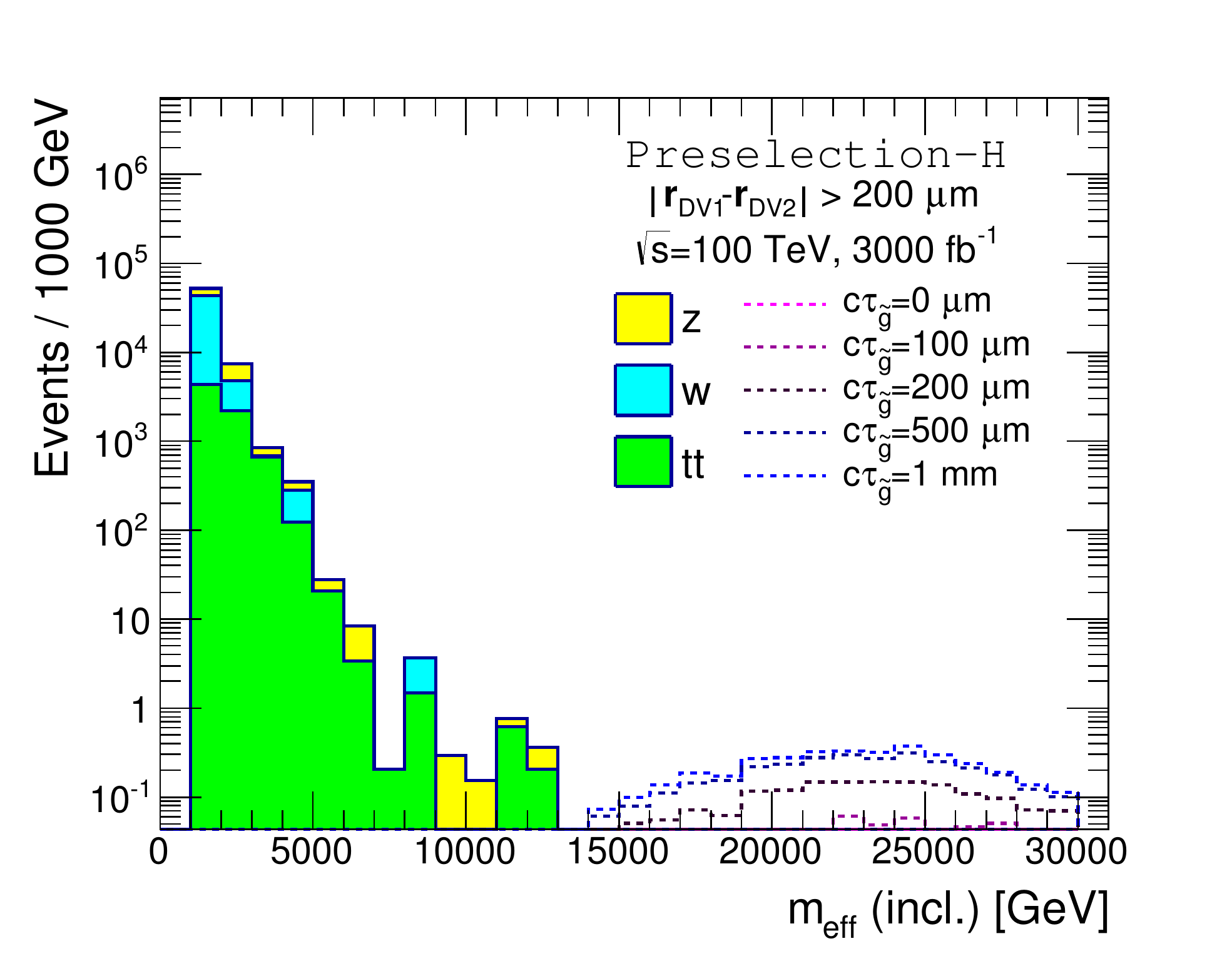}} 
  \subcaptionbox{\label{fig:meff_w_rdv_500mu_100tev} $| \bm{r}_{\rm
 DV1}-\bm{r}_{\rm DV2}| > 500~{\rm \mu
 m}$}{\includegraphics[width=0.48\columnwidth]{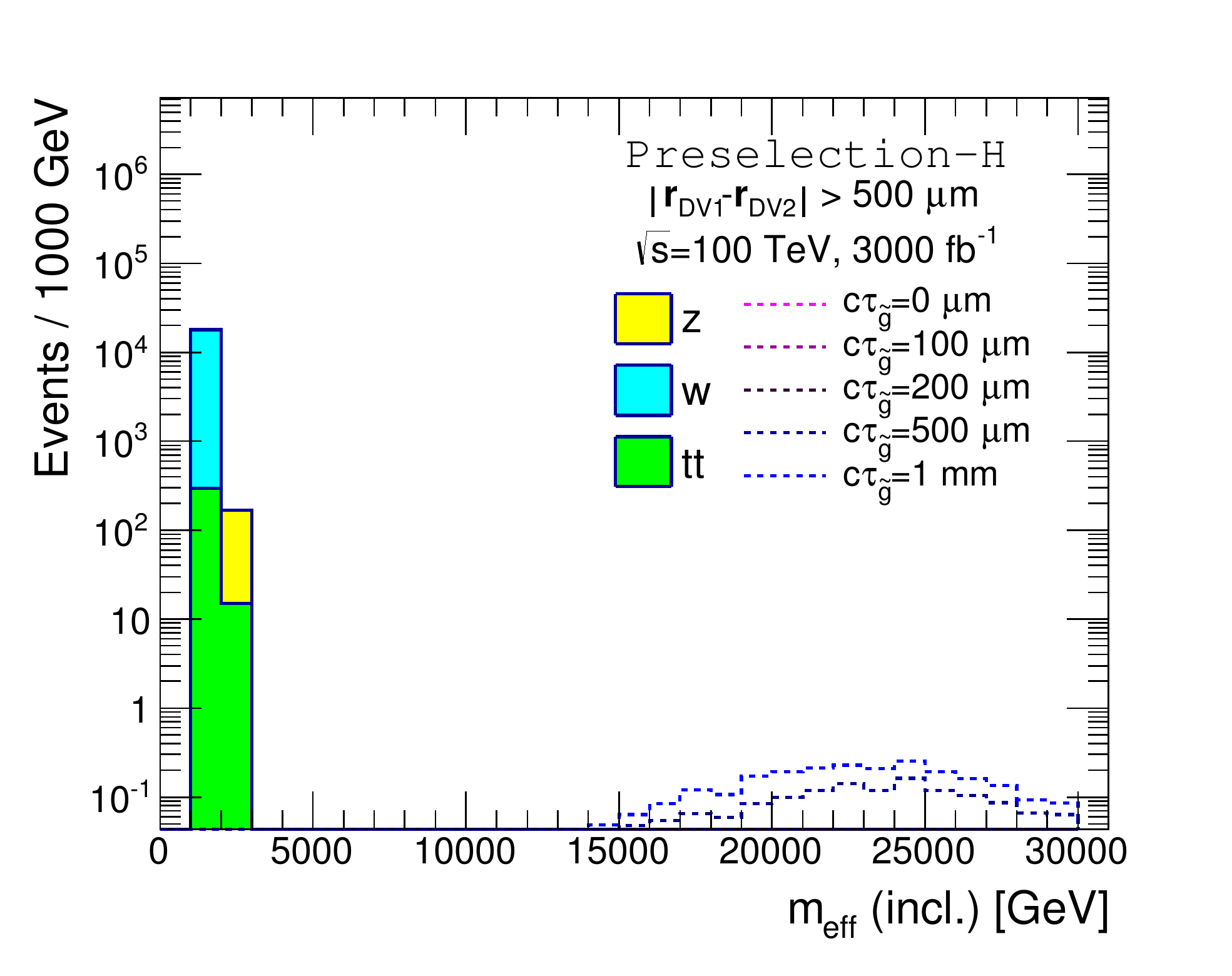}} 
\caption{Distributions of $m_{\rm eff}({\rm incl.})$ for the SM background and
signal events with different values of $c\tau_{\tilde{g}}$. The masses
of gluino and the LSP are set to be $14$~TeV and $100$~GeV, 
respectively. We have imposed {\tt preselection-H} and the vetoes given in
Table~\ref{table:selections}. }  
  \label{fig:meff_w_rdv_100tev}
\end{figure}
%%%%%%%%%%%%%%%%%%%%%%%%%%%%%%%%%%%%%%%%%%%%%%%%%%%%%%%%%%

Now we discuss the prospects of our new selection cut for a 100~TeV
collider. We again use the event selection criteria summarized in
Table~\ref{table:selections}, except that we now raise the upper limit
on $m_{\rm effcut}$ to $3\times 10^4$~GeV. In
Fig.~\ref{fig:optimal_100tev}, we show optimal cut values ($(m_{\rm
  effcut})_{\rm optimal}$, $(r_{\rm cut})_{\rm optimal}$), the
preselection ({\tt L}, {\tt M}, or {\tt H}), and the discriminator
$\Delta r_{\rm DV}$ ({\tt R}, {\tt T}, or {\tt Z}) for each sample
point with various $c\tau_{\tilde{g}}$, for a 100~TeV collider with an
integrated luminosity of $3000~{\rm fb}^{-1}$.  The expected $5\sigma$
discovery reaches for gluinos are also shown in the dashed lines. We
again find that the kinematic selection cut on $m_{\rm eff}({\rm
  incl.})$ may be highly relaxed with the help of the displaced-vertex
cut with $(r_{\rm cut})_{\rm optimal} \simeq c\tau_{\tilde{g}}$.  In
Tables~\ref{table:cut_flow_100tev} and
\ref{table:cut_flow_mod_degen_100tev}, we present the expected number
of background and signal events for an integrated luminosity of
$\mathcal{L}=3000~{\rm fb}^{-1}$.  In
Table~\ref{table:cut_flow_100tev}
(\ref{table:cut_flow_mod_degen_100tev}), we consider the case of a
light (heavy) LSP with $m_{\tilde{g}}=12~(10)$~TeV and
$m_{\tilde{\chi_1^0}}=100$~GeV (8~TeV). We then show the distributions
of $m_{\rm eff}({\rm incl.})$ for the SM background and signal events
with different values of $c\tau_{\tilde{g}}$ in
Fig.~\ref{fig:meff_w_rdv_100tev}, with {\tt preselection-H} and the
vetoes in Table~\ref{table:selections} imposed. The masses of gluino
and the LSP are set to be $14$~TeV and $100$~GeV, respectively. Both
the tables and figures demonstrate that also at a 100 TeV collider our
displaced-vertex selection cut can efficiently eliminate the SM
background while keeping the signal events.

%%%%%%%%%%%%%% FIGURE %%%%%%%%%%%%%%%%%%%%%%%%%%%%%%%%%%%%
\begin{figure}
  \centering
  \subcaptionbox{\label{fig:future_100tev} $m_{\tilde{\chi}^0_1} =
 100$~GeV}{\includegraphics[width=0.48\columnwidth]{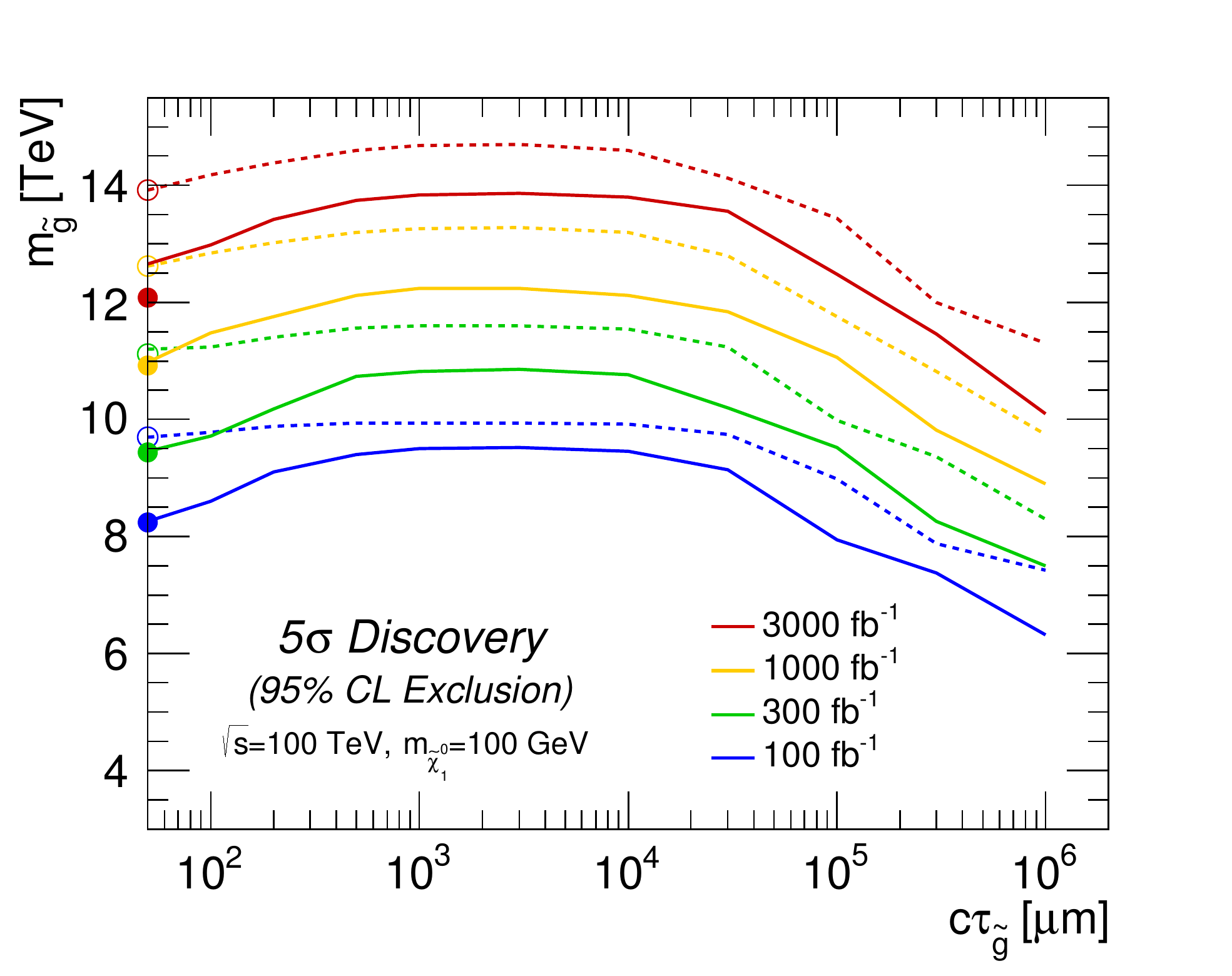}}  
  \subcaptionbox{\label{fig:future_diff100_100tev} $|m_{\tilde{g}} -m_{\tilde{\chi}^0_1}| =
 100$~GeV}{\includegraphics[width=0.48\columnwidth]{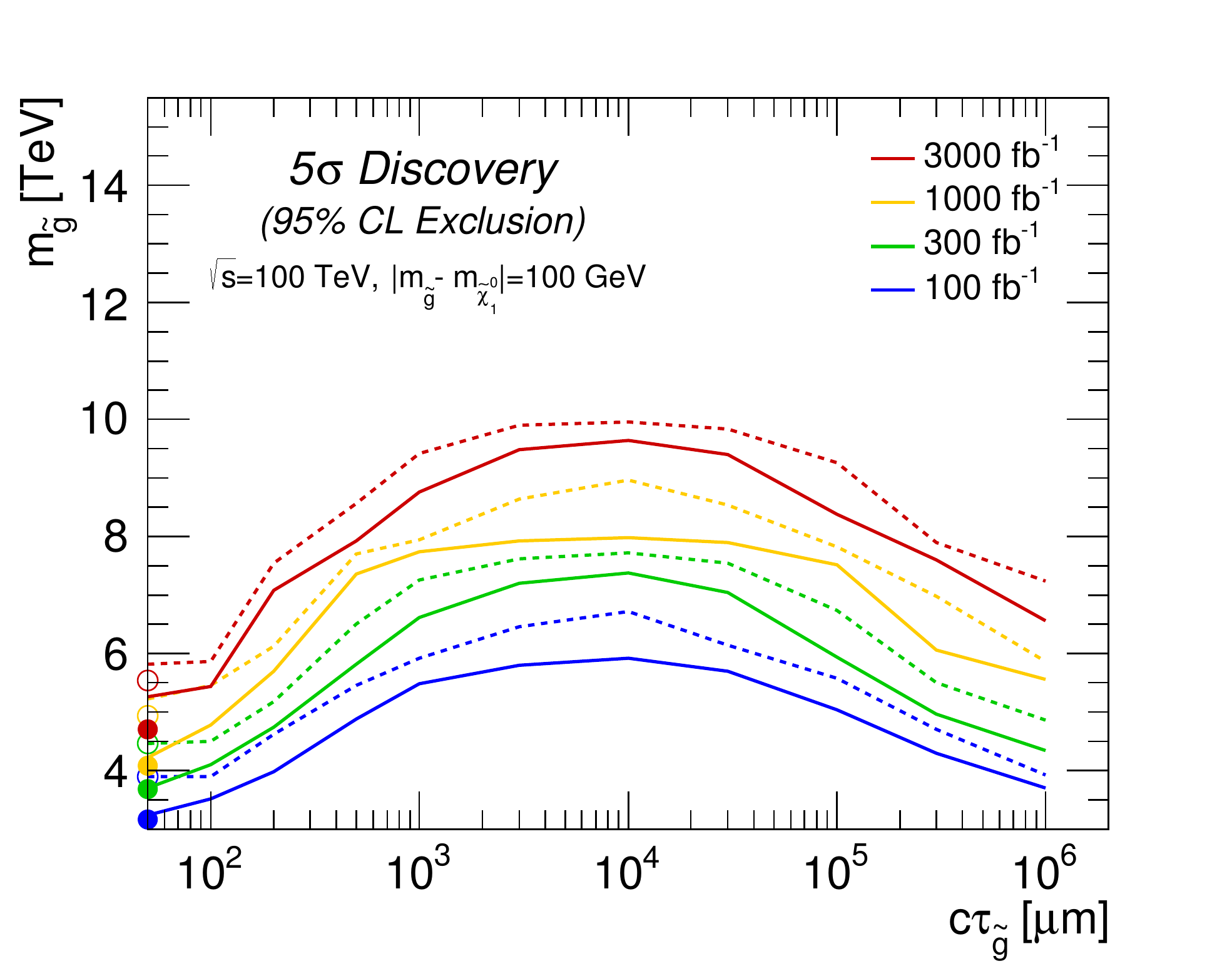}}  
\caption{The expected 95\% CL exclusion limits (dotted) and
    $5\sigma$ discovery reaches (solid) as functions of $c\tau_{\tilde
      g}$ for different values of integrated luminosity at a future 100~TeV
    $pp$ collider. The expected exclusion limit (discovery reach) for
$c\tau_{\tilde{g}}=0$ is represented by a circle (a blob).}  
  \label{fig:fu_100tev}
\end{figure}
%%%%%%%%%%%%%%%%%%%%%%%%%%%%%%%%%%%%%%%%%%%%%%%%%%%%%%%%%%

In Fig.~\ref{fig:future_100tev}, we show the expected 95\% CL exclusion limits 
(in dotted lines) and $5\sigma$ discovery reaches (in solid lines) for
gluino as functions of $c\tau_{\tilde  g}$ for different values of
integrated luminosity at a 100~TeV collider, where the LSP mass is set
to be 100~GeV. The expected exclusion limit (discovery reach) for
$c\tau_{\tilde{g}}=0$ is represented by a circle (a blob).
As can be seen from the figure, the reach for the gluino
can be extended with the help of the additional displaced-vertex cut for
$c\tau_{\tilde g} \gtrsim 100~\mu{\rm m}$; for instance, for a gluino
with $c\tau_{\tilde g} \sim \mathcal{O}$(1--10)~mm, the expected
discovery reach for the gluino mass can be extended by as large as
$\sim 1.4$~TeV ($1.8$~TeV) with an integrated luminosity of 
${\mathcal  L}=300~{\rm fb}^{-1}$ (3000~fb$^{-1}$).
These reaches for a gluino with 
$c\tau_{\tilde g} \sim \mathcal{O}$(1--10)~mm
are obtained with {\tt preselection-H},
$| \bm{r}_{\rm DV1} - \bm{r}_{\rm DV2} |$~
($| \bm{r}_{\rm DV1} - \bm{r}_{\rm DV2}|_{\rm T}$)
with 
$(m_{\rm effcut})_{\rm optimal}=10$ TeV ($12.6$ TeV), and 
$( r_{\rm cut} )_{\rm optimal}\sim120\,\mu{\rm m}$ ($160\,\mu{\rm m}$) for 
${\mathcal L}=300\,{\rm fb}^{-1}$ ($3000\,{\rm fb}^{-1}$).
Compared to a promptly decaying gluino, where the optimized values for
$(m_{\rm effcut})_{\rm optimal}$ are given by 13.8~TeV and $17.8$~TeV for
${\mathcal  L}=300~{\rm fb}^{-1}$ and 3000~fb$^{-1}$, respectively,
the $m_{\rm eff}({\rm incl.})$ selection cut is loosened
and the new vertex-based selection cut is responsible for the reduction of
background. 
We also show the degenerate case in
Fig.~\ref{fig:future_diff100_100tev}, where the mass
difference between gluino and the LSP is set to be $100$~GeV. 
We see a drastic enhancement in reaches, especially for 
$c\tau_{\tilde{g}}\sim {\mathcal O}$(1--100)~mm.

%%%%%%%%%%%%%% FIGURE %%%%%%%%%%%%%%%%%%%%%%%%%%%%%%%%%%%%
\begin{figure}
  \centering
  \subcaptionbox{\label{fig:mg_mneu1_300_disc_100tev} $5\sigma$ discovery}{\includegraphics[width=0.48\columnwidth]{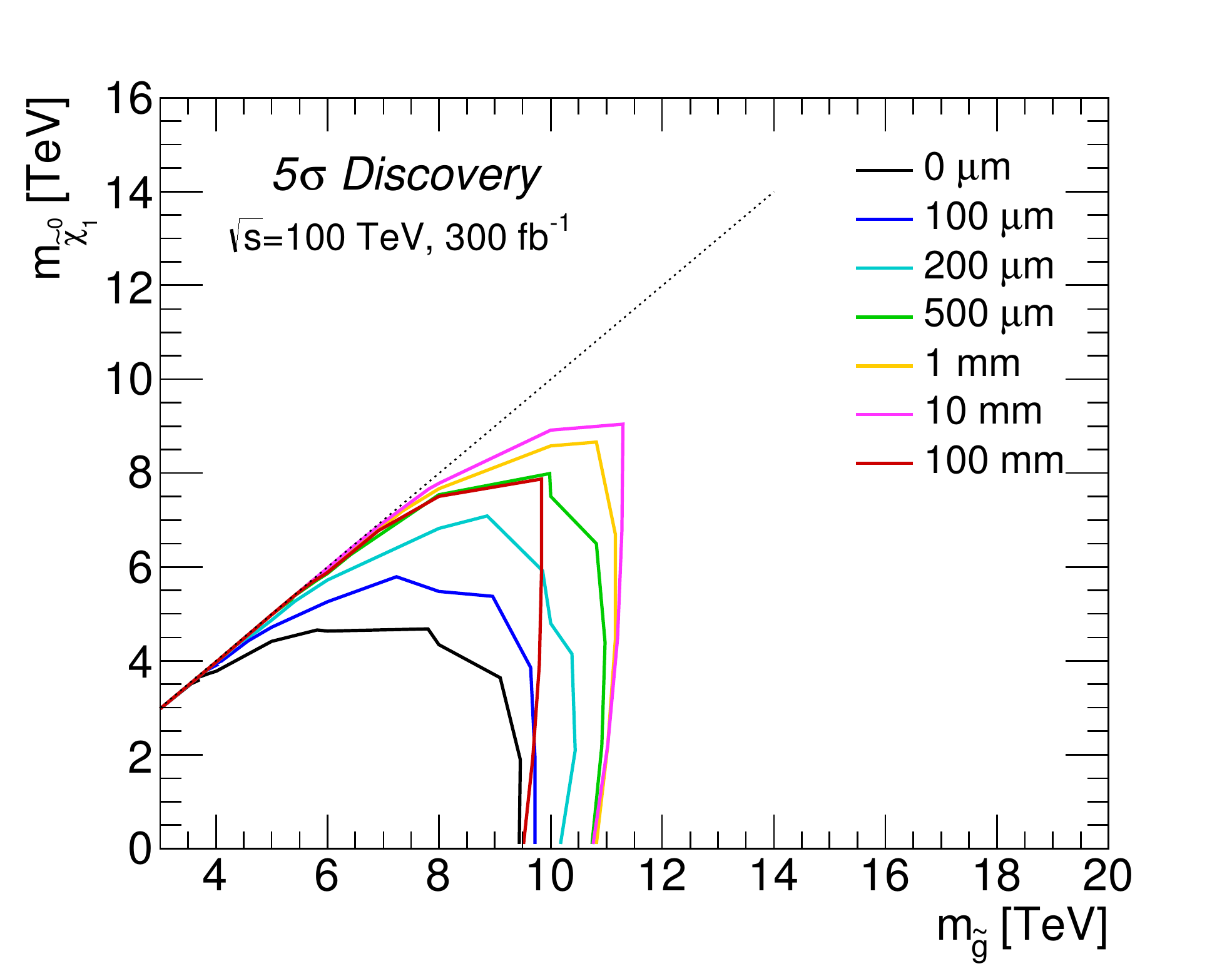}}  
  \subcaptionbox{\label{fig:mg_mneu1_300_excl_100tev} Expected 95\% CL exclusion}{\includegraphics[width=0.48\columnwidth]{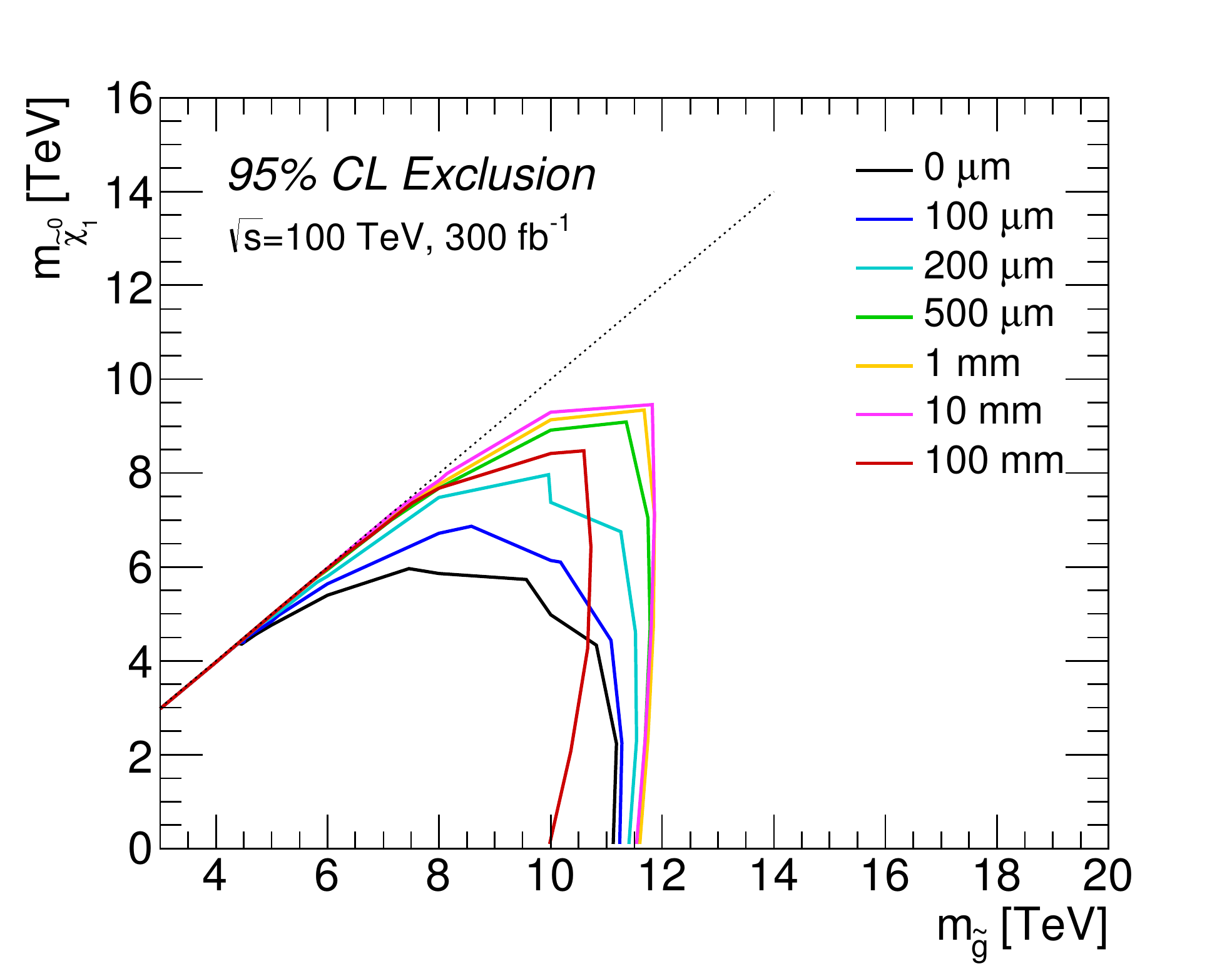}}  
\caption{$5\sigma$ discovery reaches and expected 95\% CL exclusion
 limits for gluinos with different $c\tau_{\tilde{g}}$, for an
 integrated luminosity of $300\,{\rm fb}^{-1}$. }  
  \label{fig:mg_mneu1_300_100tev}
\end{figure}
%%%%%%%%%%%%%%%%%%%%%%%%%%%%%%%%%%%%%%%%%%%%%%%%%%%%%%%%%%

%%%%%%%%%%%%%% FIGURE %%%%%%%%%%%%%%%%%%%%%%%%%%%%%%%%%%%%
\begin{figure}
  \centering
  \subcaptionbox{\label{fig:mg_mneu1_3000_disc_100tev} $5\sigma$ discovery}{\includegraphics[width=0.48\columnwidth]{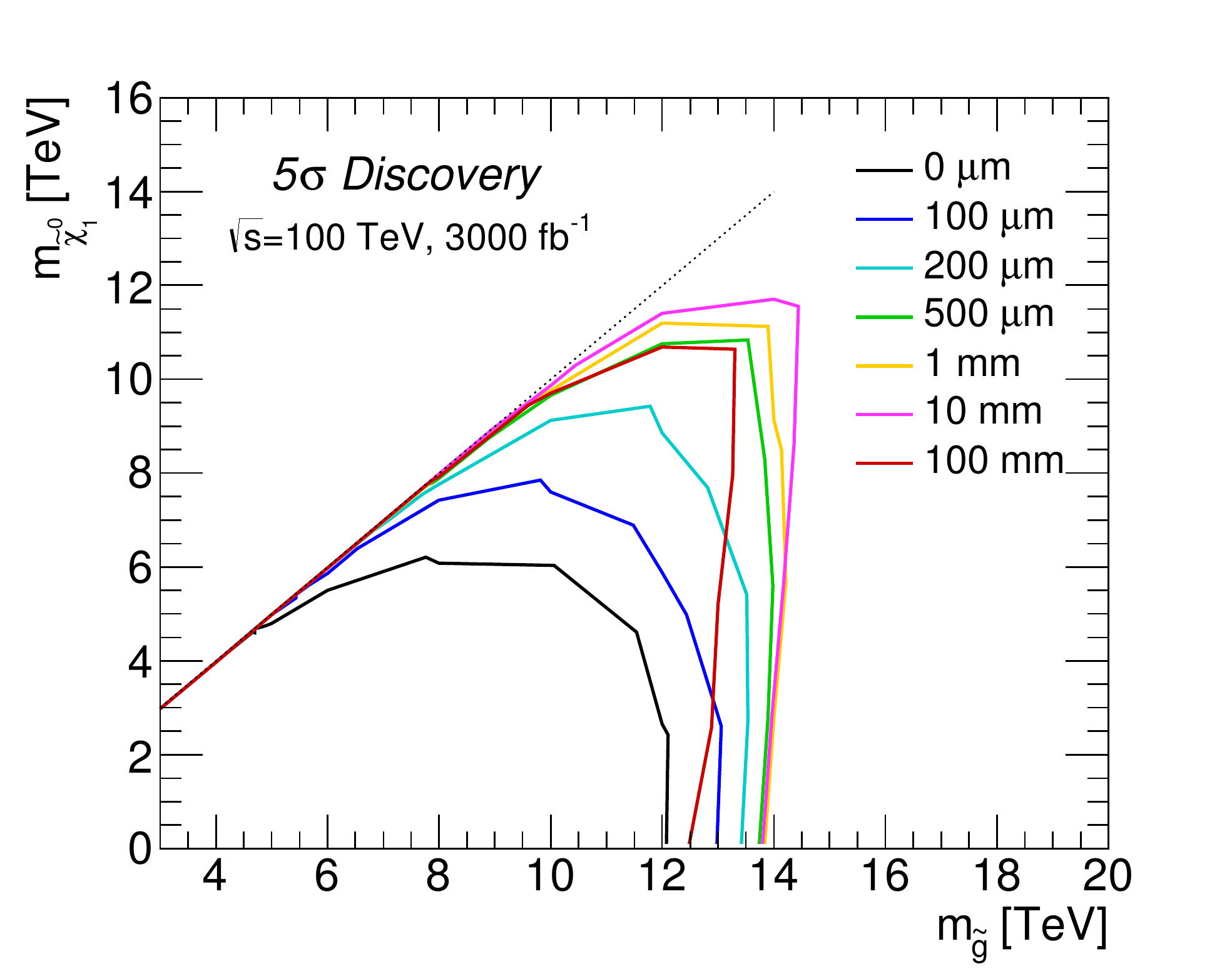}}  
  \subcaptionbox{\label{fig:mg_mneu1_3000_excl_100tev} Expected 95\% CL exclusion}{\includegraphics[width=0.48\columnwidth]{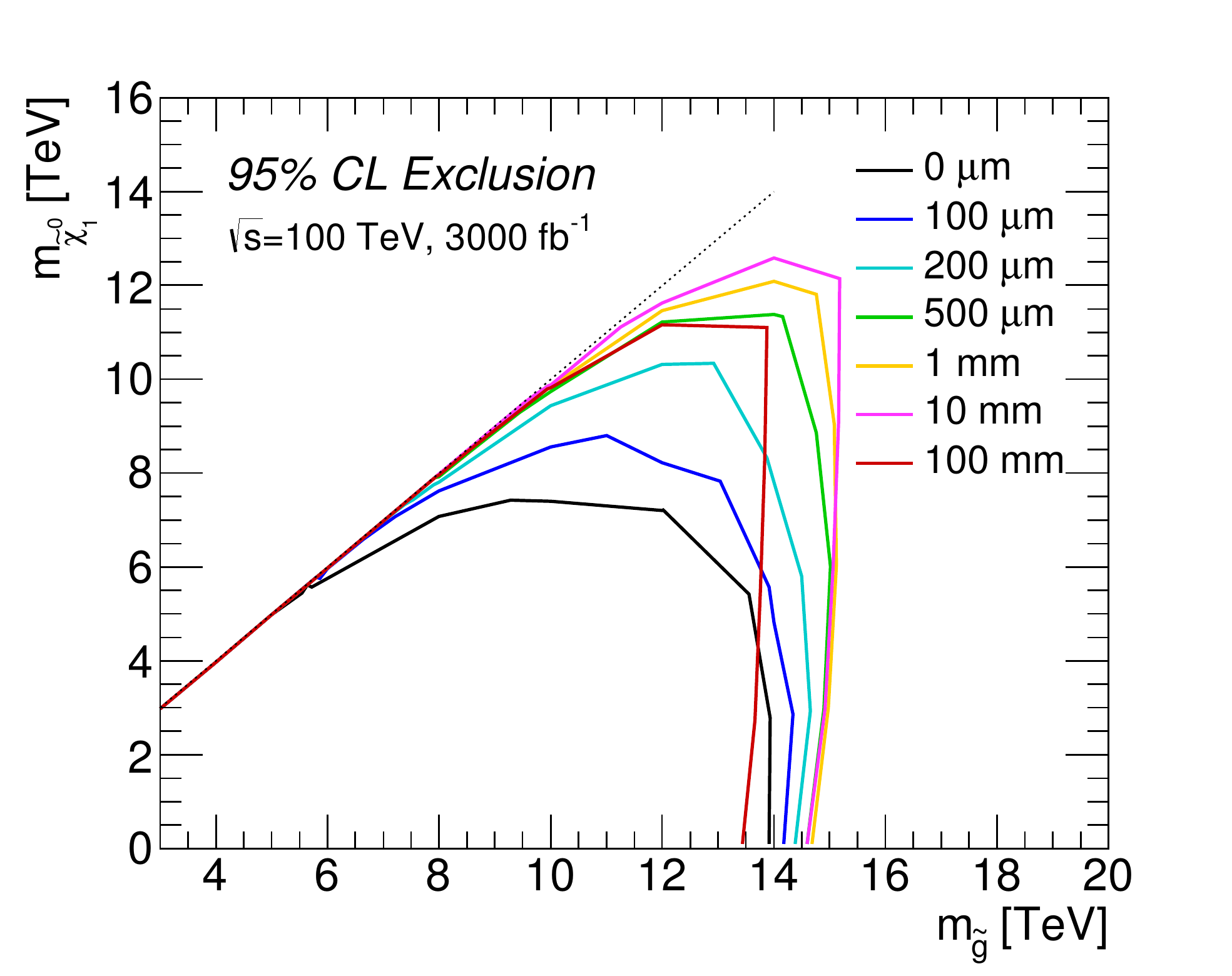}}  
\caption{Same as in Fig.~\ref{fig:mg_mneu1_300_100tev}, but for an integrated
 luminosity of $3000\,{\rm fb}^{-1}$. }  
  \label{fig:mg_mneu1_3000_100tev}
\end{figure}
%%%%%%%%%%%%%%%%%%%%%%%%%%%%%%%%%%%%%%%%%%%%%%%%%%%%%%%%%%

Finally, we show the $5\sigma$ discovery reaches and expected 95\% CL
exclusion limits for various values of $c\tau_{\tilde{g}}$ in
Figs.~\ref{fig:mg_mneu1_300_100tev} and \ref{fig:mg_mneu1_3000_100tev}
for an integrated luminosity of 300 and
3000~fb$^{-1}$, respectively. These figures illustrate that the use of
our vertex-based selection cut leads to a significant improvement in
both the discovery reach and the exclusion limit for
$c\tau_{\tilde{g}}\gtrsim 100\,\mu{\rm m}$. Notice that, compared with the
13~TeV case, searches at a 100~TeV collider
may be sensitive to gluinos with a shorter lifetime. 
The extent of the improvement
is maximized for $c\tau_{\tilde{g}}\sim {\mathcal O}$(1--10)~mm and
tends to get larger for a heavier LSP, similarly to the 13~TeV LHC
case. We also find a dramatic improvement in the degenerate mass
region; with the displaced-vertex selection cut, we may probe a gluino
degenerate with the LSP in mass up to $\sim 10$~TeV. This has
significant implications for the gluino coannihilation scenario, given
that an upper limit on the gluino mass is set for this
scenario as $m_{\tilde{g}} \lesssim 8$~TeV
\cite{Ellis:2015vaa}. Our analysis indicates that we may probe the whole
range of the gluino coannihilation scenario at a 100~TeV collider if the
gluino decay length falls into the range of $100~\mu\text{m} \lesssim
c\tau_{\tilde{g}} \lesssim 100~\text{mm}$.

%%%%%%%%%%%%%%%%%%%%%%%%%%%%%%%%%%%%%%%
\section{Conclusion and discussion}
\label{sec:conclusion}
%%%%%%%%%%%%%%%%%%%%%%%%%%%%%%%%%%%%%%%

In this paper we have discussed a strategy of improving searches for
metastable particles at hadron colliders. This strategy is based on the
reconstruction of displaced vertices caused by the decay of metastable
particles. We take account of this information as a new event-selection
cut and impose this in addition to the conventional selection criteria
based on kinematical observables. To see the significance of this new
selection cut, we consider metastable gluinos in SUSY models as an
example, whose decay length falls into the sub-millimeter range if
squark masses are around the PeV scale. Then, we have studied the
implications of this new selection cut for the gluino searches at both
the 13~TeV LHC and future 100~TeV $pp$ collider experiments. 
We have performed MC simulations for both the signal and SM background
events, and take into account the effect of the track reconstruction
performance on the resolution of vertex reconstruction. We have also
discussed optimization of kinematical selection criteria in the
presence of the new vertex selection cut. As it turns out, we can
considerably relax the kinematical selection criteria in this case,
which is of great importance especially for the cases where gluino and
the LSP are degenerate in mass.

For the 13~TeV LHC analysis, we have found that our
vertex-reconstruction method can separate out decay vertices if the
gluino decay length is $\gtrsim 100~\mu {\rm m}$. As a result, with
the displaced-vertex cut, we may considerably improve the potential of
gluino searches for a gluino with $c \tau_{\tilde{g}} \gtrsim 200~\mu
{\rm m}$.  In particular, if $c\tau_{\tilde g} \sim
\mathcal{O}$(1--10)~mm, then the exclusion and discovery reaches for
the gluino mass can be extended by about $180$~GeV and $320$~GeV,
respectively, with an integrated luminosity of 3000~fb$^{-1}$ at the
13~TeV LHC for LSP with a mass of $100$~GeV.  This improvement gets
more drastic when gluino and the LSP are degenerate in
mass. Furthermore, with an integrated luminosity of 3000~fb$^{-1}$, it
is possible to measure the gluino decay length with an ${\cal O}(1)$
accuracy for a gluino with $c\tau_{\tilde{g}}\sim{\cal O}(100)~\mu
{\rm m}$ and $m_{\tilde{g}} = 2.2$~TeV, which may allow us to probe
the PeV-scale squarks indirectly.  

%In addition, 
After the discovery of a metastable particle, its
lifetime information will become available by trying to reconstruct
displaced vertices as we have seen in Sec.~\ref{sec:lifetime}.  In the
case of gluino, such information can be used to constrain the mass
scale of squarks which mediate the decay processes.  Even though we
can only have an upper bound on the lifetime if the decay length
$c\tau_{\tilde{g}}$ is shorter than $\sim 100~\mu {\rm m}$, such an
bound is highly useful because it can provide an upper bound on the
mass scale of squarks whose direct production may not be possible at
the LHC.  For the gluino mass of $2.2\ {\rm TeV}$, for example, the
mass scale of the squarks will be known to be lower than the PeV scale
or a longevity of gluino will be observed.

We have also studied the prospects of searches for metastable gluinos
at a future $100$~TeV $pp$ collider. Since the TeV-scale gluinos tend to
be produced in a fairly boosted state at a 100~TeV collider, we expect
it is possible to probe a shorter
decay length compared with the LHC case. Indeed, we have found that the
$|\bm{r}_{\rm DV1} - \bm{r}_{\rm DV2}|$ distributions for a 3~TeV gluino
are quite different from those for the $c\tau_{\tilde{g}}=0$ case if
its decay distance is $\gtrsim 50~{\rm \mu m}$. 
By using the new selection cut, we can significantly extend the
exclusion limits and discovery reaches of gluino searches for
$c\tau_{\tilde{g}} \gtrsim 100~\mu {\rm m}$; {\it e.g.}, for
$c\tau_{\tilde g} \sim \mathcal{O}$(1--10)~mm, the exclusion and
discovery reaches of gluino mass will be extended by about $780$~GeV
and $1780$~GeV, respectively, with an integrated luminosity of
3000~fb$^{-1}$ and the LSP mass being $100$~GeV. The improvement is
found to be more drastic in the degenerate mass region.

In the analyses given in this paper, we have assumed that gluinos decay
into only the first-generation quarks and a LSP. In reality, the dominant
decay channel of a gluino depends on the mass spectrum of squarks; for
instance, if stops are much lighter than the other squarks, then 
the $\tilde{g} \rightarrow t\bar{t}\tilde{\chi}_1^0$ decay process
becomes the dominant decay channel. In this case, besides the displaced
vertices associated with the gluino decay, we may also find the secondary
vertices that originate from long-lived hadrons including $b$ quarks,
which are emitted by top quarks in the final state. Moreover, if there
is a sizable flavor violation in the sfermion mass matrices---this
possibility is experimentally allowed if the mass scale of SUSY
particles lies around the PeV scale as we discussed in
Sec.~\ref{sec:gluino}---the decay of gluinos may exhibit this
flavor mixing by containing quarks with different flavors in the final
state \cite{Sato:2013bta}. Secondary vertices due to metastable hadrons
may also appear in such cases. The presence of secondary vertices may
affect the resolution of the vertex reconstruction, whereas this
distinct signature composed of multiple displaced vertices may allow
further optimization for this class of decay processes.  
A dedicated study regarding this possibility may be worth doing.

Another possibility of optimization is related to the degenerate mass
region. As we observed in Fig.~\ref{fig:mg_mneu1_36p1_excl} in
Sec.~\ref{sec:prospects}, the limits we obtained in the degenerate mass
region are weaker than that presented in
Ref.~\cite{Aaboud:2017vwy}, since our analysis was not optimized to the
degenerate mass region. Such an optimization is however possible by
relaxing the requirements on the jet momenta. In fact, this is quite
promising given that our new selection cut can significantly reduce the
SM background and thus allow a relaxation of kinematical selection
criteria as we have seen in the above analyses. This expectation is actually
supported by the recent ATLAS result of the displaced-vertex search
\cite{Aaboud:2017iio}, where an optimization for the degenerate mass
region was successfully carried out and a stringent limit is imposed on
long-lived gluinos in a degenerate mass spectrum.

In this analysis, we do not reconstruct the position of the primary
interaction point and use only the distance of the two displaced
vertices as a discriminator. As we mentioned in Sec.~\ref{sec:eventsel},
however, the reconstruction of the primary vertex may also be possible
by using the remnants of the $pp$ collision and/or initial state
radiation emitted from the vertex. This additional information could be
useful for the further reduction of background events. For instance, by
requiring the presence of the primary vertex in-between the two
reconstructed vertices we may efficiently reject the SM background
contribution. Another, in fact simpler way of going beyond the mere use
of the vertex distance is to require both of the two decay vertices to
be reconstructed away from the beam line, with which we can avoid the
misidintification of the primary vertex as one of the two decay
vertices. Such elaboration of our new selection cut will be explored in
another occasion. 

Finally we comment here that, by reconstructing the positions of
displaced vertices as well as the momenta of the charged tracks
associated with these vertices, we may also extract the kinematical
information of both the decaying and final-state invisible particles, as
discussed in Refs.~\cite{Park:2011vw, Cottin:2018hyf}. In particular, we
may determine the masses of these particles from the above
information. This technique may be useful not only for studying the mass
spectrum of the decay chain after the discovery of the metastable
particle, but also for an additional event-selection cut to reduce the
SM background. A dedicated study is required to assess the feasibility
of this method, and thus we defer it to future work.

%%%%%%%%%%%%%%%%%%%%%%%%%%%%%%%%%%%%
\section*{Acknowledgments}
%%%%%%%%%%%%%%%%%%%%%%%%%%%%%%%%%%%%

This work was supported in part by the Grant-in-Aid for 
Scientific Research C (26400239 [TM]), 
Innovative Areas (16H06490 [TM]), and
Young Scientists B (17K14270 [NN]).

%%%%%%%%%%%%%%%%%%%%%%%%%%%%%%%%%%%%%%%%%%%%%%
%\section*{Appendix}
%\appendix
%%%%%%%%%%%%%%%%%%%%%%%%%%%%%%%%%%%%%%%%%%%%%

%%%%%%%%%%%%%%%%%%%%%%%%%%%%%%%%%%%%%%%
{\small 
\bibliographystyle{JHEP}
\bibliography{ref}

\providecommand{\href}[2]{#2}\begingroup\raggedright\begin{thebibliography}{100}

\bibitem{Fairbairn:2006gg}
M.~Fairbairn, A.~C. Kraan, D.~A. Milstead, T.~Sjostrand, P.~Z. Skands, and
  T.~Sloan, {\it {Stable massive particles at colliders}},  {\em Phys. Rept.}
  {\bf 438} (2007) 1--63, [\href{http://arxiv.org/abs/hep-ph/0611040}{{\tt
  hep-ph/0611040}}].

\bibitem{Toharia:2005gm}
M.~Toharia and J.~D. Wells, {\it {Gluino decays with heavier scalar
  superpartners}},  {\em JHEP} {\bf 02} (2006) 015,
  [\href{http://arxiv.org/abs/hep-ph/0503175}{{\tt hep-ph/0503175}}].

\bibitem{Gambino:2005eh}
P.~Gambino, G.~F. Giudice, and P.~Slavich, {\it {Gluino decays in split
  supersymmetry}},  {\em Nucl. Phys.} {\bf B726} (2005) 35--52,
  [\href{http://arxiv.org/abs/hep-ph/0506214}{{\tt hep-ph/0506214}}].

\bibitem{Sato:2012xf}
R.~Sato, S.~Shirai, and K.~Tobioka, {\it {Gluino Decay as a Probe of High Scale
  Supersymmetry Breaking}},  {\em JHEP} {\bf 11} (2012) 041,
  [\href{http://arxiv.org/abs/1207.3608}{{\tt arXiv:1207.3608}}].

\bibitem{Giudice:1998bp}
G.~F. Giudice and R.~Rattazzi, {\it {Theories with gauge mediated supersymmetry
  breaking}},  {\em Phys. Rept.} {\bf 322} (1999) 419--499,
  [\href{http://arxiv.org/abs/hep-ph/9801271}{{\tt hep-ph/9801271}}].

\bibitem{Draper:2011aa}
P.~Draper, P.~Meade, M.~Reece, and D.~Shih, {\it {Implications of a 125 GeV
  Higgs for the MSSM and Low-Scale SUSY Breaking}},  {\em Phys. Rev.} {\bf D85}
  (2012) 095007, [\href{http://arxiv.org/abs/1112.3068}{{\tt
  arXiv:1112.3068}}].

\bibitem{Evans:2016zau}
J.~A. Evans and J.~Shelton, {\it {Long-Lived Staus and Displaced Leptons at the
  LHC}},  {\em JHEP} {\bf 04} (2016) 056,
  [\href{http://arxiv.org/abs/1601.01326}{{\tt arXiv:1601.01326}}].

\bibitem{Allanach:2016pam}
B.~C. Allanach, M.~Badziak, G.~Cottin, N.~Desai, C.~Hugonie, and R.~Ziegler,
  {\it {Prompt Signals and Displaced Vertices in Sparticle Searches for
  Next-to-Minimal Gauge Mediated Supersymmetric Models}},  {\em Eur. Phys. J.}
  {\bf C76} (2016), no.~9 482, [\href{http://arxiv.org/abs/1606.03099}{{\tt
  arXiv:1606.03099}}].

\bibitem{Barbier:2004ez}
R.~Barbier et~al., {\it {R-parity violating supersymmetry}},  {\em Phys. Rept.}
  {\bf 420} (2005) 1--202, [\href{http://arxiv.org/abs/hep-ph/0406039}{{\tt
  hep-ph/0406039}}].

\bibitem{Graham:2012th}
P.~W. Graham, D.~E. Kaplan, S.~Rajendran, and P.~Saraswat, {\it {Displaced
  Supersymmetry}},  {\em JHEP} {\bf 07} (2012) 149,
  [\href{http://arxiv.org/abs/1204.6038}{{\tt arXiv:1204.6038}}].

\bibitem{Ghosh:2017yeh}
P.~Ghosh, I.~Lara, D.~E. Lopez-Fogliani, C.~Munoz, and R.~Ruiz~de Austri, {\it
  {Searching for left sneutrino LSP at the LHC}},
  \href{http://arxiv.org/abs/1707.02471}{{\tt arXiv:1707.02471}}.

\bibitem{Brandenburg:2005he}
A.~Brandenburg, L.~Covi, K.~Hamaguchi, L.~Roszkowski, and F.~D. Steffen, {\it
  {Signatures of axinos and gravitinos at colliders}},  {\em Phys. Lett.} {\bf
  B617} (2005) 99--111, [\href{http://arxiv.org/abs/hep-ph/0501287}{{\tt
  hep-ph/0501287}}].

\bibitem{Hamaguchi:2006vu}
K.~Hamaguchi, M.~M. Nojiri, and A.~de~Roeck, {\it {Prospects to study a
  long-lived charged next lightest supersymmetric particle at the LHC}},  {\em
  JHEP} {\bf 03} (2007) 046, [\href{http://arxiv.org/abs/hep-ph/0612060}{{\tt
  hep-ph/0612060}}].

\bibitem{Freitas:2011fx}
A.~Freitas, F.~D. Steffen, N.~Tajuddin, and D.~Wyler, {\it {Axinos in Cosmology
  and at Colliders}},  {\em JHEP} {\bf 06} (2011) 036,
  [\href{http://arxiv.org/abs/1105.1113}{{\tt arXiv:1105.1113}}].

\bibitem{Barenboim:2014kka}
G.~Barenboim, E.~J. Chun, S.~Jung, and W.~I. Park, {\it {Implications of an
  axino LSP for naturalness}},  {\em Phys. Rev.} {\bf D90} (2014), no.~3
  035020, [\href{http://arxiv.org/abs/1407.1218}{{\tt arXiv:1407.1218}}].

\bibitem{Redino:2015mye}
C.~S. Redino and D.~Wackeroth, {\it {Exploring the Hadronic Axion Window via
  Delayed Neutralino Decay to Axinos at the LHC}},  {\em Phys. Rev.} {\bf D93}
  (2016), no.~7 075022, [\href{http://arxiv.org/abs/1512.06822}{{\tt
  arXiv:1512.06822}}].

\bibitem{Co:2016fln}
R.~T. Co, F.~D'Eramo, and L.~J. Hall, {\it {Gravitino or Axino Dark Matter with
  Reheat Temperature as high as $10^{16}$ GeV}},  {\em JHEP} {\bf 03} (2017)
  005, [\href{http://arxiv.org/abs/1611.05028}{{\tt arXiv:1611.05028}}].

\bibitem{Co:2017orl}
R.~T. Co, F.~D'Eramo, L.~J. Hall, and K.~Harigaya, {\it {Saxion Cosmology for
  Thermalized Gravitino Dark Matter}},  {\em JHEP} {\bf 07} (2017) 125,
  [\href{http://arxiv.org/abs/1703.09796}{{\tt arXiv:1703.09796}}].

\bibitem{Fan:2011yu}
J.~Fan, M.~Reece, and J.~T. Ruderman, {\it {Stealth Supersymmetry}},  {\em
  JHEP} {\bf 11} (2011) 012, [\href{http://arxiv.org/abs/1105.5135}{{\tt
  arXiv:1105.5135}}].

\bibitem{Fan:2012jf}
J.~Fan, M.~Reece, and J.~T. Ruderman, {\it {A Stealth Supersymmetry Sampler}},
  {\em JHEP} {\bf 07} (2012) 196, [\href{http://arxiv.org/abs/1201.4875}{{\tt
  arXiv:1201.4875}}].

\bibitem{Fan:2015mxp}
J.~Fan, R.~Krall, D.~Pinner, M.~Reece, and J.~T. Ruderman, {\it {Stealth
  Supersymmetry Simplified}},  {\em JHEP} {\bf 07} (2016) 016,
  [\href{http://arxiv.org/abs/1512.05781}{{\tt arXiv:1512.05781}}].

\bibitem{Batell:2015fma}
B.~Batell, G.~F. Giudice, and M.~McCullough, {\it {Natural Heavy
  Supersymmetry}},  {\em JHEP} {\bf 12} (2015) 162,
  [\href{http://arxiv.org/abs/1509.00834}{{\tt arXiv:1509.00834}}].

\bibitem{Evans:2016htp}
J.~L. Evans, T.~Gherghetta, N.~Nagata, and Z.~Thomas, {\it {Naturalizing
  Supersymmetry with a Two-Field Relaxion Mechanism}},  {\em JHEP} {\bf 09}
  (2016) 150, [\href{http://arxiv.org/abs/1602.04812}{{\tt arXiv:1602.04812}}].

\bibitem{Feng:1999fu}
J.~L. Feng, T.~Moroi, L.~Randall, M.~Strassler, and S.-f. Su, {\it {Discovering
  supersymmetry at the Tevatron in wino LSP scenarios}},  {\em Phys. Rev.
  Lett.} {\bf 83} (1999) 1731--1734,
  [\href{http://arxiv.org/abs/hep-ph/9904250}{{\tt hep-ph/9904250}}].

\bibitem{Ibe:2006de}
M.~Ibe, T.~Moroi, and T.~T. Yanagida, {\it {Possible Signals of Wino LSP at the
  Large Hadron Collider}},  {\em Phys. Lett.} {\bf B644} (2007) 355--360,
  [\href{http://arxiv.org/abs/hep-ph/0610277}{{\tt hep-ph/0610277}}].

\bibitem{Asai:2007sw}
S.~Asai, T.~Moroi, K.~Nishihara, and T.~T. Yanagida, {\it {Testing the Anomaly
  Mediation at the LHC}},  {\em Phys. Lett.} {\bf B653} (2007) 81--87,
  [\href{http://arxiv.org/abs/0705.3086}{{\tt arXiv:0705.3086}}].

\bibitem{Asai:2008sk}
S.~Asai, T.~Moroi, and T.~T. Yanagida, {\it {Test of Anomaly Mediation at the
  LHC}},  {\em Phys. Lett.} {\bf B664} (2008) 185--189,
  [\href{http://arxiv.org/abs/0802.3725}{{\tt arXiv:0802.3725}}].

\bibitem{Asai:2008im}
S.~Asai, Y.~Azuma, O.~Jinnouchi, T.~Moroi, S.~Shirai, and T.~T. Yanagida, {\it
  {Mass Measurement of the Decaying Bino at the LHC}},  {\em Phys. Lett.} {\bf
  B672} (2009) 339--343, [\href{http://arxiv.org/abs/0807.4987}{{\tt
  arXiv:0807.4987}}].

\bibitem{Liu:2015bma}
Z.~Liu and B.~Tweedie, {\it {The Fate of Long-Lived Superparticles with
  Hadronic Decays after LHC Run 1}},  {\em JHEP} {\bf 06} (2015) 042,
  [\href{http://arxiv.org/abs/1503.05923}{{\tt arXiv:1503.05923}}].

\bibitem{Nagata:2015hha}
N.~Nagata, H.~Otono, and S.~Shirai, {\it {Probing bino-gluino coannihilation at
  the LHC}},  {\em Phys. Lett.} {\bf B748} (2015) 24--29,
  [\href{http://arxiv.org/abs/1504.00504}{{\tt arXiv:1504.00504}}].

\bibitem{Nagata:2015pra}
N.~Nagata, H.~Otono, and S.~Shirai, {\it {Probing Bino-Wino Coannihilation at
  the LHC}},  {\em JHEP} {\bf 10} (2015) 086,
  [\href{http://arxiv.org/abs/1506.08206}{{\tt arXiv:1506.08206}}].

\bibitem{Rolbiecki:2015gsa}
K.~Rolbiecki and K.~Sakurai, {\it {Long-lived bino and wino in supersymmetry
  with heavy scalars and higgsinos}},  {\em JHEP} {\bf 11} (2015) 091,
  [\href{http://arxiv.org/abs/1506.08799}{{\tt arXiv:1506.08799}}].

\bibitem{Nagata:2017gci}
N.~Nagata, H.~Otono, and S.~Shirai, {\it {Cornering Compressed Gluino at the
  LHC}},  {\em JHEP} {\bf 03} (2017) 025,
  [\href{http://arxiv.org/abs/1701.07664}{{\tt arXiv:1701.07664}}].

\bibitem{Fukuda:2017jmk}
H.~Fukuda, N.~Nagata, H.~Otono, and S.~Shirai, {\it {Higgsino Dark Matter or
  Not: Role of Disappearing Track Searches at the LHC and Future Colliders}},
  \href{http://arxiv.org/abs/1703.09675}{{\tt arXiv:1703.09675}}.

\bibitem{Chacko:2005pe}
Z.~Chacko, H.-S. Goh, and R.~Harnik, {\it {The Twin Higgs: Natural electroweak
  breaking from mirror symmetry}},  {\em Phys. Rev. Lett.} {\bf 96} (2006)
  231802, [\href{http://arxiv.org/abs/hep-ph/0506256}{{\tt hep-ph/0506256}}].

\bibitem{Burdman:2006tz}
G.~Burdman, Z.~Chacko, H.-S. Goh, and R.~Harnik, {\it {Folded supersymmetry and
  the LEP paradox}},  {\em JHEP} {\bf 02} (2007) 009,
  [\href{http://arxiv.org/abs/hep-ph/0609152}{{\tt hep-ph/0609152}}].

\bibitem{Cai:2008au}
H.~Cai, H.-C. Cheng, and J.~Terning, {\it {A Quirky Little Higgs Model}},  {\em
  JHEP} {\bf 05} (2009) 045, [\href{http://arxiv.org/abs/0812.0843}{{\tt
  arXiv:0812.0843}}].

\bibitem{Burdman:2015oej}
G.~Burdman and R.~T. D'Agnolo, {\it {Scalar Leptons in Folded Supersymmetry}},
  \href{http://arxiv.org/abs/1512.00040}{{\tt arXiv:1512.00040}}.

\bibitem{Chacko:2015fbc}
Z.~Chacko, D.~Curtin, and C.~B. Verhaaren, {\it {A Quirky Probe of Neutral
  Naturalness}},  {\em Phys. Rev.} {\bf D94} (2016), no.~1 011504,
  [\href{http://arxiv.org/abs/1512.05782}{{\tt arXiv:1512.05782}}].

\bibitem{Strassler:2006im}
M.~J. Strassler and K.~M. Zurek, {\it {Echoes of a hidden valley at hadron
  colliders}},  {\em Phys. Lett.} {\bf B651} (2007) 374--379,
  [\href{http://arxiv.org/abs/hep-ph/0604261}{{\tt hep-ph/0604261}}].

\bibitem{Strassler:2006ri}
M.~J. Strassler and K.~M. Zurek, {\it {Discovering the Higgs through
  highly-displaced vertices}},  {\em Phys. Lett.} {\bf B661} (2008) 263--267,
  [\href{http://arxiv.org/abs/hep-ph/0605193}{{\tt hep-ph/0605193}}].

\bibitem{Strassler:2006qa}
M.~J. Strassler, {\it {Possible effects of a hidden valley on supersymmetric
  phenomenology}},  \href{http://arxiv.org/abs/hep-ph/0607160}{{\tt
  hep-ph/0607160}}.

\bibitem{Nakai:2015swg}
Y.~Nakai, M.~Reece, and R.~Sato, {\it {SUSY Higgs Mass and Collider Signals
  with a Hidden Valley}},  {\em JHEP} {\bf 03} (2016) 143,
  [\href{http://arxiv.org/abs/1511.00691}{{\tt arXiv:1511.00691}}].

\bibitem{Knapen:2017kly}
S.~Knapen, H.~K. Lou, M.~Papucci, and J.~Setford, {\it {Tracking down Quirks at
  the Large Hadron Collider}},  {\em Phys. Rev.} {\bf D96} (2017), no.~11
  115015, [\href{http://arxiv.org/abs/1708.02243}{{\tt arXiv:1708.02243}}].

\bibitem{Barnard:2015rba}
J.~Barnard, P.~Cox, T.~Gherghetta, and A.~Spray, {\it {Long-Lived,
  Colour-Triplet Scalars from Unnaturalness}},  {\em JHEP} {\bf 03} (2016) 003,
  [\href{http://arxiv.org/abs/1510.06405}{{\tt arXiv:1510.06405}}].

\bibitem{Chang:2009sv}
S.~Chang and M.~A. Luty, {\it {Displaced Dark Matter at Colliders}},
  \href{http://arxiv.org/abs/0906.5013}{{\tt arXiv:0906.5013}}.

\bibitem{Co:2015pka}
R.~T. Co, F.~D'Eramo, L.~J. Hall, and D.~Pappadopulo, {\it {Freeze-In Dark
  Matter with Displaced Signatures at Colliders}},  {\em JCAP} {\bf 1512}
  (2015), no.~12 024, [\href{http://arxiv.org/abs/1506.07532}{{\tt
  arXiv:1506.07532}}].

\bibitem{Buchmueller:2017uqu}
O.~Buchmueller, A.~De~Roeck, K.~Hahn, M.~McCullough, P.~Schwaller, K.~Sung, and
  T.-T. Yu, {\it {Simplified Models for Displaced Dark Matter Signatures}},
  {\em JHEP} {\bf 09} (2017) 076, [\href{http://arxiv.org/abs/1704.06515}{{\tt
  arXiv:1704.06515}}].

\bibitem{Acharya:2017kfi}
B.~S. Acharya, S.~A.~R. Ellis, G.~L. Kane, B.~D. Nelson, and M.~Perry, {\it
  {Categorisation and Detection of Dark Matter Candidates from String/M-theory
  Hidden Sectors}},  \href{http://arxiv.org/abs/1707.04530}{{\tt
  arXiv:1707.04530}}.

\bibitem{Basso:2008iv}
L.~Basso, A.~Belyaev, S.~Moretti, and C.~H. Shepherd-Themistocleous, {\it
  {Phenomenology of the minimal B-L extension of the Standard model: Z' and
  neutrinos}},  {\em Phys. Rev.} {\bf D80} (2009) 055030,
  [\href{http://arxiv.org/abs/0812.4313}{{\tt arXiv:0812.4313}}].

\bibitem{Helo:2013esa}
J.~C. Helo, M.~Hirsch, and S.~Kovalenko, {\it {Heavy neutrino searches at the
  LHC with displaced vertices}},  {\em Phys. Rev.} {\bf D89} (2014) 073005,
  [\href{http://arxiv.org/abs/1312.2900}{{\tt arXiv:1312.2900}}]. [Erratum:
  Phys. Rev.D93,no.9,099902(2016)].

\bibitem{Izaguirre:2015pga}
E.~Izaguirre and B.~Shuve, {\it {Multilepton and Lepton Jet Probes of
  Sub-Weak-Scale Right-Handed Neutrinos}},  {\em Phys. Rev.} {\bf D91} (2015),
  no.~9 093010, [\href{http://arxiv.org/abs/1504.02470}{{\tt
  arXiv:1504.02470}}].

\bibitem{Maiezza:2015lza}
A.~Maiezza, M.~Nemev\v{s}ek, and F.~Nesti, {\it {Lepton Number Violation in
  Higgs Decay at LHC}},  {\em Phys. Rev. Lett.} {\bf 115} (2015) 081802,
  [\href{http://arxiv.org/abs/1503.06834}{{\tt arXiv:1503.06834}}].

\bibitem{Antusch:2016vyf}
S.~Antusch, E.~Cazzato, and O.~Fischer, {\it {Displaced vertex searches for
  sterile neutrinos at future lepton colliders}},  {\em JHEP} {\bf 12} (2016)
  007, [\href{http://arxiv.org/abs/1604.02420}{{\tt arXiv:1604.02420}}].

\bibitem{Antusch:2016ejd}
S.~Antusch, E.~Cazzato, and O.~Fischer, {\it {Sterile neutrino searches at
  future $e^-e^+$, $pp$, and $e^-p$ colliders}},  {\em Int. J. Mod. Phys.} {\bf
  A32} (2017), no.~14 1750078, [\href{http://arxiv.org/abs/1612.02728}{{\tt
  arXiv:1612.02728}}].

\bibitem{Accomando:2016rpc}
E.~Accomando, L.~Delle~Rose, S.~Moretti, E.~Olaiya, and C.~H.
  Shepherd-Themistocleous, {\it {Novel SM-like Higgs decay into displaced heavy
  neutrino pairs in U(1)$^\prime$ models}},  {\em JHEP} {\bf 04} (2017) 081,
  [\href{http://arxiv.org/abs/1612.05977}{{\tt arXiv:1612.05977}}].

\bibitem{Nemevsek:2016enw}
M.~Nemev\v{s}ek, F.~Nesti, and J.~C. Vasquez, {\it {Majorana Higgses at
  colliders}},  {\em JHEP} {\bf 04} (2017) 114,
  [\href{http://arxiv.org/abs/1612.06840}{{\tt arXiv:1612.06840}}].

\bibitem{Dev:2016vle}
P.~Bhupal~Dev, R.~N. Mohapatra, and Y.~Zhang, {\it {Displaced photon signal
  from a possible light scalar in minimal left-right seesaw model}},  {\em
  Phys. Rev.} {\bf D95} (2017), no.~11 115001,
  [\href{http://arxiv.org/abs/1612.09587}{{\tt arXiv:1612.09587}}].

\bibitem{Aad:2014yea}
{\bf ATLAS} Collaboration, G.~Aad et~al., {\it {Search for long-lived neutral
  particles decaying into lepton jets in proton-proton collisions at $
  \sqrt{s}=8 $ TeV with the ATLAS detector}},  {\em JHEP} {\bf 11} (2014) 088,
  [\href{http://arxiv.org/abs/1409.0746}{{\tt arXiv:1409.0746}}].

\bibitem{CMS:2014wda}
{\bf CMS} Collaboration, V.~Khachatryan et~al., {\it {Search for Long-Lived
  Neutral Particles Decaying to Quark-Antiquark Pairs in Proton-Proton
  Collisions at $\sqrt{s} =$ 8 TeV}},  {\em Phys. Rev.} {\bf D91} (2015), no.~1
  012007, [\href{http://arxiv.org/abs/1411.6530}{{\tt arXiv:1411.6530}}].

\bibitem{CMS:2014hka}
{\bf CMS} Collaboration, V.~Khachatryan et~al., {\it {Search for long-lived
  particles that decay into final states containing two electrons or two muons
  in proton-proton collisions at $\sqrt{s} =$ 8 TeV}},  {\em Phys. Rev.} {\bf
  D91} (2015), no.~5 052012, [\href{http://arxiv.org/abs/1411.6977}{{\tt
  arXiv:1411.6977}}].

\bibitem{Aaij:2014nma}
{\bf LHCb} Collaboration, R.~Aaij et~al., {\it {Search for long-lived particles
  decaying to jet pairs}},  {\em Eur. Phys. J.} {\bf C75} (2015), no.~4 152,
  [\href{http://arxiv.org/abs/1412.3021}{{\tt arXiv:1412.3021}}].

\bibitem{Aad:2015uaa}
{\bf ATLAS} Collaboration, G.~Aad et~al., {\it {Search for long-lived, weakly
  interacting particles that decay to displaced hadronic jets in proton-proton
  collisions at $\sqrt{s}=8$ TeV with the ATLAS detector}},  {\em Phys. Rev.}
  {\bf D92} (2015), no.~1 012010, [\href{http://arxiv.org/abs/1504.03634}{{\tt
  arXiv:1504.03634}}].

\bibitem{Aad:2015rba}
{\bf ATLAS} Collaboration, G.~Aad et~al., {\it {Search for massive, long-lived
  particles using multitrack displaced vertices or displaced lepton pairs in pp
  collisions at $\sqrt{s}$ = 8 TeV with the ATLAS detector}},  {\em Phys. Rev.}
  {\bf D92} (2015), no.~7 072004, [\href{http://arxiv.org/abs/1504.05162}{{\tt
  arXiv:1504.05162}}].

\bibitem{Khachatryan:2016unx}
{\bf CMS} Collaboration, V.~Khachatryan et~al., {\it {Search for R-parity
  violating supersymmetry with displaced vertices in proton-proton collisions
  at $\sqrt{s}$ = 8 TeV}},  {\em Phys. Rev.} {\bf D95} (2017), no.~1 012009,
  [\href{http://arxiv.org/abs/1610.05133}{{\tt arXiv:1610.05133}}].

\bibitem{Aaboud:2017iio}
{\bf ATLAS} Collaboration, M.~Aaboud et~al., {\it {Search for long-lived,
  massive particles in events with displaced vertices and missing transverse
  momentum in $\sqrt{s}$ = 13 TeV $pp$ collisions with the ATLAS detector}},
  \href{http://arxiv.org/abs/1710.04901}{{\tt arXiv:1710.04901}}.

\bibitem{Aad:2013yna}
{\bf ATLAS} Collaboration, G.~Aad et~al., {\it {Search for charginos nearly
  mass degenerate with the lightest neutralino based on a disappearing-track
  signature in $pp$ collisions at $\sqrt(s) = 8$~TeV with the ATLAS detector}},
   {\em Phys. Rev.} {\bf D88} (2013), no.~11 112006,
  [\href{http://arxiv.org/abs/1310.3675}{{\tt arXiv:1310.3675}}].

\bibitem{CMS:2014gxa}
{\bf CMS} Collaboration, V.~Khachatryan et~al., {\it {Search for disappearing
  tracks in proton-proton collisions at $ \sqrt{s}=8 $ TeV}},  {\em JHEP} {\bf
  01} (2015) 096, [\href{http://arxiv.org/abs/1411.6006}{{\tt
  arXiv:1411.6006}}].

\bibitem{Aaboud:2017mpt}
{\bf ATLAS} Collaboration, M.~Aaboud et~al., {\it {Search for long-lived
  charginos based on a disappearing-track signature in $pp$ collisions at
  $\sqrt{s}$ = 13 TeV with the ATLAS detector}},
  \href{http://arxiv.org/abs/1712.02118}{{\tt arXiv:1712.02118}}.

\bibitem{ATLAS:2014fka}
{\bf ATLAS} Collaboration, G.~Aad et~al., {\it {Searches for heavy long-lived
  charged particles with the ATLAS detector in proton-proton collisions at $
  \sqrt{s}=8 $ TeV}},  {\em JHEP} {\bf 01} (2015) 068,
  [\href{http://arxiv.org/abs/1411.6795}{{\tt arXiv:1411.6795}}].

\bibitem{Aad:2015qfa}
{\bf ATLAS} Collaboration, G.~Aad et~al., {\it {Search for metastable heavy
  charged particles with large ionisation energy loss in $pp$ collisions at
  $\sqrt{s} = 8$ TeV using the ATLAS experiment}},  {\em Eur. Phys. J.} {\bf
  C75} (2015), no.~9 407, [\href{http://arxiv.org/abs/1506.05332}{{\tt
  arXiv:1506.05332}}].

\bibitem{Aaboud:2016dgf}
{\bf ATLAS} Collaboration, M.~Aaboud et~al., {\it {Search for metastable heavy
  charged particles with large ionization energy loss in pp collisions at
  $\sqrt{s} = 13$ TeV using the ATLAS experiment}},  {\em Phys. Rev.} {\bf D93}
  (2016), no.~11 112015, [\href{http://arxiv.org/abs/1604.04520}{{\tt
  arXiv:1604.04520}}].

\bibitem{Aaboud:2016uth}
{\bf ATLAS} Collaboration, M.~Aaboud et~al., {\it {Search for heavy long-lived
  charged $R$-hadrons with the ATLAS detector in 3.2 fb$^{-1}$ of
  proton--proton collision data at $\sqrt{s} = 13$ TeV}},  {\em Phys. Lett.}
  {\bf B760} (2016) 647--665, [\href{http://arxiv.org/abs/1606.05129}{{\tt
  arXiv:1606.05129}}].

\bibitem{Khachatryan:2016sfv}
{\bf CMS} Collaboration, V.~Khachatryan et~al., {\it {Search for long-lived
  charged particles in proton-proton collisions at $\sqrt s=$ 13 TeV}},  {\em
  Phys. Rev.} {\bf D94} (2016), no.~11 112004,
  [\href{http://arxiv.org/abs/1609.08382}{{\tt arXiv:1609.08382}}].

\bibitem{Khachatryan:2014mea}
{\bf CMS} Collaboration, V.~Khachatryan et~al., {\it {Search for Displaced
  Supersymmetry in events with an electron and a muon with large impact
  parameters}},  {\em Phys. Rev. Lett.} {\bf 114} (2015), no.~6 061801,
  [\href{http://arxiv.org/abs/1409.4789}{{\tt arXiv:1409.4789}}].

\bibitem{CMS-PAS-EXO-16-022}
{\bf CMS} Collaboration, {\it {Search for displaced leptons in the e-mu
  channel}},  Tech. Rep. CMS-PAS-EXO-16-022, CERN, Geneva, 2016.

\bibitem{Ito:2017dpm}
H.~Ito, O.~Jinnouchi, T.~Moroi, N.~Nagata, and H.~Otono, {\it {Extending the
  LHC Reach for New Physics with Sub-Millimeter Displaced Vertices}},  {\em
  Phys. Lett.} {\bf B771} (2017) 568--575,
  [\href{http://arxiv.org/abs/1702.08613}{{\tt arXiv:1702.08613}}].

\bibitem{Arkani-Hamed:2015vfh}
N.~Arkani-Hamed, T.~Han, M.~Mangano, and L.-T. Wang, {\it {Physics
  opportunities of a 100 TeV proton-proton collider}},  {\em Phys. Rept.} {\bf
  652} (2016) 1--49, [\href{http://arxiv.org/abs/1511.06495}{{\tt
  arXiv:1511.06495}}].

\bibitem{Golling:2016gvc}
T.~Golling et~al., {\it {Physics at a 100 TeV pp collider: beyond the Standard
  Model phenomena}},  {\em CERN Yellow Report} (2017), no.~3 441--634,
  [\href{http://arxiv.org/abs/1606.00947}{{\tt arXiv:1606.00947}}].

\bibitem{ATLAS:2010lca}
{\bf ATLAS} Collaboration, {\it {Performance of primary vertex reconstruction
  in proton-proton collisions at $\sqrt{s}=$7 TeV in the ATLAS experiment}},
  Tech. Rep. ATLAS-CONF-2010-069, 2010.

\bibitem{Aaboud:2016rmg}
{\bf ATLAS} Collaboration, M.~Aaboud et~al., {\it {Reconstruction of primary
  vertices at the ATLAS experiment in Run 1 proton-proton collisions at the
  LHC}},  {\em Eur. Phys. J.} {\bf C77} (2017), no.~5 332,
  [\href{http://arxiv.org/abs/1611.10235}{{\tt arXiv:1611.10235}}].

\bibitem{ATL-PHYS-PUB-2015-026}
{{\bf ATLAS} Collaboration}, {\it {Vertex Reconstruction Performance of the
  ATLAS Detector at $\sqrt{s} = 13~\textrm{TeV}$}},  Tech. Rep.
  ATL-PHYS-PUB-2015-026, CERN, Geneva, Jul, 2015.

\bibitem{Chatrchyan:2014fea}
{\bf CMS} Collaboration, S.~Chatrchyan et~al., {\it {Description and
  performance of track and primary-vertex reconstruction with the CMS
  tracker}},  {\em JINST} {\bf 9} (2014), no.~10 P10009,
  [\href{http://arxiv.org/abs/1405.6569}{{\tt arXiv:1405.6569}}].

\bibitem{CMS-DP-2016-041}
{\bf CMS} Collaboration, {\it {Primary vertex resolution in 2016}},  Tech. Rep.
  CMS-DP-2016-041, Jul, 2016.

\bibitem{Fruhwirth:2007hz}
R.~Fruhwirth, W.~Waltenberger, and P.~Vanlaer, {\it {Adaptive vertex fitting}},
   {\em J. Phys.} {\bf G34} (2007) N343.

\bibitem{Bickel20063500}
D.~R. Bickel and R.~Fruhwirth, {\it On a fast, robust estimator of the mode:
  Comparisons to other robust estimators with applications},  {\em
  Computational Statistics \& Data Analysis} {\bf 50} (2006), no.~12 3500 --
  3530, [\href{http://arxiv.org/abs/math/0505419}{{\tt math/0505419}}].

\bibitem{ATLAS:trk}
{{\bf ATLAS} Collaboration}, ``{\it Impact Parameter Resolution}.''
  \url{https://atlas.web.cern.ch/Atlas/GROUPS/PHYSICS/PLOTS/IDTR-2015-007/},
  2015.

\bibitem{ATL-PHYS-PUB-2015-018}
{{\bf ATLAS} Collaboration}, {\it {Track Reconstruction Performance of the
  ATLAS Inner Detector at $\sqrt{s}=13$~TeV}},  Tech. Rep.
  ATL-PHYS-PUB-2015-018, CERN, Geneva, Jul, 2015.

\bibitem{PERF-2007-01}
{{\bf ATLAS} Collaboration}, {\it {The ATLAS Experiment at the CERN Large
  Hadron Collider}},  {\em JINST} {\bf 3} (2008) S08003.

\bibitem{impres2015}
{{\bf ATLAS} Collaboration}, ``{\it Impact Parameter Resolution Using 2016 MB
  Data}.''
  \url{https://atlas.web.cern.ch/Atlas/GROUPS/PHYSICS/PLOTS/IDTR-2016-018/},
  Dec, 2016.

\bibitem{Aad:2015zhl}
{\bf ATLAS, CMS} Collaboration, G.~Aad et~al., {\it {Combined Measurement of
  the Higgs Boson Mass in $pp$ Collisions at $\sqrt{s}=7$ and 8 TeV with the
  ATLAS and CMS Experiments}},  {\em Phys. Rev. Lett.} {\bf 114} (2015) 191803,
  [\href{http://arxiv.org/abs/1503.07589}{{\tt arXiv:1503.07589}}].

\bibitem{Okada:1990vk}
Y.~Okada, M.~Yamaguchi, and T.~Yanagida, {\it {Upper bound of the lightest
  Higgs boson mass in the minimal supersymmetric standard model}},  {\em Prog.
  Theor. Phys.} {\bf 85} (1991) 1--6.

\bibitem{Okada:1990gg}
Y.~Okada, M.~Yamaguchi, and T.~Yanagida, {\it {Renormalization group analysis
  on the Higgs mass in the softly broken supersymmetric standard model}},  {\em
  Phys. Lett.} {\bf B262} (1991) 54--58.

\bibitem{Ellis:1990nz}
J.~R. Ellis, G.~Ridolfi, and F.~Zwirner, {\it {Radiative corrections to the
  masses of supersymmetric Higgs bosons}},  {\em Phys. Lett.} {\bf B257} (1991)
  83--91.

\bibitem{Ellis:1991zd}
J.~R. Ellis, G.~Ridolfi, and F.~Zwirner, {\it {On radiative corrections to
  supersymmetric Higgs boson masses and their implications for LEP searches}},
  {\em Phys. Lett.} {\bf B262} (1991) 477--484.

\bibitem{Haber:1990aw}
H.~E. Haber and R.~Hempfling, {\it {Can the mass of the lightest Higgs boson of
  the minimal supersymmetric model be larger than m(Z)?}},  {\em Phys. Rev.
  Lett.} {\bf 66} (1991) 1815--1818.

\bibitem{Gabbiani:1996hi}
F.~Gabbiani, E.~Gabrielli, A.~Masiero, and L.~Silvestrini, {\it {A Complete
  analysis of FCNC and CP constraints in general SUSY extensions of the
  standard model}},  {\em Nucl. Phys.} {\bf B477} (1996) 321--352,
  [\href{http://arxiv.org/abs/hep-ph/9604387}{{\tt hep-ph/9604387}}].

\bibitem{Moroi:2013sfa}
T.~Moroi and M.~Nagai, {\it {Probing Supersymmetric Model with Heavy Sfermions
  Using Leptonic Flavor and CP Violations}},  {\em Phys. Lett.} {\bf B723}
  (2013) 107--112, [\href{http://arxiv.org/abs/1303.0668}{{\tt
  arXiv:1303.0668}}].

\bibitem{McKeen:2013dma}
D.~McKeen, M.~Pospelov, and A.~Ritz, {\it {Electric dipole moment signatures of
  PeV-scale superpartners}},  {\em Phys. Rev.} {\bf D87} (2013), no.~11 113002,
  [\href{http://arxiv.org/abs/1303.1172}{{\tt arXiv:1303.1172}}].

\bibitem{Altmannshofer:2013lfa}
W.~Altmannshofer, R.~Harnik, and J.~Zupan, {\it {Low Energy Probes of PeV Scale
  Sfermions}},  {\em JHEP} {\bf 11} (2013) 202,
  [\href{http://arxiv.org/abs/1308.3653}{{\tt arXiv:1308.3653}}].

\bibitem{Fuyuto:2013gla}
K.~Fuyuto, J.~Hisano, N.~Nagata, and K.~Tsumura, {\it {QCD Corrections to Quark
  (Chromo)Electric Dipole Moments in High-scale Supersymmetry}},  {\em JHEP}
  {\bf 12} (2013) 010, [\href{http://arxiv.org/abs/1308.6493}{{\tt
  arXiv:1308.6493}}].

\bibitem{Hisano:2013exa}
J.~Hisano, D.~Kobayashi, T.~Kuwahara, and N.~Nagata, {\it {Decoupling Can
  Revive Minimal Supersymmetric SU(5)}},  {\em JHEP} {\bf 07} (2013) 038,
  [\href{http://arxiv.org/abs/1304.3651}{{\tt arXiv:1304.3651}}].

\bibitem{Nagata:2013sba}
N.~Nagata and S.~Shirai, {\it {Sfermion Flavor and Proton Decay in High-Scale
  Supersymmetry}},  {\em JHEP} {\bf 03} (2014) 049,
  [\href{http://arxiv.org/abs/1312.7854}{{\tt arXiv:1312.7854}}].

\bibitem{Evans:2015bxa}
J.~L. Evans, N.~Nagata, and K.~A. Olive, {\it {SU(5) Grand Unification in Pure
  Gravity Mediation}},  {\em Phys. Rev.} {\bf D91} (2015) 055027,
  [\href{http://arxiv.org/abs/1502.00034}{{\tt arXiv:1502.00034}}].

\bibitem{Giudice:1998xp}
G.~F. Giudice, M.~A. Luty, H.~Murayama, and R.~Rattazzi, {\it {Gaugino mass
  without singlets}},  {\em JHEP} {\bf 12} (1998) 027,
  [\href{http://arxiv.org/abs/hep-ph/9810442}{{\tt hep-ph/9810442}}].

\bibitem{Wells:2003tf}
J.~D. Wells, {\it {Implications of supersymmetry breaking with a little
  hierarchy between gauginos and scalars}},  in {\em {11th International
  Conference on Supersymmetry and the Unification of Fundamental Interactions
  (SUSY 2003) Tucson, Arizona, June 5-10, 2003}}, 2003.
\newblock \href{http://arxiv.org/abs/hep-ph/0306127}{{\tt hep-ph/0306127}}.

\bibitem{ArkaniHamed:2004fb}
N.~Arkani-Hamed and S.~Dimopoulos, {\it {Supersymmetric unification without low
  energy supersymmetry and signatures for fine-tuning at the LHC}},  {\em JHEP}
  {\bf 06} (2005) 073, [\href{http://arxiv.org/abs/hep-th/0405159}{{\tt
  hep-th/0405159}}].

\bibitem{Giudice:2004tc}
G.~F. Giudice and A.~Romanino, {\it {Split supersymmetry}},  {\em Nucl. Phys.}
  {\bf B699} (2004) 65--89, [\href{http://arxiv.org/abs/hep-ph/0406088}{{\tt
  hep-ph/0406088}}]. [Erratum: Nucl. Phys.B706,487(2005)].

\bibitem{ArkaniHamed:2004yi}
N.~Arkani-Hamed, S.~Dimopoulos, G.~F. Giudice, and A.~Romanino, {\it {Aspects
  of split supersymmetry}},  {\em Nucl. Phys.} {\bf B709} (2005) 3--46,
  [\href{http://arxiv.org/abs/hep-ph/0409232}{{\tt hep-ph/0409232}}].

\bibitem{Wells:2004di}
J.~D. Wells, {\it {PeV-scale supersymmetry}},  {\em Phys. Rev.} {\bf D71}
  (2005) 015013, [\href{http://arxiv.org/abs/hep-ph/0411041}{{\tt
  hep-ph/0411041}}].

\bibitem{Randall:1998uk}
L.~Randall and R.~Sundrum, {\it {Out of this world supersymmetry breaking}},
  {\em Nucl. Phys.} {\bf B557} (1999) 79--118,
  [\href{http://arxiv.org/abs/hep-th/9810155}{{\tt hep-th/9810155}}].

\bibitem{Gherghetta:1999sw}
T.~Gherghetta, G.~F. Giudice, and J.~D. Wells, {\it {Phenomenological
  consequences of supersymmetry with anomaly induced masses}},  {\em Nucl.
  Phys.} {\bf B559} (1999) 27--47,
  [\href{http://arxiv.org/abs/hep-ph/9904378}{{\tt hep-ph/9904378}}].

\bibitem{Moroi:1999zb}
T.~Moroi and L.~Randall, {\it {Wino cold dark matter from anomaly mediated SUSY
  breaking}},  {\em Nucl. Phys.} {\bf B570} (2000) 455--472,
  [\href{http://arxiv.org/abs/hep-ph/9906527}{{\tt hep-ph/9906527}}].

\bibitem{Hall:2011jd}
L.~J. Hall and Y.~Nomura, {\it {Spread Supersymmetry}},  {\em JHEP} {\bf 01}
  (2012) 082, [\href{http://arxiv.org/abs/1111.4519}{{\tt arXiv:1111.4519}}].

\bibitem{Ibe:2011aa}
M.~Ibe and T.~T. Yanagida, {\it {The Lightest Higgs Boson Mass in Pure Gravity
  Mediation Model}},  {\em Phys. Lett.} {\bf B709} (2012) 374--380,
  [\href{http://arxiv.org/abs/1112.2462}{{\tt arXiv:1112.2462}}].

\bibitem{Ibe:2012hu}
M.~Ibe, S.~Matsumoto, and T.~T. Yanagida, {\it {Pure Gravity Mediation with
  $m_{3/2} = 10$--$100~{\rm TeV}$}},  {\em Phys. Rev.} {\bf D85} (2012) 095011,
  [\href{http://arxiv.org/abs/1202.2253}{{\tt arXiv:1202.2253}}].

\bibitem{Arvanitaki:2012ps}
A.~Arvanitaki, N.~Craig, S.~Dimopoulos, and G.~Villadoro, {\it {Mini-Split}},
  {\em JHEP} {\bf 02} (2013) 126, [\href{http://arxiv.org/abs/1210.0555}{{\tt
  arXiv:1210.0555}}].

\bibitem{Hall:2012zp}
L.~J. Hall, Y.~Nomura, and S.~Shirai, {\it {Spread Supersymmetry with Wino LSP:
  Gluino and Dark Matter Signals}},  {\em JHEP} {\bf 01} (2013) 036,
  [\href{http://arxiv.org/abs/1210.2395}{{\tt arXiv:1210.2395}}].

\bibitem{ArkaniHamed:2012gw}
N.~Arkani-Hamed, A.~Gupta, D.~E. Kaplan, N.~Weiner, and T.~Zorawski, {\it
  {Simply Unnatural Supersymmetry}},
  \href{http://arxiv.org/abs/1212.6971}{{\tt arXiv:1212.6971}}.

\bibitem{Evans:2013lpa}
J.~L. Evans, M.~Ibe, K.~A. Olive, and T.~T. Yanagida, {\it {Universality in
  Pure Gravity Mediation}},  {\em Eur. Phys. J.} {\bf C73} (2013) 2468,
  [\href{http://arxiv.org/abs/1302.5346}{{\tt arXiv:1302.5346}}].

\bibitem{Evans:2013dza}
J.~L. Evans, K.~A. Olive, M.~Ibe, and T.~T. Yanagida, {\it {Non-Universalities
  in Pure Gravity Mediation}},  {\em Eur. Phys. J.} {\bf C73} (2013), no.~10
  2611, [\href{http://arxiv.org/abs/1305.7461}{{\tt arXiv:1305.7461}}].

\bibitem{Pierce:1996zz}
D.~M. Pierce, J.~A. Bagger, K.~T. Matchev, and R.-j. Zhang, {\it {Precision
  corrections in the minimal supersymmetric standard model}},  {\em Nucl.
  Phys.} {\bf B491} (1997) 3--67,
  [\href{http://arxiv.org/abs/hep-ph/9606211}{{\tt hep-ph/9606211}}].

\bibitem{Pomarol:1999ie}
A.~Pomarol and R.~Rattazzi, {\it {Sparticle masses from the superconformal
  anomaly}},  {\em JHEP} {\bf 05} (1999) 013,
  [\href{http://arxiv.org/abs/hep-ph/9903448}{{\tt hep-ph/9903448}}].

\bibitem{Nelson:2002sa}
A.~E. Nelson and N.~J. Weiner, {\it {Extended anomaly mediation and new physics
  at 10-TeV}},  \href{http://arxiv.org/abs/hep-ph/0210288}{{\tt
  hep-ph/0210288}}.

\bibitem{Hsieh:2006ig}
K.~Hsieh and M.~A. Luty, {\it {Mixed gauge and anomaly mediation from new
  physics at 10-TeV}},  {\em JHEP} {\bf 06} (2007) 062,
  [\href{http://arxiv.org/abs/hep-ph/0604256}{{\tt hep-ph/0604256}}].

\bibitem{Gupta:2012gu}
A.~Gupta, D.~E. Kaplan, and T.~Zorawski, {\it {Gaugomaly Mediation Revisited}},
   {\em JHEP} {\bf 11} (2013) 149, [\href{http://arxiv.org/abs/1212.6969}{{\tt
  arXiv:1212.6969}}].

\bibitem{Nakayama:2013uta}
K.~Nakayama and T.~T. Yanagida, {\it {Anomaly mediation deformed by axion}},
  {\em Phys. Lett.} {\bf B722} (2013) 107--110,
  [\href{http://arxiv.org/abs/1302.3332}{{\tt arXiv:1302.3332}}].

\bibitem{Harigaya:2013asa}
K.~Harigaya, M.~Ibe, and T.~T. Yanagida, {\it {A Closer Look at Gaugino Masses
  in Pure Gravity Mediation Model/Minimal Split SUSY Model}},  {\em JHEP} {\bf
  12} (2013) 016, [\href{http://arxiv.org/abs/1310.0643}{{\tt
  arXiv:1310.0643}}].

\bibitem{Evans:2014xpa}
J.~L. Evans and K.~A. Olive, {\it {Universality in Pure Gravity Mediation with
  Vector Multiplets}},  {\em Phys. Rev.} {\bf D90} (2014), no.~11 115020,
  [\href{http://arxiv.org/abs/1408.5102}{{\tt arXiv:1408.5102}}].

\bibitem{Griest:1990kh}
K.~Griest and D.~Seckel, {\it {Three exceptions in the calculation of relic
  abundances}},  {\em Phys. Rev.} {\bf D43} (1991) 3191--3203.

\bibitem{Profumo:2004wk}
S.~Profumo and C.~E. Yaguna, {\it {Gluino coannihilations and heavy bino dark
  matter}},  {\em Phys. Rev.} {\bf D69} (2004) 115009,
  [\href{http://arxiv.org/abs/hep-ph/0402208}{{\tt hep-ph/0402208}}].

\bibitem{Feldman:2009zc}
D.~Feldman, Z.~Liu, and P.~Nath, {\it {Gluino NLSP, Dark Matter via Gluino
  Coannihilation, and LHC Signatures}},  {\em Phys. Rev.} {\bf D80} (2009)
  015007, [\href{http://arxiv.org/abs/0905.1148}{{\tt arXiv:0905.1148}}].

\bibitem{deSimone:2014pda}
A.~De~Simone, G.~F. Giudice, and A.~Strumia, {\it {Benchmarks for Dark Matter
  Searches at the LHC}},  {\em JHEP} {\bf 06} (2014) 081,
  [\href{http://arxiv.org/abs/1402.6287}{{\tt arXiv:1402.6287}}].

\bibitem{Harigaya:2014dwa}
K.~Harigaya, K.~Kaneta, and S.~Matsumoto, {\it {Gaugino coannihilations}},
  {\em Phys. Rev.} {\bf D89} (2014), no.~11 115021,
  [\href{http://arxiv.org/abs/1403.0715}{{\tt arXiv:1403.0715}}].

\bibitem{Ellis:2015vaa}
J.~Ellis, F.~Luo, and K.~A. Olive, {\it {Gluino Coannihilation Revisited}},
  {\em JHEP} {\bf 09} (2015) 127, [\href{http://arxiv.org/abs/1503.07142}{{\tt
  arXiv:1503.07142}}].

\bibitem{Ellis:2015vna}
J.~Ellis, J.~L. Evans, F.~Luo, and K.~A. Olive, {\it {Scenarios for Gluino
  Coannihilation}},  {\em JHEP} {\bf 02} (2016) 071,
  [\href{http://arxiv.org/abs/1510.03498}{{\tt arXiv:1510.03498}}].

\bibitem{Liew:2016hqo}
S.~P. Liew and F.~Luo, {\it {Effects of QCD bound states on dark matter relic
  abundance}},  {\em JHEP} {\bf 02} (2017) 091,
  [\href{http://arxiv.org/abs/1611.08133}{{\tt arXiv:1611.08133}}].

\bibitem{Nadolsky:2008zw}
P.~M. Nadolsky, H.-L. Lai, Q.-H. Cao, J.~Huston, J.~Pumplin, D.~Stump, W.-K.
  Tung, and C.~P. Yuan, {\it {Implications of CTEQ global analysis for collider
  observables}},  {\em Phys. Rev.} {\bf D78} (2008) 013004,
  [\href{http://arxiv.org/abs/0802.0007}{{\tt arXiv:0802.0007}}].

\bibitem{Borschensky:2014cia}
C.~Borschensky, M.~Kr$\ddot{\rm a}$mer, A.~Kulesza, M.~Mangano, S.~Padhi,
  T.~Plehn, and X.~Portell, {\it {Squark and gluino production cross sections
  in pp collisions at $\sqrt{s}$ = 13, 14, 33 and 100 TeV}},  {\em Eur. Phys.
  J.} {\bf C74} (2014), no.~12 3174,
  [\href{http://arxiv.org/abs/1407.5066}{{\tt arXiv:1407.5066}}].

\bibitem{Alwall:2014hca}
J.~Alwall, R.~Frederix, S.~Frixione, V.~Hirschi, F.~Maltoni, O.~Mattelaer,
  H.~S. Shao, T.~Stelzer, P.~Torrielli, and M.~Zaro, {\it {The automated
  computation of tree-level and next-to-leading order differential cross
  sections, and their matching to parton shower simulations}},  {\em JHEP} {\bf
  07} (2014) 079, [\href{http://arxiv.org/abs/1405.0301}{{\tt
  arXiv:1405.0301}}].

\bibitem{Sjostrand:2007gs}
T.~Sjostrand, S.~Mrenna, and P.~Z. Skands, {\it {A Brief Introduction to PYTHIA
  8.1}},  {\em Comput. Phys. Commun.} {\bf 178} (2008) 852--867,
  [\href{http://arxiv.org/abs/0710.3820}{{\tt arXiv:0710.3820}}].

\bibitem{Bjorken:1977md}
J.~D. Bjorken, {\it {Properties of Hadron Distributions in Reactions Containing
  Very Heavy Quarks}},  {\em Phys. Rev.} {\bf D17} (1978) 171--173.

\bibitem{Aaboud:2017vwy}
{\bf ATLAS} Collaboration, M.~Aaboud et~al., {\it {Search for squarks and
  gluinos in final states with jets and missing transverse momentum using 36
  fb$^{-1}$ of $\sqrt{s}$=13 TeV $pp$ collision data with the ATLAS detector}},
   \href{http://arxiv.org/abs/1712.02332}{{\tt arXiv:1712.02332}}.

\bibitem{Alwall:2007fs}
J.~Alwall et~al., {\it {Comparative study of various algorithms for the merging
  of parton showers and matrix elements in hadronic collisions}},  {\em Eur.
  Phys. J.} {\bf C53} (2008) 473--500,
  [\href{http://arxiv.org/abs/0706.2569}{{\tt arXiv:0706.2569}}].

\bibitem{Avetisyan:2013onh}
A.~Avetisyan et~al., {\it {Methods and Results for Standard Model Event
  Generation at $\sqrt{s}$ = 14 TeV, 33 TeV and 100 TeV Proton Colliders (A
  Snowmass Whitepaper)}},  in {\em {Proceedings, 2013 Community Summer Study on
  the Future of U.S. Particle Physics: Snowmass on the Mississippi (CSS2013):
  Minneapolis, MN, USA, July 29-August 6, 2013}}, 2013.
\newblock \href{http://arxiv.org/abs/1308.1636}{{\tt arXiv:1308.1636}}.

\bibitem{deFavereau:2013fsa}
{\bf DELPHES 3} Collaboration, J.~de~Favereau, C.~Delaere, P.~Demin,
  A.~Giammanco, V.~Lema\^{i}tre, A.~Mertens, and M.~Selvaggi, {\it {DELPHES 3,
  A modular framework for fast simulation of a generic collider experiment}},
  {\em JHEP} {\bf 02} (2014) 057, [\href{http://arxiv.org/abs/1307.6346}{{\tt
  arXiv:1307.6346}}].

\bibitem{Cacciari:2011ma}
M.~Cacciari, G.~P. Salam, and G.~Soyez, {\it {FastJet User Manual}},  {\em Eur.
  Phys. J.} {\bf C72} (2012) 1896, [\href{http://arxiv.org/abs/1111.6097}{{\tt
  arXiv:1111.6097}}].

\bibitem{Bjorken:1969wi}
J.~D. Bjorken and S.~J. Brodsky, {\it {Statistical Model for electron-Positron
  Annihilation Into Hadrons}},  {\em Phys. Rev.} {\bf D1} (1970) 1416--1420.

\bibitem{Chen:2011ah}
C.~Chen, {\it {New approach to identifying boosted hadronically-decaying
  particle using jet substructure in its center-of-mass frame}},  {\em Phys.
  Rev.} {\bf D85} (2012) 034007, [\href{http://arxiv.org/abs/1112.2567}{{\tt
  arXiv:1112.2567}}].

\bibitem{Capeans:1291633}
{\bf ATLAS} Collaboration, M.~Capeans, G.~Darbo, K.~Einsweiller, M.~Elsing,
  T.~Flick, M.~Garcia-Sciveres, C.~Gemme, H.~Pernegger, O.~Rohne, and
  R.~Vuillermet, {\it {ATLAS Insertable B-Layer Technical Design Report}},
  Tech. Rep. CERN-LHCC-2010-013. ATLAS-TDR-19, Sep, 2010.

\bibitem{1748-0221-9-02-C02018}
A.~Miucci, {\it The atlas insertable b-layer project},  {\em Journal of
  Instrumentation} {\bf 9} (2014), no.~02 C02018.

\bibitem{1748-0221-11-11-P11020}
G.~A. M.~Aaboud et~al., {\it A measurement of material in the atlas tracker
  using secondary hadronic interactions in 7 tev pp collisions},  {\em Journal
  of Instrumentation} {\bf 11} (2016), no.~11 P11020.

\bibitem{Cohen:2013xda}
T.~Cohen, T.~Golling, M.~Hance, A.~Henrichs, K.~Howe, J.~Loyal, S.~Padhi, and
  J.~G. Wacker, {\it {SUSY Simplified Models at 14, 33, and 100 TeV Proton
  Colliders}},  {\em JHEP} {\bf 04} (2014) 117,
  [\href{http://arxiv.org/abs/1311.6480}{{\tt arXiv:1311.6480}}].

\bibitem{Read:2002hq}
A.~L. Read, {\it {Presentation of search results: The $CL_s$ technique}},  {\em
  J. Phys.} {\bf G28} (2002) 2693--2704.

\bibitem{Junk:1999kv}
T.~Junk, {\it {Confidence level computation for combining searches with small
  statistics}},  {\em Nucl. Instrum. Meth.} {\bf A434} (1999) 435--443,
  [\href{http://arxiv.org/abs/hep-ex/9902006}{{\tt hep-ex/9902006}}].

\bibitem{Cowan:2010js}
G.~Cowan, K.~Cranmer, E.~Gross, and O.~Vitells, {\it {Asymptotic formulae for
  likelihood-based tests of new physics}},  {\em Eur. Phys. J.} {\bf C71}
  (2011) 1554, [\href{http://arxiv.org/abs/1007.1727}{{\tt arXiv:1007.1727}}].
  [Erratum: Eur. Phys. J.C73,2501(2013)].

\bibitem{ATLAS-CONF-2014-037}
{{\bf ATLAS} Collaboration}, {\it {Limits on metastable gluinos from ATLAS SUSY
  searches at 8 TeV}},  Tech. Rep. ATLAS-CONF-2014-037, CERN, Geneva, Jul,
  2014.

\bibitem{Sirunyan:2018vjp}
{\bf CMS} Collaboration, A.~M. Sirunyan et~al., {\it {Search for natural and
  split supersymmetry in proton-proton collisions at $\sqrt{s} =$ 13 TeV in
  final states with jets and missing transverse momentum}},
  \href{http://arxiv.org/abs/1802.02110}{{\tt arXiv:1802.02110}}.

\bibitem{Sirunyan:2017ezt}
{\bf CMS} Collaboration, A.~M. Sirunyan et~al., {\it {Identification of
  heavy-flavour jets with the CMS detector in pp collisions at 13 TeV}},
  \href{http://arxiv.org/abs/1712.07158}{{\tt arXiv:1712.07158}}.

\bibitem{ATLAS:2016kts}
{\bf ATLAS} Collaboration, {\it {Further searches for squarks and gluinos in
  final states with jets and missing transverse momentum at $\sqrt{s}$ =13 TeV
  with the ATLAS detector}},  Tech. Rep. ATLAS-CONF-2016-078, 2016.

\bibitem{Ellis:2015xba}
S.~A.~R. Ellis and B.~Zheng, {\it {Reaching for squarks and gauginos at a 100
  TeV p-p collider}},  {\em Phys. Rev.} {\bf D92} (2015), no.~7 075034,
  [\href{http://arxiv.org/abs/1506.02644}{{\tt arXiv:1506.02644}}].

\bibitem{Sato:2013bta}
R.~Sato, S.~Shirai, and K.~Tobioka, {\it {Flavor of Gluino Decay in High-Scale
  Supersymmetry}},  {\em JHEP} {\bf 10} (2013) 157,
  [\href{http://arxiv.org/abs/1307.7144}{{\tt arXiv:1307.7144}}].

\bibitem{Park:2011vw}
M.~Park and Y.~Zhao, {\it {Recovering Particle Masses from Missing Energy
  Signatures with Displaced Tracks}},
  \href{http://arxiv.org/abs/1110.1403}{{\tt arXiv:1110.1403}}.

\bibitem{Cottin:2018hyf}
G.~Cottin, {\it {Reconstructing particle masses in events with displaced
  vertices}},  \href{http://arxiv.org/abs/1801.09671}{{\tt arXiv:1801.09671}}.

\end{thebibliography}\endgroup
}

%%%%%%%%%%%%%%%%%%%%%%%%%%%%%%%%%%%%%%%

\end{document}